\documentclass[twoside,11pt]{article}

\usepackage{graphicx,latexsym,amssymb, epsfig}
\usepackage{multirow,amsmath,array,booktabs}
\usepackage{subfigure}
\usepackage{cite}
\usepackage[figuresright]{rotating}

\newcommand{\ben}{\begin{enumerate}}
\newcommand{\een}{\end{enumerate}}
\newcommand{\beq}{\begin{equation}}
\newcommand{\eeq}{\end{equation}}
\newcommand{\beqn}{\begin{eqnarray}}
\newcommand{\eeqn}{\end{eqnarray}}

\newcommand{\beqd}{\begin{eqnarray*}}
\newcommand{\eeqd}{\end{eqnarray*}}
\newcommand{\bea}{\begin{array}}
\newcommand{\eea}{\end{array}}
\newcommand{\bcen}{\begin{center}}
\newcommand{\ecen}{\end{center}}
\newcommand{\btab}{\begin{tabular}}
\newcommand{\etab}{\end{tabular}}
\newcommand{\bsub}{\begin{subequations}}
\newcommand{\esub}{\end{subequations}}
\newcommand{\bit}{\begin{itemize}}
\newcommand{\eit}{\end{itemize}}
\newcommand{\brule}{\begin{ruledtabular}}
\newcommand{\erule}{\end{ruledtabular}}
\newcommand{\bpm}{\begin{pmatrix}}
\newcommand{\epm}{\end{pmatrix}}

\newcommand{\cals}[1]{{\cal #1}}
\newcommand{\tr}{{\rm Tr}}

\newcommand{\ff}[1]{\frac{1}{#1}}
\newcommand{\lc}{\left<}
\newcommand{\rc}{\right>}
\newcommand{\lr}{\left|}
\newcommand{\rl}{\right|}
\newcommand{\lb}{\left(}
\newcommand{\rb}{\right)}
\newcommand{\ls}{\left[}
\newcommand{\rs}{\right]}
\newcommand{\Lb}{\left\{}
\newcommand{\Rb}{\right\}}

\newcommand{\svec}[1]{{\mbox{\boldmath${\rm #1}$}}}
\newcommand{\ivec}{\vec}
\newcommand{\re}{\nonumber\\}
\newcommand{\fm}{\text{fm}}
\newcommand{\mev}{\text{MeV}}
\newcommand{\cm}{{\text{c.m.}}}

\newcommand{\bay}{\begin{array}}
\newcommand{\eay}{\end{array}}

\begin{document}

\title{\vspace{1cm}
 {Relativistic Continuum Hartree Bogoliubov Theory for Ground
State Properties of Exotic Nuclei}}
\author{J.\ Meng,$^{1,2,3}$
        H.\ Toki,$^{4}$
        S.\ G.\ Zhou,$^{2,3}$
        S.\ Q.\ Zhang,$^{1}$ \\
 \vspace{0.5cm}
        W.\ H.\ Long,$^{1}$
        L.\ S.\ Geng,$^{1,4}$ \\
 $^1$School of Physics, Peking University, \\
     Beijing 100871, China \\
 $^2$Institute of Theoretical Physics, Chinese Academy of Sciences, \\
     Beijing 100080, China \\
 $^3$Center of Theoretical Nuclear Physics, National Laboratory of \\
     Heavy Ion Accelerator, Lanzhou 730000, China \\
 $^4$Research Center for Nuclear Physics (RCNP), Osaka University, \\
     Ibaraki, Osaka 567-0047, Japan \\
}

\maketitle

\begin{abstract}

The Relativistic Continuum Hartree-Bogoliubov (RCHB) theory, which
properly takes into account the pairing correlation and the
coupling to (discretized) continuum via Bogoliubov transformation
in a microscopic and self-consistent way, has been reviewed
together with its new interpretation of the halo phenomena
observed in light nuclei as the scattering of particle pairs into
the continuum, the prediction of the exotic phenomena --- giant
halos in nuclei near neutron drip line, the reproduction of
interaction cross sections and charge-changing cross sections in
light exotic nuclei in combination with the Glauber theory, better
restoration of pseudospin symmetry in exotic nuclei, predictions
of exotic phenomena in hyper nuclei, and new magic numbers in
superheavy nuclei, etc. Recent investigations on new effective
interactions, the density dependence of the interaction
strengthes, the RMF theory on the Woods-Saxon basis, the single
particle resonant states, and the resonant BCS (rBCS) method for
the pairing correlation, etc. are also presented in some details.
\\
  \\
 {\bf Key words} \\
 \\
 Relativistic mean field theory,
 continuum, pairing correlation, Bogoliubov transformation,
 relativistic continuum Hartree-Bogoliubov, exotic nuclei,
 halo, giant halo, hyperon halo, interaction cross section,
 charge-changing cross section, pseudospin symmetry,
 hyper nuclei, magic number, superheavy nuclei.

\end{abstract}

\tableofcontents

\section{Introduction}

Currently nuclear physics is undergoing a renaissance as evidenced
by the fact that worldwide there are lots of Radioactive Ion Beam
(RIB)~\cite{Bertulani01} facilities operating, being upgraded,
under construction or planning to be constructed, e.g., the
Cooling Storage Ring (CSR) project at HIRFL in China will be
completed in 2005~\cite{Zhan04}, the RIB Factory at RIKEN in Japan
will begin to operate  at the end of 2006~\cite{Yano04}, the FAIR
project at GSI in Germany was approved in 2003~\cite{Henning04},
and the construction of the Rare Isotope Accelerator (RIA) is
considered to be of the highest priority of all physics
disciplines in the US~\cite{riaweb}, etc. These  new facilities
together with developments in the detection techniques have
changed the nuclear physics scenario. It has come into reality to
produce and study the nuclei far away from the stability line ---
so called ``EXOTIC NUCLEI"
{\cite{Mueller93,Tanihata95,Hansen95,Casten00,Mueller01, Jonson04,
Jensen04}}.

New exciting discoveries have been made by exploring hitherto
inaccessible regions in the nuclear chart~\cite{Mueller93,
Tanihata95, Hansen95,Casten00, Mueller01, Jonson04, Jensen04}. For
examples, a new phenomenon, halo --- a state in which nucleons
spread like a thin mist around the nucleus, was first discovered
in $^{11}$Li with RIB in 1985~\cite{Tanihata85} and later in many
other light exotic nuclei as well; exotic nuclei exhibit other
interesting phenomena such as the disappearance of traditional
shell gaps and the occurrence of new ones, which result in new
magic numbers~\cite{Ozawa00}; etc. The exotic nuclei also play
important roles in nuclear astrophysics as well, as their
properties are crucial to understand stellar nucleosynthesis.

The current nuclear models are mainly based on the knowledge
obtained from the nuclei near the  $\beta$-stability line. For
instances, the cornerstones for the edifice of modern nuclear
physics include shell model~\cite{Mayer55} and collective
model~\cite{Bohr69,Bohr75}, which are respectively based on the
magic numbers (the stability of some nuclei compared with their
neighbors) and the incompressibility of the nuclear matter.
Therefore the change of magic numbers and the unprecedented low
density nuclear matter in halo nuclei have shaken the foundation
of nuclear physics. New innovative nuclear models are needed to
describe the exotic nuclei characterized by weakly binding and low
density.

The relativistic mean field (RMF) theory \cite{Walecka74} has
received wide attention due to its successful description of lots
of nuclear phenomena during the past
years~\cite{Serot86,Reinhard89,Ring96,Serot97,Bender03}. In the
framework of the RMF theory, the nucleons interact via the
exchanges of mesons and photons.  {The representations with large
scalar and vector fields in nuclei, of order of a few hundred MeV,
provide simpler and more efficient descriptions than
nonrelativistic approaches that hide these scales. The dominant
evidence is the spin-orbit splittings. Other evidence includes the
density dependence of optical potential, the observation of
approximate pseudospin symmetry, correlated two-pion exchange
strength, QCD sum rules, and more \cite{Furnstahl04}. The
relativistic Bruekner-Hartree-Fock theory and RMF theory with
density-dependent coupling constants extracted from it can
reproduce well the nuclear saturation properties (the Coester
line) in nuclear matter~\cite{Brockmann90,Brockmann92}. }
Furthermore the RMF theory can reproduce better the measurements
of the isotopic shifts in the Pb-region \cite{Sharma93b}, give
more naturally the spin-orbit potential, the origin of the
pseudospin symmetry~\cite{Arima69,Hecht69} as a relativistic
symmetry~\cite{Ginocchio97,Meng98r,Meng99prc} and spin symmetry in
the anti-nucleon spectrum~\cite{Zhou03prl}, and be more reliable
for nuclei far away from the line of $\beta$-stability, etc.
Obviously, RMF is one of the best candidates for the description
of exotic nuclei.

In order to describe exotic nuclei, the pairing correlation and
the coupling to continuum, which are extremely crucial for the
description of drip line nuclei, must be taken into account
properly  {\cite{Bulgac80}}. The continuum effect is commonly
taken into account in the Hartree-Fock-Bogoliubov (HFB)
\cite{Ring80} or relativistic-Hartree-Bogoliubov (RHB)
\cite{Ring96} approaches. In most of these calculations the
continuum is replaced by a set of positive energy states
determined by solving the HFB or RHB equations in coordinate space
and with box boundary conditions \cite{Meng96, Dobaczewski84,
Dobaczewski96}. Recently the HFB equations were also solved with
exact boundary conditions for the continuum spectrum, both for a
zero range \cite{Grasso01} and a finite range pairing forces
\cite{Grasso02}. These resonant continuum HFB or RHB results with
exact boundary conditions are generally close to those with box
boundary conditions~\cite{Grasso01,Sandulescu03}. The extension of
the RMF theory to take into account both bound states and
(discretized) continuum via Bogoliubov transformation in a
microscopic and self-consistent way, i.e., the Relativistic
Continuum Hartree-Bogoliubov (RCHB) theory has been done in
Refs.~\cite{Meng96,Meng98npa}. The RCHB theory was very successful
in describing the ground state properties of nuclei both near and
far from the $\beta$-stability line. By using a density-dependent
zero range interaction, the halo in $^{11}$Li has been
successfully reproduced in this self-consistent picture.
Remarkable successes of the RCHB theory include the new
interpretation of the halo in $^{11}$Li \cite{Meng96}, the
prediction of the exotic phenomena as giant halos in Zr
($A>122$)~\cite{Meng98prl} and Ca ($A>60$)~\cite{Meng02r}
isotopes, the reproduction of interaction cross section and
charge-changing cross sections in light exotic nuclei in
combination with the Glauber theory~\cite{Meng98plb,Meng02plb},
better restoration of pseudospin symmetry in exotic
nuclei~\cite{Meng98r,Meng99prc}, and predictions of exotic
phenomena in hyper nuclei~\cite{Lv03} and new magic numbers in
superheavy nuclei~\cite{ZhangW05}, etc.

The main purpose of the present manuscript is to review the RCHB
theory and its applications for exotic and superheavy nuclei in
spherical case. Some other related topics will also be covered
briefly. In Section \ref{sec:rmf}, the formalism, the numerical
solutions and the effective interactions for the RMF theory are
presented. In Section \ref{sec:pairing}, the pairing correlations
and the approaches for pairing are sketched. Discussion on the
continuum states follows where the resonant BCS method and several
methods to obtain single particle resonant states are briefly
reviewed. Then the detailed formalism of the RCHB theory together
with the discussion on the effective pairing interactions are
given. In Section \ref{sec:exotic}, the applications of the RCHB
theory to the properties of exotic nuclei are presented, such as
the binding energies, particle separation energies, the radii and
cross sections, the single particle levels, shell structure, the
restoration of the pseudo-spin symmetry, the halo and giant halo,
and halos in hyper nuclei, etc. In Section \ref{sec:she}, the
predictions of new magic numbers in superheavy nuclei are
presented. Finally a brief summary and perspectives are given in
Section \ref{sec:summary}.

\section{Relativistic mean field theory}
 \label{sec:rmf}

In this section, we present the formalism, the numerical solutions
and the effective interactions for the RMF theory as well as its
application for nuclear matter. In the first subsection, the
effective Lagrangian density and equations of motion for the
nucleon and the mesons are given. The numerical solutions of the
Dirac equation for finite nuclei are discussed in the second
subsection. Subsequently, the effective interactions with
nonlinear self-coupling meson fields and density dependent
meson-nucleon couplings, and their influences on properties of
nuclear matter are discussed.

\subsection{The general formalism}

The RMF theory describes the nucleus as a system of Dirac nucleons
which interact in a relativistic covariant manner via meson
fields. The meson fields are treated as classical fields. In the
simplest RMF version, i.e., the $\sigma$-$\omega$
model~\cite{Walecka74}, the mesons do not interact among
themselves, which leads to too large incompressibility in nuclear
matter. Therefore a nonlinear self-coupling of the $\sigma$-field
was proposed~\cite{Boguta77}. In order to reproduce the density
dependence of the vector and scalar potentials of the
Dirac-Brueckner calculations~\cite{Brockmann92}, the nonlinear
self-coupling of the $\omega$-meson was found to be
necessary~\cite{Sugahara94}. Recently the nonlinear self-coupling
of the isovector $\rho$-meson was also introduced to improve the
density-dependence of the isospin-dependent part of the
potentials~\cite{Long04}. Within the present scheme, the
isoscalar-scalar $\sigma$-meson provides the mid- and long-range
attractive part of the nuclear interaction whereas the short-range
repulsive part is provided by the isoscalar-vector $\omega$-meson.
The photon field $A^\mu(x)$ accounts for the Coulomb interaction
while the isospin dependence of the nuclear force is described by
the isovector-vector $\rho$-meson. The $\pi$-meson field is not
included because it does not contribute at the Hartree level. In
principle, other mesons apart from $\sigma$, $\omega$, and $\rho$
may also contribute to the nuclear interaction as well,  {e.g.,
the isovector scalar $\delta$-meson, which was suggested in
Ref.~\cite{Kubis97} and also on the basis of a relativistic
Brueckner theory in Refs~\cite{Shen97, Jong98, Hofmann01b, Ma02b},
has been used in some structure calculations for exotic
nuclei~\cite{Hofmann01b}, nuclear matter~\cite{Liu02b} and stellar
matter~\cite{Menezes04}}. However, as the RMF theory is only an
effective theory therefore it is expected that the contributions
from other mesons can be effectively taken into account by
adjusting the model parameters to the properties of nuclear matter
and finite nuclei. With various versions of the nonlinear
self-couplings of meson fields, the RMF theory has been used to
describe lots of nuclear phenomena during the past years with
great successes~\cite{Serot86, Reinhard89, Ring96}. In order to
avoid the problem of instability for the nonlinear interactions at
high densities, the RMF theories with density dependence in the
couplings are developed as well~\cite{Brockmann92, Lenske95,
Fuchs95, Typel99, Niksic02b, Long04}.


The Lagrangian density of the RMF theory can be written as
 \beq
 \label{Lagrangian}
 \begin{split}
 \cals L = ~&\bar\psi\ls i\gamma^\mu\partial_\mu - M
            - g_\sigma\sigma
            - g_\omega\gamma^\mu\omega_\mu
            - g_\rho\gamma^\mu\ivec\tau\cdot\ivec\rho_\mu
            - e \gamma^\mu A_\mu\frac{1-\tau_3}{2}\rs\psi \\
          & + \ff2\partial^\mu\sigma\partial_\mu\sigma
                - U_\sigma(\sigma)
            -\ff4\Omega^{\mu\nu}\Omega_{\mu\nu}
                + U_\omega(\omega_\mu)
            -\ff4 \ivec R^{\mu\nu}\cdot\ivec R_{\mu\nu}
                + U_\rho(\ivec\rho_\mu) \\
          & -\ff4 F^{\mu\nu}F_{\mu\nu},
 \end{split}
 \eeq
where $M$ and $m_i(g_i)$ ($i = \sigma,\omega,\rho$) in the
following are the masses (coupling constants) of the nucleon and
the mesons respectively and
 \bsub
 \beqn
\Omega^{\mu\nu}  &=& \partial^\mu\omega^\nu
                   - \partial^\nu\omega^\mu, \\
\ivec R^{\mu\nu} &=& \partial^\mu\ivec\rho^\nu
                   - \partial^\nu\ivec\rho^\mu, \\
   F^{\mu\nu}    &=& \partial^\mu A^\nu
                   - \partial^\nu A^\mu
 \eeqn
 \esub
are the field tensors of the vector mesons and the electromagnetic
field. We adopt the arrows to indicate vectors in isospin space
and bold types for the space vectors. Greek indices $\mu$ and
$\nu$ run over 0, 1, 2, 3 or $t$, $x$, $y$, $z$, while Roman
indices $i$, $j$, etc. denote the spatial components.

The nonlinear self-coupling terms $U_\sigma(\sigma)$,
$U_\omega(\omega_\mu)$, and $U_\rho(\ivec\rho_\mu)$ for the
$\sigma$-meson, $\omega$-meson, and $\rho$-meson in the Lagrangian
density (\ref{Lagrangian}) respectively have the following forms:
 \bsub
 \beqn
 U_\sigma(\sigma)    &=& \ff2m_\sigma^2\sigma^2
                       + \ff3 g_2\sigma^3 + \ff4 g_3\sigma^4, \\
U_\omega(\omega_\mu) &=& \ff2m_\omega^2\omega^\mu\omega_\mu
                       + \ff4 c_3\lb\omega^\mu\omega_\mu\rb^2, \\
U_\rho(\ivec\rho_\mu)&=&
\ff2m_\rho^2\ivec\rho~^\mu\cdot\ivec\rho_\mu
                       + \ff4 d_3 \ls \ivec\rho~^\mu\cdot\ivec\rho_\mu\rs^2.
 \eeqn
 \esub

From the Lagrangian density (\ref{Lagrangian}), the Hamiltonian
operator can be obtained by the general Legendre transformation
 \beq
 H = \int d^3x~ \cals H
   = \int d^3x\ls \sum_i\pi_i(x)\frac{\partial \phi_i(x)}{\partial t}
     - \cals L(x)\rs,
 \eeq
where the conjugate momenta of the field operators $\phi_i$
($\phi_i = \psi, \sigma, \omega_\nu, \ivec\rho_\nu, A_\nu$) are
defined as
 \beq
\pi_i(x) = \frac{\partial \cals L}
                {\partial \ls\partial\phi_i/\partial t\rs}.
 \eeq
Then the Hamiltonian density of the system can be easily obtained
as
 \beq
 \label{Hamiltonian}
 \begin{split}
 \cals H = & \cals H[\psi]
           + \cals H[\sigma] + \cals H[\omega_\nu] + \cals H[\ivec\rho_\nu]
           + \cals H[A_\nu] \\
         = & \bar\psi\ls - i\gamma^i\partial_i + M\rs\psi
           + g_\sigma\sigma\bar\psi\psi
           + g_\omega\omega_\mu\bar\psi\gamma^\mu\psi
           + g_\rho\ivec\rho_\mu\cdot\bar\psi\gamma^\mu\ivec\tau\psi
           + e A_\mu \bar\psi\gamma^\mu\frac{1-\tau_3}{2}\psi \\
           & + \ff2\partial^0\sigma\partial_0\sigma
             - \ff2\partial^i\sigma\partial_i\sigma
             + U_\sigma(\sigma)
             - \ff4\Omega^{0\nu}\Omega_{0\nu}
             + \ff4\Omega^{i\nu}\Omega_{i\nu}
             - U_\omega(\omega_\nu) \\
           & - \ff4\ivec R^{0\nu}\cdot\ivec R_{0\nu}
             + \ff4\ivec R^{i\nu}\cdot\ivec R_{i\nu}
             - U_\rho(\ivec\rho_\nu)
             - \ff4 F^{0\nu} F_{0\nu}
             + \ff4 F^{i\nu} F_{i\nu}.
 \end{split}
 \eeq

The relativistic mean field theory is formulated on the basis of
the effective Lagrangian (\ref{Lagrangian}) with the mean field
approximation, i.e. the meson fields are treated as classical
c-numbers. With this approximation, all the quantum fluctuations
of the meson fields are removed and the nucleons are described as
independent particles moving in the effective meson and photon
fields. Therefore the nucleon field operator can be expanded on a
complete set of single-particle states as,
 \beq
  \label{expansion}
  \psi(x)        = \sum_a \psi_a(x) c_a,
 \eeq
where $c_a$ is the annihilation operator for a nucleon in the
state $a$ of the Dirac fields and $\psi_a$ is the corresponding
single-particle spinor. The operator $c_a$ and its conjugate
$c^\dagger_a$ satisfy the anticommutation rules
 \beq
 \{c_a,c_b^\dagger\}= \delta_{ab} \quad\mbox{and}\quad
 \{c_a,c_b\}        = \{c^\dagger_a,c^\dagger_b\} = 0.
 \eeq
Confined to the single-particle states $i$ with positive energies,
i.e. the no-sea approximation, the ground state of the nucleus can
be constructed as,
 \beq
  \label{trial}
  \lr \Phi\rc = \prod_{i=1}^A c_i^\dag \lr 0\rc
  \quad
  \mbox{with}\quad
  \langle\Phi|\Phi\rangle=1,
 \eeq
where $|0\rangle$ is physical vacuum.

With the ground state (\ref{trial}) and the mean field
approximation, the energy functional, i.e. the expectation value
for the Hamiltonian (\ref{Hamiltonian}) is obtained as
 \beq \label{EnergyF}
\begin{split}
E_{\text{RMF}}(\rho,\phi)
   = & \langle\Phi|\cals H|\Phi\rangle \\
   = & \int d^3x \tr\ls \beta\lb \svec\gamma\cdot\svec p + M+ g_\sigma\sigma
     + g_\omega\omega^\mu\gamma_\mu  + g_\rho\ivec\tau\cdot\ivec\rho_\mu\gamma^\mu
     + \ff2 e(1-\tau_3)A_\mu\gamma^\mu\rb\rho\rs\\
     & + \int d^3x\Lb- \ff2\partial^{\mu}\sigma\partial_{\mu}\sigma
       + U_\sigma(\sigma)
       + \ff4\Omega^{{\mu}\nu}\Omega_{{\mu}\nu}
       - U_\omega(\omega_\nu)
       + \ff4\ivec R^{{\mu}\nu}\cdot\ivec R_{{\mu}\nu}
       - U_\rho(\ivec\rho_\nu)
       + \ff4 F^{{\mu}\nu}F_{{\mu}\nu}\Rb,
 \end{split}
 \eeq
where $\phi = \sigma, \omega_\nu, \ivec\rho_\nu, A_\nu$ and the
density matrix $\rho$ is defined as
 \beqn
 \rho_{ij}
   &\equiv& \langle\Phi|\psi^\dagger_j c^\dagger_j\psi_i c_i|\Phi\rangle
          = \psi^\dagger_j \psi_i \delta_{ij}.
 \eeqn

For system with time reversal symmetry, the space-like components
of the vector fields vanish. Furthermore one can assume that in
all nuclear applications the nucleon single-particle states do not
mix isospin, i.e. the single-particle states are eigenstates of
$\tau_3$, therefore only the third component of $\ivec\rho_\nu$
survives. Stationarity implies that the nucleon single-particle
wave function can be written as
 \beq\label{singwf}
 \psi_i(x)=e^{-i\epsilon_i t}\psi_i(\svec x).
 \eeq
where $\epsilon_i$ is the single-particle energy. Accordingly the
density
 matrix is reduced as,
 \beq\label{densitM}
 \rho_{ij}=\psi_j^\dagger(\svec x')\psi_i(\svec x)\delta_{ij}.
 \eeq
 Altogether there remain only the meson
fields $\sigma$, $\omega_0$, $\rho^3_0$, $A_0$ which are time
independent.

 The equations of motion for the nucleon and the
mesons can be obtained by requiring that the energy functional
(\ref{EnergyF}) be stationary with respect to the variations of
$\rho$ and $\phi$. More explicitly, the stationary condition reads
 \beq\label{stationary}
\delta \lb E_{\text{RMF}}(\rho, \phi) -\tr(\cals E \rho) \rb= 0,
 \eeq
where $\cals E$ is a diagonal matrix, whose diagonal elements are
the single particle energies $\epsilon_i (i = 1, \cdots, N)$
introduced in Eq.(\ref{singwf}), and $N$ is the number of
eigenstates.

Using the variation $\delta\rho$ with respect to $\psi_i$, the
stationary condition (\ref{stationary}) leads to the Dirac
equation
 \beq\label{Dirac}
 \left(\svec\alpha\cdot\mathbf{p} + \beta(M+S) + V\right) \psi_i(\svec x)
   = \epsilon_i \psi_i(\svec x)
 \eeq
for the nucleon and the Klein-Gordon equations for sigma, omega,
rho, and the photon:
 \bsub
 \beqn
 \label{Klein-Gordon}
 -\svec\nabla^2\sigma   + U_\sigma'(\sigma)   &=& - g_\sigma\rho_s, \label{kg-sigma}\\
 -\svec\nabla^2\omega_0 + U_\omega'(\omega_0) &=&   g_\omega\rho_v, \label{kg-omega}\\
 -\svec\nabla^2\rho^3_0 + U_\rho'(\rho^3_0)   &=&   g_\rho\rho_3  , \label{kg-rho}\\
 -\svec\nabla^2 A_0                           &=&   e\rho_c  \label{kg-photon}.
 \eeqn
 \esub
The scalar potential $S$ and vector potential $V$ in equation
(\ref{Dirac}) are respectively:
 \bsub
 \beqn
 \label{potential}
S(\svec{x})&=& g_\sigma\sigma(\svec{x}), \label{potential-scalar}\\
V(\svec{x})&=& g_\omega\omega_0(\svec{x})
             + g_\rho\tau_3\rho^3_0(\svec{x})
             + \ff2 e(1-\tau_3)A_0(\svec{x}) \label{potential-vector}.
 \eeqn
 \esub
While the scalar density ($\rho_s$), the baryon density
($\rho_v$), the isovector density ($\rho_3$), and the charge
density ($\rho_c$) in the Klein-Gordon equations
(\ref{kg-sigma}--\ref{kg-photon}) are respectively,
  {\bsub\beqn
 \label{density}
 \rho_s &=& \tr\ls \beta\rho\rs, \label{density-s}\\
 \rho_v &=& \tr \ls \rho\rs ,\label{density-v}\\
 \rho_3 &=& \tr \ls \tau_3\rho\rs,\label{density-3}\\
 \rho_c &=& \ff2\tr \ls (1-\tau_3)\rho\rs\label{density-c}.
 \eeqn\esub}

The total energy of the system can be obtained from the energy
functional (\ref{EnergyF}) as,
  {\beq\begin{split}
  E =~& \int d^3 x \Lb\tr\ls\lb\svec\alpha\cdot\svec p + \beta M\rb\rho\rs+ \ff2\tr\ls g_\sigma\beta\sigma\rho
     +  g_\omega\omega_0\rho     +  g_\rho \tau_3\rho^3_0\rho
     + \ff2 e (1-\tau_3) A_0\rho \rs\Rb \\
   ~& + \int d^3x \ls U_\sigma(\sigma)
          - U_\omega(\omega_0) - U_\rho(\rho^3_0)\rs
      - \ff2\int d^3x\ls\sigma U_\sigma'(\sigma)
      - \omega^0 U_\omega'(\omega_0)
      - \rho^3_0 U_\rho'(\rho^3_0)\rs.
 \end{split}\eeq}

In the density dependent RMF approach, where the nonlinear
self-couplings for the $\sigma$, $\omega$, and $\rho$ mesons in
the Lagrangian density are respectively replaced by the density
dependence of the coupling constants $g_\sigma(\rho)$,
$g_\omega(\rho)$, and $g_\rho(\rho)$, an additional term, i.e. the
rearrangement term, will appear in the Dirac
equation~(\ref{Dirac})~\cite{Brockmann92, Lenske95, Fuchs95,
Typel99, Niksic02b, Long04}.

\subsection{Numerical algorithm for spherical nuclei}
\label{sec:srh}

The harmonic oscillator basis has served as a very useful tool in
nuclear structure study. Normally, the equations of motion for
nucleons moving in a mean field are solved by expanding them on
the harmonic oscillator (HO)
basis~\cite{Mayer55,Nilsson55,Bohr69,Bohr75,Ring80}. However, for
exotic nuclei with large spacial extension, e.g., halo nuclei, it
is not justified to work in the conventional harmonic oscillator
basis due to its
localization~\cite{Meng96,Stoitsov98a,Stoitsov98b,Zhou00}.
Instead, one can choose to work either in the coordinate space, or
improve the asymptotic behavior of the HO wave function, or adopt
other basises which have a correct asymptotic behavior, for
example, the Woods-Saxon basis.

In this subsection, we will focus on the numerical solution of the
RMF for spherical nuclei. Due to the special spacial symmetry,
both the Dirac equation for the nucleon and the Klein-Gordon
equations for the mesons and the photon become radially dependent
only, thus facilitating much the solution of the coupled
equations. The formalism for the spherical relativistic Hartree
(SRH) theory will be briefly presented. We then review the SRH
theory in different basis, including Finite Element Method (FEM)
in coordinate space, transformed harmonic oscillator basis, and
the Woods-Saxon basis. The application of the SRH theory to doubly
magic nuclei follows.

\subsubsection{Spherical relativistic Hartree theory}

Starting from the Eqs.~(\ref{Dirac}) and
(\ref{kg-sigma}--\ref{kg-photon}) given in the previous
subsection, one derives the coupled radial equations for spherical
nuclei, i.e., the radial Dirac equation and radial Klein-Gordon
equations.

For spherical nuclei, the Dirac spinor which is the expansion
coefficient $\psi_a(\svec r)$ ($a = \Lb\alpha,\kappa,m\Rb$) in
Eq.~(\ref{singwf}) (the coordinate from $\svec x$ has been changed
to $\svec r$ to reflect the spherical symmetry) is characterized
by the angular momentum quantum numbers $\kappa$($l$,$j$), $m$,
the parity, the isospin $t = \pm 1/2$ (``+'' for neutrons and
``$-$'' for protons) and the radial quantum number $\alpha$ and
has the form
 \begin{equation}
\psi_{\alpha\kappa m}(\svec r,t)
    = \bpm i \dfrac{G_\alpha^{\kappa}(r)}{r} \\[1em]
             \dfrac{F_\alpha^{\kappa}(r)}{r}\svec\sigma\cdot\hat{\svec r}
      \epm
      Y_{jm}^{l}(\theta,\phi)\chi_{t_\alpha}(t),
 \label{eq:SRHspinor}
 \end{equation}
with $G_\alpha^{\kappa}(r) / r$ and $F_\alpha^{\kappa}(r) / r$ the
radial wave functions for the upper and lower components and $ Y^l
_{jm}(\theta,\phi)$ the spinor spherical harmonics
\cite{Varshalovich88}. Substituting Eq. (\ref{eq:SRHspinor}) into
the Dirac equation (\ref{Dirac}), one can deduce the radial Dirac
equations as
 \bsub
 \label{eq:SRHDirac}
 \beqn
\epsilon_\alpha G_\alpha^{\kappa}
    & = & \left( -\dfrac{d}{d r} + \dfrac{\kappa}{r} \right) F_\alpha^{\kappa}
        + \left( M + S(r) + V(r) \right) G_\alpha^{\kappa} ,    \\
\epsilon_\alpha F_\alpha^{\kappa}
    & = & \left( +\dfrac{d}{d r} + \dfrac{\kappa}{r} \right) G_\alpha^{\kappa}
        - \left( M + S(r) - V(r) \right) F_\alpha^{\kappa},
 \eeqn
 \esub
with the scalar and vector potentials
 \bsub\label{eq:sources}\beqn
S(r) &=& g_\sigma\sigma,\\
V(r) &=& g_\omega\omega_0 + g_\rho \tau_3 \rho^3_0
       + \ff2 e (1-\tau_3) A_0.
 \eeqn\esub
 The meson field equations (\ref{kg-sigma}--\ref{kg-photon}) simply become radial Laplace equations of the
form
\begin{equation}
 \left( - \frac{d^2}{d r^2}
        - \frac{2}{r}\frac{d}{d r} + m_{\phi}^2
 \right) \phi(r)
 = s_{\phi}(r).
 \label{eq:SRHmesonmotion}
\end{equation}
where $m_{\phi}$ are the meson masses for $\phi = \sigma,
\omega,\rho$ and zero for the photon. The source terms are
 \begin{eqnarray}
 s_{\phi}(r) =
  \left\{
    \begin{array}{ll}
      - g_\sigma\rho_s(r) - g_2 \sigma^2(r)  - g_3 \sigma^3(r),
     & \mathrm{for \ the} \ \sigma\mathrm{-field}, \\
        g_\omega \rho_v(r) - c_3 \omega_0^3(r),
     & \mathrm{for \ the} \ \omega\mathrm{-field}, \\
        g_{\rho} \rho_3(r) - d_3 [\rho_0^3(r)]^3,
     & \mathrm{for \ the} \ \rho\mathrm{-field}, \\
        e \rho_c(r),
     & \mathrm{for \ the \ Coulomb \ field},
    \end{array}
  \right.
 \label{eq:mesonsourceS}
 \end{eqnarray}
with
 \bsub
 \beqn
 \rho_s(r) & = &\ff{4\pi r^2} \sum_{i=1}^A (|G_i(r)|^2 - |F_i(r)|^2), \\
 \rho_v(r) & = &\ff{4\pi r^2} \sum_{i=1}^A (|G_i(r)|^2 + |F_i(r)|^2), \\
 \rho_3(r) & = &\ff{4\pi r^2} \sum_{i=1}^A \tau_3 (|G_i(r)|^2 + |F_i(r)|^2), \\
 \rho_c(r) & = &\ff{4\pi r^2} \sum_{i=1}^A \ff2 (1-\tau_3) (|G_i(r)|^2 + |F_i(r)|^2).
 \eeqn
 \label{eq:SRHdensity}
 \esub

The procedures of solving these coupled equations are as the
following: a) with a set of estimated meson and photon fields, the
scalar and vector potentials in Eqs.(\ref{eq:sources}) are
calculated and the radial Dirac equation is solved; b) the so
obtained nucleon wave functions are used to give the source term
of each radial Laplace equation for mesons and the photon; and c)
the new meson and photon fields obtained from solving these
Laplace equations will be used to replace the fields in step a).
This procedure is iterated until a demanded accuracy is achieved.

With the above procedures, there are two methods to solve the
coupled equations (\ref{eq:SRHDirac}--\ref{eq:SRHdensity}). One is
in coordinate space with the shooting method~\cite{Horowitz81},
i.e. spherical relativistic Hartree theory in $r$ space (SRHR),
and the finite element
method~\cite{Poschl97a,Meng97,Poschl97b,Meng98npa} (SRHFEM).
Another is in configuration space, e.g., the harmonic oscillator
basis~\cite{Gambhir90} (SRHHO), the transformed harmonic
oscillator basis~\cite{Stoitsov98a} (SRHTHO) and the Woods-Saxon
basis~\cite{Zhou03prc} (SRHWS). The readers are referred to
Ref.~\cite{Horowitz81} and Ref.~\cite{Gambhir90} for SRHR and
SRHHO, respectively. In the following the numerical techniques for
the solution of the Dirac equation in the finite element method,
the transformed harmonic oscillator basis and Woods-Saxon basis
will be briefly introduced.

\subsubsection{Finite element method} \label{subsec:FEM}

A convenient procedure for the coordinate space discretization of
the Dirac equations (\ref{eq:SRHDirac}) is provided by Finite
Element Method (FEM)~\cite{Poschl97a,Meng97,Poschl97b,Meng98npa}.
Similar as in the $r$ space~\cite{Horowitz81}, the Dirac equations
(\ref{eq:SRHDirac}) are solved in a box with a box size $R$ and
proper boundary condition. The radial wave functions
$G_\alpha^{\kappa}$ and $F_\alpha^{\kappa}$ are discretized at
$N+1$ points, i.e., $r_1=0$, $r_2=\Delta r$, ... , $r_{N+1} = N
\cdot \Delta r$, where $\Delta r= R/N$. The wave functions thus
obtained become tabulated values:
$G_\alpha^{\kappa}(r_0)$,$G_\alpha^{\kappa}(r_1)$, ... ,
$G_\alpha^{\kappa}(r_{N+1})$ and
$F_\alpha^{\kappa}(r_0)$,$F_\alpha^{\kappa}(r_1)$, ... ,
$F_\alpha^{\kappa}(r_{N+1})$.

The idea of FEM is as following: if the $\Delta r= R/N$ is small
enough, or equivalently $N$ is large enough, the wave function
$G_\alpha^{\kappa}(r)$ in the element $r \in [r_i, r_{i+1}]$ can
be well approximated by $G_\alpha^{\kappa}(r_{i})$ and
$G_\alpha^{\kappa}(r_{i+1})$ together with some simple analytic
function, e.g., the linear, quadratic, cubic, or 4th order shape
function~\cite{Poschl97a}.

Taking the linear shape function,
\begin{eqnarray}
\left\{
\begin{array}{lll}
    \eta_a(r) &=& 1 - {\rho}, \\
    \eta_b(r) &=& {\rho},
\end{array}
\right. \label{ShEq}
\end{eqnarray}
with $\rho = \dfrac {r-r_i} {\Delta r}$ and $r \in [r_i, r_{i+1}]$
as an example, the wave function $G_\alpha^{\kappa}(r)$ can be
well approximated by, \beqn
    G_\alpha^{\kappa}(r) = G_\alpha^{\kappa}(r_{i}) \eta_a(r)
                         + G_\alpha^{\kappa}(r_{i+1})\eta_b(r), ~~
    r \in [r_i, r_{i+1}].
\eeqn Similarly the expression for $F_\alpha^{\kappa}(r)$ can also
be obtained. For $r \in [r_i, r_{i+1}]$, the Dirac equations
(\ref{eq:SRHDirac}) can now be written as:
\begin{eqnarray}
    \left(
    \begin{array}{ll}
       M + S(r) + V(r)-\epsilon_\alpha  &  -\dfrac{d}{d r} + \dfrac{\kappa}{r}  \\
       \dfrac{d}{d r} + \dfrac{\kappa}{r} & - M - S(r) + V(r) -\epsilon_\alpha
    \end{array}
    \right)
    \left(
    \begin{array}{l}
       G_\alpha^{\kappa}(r_{i}) \eta_a(r)
                         + G_\alpha^{\kappa}(r_{i+1})\eta_b(r) \\
       F_\alpha^{\kappa}(r_{i}) \eta_a(r)
                         + F_\alpha^{\kappa}(r_{i+1})\eta_b(r)
    \end{array}
    \right)
     = 0
\label{FEMDEq}
\end{eqnarray}
Multiplying
       $\displaystyle
     ( G_\alpha^{\kappa}(r_{i}) \eta_a(r)
                         + G_\alpha^{\kappa}(r_{i+1})\eta_b(r) ~~~
       F_\alpha^{\kappa}(r_{i}) \eta_a(r)
                         + F_\alpha^{\kappa}(r_{i+1})\eta_b(r) )$
from left and integrating with $\displaystyle \int^{r_{i+1}}_{r_i}
r^2 dr$, algebra equations for $G_\alpha^{\kappa}(r_{i})$,
$G_\alpha^{\kappa}(r_{i+1})$ and $F_\alpha^{\kappa}(r_{i})$,
$F_\alpha^{\kappa}(r_{i+1})$ are obtained. Repeating the same
procedure for all the $N$ elements, the Dirac equations
(\ref{eq:SRHDirac}) are transformed into a generalized eigenvalue
problem. This generalized eigenvalue problem can be solved by the
standard diagonaliztion algorithms and all the energies and wave
functions discretized at $r=r_1,r_2, ... , r_{N+1}$ can be
obtained. It has been shown that FEM can provide very accurate
solutions for the relativistic eigenvalue problem in the
self-consistent mean-field
approximation~\cite{Poschl97a,Meng97,Poschl97b,Meng98npa}.

Comparing with the shooting method in the $r$
space~\cite{Horowitz81}, the FEM has the advantage that one can
get all the energy $\epsilon_\alpha$ and wave functions
$G_\alpha^{\kappa}$ and $F_\alpha^{\kappa}$ by a single
diagonalization and it is straight forward to be generalized to
the cases with nonlocal interaction in the pairing channel.
However, in order to obtain the same accuracy as the shooting
method, a huge matrix has to be constructed and it becomes more
time consuming. One can also replace the linear shape function
$\eta_a(r)$ and $\eta_b(r)$ in Eq.(\ref{ShEq}) by the quadratic,
cubic, or 4th order shape function, the procedures are the same
and in principle the same accuracy can be achieved by the linear
shape function with larger $N$.

\subsubsection{\label{subsec:THO}Transformed harmonic oscillator basis}

In order to modify the asymptotic behavior of the harmonic
oscillator wave function at large $r$, the local-scaling
transformation method \cite{Petkov81, Petkov83a, Petkov83b} has
been introduced to construct the so called local-scaling
transformed harmonic oscillator basis (THO) in
Refs.~\cite{Stoitsov98a, Stoitsov98b}.

A local-scaling point coordinate transformation (LST) is defined
as
\begin{equation}
{\bf r}^{\prime }={\bf f}({\bf r})\equiv \hat{\bf r}f({\bf r)}.
\end{equation}
The transformed radius vector has the same direction $\hat{\bf r}
\equiv {\bf r}/|{\bf r}|$, while its magnitude $r^{\prime }=f({\bf
r)}$ depends on the scalar LST function. $f({\bf r)}$ is assumed
to be an increasing function of $r$, and $f(\bf{0})$=0. The
corresponding LST wave function can be expressed as
\begin{equation}
 \Psi_{f}({\bf r}_{1},{\bf r}_{2},...,{\bf r}_{A})
 = \left[ \prod_{i=1}^{A} \frac{f^2({\bf r}_i)}{r^2_i}
          \frac{\partial f({\bf r}_i)}{\partial r_i}
   \right]^{1/2}
   \bar{\Psi}(f({\bf r}_{1}), f({\bf r}_{2}), ..., f({\bf r}_{A})),
 \label{eq:lstwf}
\end{equation}
where $\bar{\Psi}({\bf r}_{1}, {\bf r}_{2}, ..., {\bf r}_{A})$ is
an $A$-particle wave function normalized to unity. The local
one-body density corresponding to an $A$-body wave function
${\Psi}$ is
\begin{equation}
 \rho({\bf r}) =
 A \int | \Psi({\bf r}, {\bf r}_{2}, ..., {\bf r}_{A}) |^{2}
        d{\bf r}_{2}...d{\bf r}_{A}.
 \label{eq:modden}
\end{equation}
There exists a simple relation between the local density
$\rho_f(\bf r)$ associated with the LST function ${\Psi}_f$, and
the density $\bar{\rho}(\bf r)$ which corresponds to the
prototypical model function $\bar{\Psi}$:
\begin{equation}
 \rho_f({\bf r}) =
 \frac{f^{2}({\bf r})}{r^{2}} \frac{\partial f({\bf r})} {\partial r}
 \bar{\rho}(f({\bf r)}).
 \label{eq:lstden}
\end{equation}
When the form of the density $\rho_f(\bf r)$ is known,
Eq.~(\ref{eq:lstden}) becomes a first order nonlinear differential
equation for the LST function $f$ and can be solved easily. For a
system with spherical symmetry, $\rho_f$, $\bar{\rho}$, and $f$
depend only on $r$=$|{\bf r|}$, and Eq.~(\ref{eq:lstden}) can be
reduced to a nonlinear algebraic equation
\begin{equation}
 \int_{0}^{r} \rho_f(u) u^{2} du =
 \int_{0}^{f(r)} \bar{\rho}(u) u^{2} du.
 \label{eq:lsteq}
\end{equation}
The solution can be found subject to the boundary condition
$f(0)=0$.

For shell-model or mean-field applications, one has to consider
the case when the model many-body wave function is a Slater
determinant
\begin{equation}
 \label{eq:ssmod}
 \bar{\Psi}({\bf r}_{1}, {\bf r}_{2}, ..., {\bf r}_{A}) =
 \frac{1}{\sqrt{A!}} \det |\bar{\phi}_{i}({\bf r}_{j})|.
\end{equation}
The single-particle functions $\bar{\phi}_{i}({\bf r})$ form a
complete set. The LST wave function is defined by the
transformation (\ref{eq:lstwf}), and is written as a product state
\begin{equation}
 \label{eq:ssf}
 \Psi_{f}({\bf r}_{1}, {\bf r}_{2}, ..., {\bf r}_{A}) =
 \frac{1}{\sqrt{A!}} \det |\phi_{i}({\bf r}_{j})| ,
\end{equation}
of the transformed basis state
\begin{equation}
 \label{eq:lstspwf}
 \phi_{i}({\bf r}) =
 \left[ \frac{f^{2}({\bf r})}{r^{2}} \frac{\partial f({\bf r})}{\partial r}
 \right]^{1/2}
 \bar{\phi}_{i}({\bf f}({\bf r)}).
\end{equation}

\begin{figure}[tbp]
\centerline{\includegraphics[width=8.5cm]{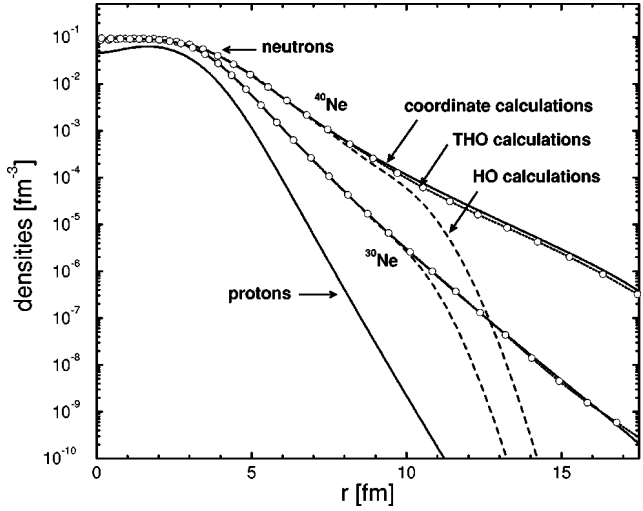}}
\caption{Self-consistent RHB proton and neutron densities for the
ground-states of $^{30}$Ne and $^{40}$Ne. Density profiles
calculated in THO are compared with FEM. Taken from Stoitsov et
al.~\protect\cite{Stoitsov98a}.} \label{FigB1}
\end{figure}

In Refs.~\cite{Stoitsov98a, Stoitsov98b}, the transformed harmonic
oscillator basis (THO) has been derived by a local scaling-point
transformation of the spherical harmonic-oscillator radial wave
functions. The unitary scaling transformation produces a basis
with improved asymptotic properties. The THO basis is employed in
the solution of the relativistic Hartree-Bogoliubov (RHB)
equations in configurational space~\cite{Stoitsov98a}. As shown in
Fig.~\ref{FigB1}, an expansion of nucleon spinors and mean-field
potentials in the THO basis reproduces the asymptotic properties
of neutron densities calculated by FEM in the coordinate space.

\subsubsection{Woods-Saxon basis} \label{subsec:WS}

\paragraph{Woods-Saxon basis from the
Schr\"odinger equation (SWS basis)}

For the Schr\"odinger equation with a spherical Woods-Saxon
potential
\begin{equation}
 V_\mathrm{WS}(r)
  = \left\{
     \begin{array}{ll}
      \dfrac{V_0}{1+e^{(r-R_0)/a_0}},\ \ & r <    R_\mathrm{max}, \\
      \infty,                            & r \geq R_\mathrm{max}, \\
     \end{array}
    \right.
\end{equation}
where $R_\mathrm{max}$ is introduced for practical reasons to
define the box boundary. The eigenfunction can be written as
$\phi_{nlm_l}(\svec{r}) = R_{nl}(r) Y_{lm_l}(\theta,\phi)$ and its
radial Schr\"odinger equation is derived as
\begin{equation}
 \left[ -\dfrac{1}{2M}
  \left( \dfrac{1}{r^2} \dfrac{d}{d r}
         r^2 \dfrac{d}{d r} - \dfrac{l(l+1)}{r^2}
  \right)
  + V_\mathrm{WS}(r)
 \right] R_{nl}(r)
 = \ E_{nl} R_{nl}(r) .
 \label{eq:schws}
\end{equation}

Equation~(\ref{eq:schws}) is solved on a discretized radial mesh
with a mesh size $\Delta r$. $R_\mathrm{max}$ ($\Delta r$) is
chosen large (small) enough to make sure that the final results do
not depend on it. The radial wave functions thus obtained form a
complete basis,
\begin{equation}
 \left\{ R_{nl}(r); n=0, 1, \cdots; l = 0,1,\cdots,n \right\},
\end{equation}
in terms of which the radial part of the upper and the lower
components of the Dirac spinor in Eq.~(\ref{eq:SRHDirac}) are
expanded respectively as
\begin{equation}
 \left\{
  \begin{array}{l}
   \displaystyle
   G_\alpha^\kappa(r) = -i \sum_{n=0}^{n_\mathrm{max}}
                             g_{\alpha n} r R_{nl}(r), \\
   \displaystyle
   F_\alpha^\kappa(r) = -i \sum_{\tilde n=0}^{\tilde n_\mathrm{max}}
                             f_{\alpha \tilde n}
                             r R_{\tilde n\tilde l}(r). \\
  \end{array}
 \right. \label{eq:DiracSWS}
\end{equation}
The radial Dirac equation (\ref{eq:SRHDirac}) is transformed into
the WS basis as
\begin{equation}
 \left(
  \begin{array}{cc}
   {\cal A}_{mn}        & {\cal B}_{m\tilde n}        \\
   {\cal C}_{\tilde mn} & {\cal D}_{\tilde m\tilde n} \\
  \end{array}
 \right)
 \left(
  \begin{array}{c}
   g_{\alpha n} \\
   f_{\alpha \tilde n} \\
  \end{array}
 \right)
 = \epsilon_\alpha
 \left(
  \begin{array}{c}
   g_{\alpha n} \\
   f_{\alpha \tilde n} \\
  \end{array}
 \right),
\end{equation}
where the matrix elements are calculated as follows
\begin{equation}
 \left\{
  \begin{array}{l}
   \displaystyle
   {\cal A}_{mn}
     = \int_0^{R_\mathrm{max}} r^2 dr
        R_{ml}(r)
        \left( V(r)+S(r)+M \right)
        R_{nl}(r),
   \\
   \displaystyle
   {\cal B}_{m\tilde n}
     = \int_0^{R_\mathrm{max}} r^2 dr
        R_{ml}(r)
        \left( +\frac{d}{d r} - \frac{\kappa_\alpha-1}{r}
        \right)
        R_{\tilde n\tilde l}(r),
   \\
   \displaystyle
   {\cal C}_{\tilde mn}
     = \int_0^{R_\mathrm{max}} r^2 dr
        R_{\tilde m\tilde l}(r)
        \left( -\frac{d}{d r} - \frac{\kappa_\alpha+1}{r}
        \right)
        R_{nl}(r),
   \\
   \displaystyle
   {\cal D}_{\tilde m\tilde n}
     = \int_0^{R_\mathrm{max}} r^2 dr
        R_{\tilde m\tilde l}(r)
        \left( V(r)-S(r)-M \right)
        R_{\tilde n\tilde l}(r).
   \\
  \end{array}
 \right.
 \label{eq:DiracMatrix}
\end{equation}

In practical calculations, an energy cutoff $E_\mathrm{cut}$
(relative to the nucleon mass $M$) is used to determine the cutoff
of the radial quantum number $n_\mathrm{max}$~\cite{Zhou03prc}.

\paragraph{Woods-Saxon basis from the Dirac
equation (DWS basis)}

The radial Dirac equation (\ref{eq:SRHDirac}) may be solved in $r$
space~\cite{Horowitz81} with Woods-Saxon-like potentials for
$V_0(r) \pm S_0(r)$~\cite{Koepf91a} within a spherical box of the
size $R_\mathrm{max}$, together with the spherical spinor which
gives a complete WS basis
\begin{equation}
 \left\{
  \left[ \epsilon_{n\kappa m}^0, \psi_{n\kappa m}^0(\svec{r},s,t) \right];
   \epsilon_{n\kappa m}^0 \ {}^{<}_{>} \ 0
 \right\},
\end{equation}
with $n=0,1,\cdots$, $\kappa=\pm1,\pm2,\cdots$, and $m =
-j_\kappa,\cdots,j_\kappa$. $\psi_{n\kappa m}^0(\svec{r},s,t)$
takes the form of Eq.~(\ref{eq:SRHspinor}). In such cases, states
both in the Fermi sea and in the Dirac sea should be included in
the basis for completeness. The nucleon wave function
(\ref{eq:SRHspinor}) can be expanded in terms of this set of basis
as
\begin{equation}
 \psi_{\alpha\kappa m}(\svec{r},s,t)
  = \sum_{n=0}^{n_\mathrm{max}}
     c_{\alpha n} \psi_{n\kappa m}^0(\svec{r},s,t),
 \label{eq:DSexpansion}
\end{equation}
where $n_\mathrm{max} = n_\mathrm{max}^+ + n_\mathrm{max}^- + 1$
and the summation runs over positive energy levels in the Fermi
sea for $0 \le n \le n_\mathrm{max}^+$ and over negative energy
levels in the Dirac sea for $n_\mathrm{max}^+ + 1 \le n \le
n_\mathrm{max}$. The negative energy states are obtained with the
same method as the positive energy ones. In this WS basis, the
Dirac equation (\ref{eq:SRHDirac}) turns out to be
\begin{equation}
 c_{\alpha m} \epsilon_m^0 + \sum_{n=0}^{n_\mathrm{max}}
                              c_{\alpha n} H'_{mn}
 = \epsilon_\alpha c_{\alpha m},\
 m = 1, \cdots, n_\mathrm{max},
\label{eq:matrix}
\end{equation}
with
\begin{eqnarray}
 H'_{mn}
 & = &
  \left\langle \psi_m^0(\svec{r}) \right|
   \left[ \Delta V(\svec{r})+\beta\Delta S(\svec{r})\right]
  \left| \psi_n^0(\svec{r}) \right\rangle
 \nonumber \\
 & = &
  \int_0^{R_\mathrm{max}} dr
    G_m^0(r) \left[ \Delta V(r)+\Delta S(r) \right] G_n^0(r)
 \nonumber \\
 & + &
  \int_0^{R_\mathrm{max}} dr
    F_m^0(r) \left[ \Delta V(r)-\Delta S(r) \right] F_n^0(r)
 ,
 \label{eq:matrixelement}
\end{eqnarray}
where $\Delta V(\svec{r}) = V(\svec{r})-V_0(\svec{r})$, $\Delta
S(\svec{r}) = S(\svec{r})-S_0(\svec{r})$ and the angular, spin,
and isospin quantum numbers are omitted for brevity.


In the expansion (\ref{eq:DSexpansion}) of the nucleon wave
function in the SRHDWS theory, one has to take into account not
only the states in the Fermi sea but also those in the Dirac sea
because these states form a complete basis together. The
contribution from negative energy states for $^{16}$O is given in
Table~\ref{TabB1}. It is found that, without including the
negative energy levels, the calculated results will depend on the
potentials for the basis.

\begin{table}
\caption{\label{TabB1} Effects of negative energy levels on bulk
properties in SRHDWS for $^{16}$O. The effective interaction for
the Lagrangian is NLSH, $R_{\rm{max}} = 20$ fm, $\Delta r = 0.1$
fm, and $N^+_{\rm{max}} = 25$. For the initial Woods-Saxon like
potentials, parameters in Ref.~\cite{Koepf91a} are used except for
$V_0$ which is specified in the table. The left value in each
entry gives the result without negative energy levels included and
the right one that with $N^-_{\rm{max}}$ = 25. Energy is in MeV
and radius in fm.  {Taken form Ref.~\cite{Zhou03prc}}.}

\begin{center}
\begin{tabular}{c|c|c|c|c|c|c}
\hline\hline
 $V_0$         & \multicolumn{2}{c|}{$54$ MeV}  &\multicolumn{2}{c|}{$72$ MeV}&\multicolumn{2}{c}{$90$
 MeV}\\ \hline
 $N^-_{\rm{max}}$&0&25&0&25&0&25\\ \hline
 $E/A$          & 8.547 & 8.013 & 8.117 & 8.015 & 8.427 & 8.012 \\
 $r_{\rm rms}$  & 2.385 & 2.568 & 2.531 & 2.567 & 2.610 & 2.567 \\
\hline\hline
\end{tabular}
\end{center}
\end{table}

The contribution of negative energy states in the Dirac sea to the
wave function can be calculated by $\sum_n |c_{n}^-|^2$ in the
expansion (\ref{eq:DSexpansion}) and it is around $10^{-4\sim -5}$
(Note that the nucleon wave function is normalized to one).
However, such a small component from negative energy states in the
wave functions contributes to the physical observables such as
$E/A$ and $r_{\rm{rms}}$ by magnitudes of 1$\sim$10 \% as can be
seen from Table~\ref{TabB1}.

\paragraph{Comparisons between $r$ space, harmonic oscillator basis and
Woods-Saxon basis}
 It is found that for stable nuclei, the SRHR,
SRHSWS, SRHDWS, and SRHHO approaches are all valid and results
from them are in excellent agreement with each
other~\cite{Zhou03prc}. However, for unstable nuclei near the
neutron drip line, these methods differ from each other to some
extent.

\begin{figure}\centering
\includegraphics[width=6.0cm]{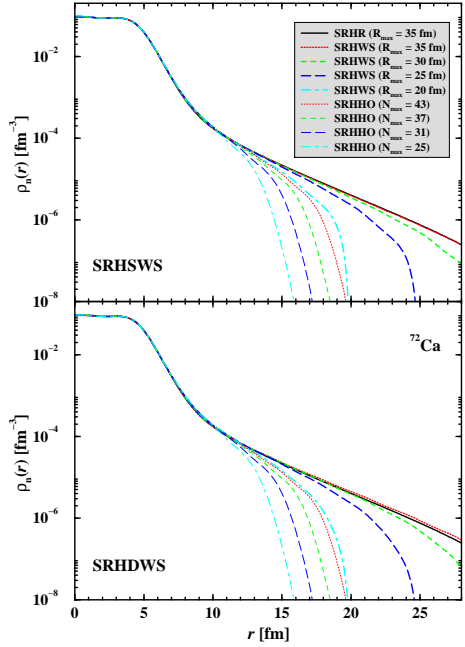}
\caption{\label{FigB2} Comparison of density distributions for
$^{72}$Ca from SRHR, SRHWS and SRHHO with NLSH. $\Delta r$ = 0.1
fm for SRHR and SRHWS. $E_{\rm cut}$ = 75 MeV and $R_{\rm max}$ =
20 (thick dot-dashed curve), 25 (thick long-dashed curve), 30
(thick dashed curve) and 35 fm (thick dotted curve) for SRHWS.
These sets of cutoff's correspond to cutoff's in principal quantum
number $N_{\rm max}$ = 25 (thin dot-dashed curve), 31 (thin
long-dashed curve), 37 (thin dashed curve) and 43 (thin dotted
curve) which are used in SRHHO calculations. Taken form
Ref.~\cite{Zhou03prc}.}
\end{figure}

In Fig.~\ref{FigB2}, the neutron density distribution of $^{72}$Ca
from different SRH approaches are compared. With the same box
size, the density distribution from SRHR are almost identical with
those from SRHWS, which indicates the equivalence between SRHWS
and SRHR. For brevity, only $\rho_{\rm n}(r)$ from SRHR with
$R_{\rm max}$ = 35 fm is displayed which covers the curve
corresponding to $\rho_n(r)$ from SRHWS with $R_{\rm max}$ = 35 fm
in Fig.~\ref{FigB2}. On the other hand, $\rho_{\rm n}(r)$ from
SRHHO even with $N_{\rm max}$ = 43 fails to reproduce the result
of SRHR due to the well known localization property of HO
potential~\cite{Stoitsov98a}.

These results indicate that even the long tail (or halo) behavior
in neutron density distribution for nuclei near the drip line can
be reproduced quite well by the expansion method in the
Woods-Saxon basis in the Schr\"{o}dinger framework (SRHSWS) and
that in the Dirac framework (SRHDWS). This can be seen that the
neutron density distribution obtained in SRHR is reproduced well
in both SRHSWS and SRHDWS when enough box size is taken.

\subsection{Effective interactions}

In the Lagrangian density (\ref{Lagrangian}), there are meson
masses $m_\sigma, m_\omega, m_\rho$ and meson-nucleon coupling
constants $g_\sigma, g_\omega, g_\rho$ together with the nonlinear
self-couplings of the meson fields left to be determined. They are
the nucleon-nucleon interactions in the RMF theory and in
principle should be determined either by more fundamental theories
or by experiments. However, as the relativistic mean field theory
is formulated on the basis of the above effective Lagrangian in
connection with the mean-field and the no-sea approximations, it
is difficult to determine these interactions microscopically.
Instead, the masses and coupling strengthes of the mesons and the
nonlinear self-couplings of the meson fields are determined by
reproducing the properties of nuclear matter and a few doubly
magic nuclei. Namely, they are effective interactions in the
similar sense as their conventional counterparts. The effective
interactions in the RMF theory could be determined by minimizing
the least square error as follows,
 \beq \label{chisquare}
\chi^2(\svec a)
    = \sum_{i=1}^N
      \ls \frac{y^{exp}_i-y(x_i;\svec a)}{\sigma_i}\rs^2,
 \eeq
where $\svec a$ is the ensemble of the meson masses $m_\sigma,
m_\omega, m_\rho$ and the meson-nucleon coupling constants
$g_\sigma, g_\omega, g_\rho$ together with the nonlinear
self-couplings of the meson fields to be fitted, and $y^{exp}_i$
and $\sigma_i$ are the experimental observables and corresponding
weights. In general, the masses and charge radii of spherical
nuclei near the $\beta$-stability line are adopted as observables
in the least-square fitting procedure.

Among the existing effective interactions for the RMF theory, the
most frequently used are NL1~\cite{Reinhard86},
NLSH~\cite{Sharma93a}, TM1~\cite{Sugahara94} and
NL3~\cite{Lalazissis97} with nonlinear self-couplings of mesons.
Along the $\beta$-stability line NL1 gives excellent results for
binding energies and charge radii, in addition it provides an
excellent description of the superdeformed bands
~\cite{Afanasjev96a,Afanasjev96b}. However, when moving away from
the stability line the results are less satisfactory. This can be
partly attributed to the large asymmetry energy $J\simeq$ 44 MeV.
Moreover, the neutron skin thicknesses calculated with NL1 show
systematic deviations from the experimental data \cite{Schmid92}.

In NLSH this problem was treated in a better way and the improved
isovector properties have been obtained with an asymmetry energy
of $J\simeq$ 36 MeV. Furthermore, NLSH seems to describe the
deformation properties in a better way than NL1 does. However,
NLSH produces a slight over-binding along the line of
$\beta$-stability and in addition it fails to reproduce
successfully the superdeformed minima in Hg-isotopes in
constrained calculations for the energy landscape. A remarkable
difference between the two effective interactions are the quite
different values predicted for the nuclear matter
incompressibility ~\cite{Ma03}, i.e., $K$ = 212 MeV for NL1 while
$K$ = 355 MeV for NLSH~\cite{Ma01, Ma02a}. As an improvement, NL3
and TM1 provide reasonable compression modulus
($K_{\text{NL3}}=271.7 \ \mev, K_{\text{TM1}}=281.16\ \mev$) and
asymmetry energy ($J_{\text{NL3}}=37.42 \ \mev,
J_{\text{TM1}}=36.89\ \mev$) but fairly small baryonic saturation
density ($\rho_{\text{TM1}}=0.145$). In order to improve the
description of these quantities, two nonlinear self-coupling
effective interactions, PK1 with nonlinear $\sigma$- and
$\omega$-meson self-couplings and PK1R with nonlinear $\sigma$-,
$\omega$- and $\rho$-meson self-couplings were
developed~\cite{Long04} (see Table~\ref{TabB2}).

In order to reproduce better the experimental quantities such as
the binding energies and nuclear radii, etc., an additional
correction should be added in calculating the energy, i.e., the
center-of-mass correction. Conventionally a phenomenological
center-of-mass correction as $-\dfrac{3}{4}\,41 A^{-1/3}$ is used.
Microscopically the correction can be calculated by the
projection-after-variation in first-order
approximation~\cite{Bender00}:
 \beq \label{ecm1}
E^{\text{mic.}}_\cm = -\frac{1}{2MA}\lc {\svec P}_\cm^2\rc,
 \eeq
where the center-of-mass momentum $\svec P_\cm =\sum_i^A\svec p_i$
and the expectation value of its square $\lc \svec P_\cm^2\rc$
reads
 \beq\label{ecm2}
 \lc {\svec P}_\cm^2\rc
    =  \sum_a v_a^2 p_{aa}^2
     - \sum_{a,b} v_a^2 v_b^2\svec p_{ab}\cdot\svec p^*_{ab}
     + \sum_{a,b}v_a u_a v_b u_b \svec p_{ab}\cdot\svec p_{\bar a\bar b}
 \eeq
with occupation probabilities $v_a^2$ and $u_a^2 = 1-v_a^2$
accounting for the pairing effects, where $a,b$ denote the BCS
states (see the following section).

\begin{figure}[htbp]
\setlength{\abovecaptionskip}{-0.3cm} \centering
\includegraphics[width=10.0cm]{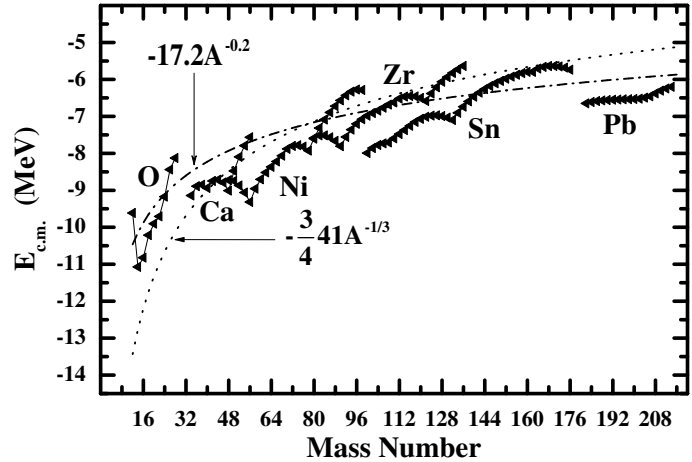}
\caption{The microscopic center-of-mass correction in comparison
with two phenomenological cases. Taken from Ref.~\cite{Long04}.
}\label{FigB3}
\end{figure}

Here it should be mentioned that the prescription (\ref{ecm1}) is
based on non-relativistic considerations. It does not preserve
Lorentz invariance. Furthermore, it also breaks the complete
self-consistence of the variational scheme. However, as it is not
included in the self-consistent procedure and only presents an
additional correction term to the binding energy, one can be
satisfied with it for the moment. Compared with the binding
energy, this center-of-mass correction is sizable in light nuclei
(about 9\% in $^{16}$O) but much less important in medium and
heavy nuclei (about 0.4\% in $^{208}$Pb) as seen in Fig.
\ref{FigB3}.

\begin{table}[htbp]
 \centering\setlength{\tabcolsep}{1.0em}
 \caption{ The nonlinear effective interactions PK1, PK1R and
density-dependent effective interaction PKDD ~\cite{Long04} in
comparison with TM1~\cite{Sugahara94}, NL3~\cite{Lalazissis97},
TW99~\cite{Typel99}, and DD-ME1~\cite{Niksic02b}.}
 \label{TabB2}
 \begin{tabular}{cccccccc}\hline\hline
&PK1&PK1R&PKDD&TM1&NL3&TW99&DD-ME1\\ \hline
$M_n$&939.5731&939.5731&939.5731&938&939&939&938.5000\\
$M_p$&938.2796&938.2796&938.2796&938&939&939&938.5000\\
$m_\sigma$&514.0891&514.0873&555.5112&511.198&508.1941&550&549.5255\\
$m_\omega$&784.254&784.2223&783&783&782.501&783&783.0000\\
$m_\rho$&763&763&763&770&763&763&763.0000\\
$g_\sigma$&10.3222&10.3219&10.7385&10.0289&10.2169&10.7285&10.4434\\
$g_\omega$&13.0131&13.0134&13.1476&12.6139&12.8675&13.2902&12.8939\\
$g_\rho$&4.5297&4.55&4.2998&4.6322&4.4744&3.661&3.8053\\
$g_2$&$-$8.1688&$-$8.1562&0&$-$7.2325&$-$10.4307&0&0.0000\\
$g_3$&$-$9.9976&$-$10.1984&0&0.6183&$-$28.8851&0&0.0000\\
$c_3$&55.636&54.4459&0&71.3075&0&0&0.0000\\
$d_3$&0&350&0&0&0&0&0 \\ \hline\hline
\end{tabular}
\setlength{\tabcolsep}{0.30em}

\caption{Density-dependent parameters of PKDD~\cite{Long04} for
meson-nucleon couplings in comparison with TW99~\cite{Typel99} and
DD-ME1~\cite{Niksic02b}.} \label{TabB3}
\begin{tabular}{cccccccccc}\hline\hline
&$a_\sigma$&$b_\sigma$&$c_\sigma$&$d_\sigma$&$a_\omega$&$b_\omega$
&$c_\omega$&$d_\omega$&$a_\rho$\\
\hline
PKDD&1.327423&0.435126&0.691666&0.694210&1.342170&0.371167&0.611397&0.738376&0.183305\\
TW99&1.365469&0.226061&0.409704&0.901995&1.402488&0.172577&0.344293&0.983955&0.515000\\
DD-ME1&1.3854&0.9781&1.5342&0.4661&1.3879&0.8525&1.3566&0.4957&0.5008\\
\hline\hline\end{tabular}
\end{table}

For the nuclear radii, the effects from the center-of-mass motion
can also be taken into account as follows. Because of its fairly
small effects, a rather rough correction is adopted for protons,
 \beq\label{radii-cm}
\delta R_p^2
    = - \frac{2}{Z}\sum_a^A
        \lc\phi_a\rl \svec R_\cm\cdot\sum_{i}^Z \svec r_i \lr \phi_a\rc
      + \sum_a^A
        \lc\phi_a\rl \svec R_\cm^2\lr \phi_a\rc,
 \eeq
where the center-of-mass coordinate  $\svec R_\cm =1/A \sum_i ^A
\svec r_i$. Then one obtains,
 \beq\label{cmc-r}
\delta R_p^2 = -\frac{2}{A} R_p^2 + \ff A R_M^2,
 \eeq
where $R_p $ and $R_M$ denote the proton and matter radii. Here we
only consider the direct-term contributions to
Eq.~(\ref{radii-cm}) to conform to the spirit of the RMF theory.
For the neutron radii, one can follow the same procedure as that
for protons. The charge radius is obtained from the proton radius
combining with the proton and neutron size and the center-of-mass
correction (\ref{cmc-r}) included in $R_p^2$~\cite{Sugahara94}
 \beq\label{rch}
R_{\text{ch}}^2 = R_p^2 + (0.862\ \fm)^2 - (0.336\ \fm)^2 N/Z.
 \eeq

With the microscopic center-of-mass motion, a multi-parameter
fitting can be performed using the Levenberg-Marquardt
method~\cite{Press1992}. In Ref.~\cite{Long04}, the masses of
$^{16}$O, $^{40}$Ca, $^{48}$Ca, $^{56}$Ni, $^{68}$Ni, $^{90}$Zr,
$^{116}$Sn, $^{132}$Sn, $^{194}$Pb and $^{208}$Pb and the bulk
quantities of nuclear matter are chosen as observables to
determine the effective interactions. The radii are excluded
because the proper values of the compression modulus $K$ and the
baryonic saturation density $\rho_0$ are sufficient to give a good
description of the radii. For a fixed value of the compression
modulus $K$, a large baryonic saturation density $\rho_0$ will
give a small charge radius and vice versa. Therefore a proper
description of both masses and radii of finite nuclei could be
obtained by carefully adjusting the values of these two quantities
$K$ and $\rho_0$. To give a fairly precise description of the
masses, the center-of-mass correction is also essential for both
light and heavy nuclei. As it can be seen in Fig.~\ref{FigB3}, the
deviation between the microscopic and phenomenological results is
considerably large not only for the light nuclei but also for the
heavy ones. And there exist very remarkable shell effects in the
microscopic results which are impossible to obtain with the
phenomenological methods. The microscopic center-of-mass
correction~\cite{Bender00}, therefore, is chosen to deal with the
center-of-mass motion.

 Because the contribution to the nuclear masses from the nonlinear
$\rho$-meson term is found to be fairly small, the effective
interaction PK1R is obtained by fixing the nonlinear self-coupling
constant $d_3$ to 350.0 and adjusting other parameters.

In the RMF theory with density-dependent meson-nucleon couplings,
the density-dependence of the coupling constants $g_\sigma$ and
$g_\omega$ can be parameterized as,
 \beq
g_i(\rho_v)
    = g_i(\rho_{\text{sat}}) f_i(x)
    ~~~~~~~~\text{ for } i = \sigma, \omega,
 \eeq
where
 \beq
f_i(x) = a_i\frac{1 + b_i(x+d_i)^2}{1+ c_i(x+d_i)^2}
 \eeq
is a function of $x =\rho_v/\rho_{\text{sat}}$, and
$\rho_{\text{sat}}$ denotes  the baryonic saturation density of
nuclear matter. For the $\rho$ meson, an exponential dependence is
utilized as
 \beq
g_\rho  = g_\rho(\rho_{\text{sat}}) \exp[-a_\rho(x-1)].
 \eeq
For the functions $f_i(x)$, one has five constraint conditions
$f_i(1) = 1, f_\sigma'' (1) = f_\omega''(1)$ and $f_i''(0) =0$.
Then 8 parameters related to density dependence for $\sigma$-N and
$\omega$-N couplings are reduced to 3 free parameters. In general,
the masses of the nucleons and the $\rho$-meson are fixed and the
nonlinear self-coupling constants $g_2, g_3, c_3$ and $d_3$ are
set to zero. With 4 free parameters for density dependence,
totally there are 8$\sim$9 parameters left free in the Lagrangian
density (\ref{Lagrangian}) for the density-dependent meson-nucleon
coupling RMF theory. A density-dependent meson-nucleon coupling
effective interaction PKDD has also been obtained in
Ref.~\cite{Long04}  (see Tables~\ref{TabB2} and \ref{TabB3}).

Tables~\ref{TabB2} and \ref{TabB3} tabulate the new effective
interactions PK1, PK1R and PKDD \cite{Long04} in comparison with
the old ones TM1~\cite{Sugahara94}, NL3~\cite{Lalazissis97},
TW99~\cite{Typel99} and DD-ME1~\cite{Niksic02b}. The newly
obtained ones reproduce better the experimental
masses~\cite{Audi95}. PK1, PK1R and PKDD also describe the charge
radii very well, especially for those of the Pb isotopes. More
comprehensive comparisons between Hartree-Fock-Bogoliubov,
 extended Thomas-Fermi model with Strutinski integral,
 RMF, and macroscopic-microscopic approaches with
different forces have been performed for the description of
nuclear masses and charge radii of spherical even-even nuclei (116
nuclides), from light (A=16) to heavy (A=220) ones in
Ref.~\cite{Patyk99}.

Table~\ref{TabB4} lists the nuclear matter quantities calculated
with the newly obtained effective interactions PK1, PK1R and PKDD,
in comparison with those from the other interactions. All the new
effective interactions give a proper value for the compression
modulus $K$.

\begin{table}[htbp]
 \centering\setlength{\tabcolsep}{1em}
 \caption{Nuclear matter properties calculated with PK1, PK1R and
PKDD~\cite{Long04} in comparison with those with
TM1~\cite{Sugahara94}, NL3~\cite{Lalazissis97},
TW99~\cite{Typel99}, DD-ME1~\cite{Niksic02b}.}
 \label{TabB4}
 \begin{tabular}{lcccccc}\hline\hline
Interaction& $\rho_{\text{sat.}}(\fm^{-3})$&$E_b$ [MeV]&K [MeV]&$J$ [MeV]& $M^*/M$(n)& $M^*/M$(p)\\
\hline
PK1    &0.1482 &$-$16.268    &282.644   &37.641    &0.6055    &0.6050\\
PK1R   &0.1482 &$-$16.274    &283.674   &37.831    &0.6052    &0.6046\\
PKDD   &0.1496 &$-$16.267    &262.181   &36.790    &0.5712    &0.5706\\
NL3    &0.1483   &$-$16.250    &271.729   &37.416    &0.5950    &0.5950\\
TM1    &0.1452 &$-$16.263    &281.161   &36.892    &0.6344    &0.6344\\
TW99   &0.1530 &$-$16.247    &240.276   &32.767    &0.5549    &0.5549\\
DD-ME1 &0.1520 &$-$16.201    &244.719   &33.065    &0.5780    &0.5780\\
\hline\hline
 \end{tabular}\end{table}

\subsection{Density and isospin dependence of effective interactions}

There are so far quite a number of effective interactions, PK1,
PK1R, PKDD~\cite{Long04} together with NL1, NL2~\cite{Lee86},
NL3~\cite{Lalazissis97}, NLSH~\cite{Sharma93a}, TM1,
TM2~\cite{Sugahara94}, GL-97~\cite{Glendenning97} and the
density-dependent effective interactions TW-99~\cite{Typel99},
DD-ME1~\cite{Niksic02b}, etc.. It is very interesting to
investigate the density and isospin dependence of the interaction
strengthes of various effective interactions in the RMF theory and
study their effects on nuclear matter~\cite{Ban04}. Although for
the nonlinear self-coupling effective interactions, the density
dependencies are only embodied in the Klein-Gordon equations, it
is still worthwhile to obtain a quantitative understanding of the
coupling constants.

\begin{figure}\centering
\includegraphics[width = 10.0cm]{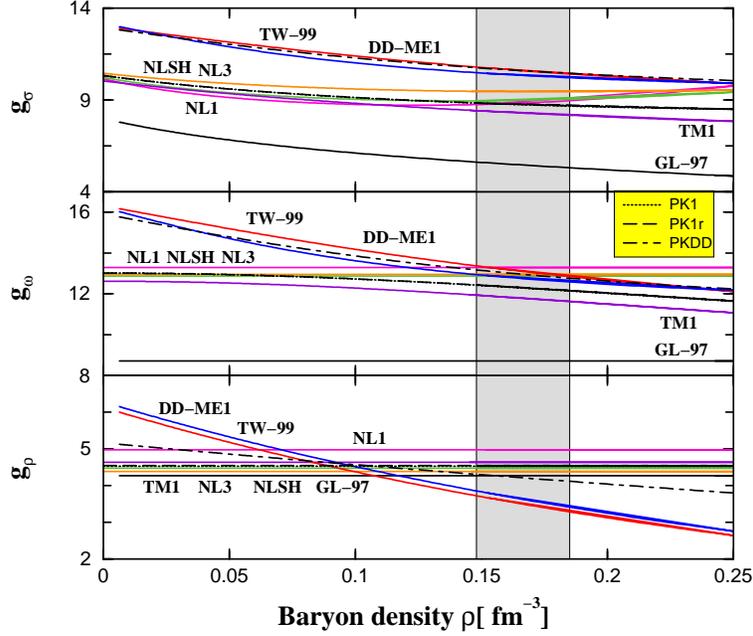}
\caption{Density dependence of the interaction strengths for
$g_\sigma$(top), $g_\omega$(middle) and $g_\rho$(bottom) in
nuclear matter as functions of the nuclear density. The shadowed
area corresponds to the empirical value of saturation point in
nuclear matter (Fermi momentum $k_F=1.35\pm0.05$ fm$^{-1}$ or
density $\rho=0.166\pm0.018$ fm$^{-3}$). Taken from
Ref.~\cite{Long04}. }\label{FigB4}
\end{figure}

In Fig. \ref{FigB4}, the density dependencies of the interaction
strengthes for $g_\sigma$(top), $g_\omega$(middle) and
$g_\rho$(bottom) in nuclear matter as functions of the nuclear
density are shown. The curves in the figures from top to bottom
are labelled in the order of from left to right. The shadowed area
corresponds to the empirical value of the saturation point in
nuclear matter, i.e., Fermi momentum $k_F=1.35\pm0.05$ fm$^{-1}$
or density $\rho=0.166\pm0.018$ fm$^{-3}$. For the nonlinear
effective interactions, the ``equivalent" density dependence of
the interaction strengthes for $\sigma$, $\omega$ and $\rho$ is
extracted from the meson field equations according
to~\cite{Long04}:
 \bsub\label{couplings}\beqn
g_\sigma &\sim&  g_\sigma + \ls g_2\sigma^2 + g_3\sigma^3\rs/\rho_s,\\
g_\omega &\sim&  g_\omega - c_3\omega_0^3/\rho_v,\\
g_\rho   &\sim&  g_\rho   - d_3({\rho_0^3})^3/{\rho_3}.
 \eeqn\esub
For the $\sigma$-meson, TW-99 and DD-ME1 exhibit quite different
behaviors for $g_\sigma$ compared with those of the other
effective interactions in either magnitude or slope. In
particular, the strengthes in TW-99 and DD-ME1 for the density
interval in Fig. \ref{FigB4} are almost two times larger than that
of GL-97. Quite different results can also be seen at the
empirical nuclear matter densities. For the $\omega$-meson, except
for TW-99, DD-ME1, TM1, and GL-97, the strengthes in all the other
effective interactions are density-independent. However, the
strengthes are closer to each other at the empirical saturation
density than those of the $\sigma$-meson although large
differences can also be seen at low densities. For the
$\rho$-meson which describes the isospin asymmetry, the strengthes
for TW-99 and DD-ME1 show strong density dependence in contrast
with those of the other effective interactions; while those of
PK1, PK1R and PKDD are just in between \cite{Long04}.

\begin{figure}[htbp]
\centering
\begin{minipage}[t]{8.0cm}
\includegraphics[width=8.0cm]{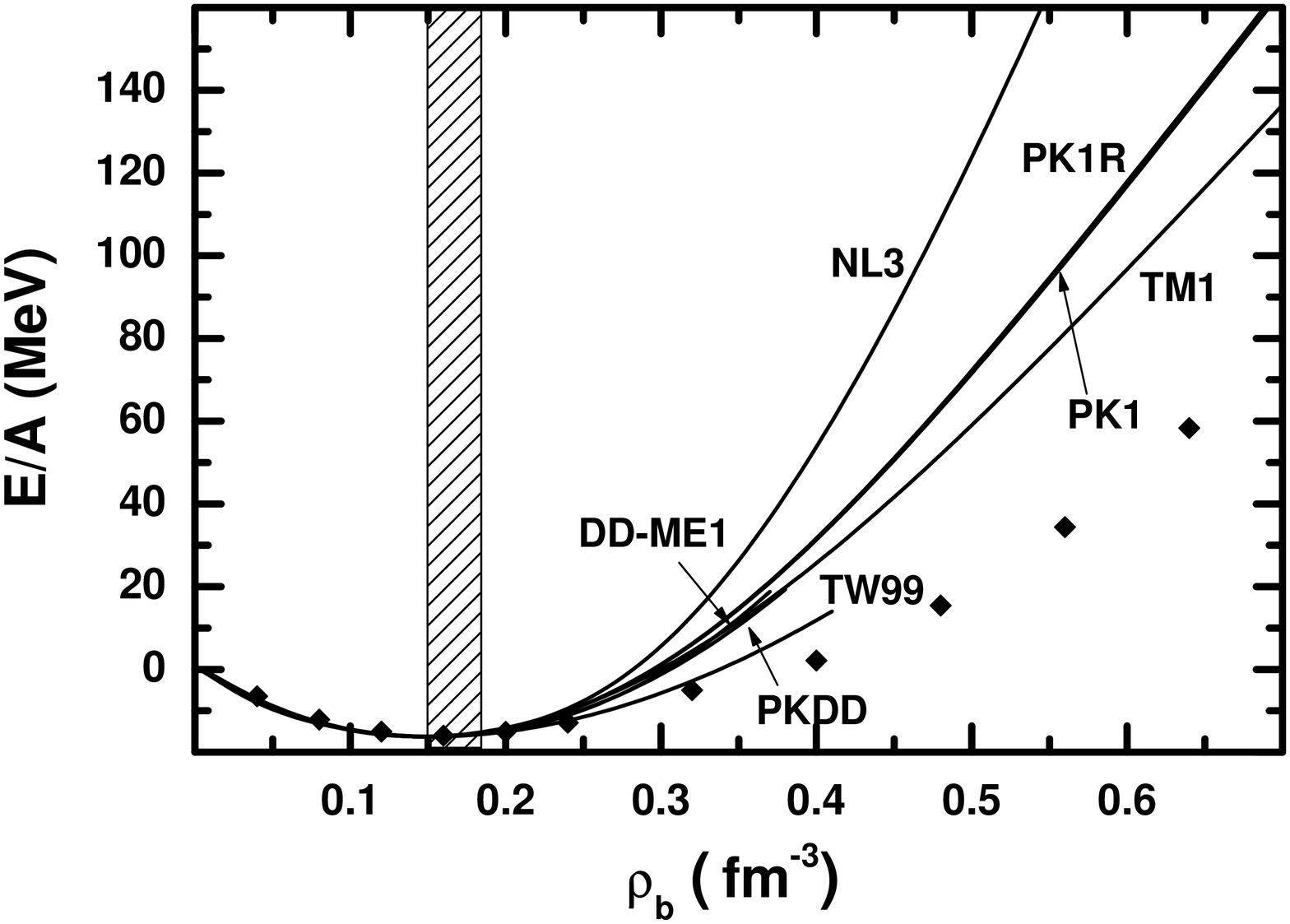}
\end{minipage}
\begin{minipage}[t]{8.0cm}
\includegraphics[width=8.0cm]{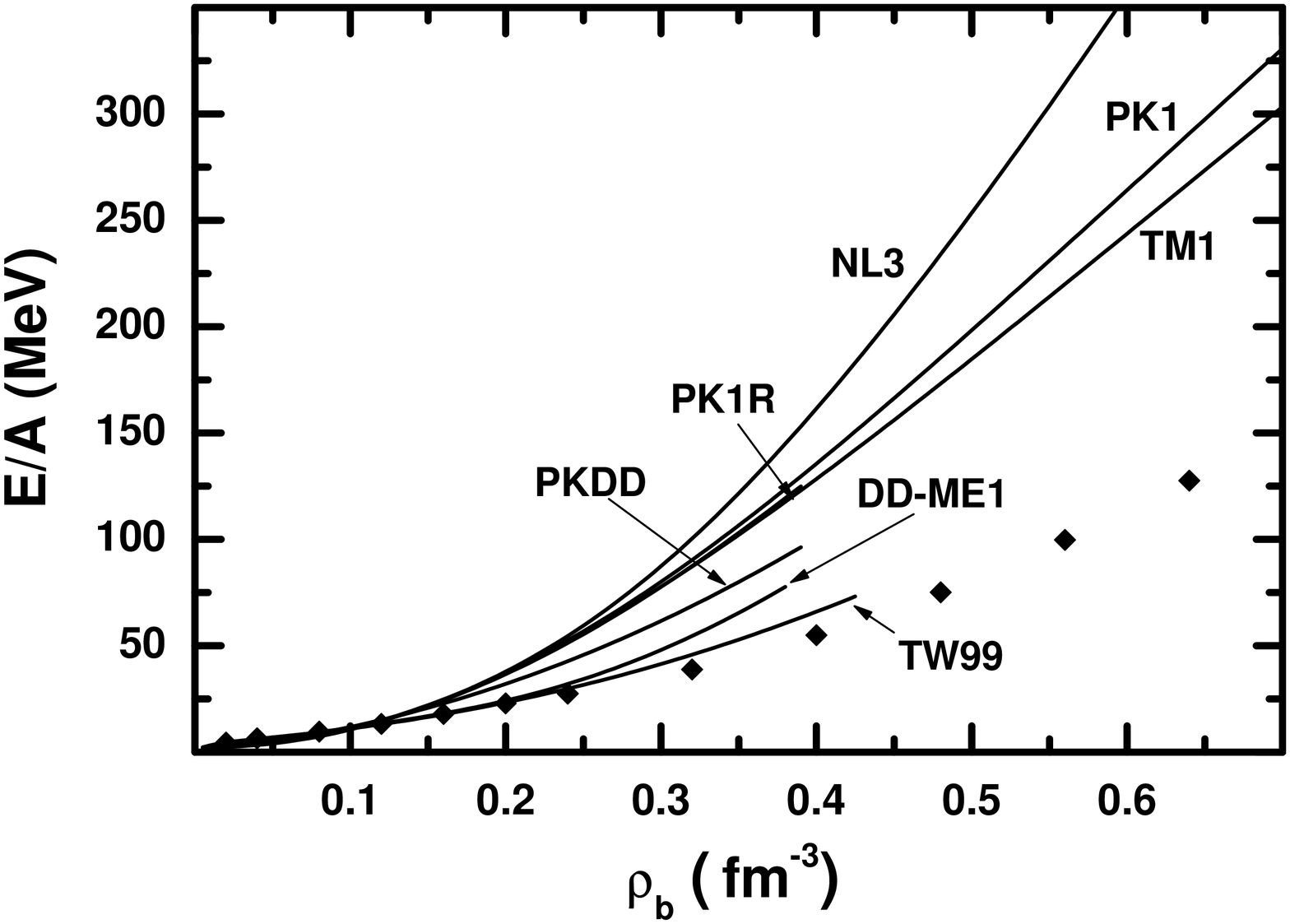}
\end{minipage}
\caption{The energy per particle $E/A$ in symmetric nuclear matter
(left) and neutron matter (right) as functions of the baryon
density $\rho$. The shaded area indicates the empirical saturation
region. The filled diamond presents the data taken from Ref.
\cite{Akmal98} as comparison.} \label{FigB5}
\end{figure}

In Fig.~\ref{FigB5}, the behavior of the binding energy per
particle $E/A$ as a function of the baryonic density $\rho$ is
shown for symmetric nuclear matter (left) and neutron matter
(right). It is seen that all the density-dependent meson-nucleon
coupling effective interactions give softer results than the
nonlinear self-coupling ones , especially for neutron matter. The
behaviors predicted by PK1, PK1R are much softer than that by NL3
and a little harder than that by TM1. The results from PKDD are
slightly softer than that from DD-ME1 and much harder than that
from TW99 at high densities. All these behaviors can be explained
in the density-dependent meson-nucleon coupling framework.

\begin{figure}[htbp]\centering
\includegraphics[width=8.0cm]{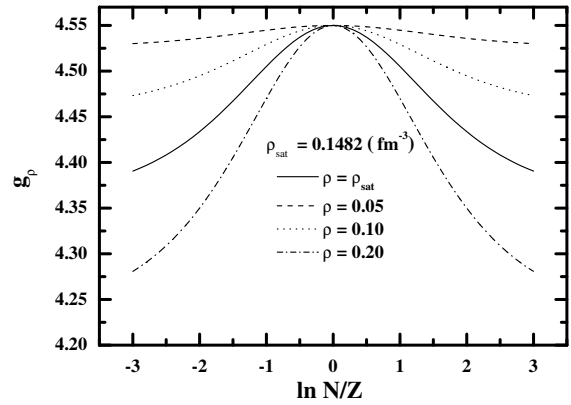}
\caption{The density-dependence of $g_\rho$ for PK1R with respect
to the neutron-proton ratio N/Z. Taken from Ref.~\cite{Long04}.}
\label{FigB6}
\end{figure}

As what have been mentioned in expressions (\ref{couplings}), the
meson-nucleon coupling constants in the nonlinear self-coupling of
mesons can be expressed as some kind of density-dependence. Fig.
\ref{FigB4} shows the density-dependence of the coupling constants
for the nonlinear self-coupling effective interactions and for the
density-dependent version, where almost all the density-dependent
coupling constants decrease with increasing density except for
$g_\sigma$ of NL3, NLSH, NL1, which has strong $\sigma^4$
self-couplings. On the other hand, the coupling constants
$g_\sigma$ and $g_\omega$ of TM1, which has relatively weak
$\sigma$ self-coupling ($g_3 = 0.6183$) and strong $\omega$
self-coupling ($c_3 = 71.3075$), are smaller than the others,
which means that TM1 provides relatively weaker scalar and vector
potentials. This is the reason why TM1 presents the softer
behavior than the other nonlinear self-coupling effective
interactions. In Fig.~\ref{FigB5}, TW99 predicts the softest
results because of its relatively small $g_\omega$ as compared
with DD-ME1, PKDD and NL3, and large $g_\sigma$ as compared with
PK1, PK1R and TM1 in Fig.~\ref{FigB4}. As it is known, the
repulsive potential would be dominant at high densities. In
Fig.~\ref{FigB5}, NL3 gives the hardest results because of its
constant and large $g_\omega$ even though its $\sigma$-N coupling
constant $g_\sigma$ increases with the density. For PK1, PK1R and
PKDD, which present the mid soft behaviors in Fig.\ref{FigB5}, the
coupling constants also lie between the largest and the smallest
in Fig. \ref{FigB4}.  There exists a significant difference
between symmetric nuclear matter and neutron matter. The
density-dependent effective interactions PKDD and DD-ME1 present
similar trends as PK1, PK1R and TM1 in symmetric nuclear matter
but much softer in neutron matter, which may be interpreted by the
density-dependence of $g_\rho$. For the effective interaction
PK1R, the density-dependence of the $g_\rho$ is fairly weak as
compared with that of the density-dependent meson-nucleon coupling
effective interactions. It can be explained by a very weak
$\rho$-field, which generates neutron-proton asymmetry field. The
behavior of $g_\rho$ with respect to the neutron-proton ratio is
shown in Fig.\ref{FigB6}. As expected, the behavior is symmetric
with respect to $\ln(N/Z)$ and the density-dependence becomes more
remarkable with the increase of the baryonic density and the
neutron-proton asymmetry.

\section{Relativistic continuum Hartree-Bogoliubov theory}
\label{sec:pairing}

Pairing correlations are due to the short range part of the
nucleon-nucleon interaction and play important roles in open shell
nuclei. A simple and commonly used method of dealing with pairing
correlations is the BCS approach under the constant gap
approximation. The BCS method can be combined easily with the
relativistic mean field theory as shown in
Refs.~\cite{Reinhard86,Ring96}. However, the conventional BCS
method is not justified for exotic nuclei because it could not
include properly the contribution of continuum states. Bogoliubov
transformation is the generalization of the BCS scheme. By
quantizing the meson field and making Gorkov factorization, the
relativistic Hartree Bogoliubov (RHB) formalism was
derived~\cite{Kucharek91}. In order to describe the exotic nuclei,
the relativistic continuum Hartree Bogoliubov (RCHB) theory was
developed, in which the continuum states are discretized and the
RHB equations are solved in the coordinate
space~\cite{Meng96,Meng98npa}.

In this section, the pairing correlations, the conventional BCS
approach, the continuum states, the resonant BCS method and
several methods for obtaining single particle resonant states are
briefly reviewed. Finally, the detailed formalism of the RCHB
theory together with the discussion on the effective pairing
interactions are given.

\subsection{Pairing correlations}
\label{subsec:BCS}

\subsubsection{BCS approximation}
For Fermion systems like atomic nuclei, Kramers degeneracy ensures
the existence of pairs of degenerate, mutually time-reversal
conjugate states, which could be coupled strongly by a short-range
force, namely a $pp$-channel in the framework of mean field.
Physically, such pairing correlations lead to the superfluidity.
In Hartree approximation, the ground state properties could be
described by filling the single particle levels from the bottom up
to the Fermi level. The occupation probability of each single
particle level is either zero or one. Due to pairing interaction,
pairs of nucleons are scattered from the levels below the Fermi
level to those above. Thus for the open shell nuclei, one has to
deal with the occupation probabilities ranging from zero to one.
Mathematically, this could be treated by introducing the concept
of quasi-particles. For stable nuclei, the pairing gap can be
extracted from the experimental odd-even mass difference and the
BCS approximation is very simple and useful. For $M$-single
particle orbitals in the Hartree approximation, for each particle
orbital $k$ ($k=1,\cdots,M$), one can introduce the quasi-particle
energy:
\begin{equation}
E_k = \left[  ( \epsilon_k - \lambda )^2 + \Delta^2 \right]^{1/2},
\end{equation}
and the corresponding occupation probability:
\begin{equation}
v_{k}^2 = \dfrac 1 2 \left[ 1 - \dfrac {\epsilon_k - \lambda}{E_k}
\right],
\end{equation}
where $\lambda$ is the Fermi level determined by the particle
number $N=\displaystyle\sum_{k=1}^M v_{k}^2$ and $\Delta$ the
pairing gap determined from the experimental odd-even mass
difference
\begin{equation}
\Delta=\frac{1}{2}\left[E(N+2)-E(N+1)-(E(N+1)-E(N))\right],
\end{equation}
or deduced from certain pairing Hamiltonian $H_{\rm pair}$ when
the experimental binding energies are unknown. For monopole
pairing interaction $H_{\rm pair} = G
\displaystyle\sum_{kk'>0}a_k^+a_{\bar k}^+a_{\bar k'}a_{k'}$,
there lies the pairing gap equation with classical variation:
\begin{equation}
\Delta = G \sum_{k>0} u_k v_k
       = \frac{G}{2}\sum_{k>0} \frac{\Delta}{E_k},
\end{equation}
where $u_k$ satisfies $u_k^2=1-v_k^2$. For fixed values of the gap
parameters $\Delta$ or pairing interaction strengthes $G$ for
neutrons and protons, the self-consistent solution is obtained by
iteration.

\begin{figure}[ht]
\centering
\includegraphics[width=6cm]{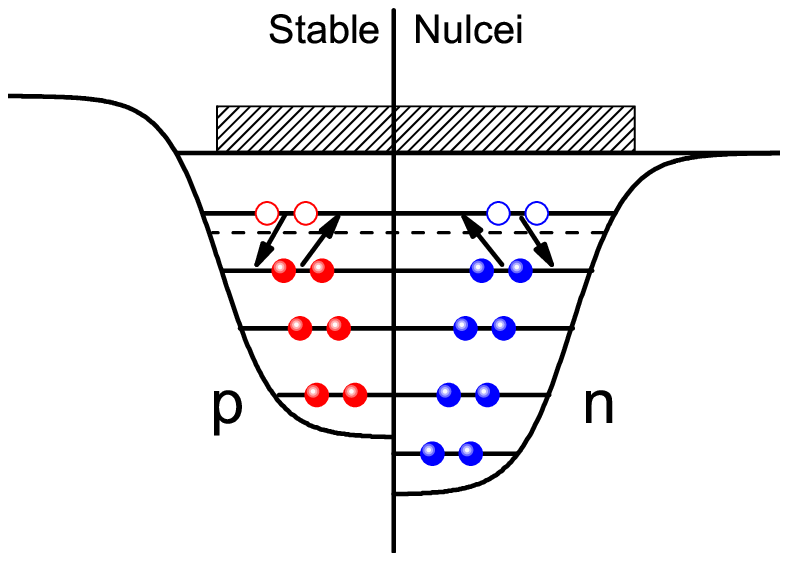}
\includegraphics[width=6cm]{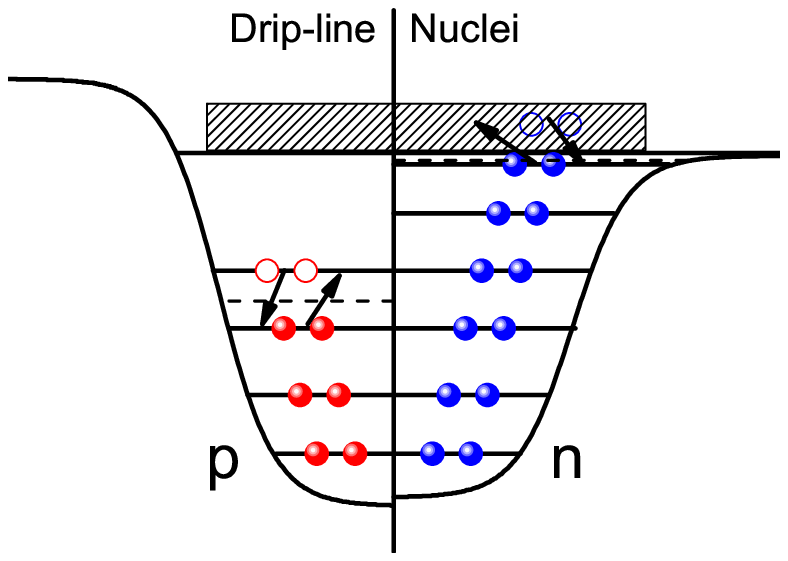}
\caption{The schematic picture for the difference in the pairing
correlation of stable and drip line nuclei. Normally the last
filled nucleon in stable nuclei has $8$ MeV binding, while it is
near the threshold in drip line nuclei. The pairing correlation in
exotic nuclei provides the possibility to scatter valence nucleons
back and forth in the continuum.} \label{FigC1} \end{figure}

The BCS method can be easily implemented into the relativistic
mean field description with the single particle orbitals in the
relativistic Hartree level. However, it is only valid for bound
states. For the extremely neutron-rich (or proton-rich) nuclei
near the drip lines, the continuum states begin to contribute as
schematically shown in Fig.~\ref{FigC1}. The conventional BCS
approach will involve unphysical states and lead to the
divergence. For these nuclei, one must either investigate the
detailed properties of continuum states and include the coupling
between the bound state and the continuum by extending the BCS
method to resonant BCS theory (see Section~\ref{subsec:resonant})
or generalize the BCS model by unifying the Hartree (or
Hartree-Fock) scheme and the Bogoliubov transformation (see
Section~\ref{subsec:RHB}).

\subsection{Continuum states}

\subsubsection{Single particle resonant states}

For stable open shell nuclei, the pairing gaps are of the order of
MeV~\cite{Bohr69}. There are some evidences that the pairing gaps
will increase towards the drip lines. For nuclei close to drip
lines, the one or two particle separation energies are usually
very small. For example, the one proton separation energies
$S_{p}$ are 0.138 MeV, 0.60 MeV and 0.29 MeV in $^{8}$B, $^{12}$N
and $^{31}$Cl; the two neutron separation energies $S_{2n}$ are
0.30 MeV and 1.33 MeV in $^{11}$Li and $^{14}$Be~\cite{Audi95}. A
small particle separation energy means that the Fermi level is
very close to the threshold. Therefore, pairing correlations can
scatter valence nucleons between bound and continuum states as
seen from Fig.~\ref{FigC1}. An unphysical continuum state
distributes over the whole space while a physical resonance state
with positive energy stays inside the nucleus for a long time and
has better asymptotic behavior. For the cases where the pairing
correlations involve the continuum, such as exotic nuclei, it has
been demonstrated that the largest contribution to the pairing
correlations is expected to come from the resonant continuum
part~\cite{Dobaczewski84,Meng96,Meng98npa,Bennett96,
Sandulescu00b, Sandulescu03}. Therefore it's very important to
investigate single particle resonant states in exotic nuclei.

Various techniques have been developed to study resonant states in
the continuum such as the R-matrix theory~\cite{Wigner47} and the
extended R-matrix theory~\cite{Hale87}, the K-matrix
theory~\cite{Humblet91} and the conventional scattering
theory~\cite{Taylor72}. By discretizing the continuum, the
contribution of the resonant states can be self-consistently taken
into account via a Bogoliubov transformation in coordinate
space~\cite{Dobaczewski84,Meng96,Meng98npa}. The bound-state-type
methods have also been developed, including the real stabilization
method~\cite{Hazi70}, the complex scaling method
(CSM)~\cite{Ho83}, and the analytic continuation in the coupling
constant (ACCC) method~\cite{Kukulin89}. Within the framework of
the non-relativistic Schr\"{o}dinger equation, the ACCC method has
been applied to investigate the energies and widths for resonant
states in light nuclei combined with few-body
methods~\cite{Tanaka99,Aoyama02}, and to study single-particle
resonant states in spherical and deformed nuclei by solving the
Schr\"{o}dinger equation with Woods-Saxon
potentials~\cite{Cattapan00}. Recently the ACCC method and phase
shift scattering method have been combined with the RMF model.

The philosophy of the ACCC method is that a resonant state becomes
a bound one if one increases the attractive potential. Then the
energy, width, and wave function of the resonant state can be
obtained by an analytic continuation carried out via a
$\mbox{Pad$\acute{\text{e}}$}$ approximant (PA) from the
bound-state solutions. Compared with other bound-state-type
methods, the ACCC approach is very effective and numerically quite
simple because many methods available for bound-state problems can
be used and the PA for analytic continuation can be implemented in
easily.

By solving the spherical relativistic Hartree equations in a
meshed box of size $R_0$ self-consistently, one can calculate the
ground state properties of a nucleus. The vector potential $V(r)$
and the scalar potential $S(r)$, energies and wave functions for
bound states are also obtained. By increasing the attractive
potential $(V(r)+S(r))$ to $\lambda (V(r)+S(r))$ , a resonant
state will be lowered and becomes bound if the coupling constant
$\lambda$ is large enough. Near the branch point $\lambda_0$,
defined by the scattering threshold
$k(\lambda_0)=0$~\cite{Kukulin89}, the wave number $k(\lambda)$
behaves as
\begin{eqnarray}
   k(\lambda) \sim \left\{
   \begin{array}{l@{\mbox{~~~}}l}
         i\sqrt{\lambda-\lambda_0},        &  l>0 ,\\
         i(\lambda-\lambda_0),            &  l=0.
   \end{array}\right.
\end{eqnarray}
These properties suggest an analytic continuation of the wave
number $k$ in the complex $\lambda$ plane from the bound-state
region into the resonance region by $\mbox{Pad$\acute{\text{e}}$}$
approximant of the second kind (PAII)~\cite{Kukulin89}
\begin{equation}
\label{pade-e}
 k(x)\approx k^{[L,N]}(x)
    = i \dfrac{c_0+c_1 x+c_2x^2+\ldots +c_Lx^L}{ 1+d_1 x+d_2 x^2+\ldots+d_N x^N},
\end{equation}
where $x\equiv\sqrt{\lambda-\lambda_0}$, and $c_0, c_1,\ldots,
c_L, d_1, d_2,\ldots,d_N$ are the coefficients of PA. These
coefficients can be determined by a set of reference points $x_i$
and $k(x_i)$ obtained from the Dirac equation with
$\lambda_i>\lambda_0,~i=1,2,...,L+N+1$. With the complex wave
number $k(\lambda=1)= k_r + i k_i$, the resonance energy $E$ and
the width $\Gamma$ can be extracted from the relation $
\varepsilon=E-i \dfrac{\Gamma}{2} ~~ (E,\Gamma\in \mathbb{R})$ and
$k^2=\varepsilon^2-M^2$, i.e.,
\begin{eqnarray}
    \label{E-W}
    E      &=& \sqrt{\dfrac{\sqrt{(M^2+k_r^2-k_i^2)^2+4k_r^2k_i^2}+(M^2+k_r^2-k_i^2)}{2}}- M , \re
    \Gamma &=& \sqrt{2\sqrt{(M^2+k_r^2-k_i^2)^2+4k_r^2k_i^2}-2(M^2+k_r^2-k_i^2)}.
\end{eqnarray}
Based on the Schr\"{o}dinger and Dirac equations, the stability
and convergence of the energies and widths for single-particle
resonant states with square-well, harmonic-oscillator and
Woods-Saxon potentials have been investigated and their dependence
on the coupling constant interval and the order of the PA have
been discussed~\cite{ZhangSS03,ZhangSS04a}. Combined with the RMF
theory, the ACCC method was employed to study energies and widths
for resonant states in stable nuclei $^{16}$O and $^{48}$Ca in
Ref.~\cite{Yang01} and to explore the single-particle resonant
states, including resonance parameters and wave functions, in
exotic nuclei in Ref.~\cite{ZhangSS04b}.

The continuation in the coupling constant can be replaced by the
continuation in $k$ plane along the $k(\lambda)$ trajectory
determined by Eq.~(\ref{pade-e}) to the point $k_R$ corresponding
to the wave number for Gamow state, i.e.,
$k_R=k^{[L,N]}(\lambda=1)$~\cite{Kukulin89}. Similarly, the wave
function $\varphi(k_R,r)$ for a resonant state can be obtained by
an analytic continuation of the bound-state wave function
$\varphi(k_i,r)$ in the complex $k$ plane. One can also prove that
the wave function $\varphi(k,r)$ is an analytic function of the
wave number $k$ in the inner region $r<R_0$ where the Jost
function analyticity dominates~\cite{Kukulin89}. Therefore, the
technique, which has been adopted to find the complex resonance
energy, has been used to determine the resonance wave function
$\varphi(k_R,r)$. Firstly, the PA are constructed to define the
resonance wave function at any point $r$ in the inner region
($r<R_0$)~\cite{Kukulin89}
\begin{equation}
    \label{inner}
    \varphi^{[L,N]}(k,r) = \dfrac{P_L(k,r)}{Q_N(k,r)}
                         = \dfrac{a_0(r)+a_1(r)k+a_2(r)k^2+\ldots +a_L(r)k^L}
                                 { 1+b_1(r) k+b_2(r)k^2+\ldots+b_N(r) k^N},
\end{equation}
where the coefficients ${a_i(r)}~(i=0,1,\ldots, L)$ and
${b_j(r)}~(j=1,2,\ldots, N)$ are dependent on $r$. These
coefficients can be determined by a set of reference points $k_i$
and $\varphi(k_i,r)$ obtained from the Dirac equation with
$\lambda_i>\lambda_0,~(i=1,2,...,L+N+1)$. The resonance wave
function $\varphi(k_R,r)=\varphi^{[L,N]}(k_R,r)~(r<R_0)$ can be
extrapolated in this way.

\begin{figure}[htb!]
\centering
\includegraphics[width=8.5cm]{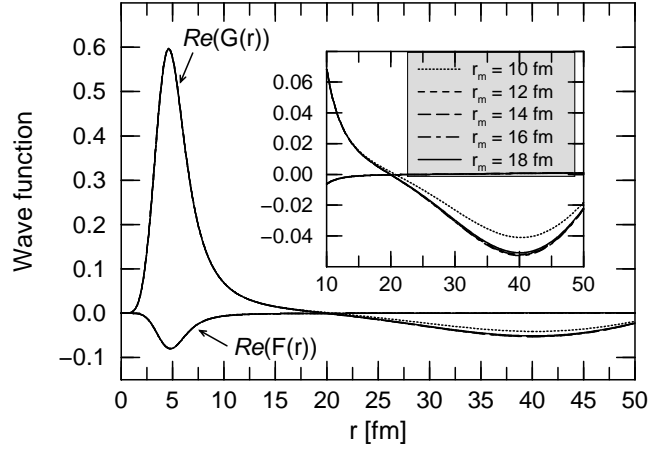}
\caption{Real parts of the upper and lower components of the
radial wave function for the neutron $\nu 1g_{9/2}$ state in
$^{60}$Ca calculated by the ACCC method with different matching
points $r_m$. Taken from Ref.~\cite{ZhangSS04b}.} \label{FigC2}
\end{figure}

The Dirac equation can be rewritten as two decoupled
Schr\"odinger-like equations for the upper and the lower
components respectively~\cite{Meng99prc}. For neutrons, the
Schr\"odinger-like equation for the upper component reads
\begin{equation}
  \label{G-diff}
   \dfrac{d^2G_i^{lj}(r)}{d r^2}
 - \dfrac{\kappa_i(\kappa_i+1)}{r^2}G_i^{lj}(r)
 + (E_i^2-M^2)G_i^{lj}(r)
 = 0,
\end{equation}
in the outer region where $V(r)\simeq 0$ and $S(r)\simeq 0$. Its
solution is the well known Riccati-Hankel function
\begin{equation}
    \hat{h}^{\pm}_\kappa (z)=\hat{n}_\kappa (z)\pm i\hat{j}_\kappa(z), ~~~~~z=kr,
\end{equation}
where $ \hat{j}_\kappa(z)=z{j}_\kappa (z)$ is the usual regular
solution, i.e. Riccati-Bessel function, and
$\hat{n}_\kappa(z)=z{n}_\kappa (z)$ is the irregular solution,
i.e. Riccati-Neumann function. The outer wave function is matched
to the inner wave function at $r = r_m < R_0$
\begin{equation}
   \label{match-G}
   \varphi^{[L,N]}(k_R,r_m)
   = C(k_R)\ls\hat{j}_\kappa(k_Rr_m) + D(k_R)\hat{n}_\kappa(k_Rr_m)\rs,
\end{equation}
where $C(k_R)$ is the coefficient for matching and
$D(k_R)=\tan\delta_\kappa(k_R)$ with $\delta_\kappa(k_R)$ the
phase shift. It has been found that $\delta_\kappa(k_R)$ is almost
a constant when $r_m$ is large enough. Given the upper component,
the lower component $F_{\kappa}(r)$ can be calculated from the
relationship between the upper and the lower components and the
resonance wave function is finally normalized according to the
Zel'dovich procedures~\cite{Kukulin89}. The wave function for the
neutron resonant state $\nu 1g_{9/2}$ in $^{60}$Ca with different
matching points is given in Fig.~\ref{FigC2} where one finds the
convergence of the wave function with respect to the matching
point $r_m$ when $r_m$ changes from 14 fm to 18 fm.

The resonant states can be also investigated in the scattering
phase shift method~\cite{Sandulescu00b,Sandulescu03,Cao02} where
the RMF equations are solved with the scattering-type boundary
conditions. At large distances, where both the scalar and the
vector potentials are zero, the RMF radial equations can be
written in the form:
\begin{equation}
 \frac{d^2G}{dr^2} +(\alpha^2- \frac{\kappa(\kappa+1)}{r^2})G  =
 0,
\end{equation}
\begin{equation}
 F  =  \frac{1}{E+M}(\frac{dG}{dr}+\frac{\kappa}{r}G) ,
\end{equation}
where $\alpha^2=E^2-M^2$. These equations are suited for fixing
the scattering-type boundary conditions for the continuum
spectrum. They are given by:
\begin{equation}
 G =  C \alpha r [\cos(\delta) j_l(\alpha r)-\sin(\delta) n_l(\alpha
 r)],
\end{equation}
\begin{equation}
 F =  \frac{C \alpha^2r}{E+M}
      [\cos(\delta) j_{l-1}(\alpha r)-\sin(\delta) n_{l-1}(\alpha r)] ,
\end{equation}
where $j_{l}$ and $n_{l}$ are the Bessel and  Neumann functions
and $\delta$ is the phase shift associated to the relativistic
mean field. The constant $C$ is fixed by the normalisation
condition of the scattering wave functions and the phase shift
$\delta$ is calculated from the matching conditions. In the
vicinity of an isolated resonance the derivative of the phase
shift has a Breit-Wigner form, i.e.
\begin{equation}
\frac{d\delta(E)}{dE} = \frac{\Gamma/2}{(E_r-E)^2+\Gamma^2/4},
\end{equation}
from which one estimates the energy and the width of the
resonance. In the vicinity of a resonance the radial wave
functions of the scattering states have a large localisation
inside the nucleus. Close to a resonance the energy dependence of
both components of the Dirac wave functions can be factorized
approximatively by a unique energy dependent
function~\cite{Migdal72b}. As in the non-relativistic
case~\cite{Unger67}, this energy dependent factor is the square
root of the Breit-Wigner function written above, or, equivalently,
the square root of the derivative of the phase shift.

\begin{figure}[htb!]
\centering
\includegraphics[width=8.5cm]{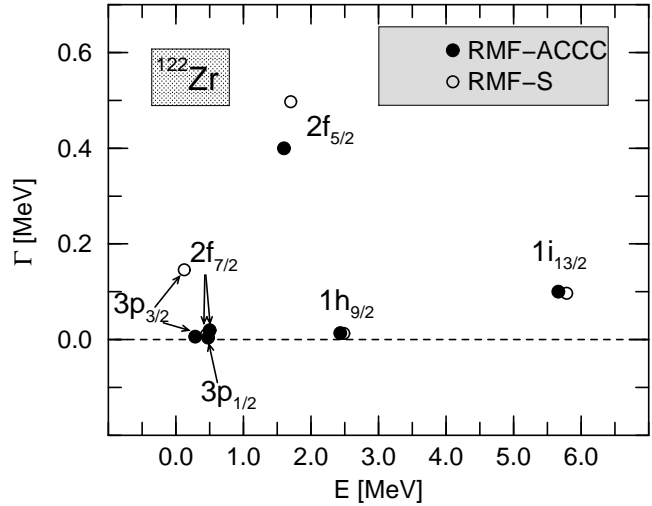}
\caption{Energies and widths for the neutron states $\nu
3p_{3/2}$, $\nu 3p_{1/2}$, $\nu 2f_{7/2}$, $\nu 2f_{5/2}$, $\nu
1h_{9/2}$, and $\nu 1i_{13/2}$ in $^{122}$Zr. Solid circles
represent the results of the ACCC method, while open circles
denote the results of the scattering method. Taken from
Ref.~\cite{ZhangSS04b}.} \label{FigC3}
\end{figure}

The energies and widths for the neutron resonant states $\nu
3p_{3/2}$, $\nu 3p_{1/2}$, $\nu 2f_{7/2}$, $\nu 2f_{5/2}$, $\nu
1h_{9/2}$, and $\nu 1i_{13/2}$ in $^{122}$Zr are shown in
Fig.~\ref{FigC3}. The results of the ACCC and the scattering
methods are in good agreement with each other for most of these
states. From both methods, $\nu 2f_{5/2}$ and $\nu 1i_{13/2}$ have
large widths, while $\nu 2f_{7/2}$ and $\nu 1h_{9/2}$ are very
narrow. The width for $\nu 3p_{3/2}$ from the scattering method is
slightly larger than that from the ACCC calculation. For the
resonant state $\nu 3p_{1/2}$, neither the energy nor the width
can be extracted from the scattering calculation. From a quantum
mechanical point of view, these single-particle resonant states
are quasi-stationary ones captured by centrifugal barriers. The
decay width for a resonant state can be roughly explained by the
penetration through the barrier. For those states with the same
$l$, i.e., the same centrifugal barrier, the higher state has a
larger width, e.g., for the two $f$ states $\nu 2f_{5/2}$ and $\nu
2f_{7/2}$. Although $\nu 1h_{9/2}$ is higher than $\nu 2f_{5/2}$,
its width is smaller because of higher centrifugal barrier. A
similar argument also holds for a much higher but narrow state
$\nu 1i_{13/2}$.

\subsubsection{BCS approximation with resonant states}
\label{subsec:resonant}
 The conventional BCS method is incapable of describing weakly bound nuclei
 due to the oscillating asymptotic behavior of single particle wave functions
 in the continuum~\cite{Dobaczewski84}. Recently, many authors
 demonstrate that by picking up only those low-lying resonant states in the
 continuum, the BCS method can be successfully applied to weakly bound
 nuclei~\cite{Sandulescu00b,Sandulescu03,Yadav04,Hagino05,Geng03b}.
 This is because the resonant states are well localized inside the
 nucleus and there is a large region outside the nucleus where the
 resonant wave functions have values close to zero before they
 start oscillating~\cite{Sandulescu00b}. Therefore,the wave
 functions of the resonant states can be taken to be zero beyond a
 cutoff radius~\cite{Sandulescu00b}. Meanwhile a realistic
 pairing potential, such as a zero-range $\delta$-force, prefer to
 pick up the resonant states,
 whose wave functions have a large overlap with those of
 bound states below the Fermi surface, rather than the
 continuum~\cite{Yadav04,Geng03b}. In such a way,
 the HF+rBCS and RMF+rBCS results do not depend sensitively on
 the cutoff radius~\cite{Sandulescu00b}.

The BCS method can be easily extended to study the width effects
of resonant states. The extended BCS equations for a general
(finite range) pairing interaction including the contribution of
the resonant continuum, referred as the resonant-BCS (rBCS)
equations~\cite{Sandulescu00b, Sandulescu03}, are :

\begin{equation}\label{eq:gapr1}
\Delta_i
     = \sum_{j}V_{i\overline{i}j\overline{j}} u_j v_j
     + \sum_\nu V_{i\overline{i},\nu\epsilon_\nu\overline{\nu\epsilon_\nu}}
       \int_{I_\nu} g_{\nu}(\epsilon) u_\nu(\epsilon) v_\nu(\epsilon)d\epsilon~,
\end{equation}
\begin{equation}\label{eq:gapr2}
\Delta_\nu
    \equiv \sum_{j} V_{\nu\epsilon_\nu\overline{\nu\epsilon_\nu},j\overline{j}} u_j v_j
         + \sum_{\nu^\prime} V_{\nu\epsilon_\nu\overline{\nu\epsilon_\nu},
           \nu^\prime\epsilon_{\nu^\prime}\overline{\nu^\prime\epsilon_{\nu^\prime}}}
           \int_{I_{\nu^\prime}} g_{\nu^\prime}(\epsilon^\prime)
                 u_{\nu^\prime}(\epsilon^\prime)
                 v_{\nu^\prime}(\epsilon^\prime)
                d\epsilon^\prime~,
\end{equation}
\begin{equation}\label{eq:gapr3}
N = \sum_i v_i^2
   + \sum_\nu \int_{I_\nu} g^c_{\nu}(\epsilon) v^2_\nu (\epsilon) d\epsilon~.
\end{equation}
Here $\Delta_i$ are the gaps for the bound states and $\Delta_\nu$
are the averaged gaps for the resonant states. The quantity
$\displaystyle g^c_\nu(\epsilon) = \frac {2j_\nu +1}{\pi}
\frac{d\delta_\nu}{d\epsilon}$ is the total level density and
$\delta_\nu$ is the phase shift of the state with the angular
momentum $(l_{\nu} j_{\nu})$. The factor $g^c_\nu(\epsilon)$ can
take into account the effect of the width and it is approximately
a delta function for a very narrow resonance. The interaction
matrix elements are calculated with the scattering wave functions
at resonance energies and normalized inside a volume where the
pairing interaction is active (see last paragraph).

\begin{figure}[t]
\centering
\includegraphics[width=8.5cm]{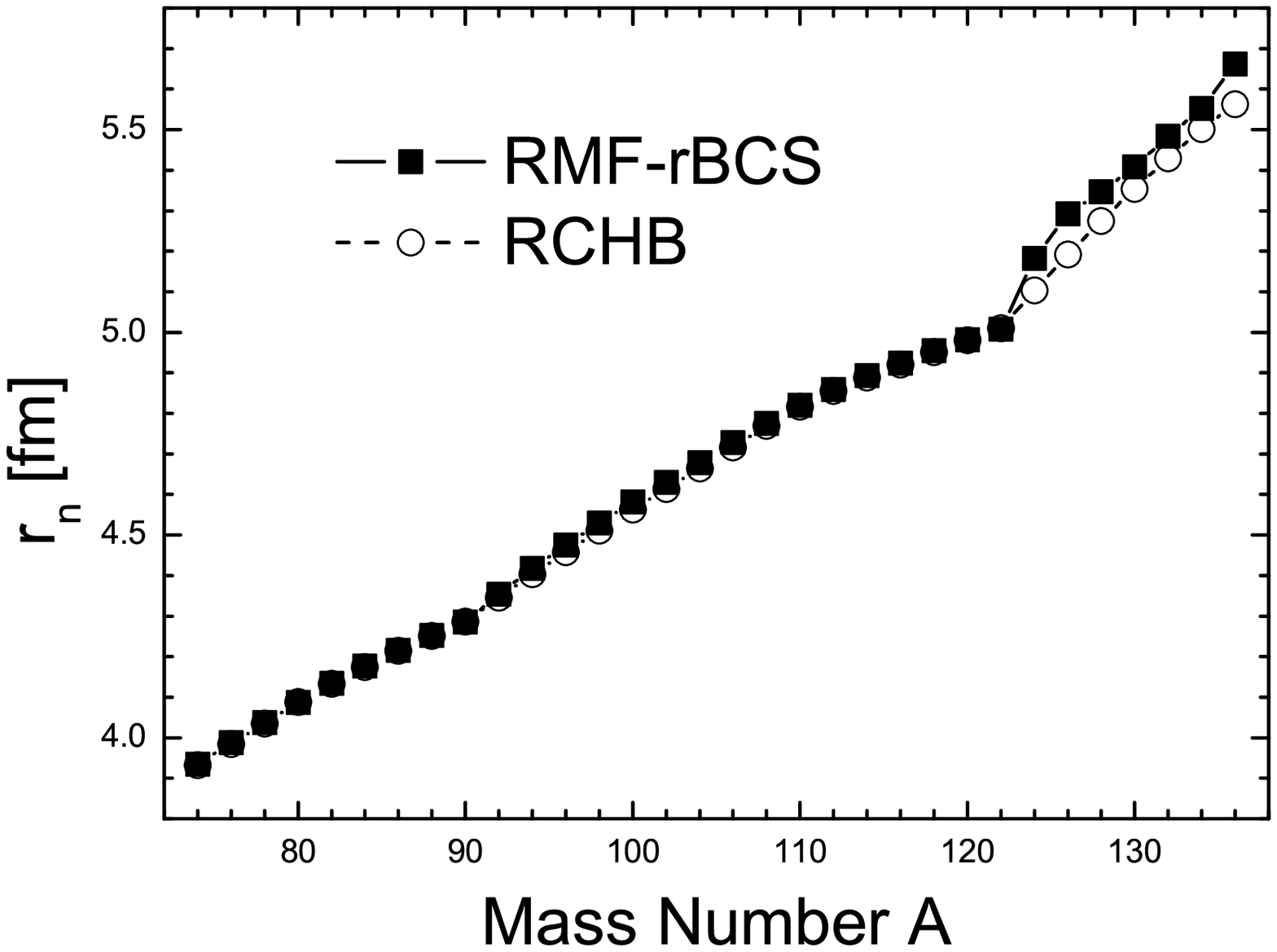}
\includegraphics[width=8.5cm]{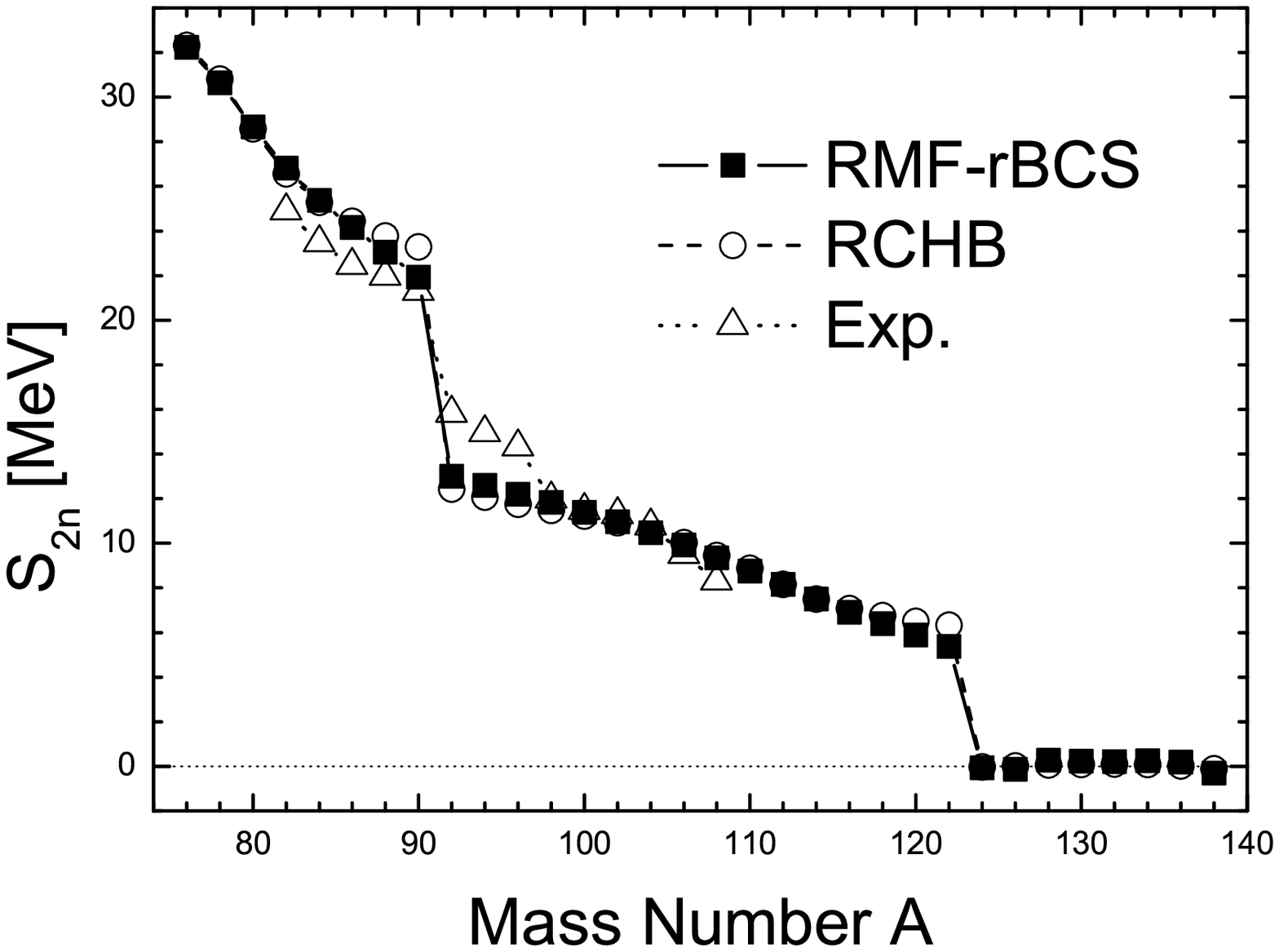}
\caption{The rms neutron radii (left panel) and the two-neutron
separation energies (right panel) of even Zr isotopes as a
function of the mass number $A$. Taken from
Ref.~\cite{Sandulescu03}.}\label{FigC4}
\end{figure}

The BCS equations in the conventional RMF+BCS formalism can be
easily substituted by these rBCS equations
(\ref{eq:gapr1},\ref{eq:gapr2},\ref{eq:gapr3}). To make this
implementation, one can solve the RMF equations with the
scattering-type boundary conditions or any other methods like the
ACCC method described in the last subsection and find the
energies, widths and wave functions for the resonant states. This
approximation scheme, called the RMF+rBCS method, has been applied
to study neutron-rich Zr isotopes~\cite{Sandulescu03}. It was
demonstrated that the sudden increase of neutron radii close to
the neutron drip line depends on a few resonant states close to
the continuum threshold. Including these resonant states, the
RMF-BCS calculations give practically the same neutron radii and
neutron separation energies as  the RCHB calculations (see
Fig.\ref{FigC4}).

\subsection{Relativistic continuum Hartree-Bogoliubov theory}
\label{subsec:RHB}

\subsubsection{Bogoliubov transformation and continuum spectra}

\begin{figure}
 \begin{center}
 \includegraphics[width=10.cm]{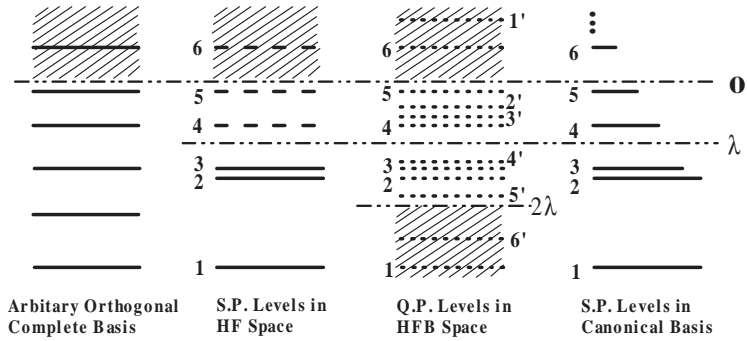}
 \caption{The schematic pictures of different basis. The first column
 is an arbitrary $M$-dimension orthogonal and complete basis to
 diagonalize the system Hamiltonian. The second column is the single
 particle levels in the HF space. The third column is quasi-particle
 levels in the quasi-particle HFB space which is enlarged into
 $2M$-dimension. The quasi-particle orbitals has a reflection symmetry
 with respect to the Fermi surface. The physical orbital in the
 canonical basis is given in the forth column. } \label{FigC5}
 \end{center}
 \end{figure}

The contribution of the resonant states can be self-consistently
taken into account in the Hartree-Fock Bogoliubov (HFB) theory
solved in the coordinate space~\cite{Dobaczewski84}. The advantage
of the general HFB theory is that the variation method based on
quasi-particle transformation unifies the self-consistent
description of nuclear orbitals and the BCS pairing theory into a
single variation theory. In Fig.~\ref{FigC5}, schematic pictures
for different bases in the HF, HFB and the canonical basis have
been given. For a given many-body system, one can choose any
$M$-dimension orthogonal and complete basis to diagonalize the
Hamiltonian. That is shown in the first column of
Fig.~\ref{FigC5}. If only HF approach are considered, the
Hamiltonian is diagonal in the HF basis. We get a $M$-dimension
basis, with a Fermi surface which divides the full open orbitals
and full occupied orbitals. They are given in the second column.
After the Bogoliubov transformation, we are working in the
quasi-particle basis, which is enlarged into $2M$-dimension. The
quasi-particle orbitals are reflection symmetric with respect to
the Fermi surface, as shown in the third column. They must be
transformed into the physical space
--- canonical basis~\cite{Ring80,Meng98npa}, in order to have a
similar explanation as the HF space in the second column.  For the
Bogoliubov transformation involving the continuum, as the Fermi
energy is negative for a bound nucleus, only those resonant states
in the canonical basis with proper asymptotic behavior could be
picked up. In the forth column, a schematic picture for the
orbitals in the canonical basis is given, now the orbitals could
be partially occupied, which is another essential difference
between the HF basis and the canonical basis.

\subsubsection{The formalism for the RCHB theory}

In the BCS approach, the gap equation takes into account only
pairing between time-reversed continuum states. A more general
pairing between continuum states at neighboring energies can be
taken care of by a continuum Bogoliubov approach. The improvement
to the BCS approximation is to introduce the concept of
quasi-particle by the Bogoliubov transformation. Instead of
introducing a mapping factor, the Bogoliubov transformation
transforms the equation of motion into quasi-particle space. In
the following we shall consider the transformation from the single
particle basis to the quasi-particle basis. The single particle
operators in coordinate space
 $a_{\mbox{r}}$,$a_{\mbox{r}}^\dagger$, are connected to the
operators in the quasi-particle basis, $\beta_{i}$,
$\beta_{i}^\dagger$, $i=1,M$, by the following transformation:
\begin{equation}
    \left( { \beta_i \atop \beta_i^\dagger } \right)
        = \int d^3 \mbox{r} {\cal W}^\dagger
          \left( { a_{\mbox{r}} \atop a_{\mbox{r}}^\dagger } \right)
        = \int d^3 r \left( \begin{array}{cc}
        U_{i,\mbox{r}}^*   & V_{i,\mbox{r}}^* \\
        V_{i,\mbox{r}}     & U_{i,\mbox{r}}
    \end{array} \right)
    \left( { a_{\mbox{r}} \atop a_{\mbox{r}}^\dagger } \right),
\label{qptrsf}
\end{equation}
where $i$ and $\mbox{r}$ represents the row and column indices
respectively. The basis of the quasi-particle states,
$\psi_{U,i}(\svec r)$ and $\psi_{V,i}(\svec r)$, are defined via
the single particle coordinate basis of $\delta$-function as:
\begin{equation}
    \left( \psi_{U,i}(\svec r) \atop \psi_{V,i}(\svec r) \right)
        =  \int d^3 \mbox{r}' \ \delta{(\mbox{r}-\mbox{r}')} \
    \left( U_{i,\mbox{r}'}^* \atop V_{i,\mbox{r}' } \right).
\label{qpstrsf}
\end{equation}

In spherical case, the Dirac spinor wave functions
$\psi_{U,i}(\svec r)$ and $\psi_{V,i}(\svec r)$ are similar to
Eq.(\ref{eq:SRHspinor}) and satisfy the following normalization
relation:
\begin{equation}
   \int d^3r \cal W^\dagger \cal W = I.
\end{equation}
The variation method based on quasi-particle transformation can
unify the self-consistent description of nuclear orbitals, as
given by the HF approach, and the BCS pairing theory into a single
variation theory. As the equations of motion are self-consistently
solved in the quasi-particle space, the convergence is guaranteed
automatically. The unphysical continuum are excluded and the
contribution from the resonance states with positive energies can
be taken into account. From now on, the concept of continuum in
Bogoliubov transformation is just the resonance states with
positive energy as the unphysical continuum are excluded already.

Following the standard procedure of Bogoliubov transformation, a
Dirac Hartree-Bogoliubov equation could be derived and the unified
description of the mean field and pairing correlation in nuclei
could be achieved. Using Green's function techniques it has been
shown in Refs.~\cite{Kucharek91, Ring96} how one can derive a
relativistic Hartree-Fock-Bogoliubov theory: after a full
quantization of the system the mesonic degrees of freedom are
eliminated and, in full analogy to the non-relativistic case, the
higher order Green's functions are factorized in the sense of
Gorkov~\cite{Gorkov58}. Finally, neglecting retardation effects
and the Fock term, as it is mostly done in relativistic mean field
theory, one ends up with RHB equations as the following:
\begin{equation}
   \left( \begin{array}{cc}
          h-\lambda &   \Delta \\
          -\Delta^*    &  - h^*+\lambda
          \end{array} \right)
   \left( { \psi_U \atop\psi_V } \right)  ~
   = ~ E ~ \left( { \psi_U \atop \psi_V } \right),
\label{ghfb}
\end{equation}
where
\begin{equation}
   h(\svec r,\svec r') =  \left[ {\svec \alpha} \cdot {\svec p} +
      V( {\svec r} ) + \beta ( M + S ( {\svec r} ) ) \right]
\label{NHamiltonian}
\end{equation}
is the Dirac Hamiltonian and the Fock term has been neglected. The
pairing potential is
\begin{eqnarray}
    \Delta_{kk'}(\svec r, \svec r')
    &=& - \int d^3 \mbox{r}_1 \int d^3\mbox{r}_1' \sum_{\tilde k \tilde k'}
    V_{kk',\tilde k \tilde k'} ( \svec r \svec r'; \svec r_1 \svec r_1' )
    \kappa_{\tilde k \tilde k'}  (\svec r_1, \svec r_1' ),
\label{gap}
\end{eqnarray}
where $k$, $k'$, $\tilde k$, or $\tilde k'$ represent the other
quantum numbers which together with the coordinate specify the
single particle states. It is obtained from the pairing tensor,
\begin{eqnarray}
   \kappa_{k k'}( \svec r, \svec r') = \langle | a_{k,i} a _{k',i'} | \rangle
   = \sum_{E_i > 0}
          \psi_U^{k,i}(\svec r) ^*
          \psi_V^{k',i}(\svec r'),
\end{eqnarray}
and one-meson exchange interaction $V_{kk',\tilde k \tilde k'}
(\svec r \svec r'; \svec r_1 \svec r_1' )$ in the $pp$-channel.
The nuclear density is given as
\begin{eqnarray}
   \rho(\svec r ,\svec r' )
   = \sum_{k,E_i>0}  \psi_V^{k,i} (\svec r) ^* \psi_V ^{k,i} (\svec r '
   ).
\end{eqnarray}

In the RHB theory, the ground state $|\Psi\rangle$ of the even
particle system is defined as the vacuum with respect to the
quasi-particle: $\beta_{i} |\Psi\rangle=0$ for all $i$, i.e.,
$|\Psi\rangle = \prod_i \beta_{i} |-\rangle$, where $|-\rangle$ is
the bare vacuum. For an odd particle system, the ground state can
be correspondingly written as: $|\Psi \rangle_{j} =
\beta_{j}^\dagger\prod_{i \ne j} \beta_{i} | - \rangle$, where $j$
is the level which is blocked. The exchange of the quasiparticle
creation operator $\beta_{j}^\dagger$ with the corresponding
annihilation operator $\beta_{j}$ means the replacement of the
column $j$ in the  $U$ and $V$ matrices by the corresponding
column in the matrices $V^*$, $U^*$~\cite{Ring80}.

As the one-meson exchange interactions are not able to reproduce
even in a semi-quantitative way the proper pairing in the
realistic nuclear many-body problem, the interaction $V$ in
Eq.~(\ref{gap}) is replaced by the phenomenological interaction
which has been generally used in the pairing channel of the
conventional HFB theory. As in Ref.~\cite{Meng96} $V$ used for the
pairing potential (\ref{gap}) is either the density-dependent
two-body force of zero range:
\begin{equation}
   V(\mbox{\boldmath $r$}_1,\mbox{\boldmath $r$}_2) = V_0
     \delta(\mbox{\boldmath $r$}_1-\mbox{\boldmath $r$}_2)
     \frac{1}{4}\left[1-
     \mbox{\boldmath $\sigma$}_1\mbox{\boldmath $\sigma$}_2\right]
     \left(1 - \frac{\rho(r)}{\rho_0}\right)
\label{vpp}
\end{equation}
with the interaction strength $V_0$ and the nuclear matter density
$\rho_0$ or  Gogny-type finite range force:
\begin{equation}
   V(\mbox{\boldmath $r$}_1,\mbox{\boldmath $r$}_2)
      ~=~\sum_{i=1,2}
       e^{((\mbox{\boldmath $r$}_1-\mbox{\boldmath$r$}_2) / \mu_i)^2}
       (W_i + B_i P^{\sigma} - H_i P^{\tau} - M_i P^{\sigma} P^{\tau})
\label{vpp2}
\end{equation}
with the parameter $\mu_i$, $W_i$, $B_i$, $H_i$ and $M_i$
($i=1,2$) as the finite range part of the Gogny force
D1S~\cite{Berger84}.

The RHB equations can be solved in different bases as well. In
many applications an expansion of the wave functions in an
appropriate harmonic-oscillator basis of spherical or axial
symmetry provides a satisfactory level of accuracy. For example,
in Ref.~\cite{Gonzalez96} the RHB equations were solved by
expanding the nucleon spinors $U_k(r)$ and $V_k(r)$ and the meson
fields in a basis of spherical harmonic oscillators. However, for
nuclei near the drip lines the expansion in the localized
oscillator basis presents only a poor approximation for the
continuum states. Oscillator expansions produce densities which
decrease too steeply in the asymptotic region at large distance
from the center of the nucleus. The calculated rms radii cannot
reproduce the experimental values, especially for halo nuclei. The
transformed harmonic oscillator (THO) basis has also been employed
in the solution of the RHB equations in configurational
space~\cite{Stoitsov98a}. Solving the RHB equations in Woods-Saxon
basis is another alternative.

In order to describe the coupling between bound and continuum
states more exactly, the RHB equations and the equations for the
meson fields should be solved in coordinate
space~\cite{Meng98npa,Meng96,Poschl97} and one arrives at the RCHB
theory. When spherical symmetry is imposed, the wave function can
be conveniently written as
\begin{equation}
   \psi^i_U = \left( {\displaystyle {i  \frac {G_U^{i\kappa}(r)} r }  \atop
     {\displaystyle \frac {F_U^{i\kappa}(r)} r
        (\svec\sigma \cdot \hat {\svec r} )  } }
              \right) {Y^l _{jm} (\theta,\phi)}  \chi_{t}(t) ,
   ~~~
   \psi^i_V =
       \left( {\displaystyle {i  \frac {G_V^{i\kappa}(r)} r }  \atop
          {\displaystyle \frac {F_V^{i\kappa}(r)} r
             (\svec\sigma \cdot \hat {\svec r} )
       } } \right)  {Y^l _{jm} (\theta,\phi)}  \chi_{t}(t).
\end{equation}
The above equation (\ref{ghfb}) only depends on radial coordinates
and can be derived as the following integro-differential
equations:
\begin{eqnarray}
\left\{
   \begin{array}{lll}
      \displaystyle
      \frac {d G_U(r)} {dr} + \frac {\kappa} r G_U(r) -
       ( E + \lambda-V(r) + S(r) ) F_U(r) +
         r \int r'dr' \Delta(r,r')  F_V(r') &=& 0 , \\
      \displaystyle
      \frac {d F_U(r)} {dr} - \frac {\kappa} r F_U(r) +
       ( E + \lambda-V(r)-S(r) ) G_U(r) +
         r \int r'dr' \Delta(r,r') G_V(r') &=& 0 ,\\
      \displaystyle
      \frac {d G_V(r)} {dr} + \frac {\kappa} r G_V(r) +
       ( E - \lambda+V(r)-S(r) ) F_V(r) +
         r \int r'dr' \Delta(r,r') F_U(r') &=& 0 ,\\
      \displaystyle
      \frac {d F_V(r)} {dr} - \frac {\kappa} r F_V(r) -
       ( E - \lambda+V(r)+S(r) ) G_V(r) +
         r \int r'dr' \Delta(r,r')  G_U(r') &=& 0 ,\\
\end{array}
\right. \label{eq:RCHBCoupEq}
\end{eqnarray}
where the nucleon mass has been included in the scalar potential
$S(r)$. Instead of solving Eqs. (\ref{eq:SRHDirac}) and
(\ref{eq:SRHmesonmotion}) self-consistently for the RMF case, now
one has to solve Eqs. ({\ref{eq:RCHBCoupEq}) and
(\ref{eq:SRHmesonmotion}) self-consistently for the RCHB case. The
densities are calculated as
\begin{eqnarray}
 \left\{
  \begin{array}{lll}
   4\pi r^2 \rho_s(r) & = & \sum_{i} (|G^i_V(r)|^2 - |F^i_V(r)|^2), \\
   4\pi r^2 \rho_v(r) & = & \sum_{i} (|G^i_V(r)|^2 + |F^i_V(r)|^2), \\
   4\pi r^2 \rho_3(r) & = & \sum_{i} \tau_3
                                         (|G^i_V(r)|^2 + |F^i_V(r)|^2), \\
   4\pi r^2 \rho_c(r) & = & \sum_{i} \ff2 (1-\tau_3)
                                         (|G^i_V(r)|^2 + |F^i_V(r)|^2), \\
  \end{array}
 \right.
 \label{eq:RCHBdensity}
\end{eqnarray}
where the summations are over all quasi particle states. From the
densities given above, one can calculate the rms radii and charge
radius of the nucleus. The total binding energy is calculated as
\begin{eqnarray}
 E & = & E_{\rm nucleon} + E_\sigma + E_\omega + E_\rho + E_c + E_{\rm CM},
\end{eqnarray}
where except for $E_{\rm nucleon}$, the other terms are the same
as those in the spherical RH theory. The energy of nucleons
$E_{\rm nucleon}$ is calculated as
\begin{equation}
 E_{\rm nucleon} = \sum_{i}\int dr (\lambda-E^i)
 \left[|G^i_V(r)|^2 + |F^i_V(r)|^2\right] - 2E_{\rm pair},
\end{equation}
and $E_{\rm pair} = -\frac{1}{2}{\rm Tr} \Delta\kappa$. For the
details of the transformation from the quasi particle space to the
canonical basis, the reader is referred to Ref.~\cite{Meng98npa}.

To solve RCHB equations (\ref{eq:RCHBCoupEq}), one has to
discretize the RCHB continuum. The discretization can be done by
solving the RCHB equation in a spherical box of radius $R$, i.e.,
by imposing the boundary conditions
\begin{equation}
   \psi_U(E,|\svec r| = R, s) = \psi_V(E,|\svec r| = R, s) = 0.
\end{equation}
By increasing $R$, one can make the RCHB spectrum dense and better
approximate the continuum, while increasing $E_{max}$ allows one
to take into account coupling to highly excited quasi-particle
state. Both $R$ and $E_{max}$ must be reasonably taken in such a
way that the final results do not depend on the chosen $R$ and
$E_{max}$. Only those solutions with $E>0$ are necessary due to
the special symmetry of RCHB equation \cite{Meng98npa}. If $E$ is
the solution of Eq.~(\ref{ghfb}), $-E$ must be another solution,
but they are not independent.  Both solutions are connected as
following:
\begin{equation}
  \psi_U^i ( -E,\svec r ) =    \psi_V^i (E,\svec r ), ~ ~
  \psi_V^i ( -E,\svec r ) =  - \psi_U^i (E,\svec r ).
\end{equation}
So one can concentrate only on the solution with either positive
energies or negative energies.

Equations~(\ref{eq:RCHBCoupEq}), in the case of $\delta$-force
~Eq.(\ref{vpp}), is reduced to ordinary coupled differential
equations and can be solved with shooting method combined with
Runge-Kutta algorithms, which is so far the most elegant method
for coupled differential equations in coordinate representation.
For each fixed value of the energy $E$ the RCHB equations
(\ref{eq:RCHBCoupEq}) are solved in the following steps:
\begin{enumerate}
\item Discretizing the whole space $[0,\infty]$;
\item Choosing a value for the energy $E$ in Eq.~(\ref{eq:RCHBCoupEq}) with $E \ge 0$
      for the given spin-parity channel;
\item Finding the proper boundary condition for $\psi(0)$, where $\psi$
      represents  $G_V$,$F_V$,$G_U$,$F_U$ and integrated Eq.~(\ref{eq:RCHBCoupEq})
      outward from $r=0$ to proper matching point $r=R_{match}$ by
      Runge-Kutta algorithms;
\item Finding the proper boundary condition for $\psi(\infty)$ and
      integrated Eq.~(\ref{eq:RCHBCoupEq}) from $r=\infty$ inward to proper matching
      point $r=R_{match}$ by Runge-Kutta algorithms;
\item Requiring the $\psi(r)$ derived from the former two steps
      to be the same at $r=R_{match}$ will lead to the correct energy $E$;
\item Repeating the process till all the energies in the given
      spin-parity channel lying in $[0,E_{cut}]$
      have been found.
\end{enumerate}
Equations (\ref{eq:RCHBCoupEq}) for Gogny forces~(\ref{vpp2}) are
a set of four coupled integro-differential equations. They are
discretized in the space and solved by the finite element method
as the following:
\begin{enumerate}
\item Discretizing the whole space $[0,\infty]$ with $N+1$ points
      at $r_1=0,r_2=\rho, ... , r_{N+1} = N \times \rho$, where $\rho = R/N$.
\item Assuming $\psi(r)$ at $r_1=0,r_2=\rho, ... , r_{N+1} =
      N \times \rho$ are known, if $\rho = R/N$ is small enough, $\psi(r)$ in the
      element $(i-1)\times \rho \le r \le i\times \rho$ could be written as a
      simple linear combination of $\psi(r_i)$ and $\psi(r_{i+1})$;
\item Requiring Eqs.~(\ref{eq:RCHBCoupEq}) is valid in the element
      $(i-1)\times \rho \le r \le i\times \rho$, which could be easily
      integrated analytically and lead to a simple relation for
      $E$, $\psi(r_i)$ and $\psi(r_{i+1})$;
\item Repeating the process for all the $N$ element, one then is lead
      to a $4 \times (N+1)$ linear algebra equations for $E$,
      $G_V$,$F_V$,$G_U$,$F_U$ for $r=r_1,r_2, ... , r_{N+1}$, which
      is nothing but a generalized eigenvalue problem;
\item Diagonalizing this generalized eigenvalue problem one can get
      all the energies and their wave function discretized at
      $r=r_1,r_2, ... , r_{N+1}$.
\end{enumerate}
Normally, equations (\ref{eq:RCHBCoupEq}) and
(\ref{eq:SRHmesonmotion}) are solved in a self-consistent way by
the shooting method or finite element methods with a step size of
$0.1$ fm using proper boundary conditions in a spherical box of
radius $R \le 20$ fm. The Fermi-surface $\lambda$ is determined
self-consistently by the particle number~\cite{Meng98npa}. The
results do not depend on the box size for $R>15$
fm~\cite{Meng98prl}. For $\delta$-force~(\ref{vpp}), the number of
continuum levels is limited by a cut-off energy, which must be
larger than the depth of the potentials.

\subsubsection{Pairing correlation: finite range vs. zero range}

In the vicinity of the drip line, as the contribution from the
continuum is switched on, it is very challenging to have the
proper single particle levels and treat the coupling of the
continuum without any ambiguity. Therefore, a proper interaction
for the coupling between the bound state and the continuum is also
essential. In principle, the effective nucleon-nucleon interaction
should be obtained by means of Bruckner renormalization which
gives the correct interaction after modifying the free interaction
for the effect of the nuclear medium. In practice, however, the
effective interactions are approximated by some phenomenological
forces, such as Skyrme-type $\delta$-force or finite range Gogny
force. Delta force is relatively simple for numerical calculation
and has a realistic density-dependent behavior, but it allows a
coupling to the very highly excited states. Normally an energy
cutoff has to be introduced and the interaction strength has to be
properly justified. The Gogny force has a better treatment for the
coupling to the highly excited states, but it involves more
sophisticated numerical techniques.

\begin{figure}
\begin{center}
\includegraphics[width=6.0cm,,angle=270]{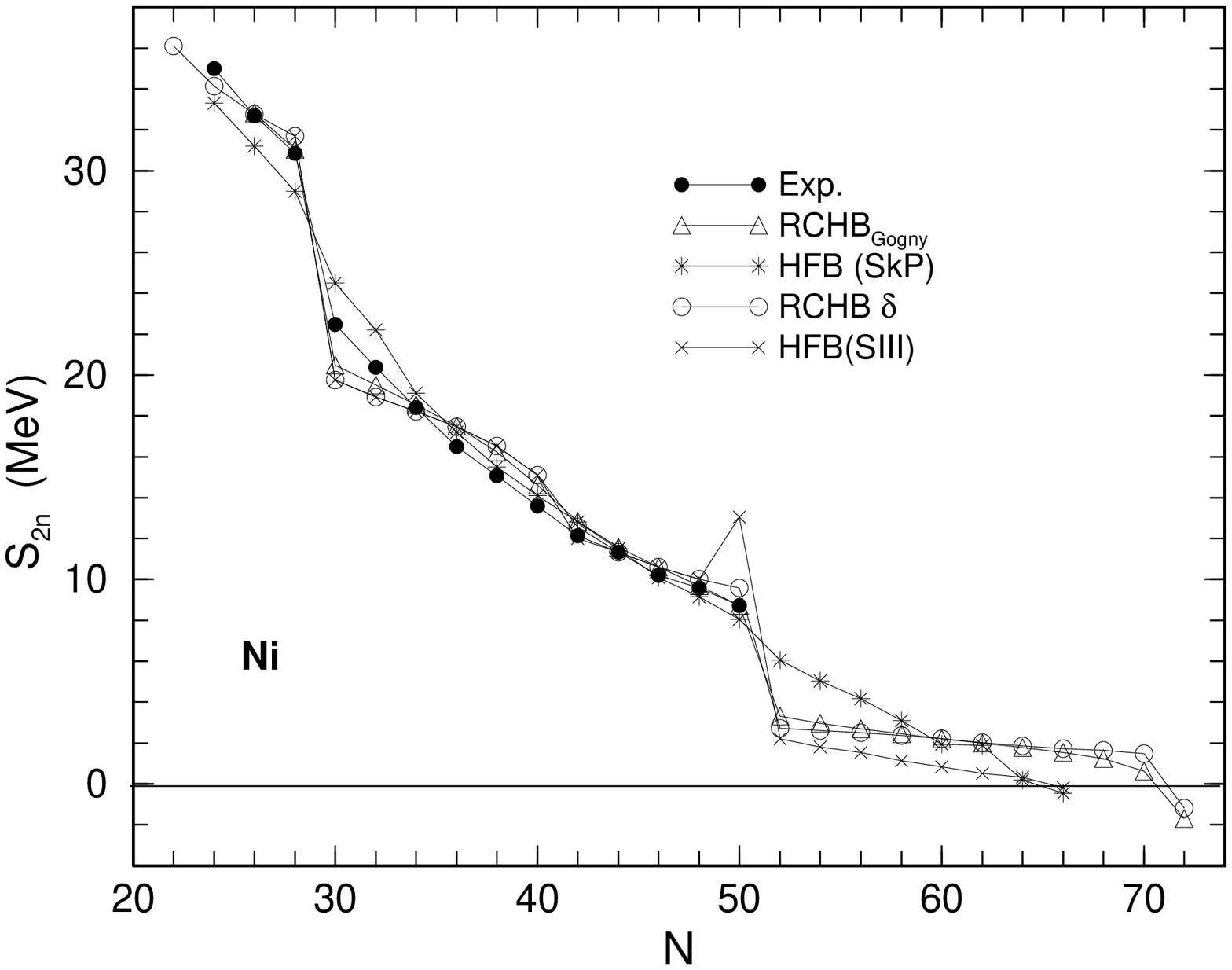}
\includegraphics[width=6.1cm,angle=270]{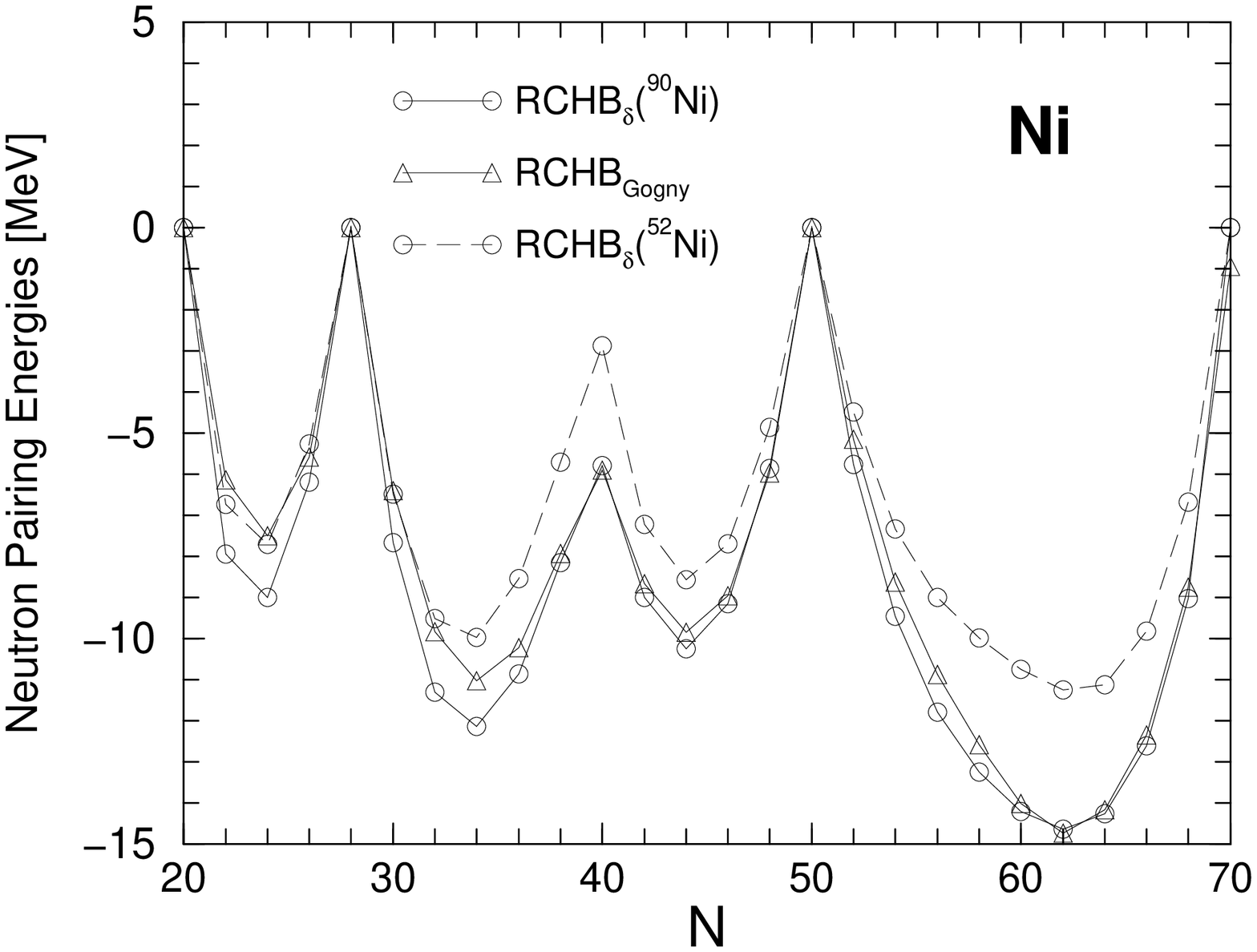}
\caption{Left: Two-neutron separation energies $S_{2n}$ of even Ni
isotopes as a function of $N$, including the experimental data
(solid points), RCHB with $\delta$-force (open circles), RCHB with
Gogny force (triangles) ,HFB with SkP interaction (stars) and HFB
with SIII interaction (pluses); Right: The neutron pairing
energies for RCHB with Gogny force (triangles) and $\delta$-force
(open circles). The open circles connected by dashed line or solid
line are respectively the pairing energies for RCHB with
$\delta$-force by fitting that of RCHB with Gogny force at
$^{52}$Ni or $^{90}$Ni. Taken from~\cite{Meng98prc}.}
\label{FigC6}
\end{center}
\end{figure}

The Skyrme- and Gogny-type pairing forces have been discussed
before within the conventional HFB formalism, but the HFB
equations for Gogny force are solved on a harmonic basis, which is
valid only for stable nuclei. In Ref.~\cite{Meng98prc}, the proper
form of the pairing interaction near the drip line is discussed in
the framework of the RCHB theory. The chain of even-even nickel
isotopes ranging from the proton drip line to the neutron drip
line are taken as examples. The pairing correlations are taken
into account by either a density-dependent force of zero range or
the finite range Gogny force. Through the comparisons of the two
neutron separation energies $S_{2n}$, the neutron, proton, and
matter rms radii, excellent agreements have been found between the
calculations with both interactions and the empirical values. In
Fig.~\ref{FigC6}, for example, the two neutron separation energies
$S_{2n}$ of even Ni isotopes are shown as a function of the
neutron number $N$ from the proton drip line to the neutron drip
line, including the experimental data (solid points), RCHB with
$\delta$-force (open circles), RCHB with Gogny force (triangles),
HFB with SkP interaction~\cite{Dobaczewski96} (stars) and SIII
interaction~\cite{Terasaki96} (pluses). The RCHB results with
$\delta$-force and Gogny force are almost identical. They show a
strong kink at $N$=28 and a less one at $N$=50. The drip line
nucleus is predicated at $^{100}$Ni in both calculations. The
empirical data is known only up to $N$=50. Comparing with the
available empirical data, the general trend and gradual decline of
$S_{2n}$ have been well reproduced.

The neutron pairing energies $-\frac{1}{2}\mbox{Tr}\Delta\kappa$
for the calculation of RCHB with Gogny force are given in
Fig.~\ref{FigC6} as triangles. The pairing energies demonstrate
the shell structure revealed by the $S_{2n}$ in Fig.~\ref{FigC6}.
They vanish at the closure shell and have a maximum values in the
middle of two neighboring closure shells. A tendency of the
enhancement of pairing energies appears for the neutron-rich
nuclei. The pairing energies for RCHB with $\delta$-force are
given by open circles. The circles connected by the solid line and
dashed line are calculations with $V_0 = 742$ MeV$\times$fm$^{-3}$
(by fitting the pairing energy of the Gogny force at $^{90}$Ni)
and $715$ MeV$\times$fm$^{-3}$ (by fitting at $^{52}$Ni),
respectively. The variation behavior of the pairing energies with
$N$ is quite similar, although $V_0$ fitted at $^{90}$Ni gives 2
MeV more for stable nuclei and $V_0$ fitted at $^{52}$Ni gives 4
MeV less for neutron rich nuclei. But these differences have
little influence on the two-neutron separation energies and the
rms radii, as one observes in Fig.~\ref{FigC6}, for example. This
clearly demonstrated the reliability of the result against the
pairing strength.

\begin{figure}
\begin{center}
\includegraphics[width=5.0cm,angle=270]{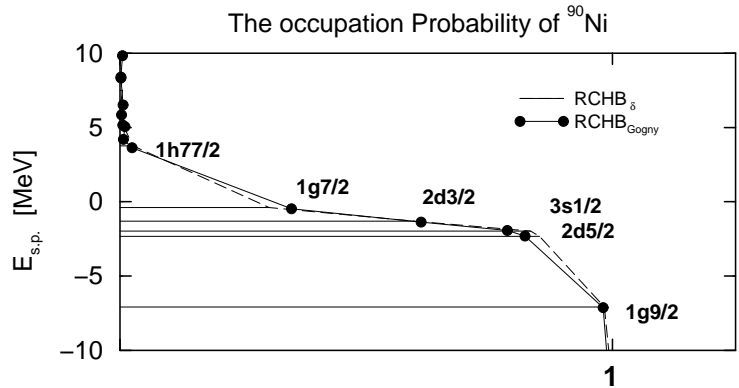}
\caption{The occupation probabilities in the canonical basis for
$^{90}$Ni as a function of the single particle energy around the
threshold, for RCHB with $\delta$-force (dashed line) and Gogny
force (solid points). Taken from~\cite{Meng98prc}.} \label{FigC7}
\end{center}
\end{figure}

Another interesting feature of the exotic nuclei is the
contribution from the continuum. Definitely it is very interesting
to see how different the contribution from the continuum is for
the calculations with Gogny force and $\delta$-force. In
Fig.~\ref{FigC7} the occupation probabilities of $^{90}$Ni in the
canonical basis is shown~\cite{Ring80} for the neutron levels near
the Fermi surface, i.e. in the interval $-10 \le E_{s.p.} \le 10 $
MeV. The occupation probabilities for RCHB $\delta$-force are
represented by the dashed line and RCHB with Gogny force by the
solid points. For $V_0 = 715$ MeV$\times$fm$^{-3}$ or  $V_0 = 742$
MeV$\times$fm$^{-3}$, the occupation probabilities are the same.
Even the difference from the Gogny and the $\delta$-force is too
small to have any influence on the physical observables like
$S_{2n}$ and rms radii.

It has been shown that after proper renormalization (e.g., fixing
the corresponding pairing energies), physical observables such as
$S_{2n}$ and rms radii remain the same, even within a reasonable
region of the interaction strength~\cite{Meng98prc}.


\section{Exotic nuclei}
\label{sec:exotic}

The study of exotic nuclei has attracted world wide attention due
to their large $N/Z$ ratios (isospin) and interesting properties
such as halo and skin. It is also important for other fields
including astrophysics, e.g., the properties of exotic nuclei are
essential to understand the nucleosynthesis in the {\it r}
process. Since the first case of halo in an exotic nucleus
$^{11}$Li was observed with RIB in 1985~\cite{Tanihata85}, more
and more exotic nuclei have been investigated with various modern
experimental methods to understand this attractive phenomenon
better~\cite{Mueller93, Tanihata95, Hansen95, Jonson04, Jensen04}.
For nuclei far from the $\beta$-stability valley and with small
nucleon separation energy, the valence nucleons in exotic nuclei
extend over quite a wide space to form low density nuclear matter.
Furthermore, the Fermi surface for exotic nuclei is usually very
close to the continuum threshold. The valence nucleons could be
easily scattered to the continuum states due to the pairing
correlation. Thus, theories which can properly handle the pairing
and continuum states are needed to describe the properties of
exotic nuclei. One of such theories is the RCHB
theory~\cite{Meng96, Meng98npa} introduced in the previous
section. In this section, the RCHB theory will be applied to study
the properties of exotic nuclei, e.g., the binding energies,
particle separation energies, the radii and cross sections, the
single particle levels, shell structure, the restoration of the
pseudo-spin symmetry, the halo and giant halo, and halos in hyper
nuclei, etc.

\subsection{Binding and separation energies}

The self-consistent microscopic RCHB theory has been used to study
the ground state properties systematically such as binding energy,
separation energy, radius, density distribution, single particle
spectrum and shell structure etc. for a large number of isotope
chains from the proton drip line to the neutron drip line
throughout the periodic table including Li, C, N, O, F, Na, Ca,
Ni, Zr, Sn and Pb~\cite{Meng96, Meng98prl, Meng98npa, Meng98plb,
Meng98prc, Meng99npa, Meng99prc, Meng02plb, Meng02r, Zhang02a,
Zhang03, Lv03}. In this subsection, we will briefly present the
description of the binding energies and separation energies for
some typical spherical nuclei in RCHB theory.

Ground state properties of all the even-even O, Ca, Ni, Zr, Sn, Pb
isotopes ranging from the proton drip line to the neutron drip
line were investigated with the RCHB theory~\cite{Zhang03}. The
binding energy $E_b$ calculated from the RCHB theory for these
isotope chains and the corresponding data available~\cite{Audi95}
are compared. Generally speaking, the calculations reproduce the
available data of binding energy quite well. The errors are
usually within 3-4 MeV which is less than $1\%$ of the
experimental values. Since the deformation effects are not
included in the RCHB calculations yet, for some nuclei with large
deformation, larger discrepancies ($\Delta E_b\sim 10$ MeV) of
binding energy from experimental values are found, e.g., in some
Zr isotopes.

\begin{figure}
\centering
\includegraphics[width=8.0cm]{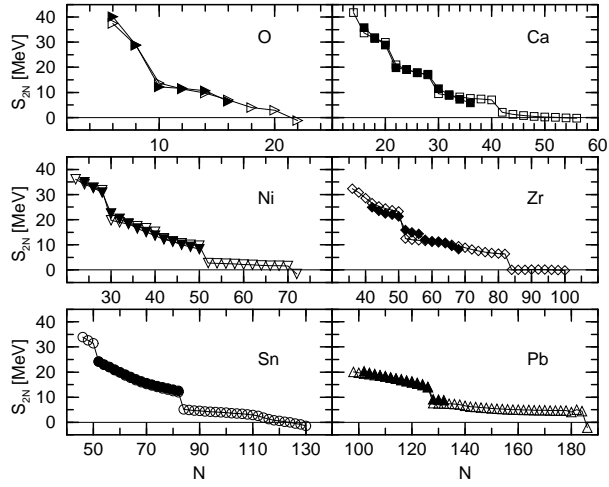}
\caption{The two neutron separation energies $S_{2n}$ for the
proton magic isotopes plotted as a function of the neutron number
in RCHB with NLSH. The experimental values are denoted by the
solid symbols and the calculated ones by the open symbols. Taken
from Ref.~\cite{Zhang03}.} \label{FigD1}
\end{figure}

In Table~\ref{TabD1}, the binding energies for Pb isotopes
calculated with different effective interactions PK1, PK1R, PKDD,
TM1, NL3, TW99 and DD-ME1 are given in comparison with the
experimental values~\cite{Long04}. Many effective interactions,
particularly PK1, PK1R and PKDD provide good descriptions of the
available data. All the effective interactions overestimate the
binding energy in the beginning of the isotopic chain. However,
from $^{190}$Pb to $^{210}$Pb, the effective interactions PK1,
PK1R and PKDD give better descriptions than all the others. There
exist however systematic deviations out of the neutron magic
numbers, e.g., in $^{214}$Pb. For the effective interactions TM1,
NL3, TW99 and DD-ME1, the deviations in $^{214}$Pb are smaller but
fairly large in $^{208}$Pb.

The one or two nucleon separation energies $S_{p(n)}$ or
$S_{2p(2n)}$ are quite sensitive quantities to test a microscopic
theory. In Fig.~\ref{FigD1} both the theoretical and the available
experimental $S_{2n}$ are presented as a function of neutron
number $N$ for the O, Ca, Ni, Zr, Sn, Pb isotope chains,
respectively. The good coincidence between experiment and
calculation is clearly seen. Seen along the $S_{2n}$ versus $N$
curve in Fig.~\ref{FigD1}, strong kinks can be clearly seen at
$N=8$, $N=20, 28, 40$, $N=28,50$, $N=50,82$ and $N=126$ for O, Ca,
Ni, Zr, Sn, Pb isotope chains, respectively. All these numbers
correspond to the neutron magic numbers, which come from the large
gaps of single particle energy levels.

\begin{table}[htbp]
\begin{minipage}[t]{1.0\textwidth}
\renewcommand{\footnoterule}{\rule{0pt}{0pt}\vspace{-7.5pt}}
\centering \caption{Total binding energies of Pb isotopes (in MeV)
calculated with the nonlinear self-coupling effective interactions
PK1, PK1R and the density-dependent meson-nucleon coupling
effective interaction PKDD~\cite{Long04}, in comparison with the
experimental values~\cite{Audi95} and the results of TM1, NL3,
TW99 and DD-ME1. }\label{TabD1}
\begin{tabular}{ccccccccc}
\hline\hline
$A$ &  Exp.       & PK1         &      PK1R   &   PKDD      &         NL3 & TM1         &   TW99      &  DD-ME1     \\
\hline
182 & $$1411.650 & $$1416.431 & $$1416.619 & $$1416.309 & $$1415.184 & $$1420.872 & $$1417.202 & $$1415.216 \\
184 & $$1431.960 & $$1435.548 & $$1435.706 & $$1435.478 & $$1434.569 & $$1439.768 & $$1436.761 & $$1434.569 \\
186 & $$1451.700 & $$1454.258 & $$1454.382 & $$1454.192 & $$1453.511 & $$1458.222 & $$1455.879 & $$1453.306 \\
188 & $$1470.900 & $$1472.586 & $$1472.673 & $$1472.504 & $$1472.056 & $$1476.272 & $$1474.640 & $$1471.639 \\
190 & $$1489.720 & $$1490.555 & $$1490.603 & $$1490.468 & $$1490.247 & $$1493.958 & $$1493.109 & $$1489.657 \\
192 & $$1508.120 & $$1508.197 & $$1508.201 & $$1508.116 & $$1508.130 & $$1511.317 & $$1511.329 & $$1507.411 \\
194 & $$1525.930 & $$1525.536 & $$1525.494 & $$1525.474 & $$1525.733 & $$1528.378 & $$1529.309 & $$1524.937 \\
196 & $$1543.250 & $$1542.592 & $$1542.502 & $$1542.545 & $$1543.085 & $$1545.163 & $$1547.012 & $$1542.262 \\
198 & $$1560.070 & $$1559.378 & $$1559.236 & $$1559.329 & $$1560.199 & $$1561.685 & $$1564.338 & $$1559.403 \\
200 & $$1576.365 & $$1575.893 & $$1575.697 & $$1575.831 & $$1577.076 & $$1577.942 & $$1581.344 & $$1576.365 \\
202 & $$1592.202 & $$1592.095 & $$1591.843 & $$1592.022 & $$1593.679 & $$1593.905 & $$1598.084 & $$1593.133 \\
204 & $$1607.520 & $$1607.851 & $$1607.545 & $$1607.770 & $$1609.906 & $$1609.477 & $$1614.197 & $$1609.676 \\
206 & $$1622.340 & $$1623.126 & $$1622.765 & $$1623.167 & $$1625.725 & $$1624.530 & $$1630.122 & $$1625.966 \\
208 & $$1636.446 & $$1637.443 & $$1637.024 & $$1637.387 & $$1640.584 & $$1638.777 & $$1644.790 & $$1641.415 \\
210 & $$1645.568 & $$1644.844 & $$1644.345 & $$1643.643 & $$1647.969 & $$1647.245 & $$1650.888 & $$1648.312 \\
212 & $$1654.524 & $$1652.155 & $$1651.573 & $$1649.873 & $$1655.289 & $$1655.549 & $$1657.020 & $$1655.174 \\
214 & $$1663.298 & $$1659.382 & $$1658.718 & $$1656.084 & $$1662.551 & $$1663.706 & $$1663.196 & $$1662.011 \\
\hline
 $\sigma$
&& 0.00033 & 0.00036 & 0.00042 & 0.00031 & 0.00068 & 0.00078 & 0.00033 \\
\hline\hline
\end{tabular}
\end{minipage}
\end{table}

For the proton magic even mass nuclei studied in
Ref.~\cite{Zhang03}, the neutron drip line nuclei are predicted as
$^{28}$O, $^{72}$Ca, $^{98}$Ni, $^{136}$Zr, $^{174}$Sn, and
$^{266}$Pb, respectively. In Refs.~\cite{Fayans00,Im00}, the last
bound nucleus in neutron rich Ca isotopes is predicted as
$^{70}$Ca with the HFB and Skyrme HF methods, respectively. The
reason for that $^{72}$Ca is bound is caused by the halo effect,
which also results in the disappearance of the normal $N=50$ magic
number in the RCHB calculation as shown in Fig.~\ref{FigD1} where
no kink appears at $N=50$ for Ca isotopes. In stable nuclei, the
$N=50$ magic number is given by the big energy gap between the
$1g_{9/2}$ and $3s,~2d$ levels. However in the vicinity of the
drip line region of Ca isotopes, the single particle energy of
$3s_{1/2}$ would decrease with the halo effect and its zero
centrifugal potential barrier, the large gap between $1g_{9/2}$
and $3s_{1/2}$ disappears. Therefore, the nucleus $^{70}$Ca is no
more a doubly magic nucleus and $^{72}$Ca is bound. This
disappearance of the $N=50$ magic number at the neutron drip line
is due to the halo property. This is different from the
disappearance of the $N=20$ magic number which is due to
deformation. Recently a new magic number $N=16$ has been
discovered in neutron drip line light-nuclei region which is also
considered as owing to the halo effect~\cite{Ozawa00}. It can be
expected that near drip line, more traditional magic numbers would
disappear and new magic numbers would be found with the same
mechanism.

\begin{figure}
\centering
\includegraphics[width=8.0cm]{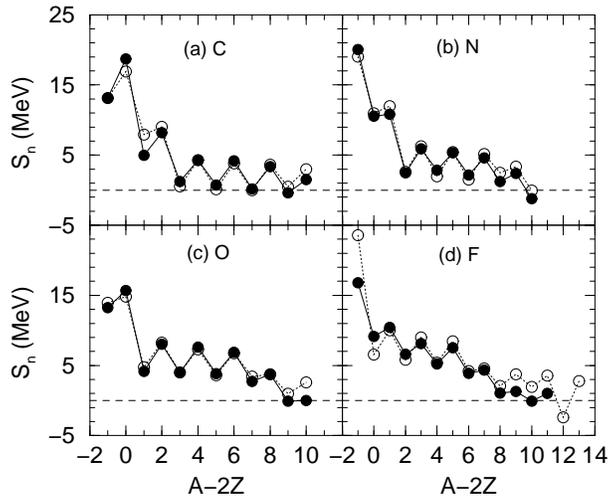}
\caption{The one neutron separation energies $S_{\rm n}$ for the
nuclei $^{11-22}$C,$^{13-24}$N,$^{15-26}$O and $^{17-25}$F by RCHB
theory(open circles) and their experimental counterparts (solid
circles). Taken from Ref.~\cite{Meng02plb}.} \label{FigD2}
\end{figure}

Systematic RCHB calculations with NLSH have been carried out for
the C, N, O and F isotopes~\cite{Meng02plb}. The one neutron
separation energies $S_{\rm n}$ predicted by RCHB and the
data~\cite{Audi95} for the nuclei $^{11-22}$C, $^{13-24}$N,
$^{15-26}$O and $^{17-25}$F are shown in Fig.~\ref{FigD2} as open
and solid circles respectively. For carbon isotopes, the
theoretical one neutron separation energies for $^{11-18,20,22}$C
are in agreement with the data. The calculated $S_{\rm n}$ is less
than 0 ($-$0.003 MeV) for the odd-$A$ nucleus $^{19}$C which is
bound in experiment. While for the experimentally unbound nucleus
$^{21}$C, the predicted value of $S_{\rm n}$ is positive. From the
neutron-deficient side to the neutron drip line, excellent
agreement has been achieved for the nitrogen isotopes. As almost
all relativistic mean field approaches do, the RCHB calculations
overestimate binding for $^{25,26}$O which are unstable in
experiment. For fluorine isotopes, the $S_{\rm n}$ in
$^{17,26-29}$F are overestimated in contrast with the
underestimated one in $^{18}$F. The last bound neutron-rich
nucleus is predicted as $^{30}$F. In general, the RCHB theory
reproduces the $S_{\rm n}$ data well considering that this is a
microscopic and almost parameter free model. As discussed above,
there are some discrepancies between the calculations and the
empirical values for some of the studied isotopes. This may be due
to deformation effects, which has been neglected in
Ref.~\cite{Meng02plb}.

\subsection{Radius, nuclear density distribution, and cross section}

\subsubsection{Nuclear root mean square radii}

The nuclear radii are important basic physical quantities to
describe atomic nuclei. The calculated neutron, proton, matter,
and charge radii for all even Ca isotopes against mass number $A$
are plotted in Fig.~\ref{FigD3}. For nuclear charge radii $r_c$,
the well-known parabolic behavior along $^{40-48}$Ca is not
reproduced in RCHB calculation. It is mainly due to the improper
spherical supposition to the real deformed $^{42, 44, 46}$Ca
nuclei. It can be clearly seen that the radii $r_p$, $r_n$, $r_m$
and $r_c$ all increase with the mass number $A$. The mass radii
$r_m$ as well as $r_n$ increase much faster than $r_p$ and $r_c$.
In addition to the normal increase of $r_n$ and $r_m$ with $A$, a
slightly up-bend tendency occurs in the neutron drip line region
from $A= 62$ up to $A=72$.

\begin{figure}
\centering
\includegraphics[width=7.0cm]{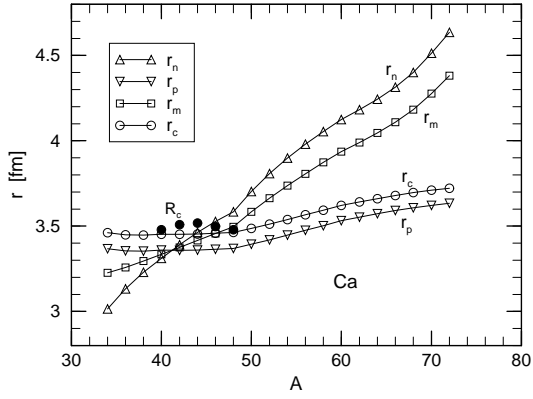}
\caption{The rms radii for the proton, neutron, charge and matter
distributions are plotted as a function of the neutron number for
the Ca isotope.  The available experimental values are denoted by
solid symbols. Taken from Ref.~\cite{Zhang03}.} \label{FigD3}
\end{figure}

For light nuclei, the RCHB calculations predict a slight increase
of proton or charge radii when the proton drip line is
approached~\cite{Meng02plb}. The neutron and proton rms radii
predicted by RCHB for the nuclei $^{10-22}$C,
$^{12-24}$N,$^{14-26}$O and $^{16-25}$F are given in
Fig.~\ref{FigD4}. The neutron radii for nuclei in each isotope
chain increase steadily. While the corresponding proton radii
remains almost constant for nuclei in each isotope chain except
for proton rich ones. The increase of proton radii with neutron
number decreasing in the proton rich side in each isotope chain is
responsible for the slightly increase in the charge changing cross
section of these nuclei on the target $^{12}$C as discussed in
Section~\ref{subsec:crosssection}.

\begin{figure}
\centering
\includegraphics[width=8.0cm]{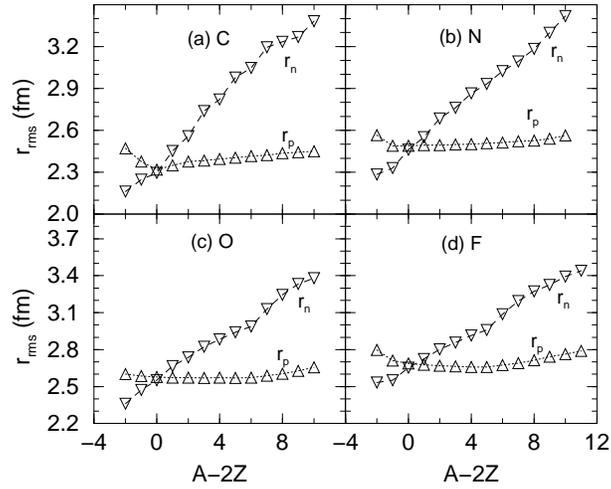}
\caption{The neutron and proton rms radii predicted by RCHB for
the nuclei $^{10-22}$C, $^{12-24}$N,$^{14-26}$O and $^{16-25}$F.
Taken from Ref.~\cite{Meng02plb}.} \label{FigD4}
\end{figure}

In Fig.~\ref{FigD5} the neutron radii $r_n$ from the RCHB
calculation for the even-even nuclei of the O, Ca, Ni, Zr, Sn and
Pb isotope chains are ploted. The predicted $r_n$ curve using the
simple empirical equation $r_n=r_0\cdot N^{1/3}$ with $r_0=1.139$
normalizing to $^{208}$Pb is also represented in the figure. The
simple formula for $r_n$ agrees with the calculated neutron radii
except for two anomalies. One anomaly appears in the Ca chain
above $N=40$, and the other in Zr chain above $N=82$. The increase
of exotic Ni and Sn nuclei is not as rapid as that in Ca and Zr
chains. The regions of the abnormal increases of the neutron radii
are just the same as those for $S_{2n}$.  Both behaviors are
connected with the formation of giant halo (see
Section~\ref{subsec:halo} for details).

\begin{figure}
\centering
\includegraphics[width=10.0cm]{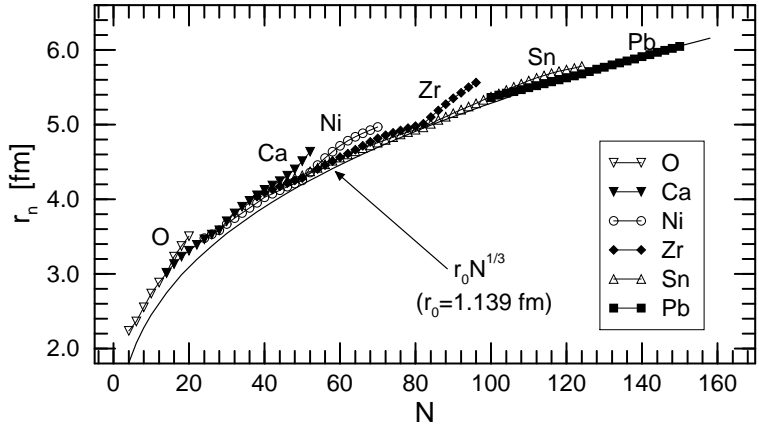}
\caption{The rms neutron radii for the proton magic even mass
nuclei plotted as a function of the neutron number. The solid
curve corresponds to the radii calculated with the $N^{1/3}$
formula.} \label{FigD5}
\end{figure}

It is interesting to investigate the neutron number dependence of
the proton radii (or charge radii) and the neutron radii for these
proton-magic isotope chains. In Fig.~\ref{FigD6}, the rms charge
radii $r_c$ obtained from the RCHB theory (open symbols) and the
data available (solid symbols) for the even-even Ca, Ni, Zr, Sn
and Pb isotopes are shown. The RCHB calculations reproduce the
data very well (within 1.5$\%$). For a given isotopic chain, an
approximate linear $N$ dependence of the calculated rms charge
radii $ r_c$ is clearly seen. However, the variation of $r_c$ for
a given isotopic chain deviates from the general accepted simple
$A^{1/3}$ law (denoted by dashed lines in Fig.~\ref{FigD6}), which
shows a strong isospin dependence of nuclear charge radii for
nuclei with extreme $N/Z$ ratio. Recently, these charge radii as
well as the large amount of experimental data are investigated. A
formula with the $Z^{1/3}$ dependence and isospin dependence is
proposed which describes well the charge radii all over the
nuclear isotope chart, see subsection
\ref{Z-law}.~\cite{Zhang02b}.

\subsubsection{Neutron skin thickness}

The measurement of nucleon density distributions provides basic
and important information in nuclear structure. The proton
distribution of stable nuclei can be determined accurately by
elastic electron or muon scattering. However, there is not
comparable measurement of the neutron density distribution up to
now. Although it is difficult to obtain the neutron density
distribution, the difference in radii of the neutron and proton
density distributions can be more easily
determined~\cite{Krasznahorkay99}. Theoretically, the neutron skin
thickness, $r_{\mathrm n} - r_{\mathrm p}$ could place important
constraints on effective interactions used in nuclear
models~\cite{Horowitz01, Horowitz01PRL, Shen05}.

Many investigations have been done aiming at the description of
the neutron skin thickness in the framework of the RMF model. With
the pairing correlation treated by a constant BCS approach, the
neutron rms radii of some closed and open shell nuclei are
investigated with the spherical RMF model~\cite{Sharma92}. It has
been demonstrated that with an increase in isospin of nuclei, the
neutron rms radii in the relativistic mean-field approach with the
parameter sets NL1 and NL2 are larger than both the empirical data
and the predictions of the Skyrme mean field~\cite{Sharma92b}.
This overestimation of the neutron skin thickness remains true for
the effective interaction NL3, although the relativistic
Hartree-Bogoliubov model with NL3+D1S give a good result of the
neutron-skin thickness along the Sn isotopic chain compared to the
available data~\cite{Vretenar03b, Vretenar03a}.

The overestimation of the neutron skin thickness is attributed to
a poor description of the isovector channel of the nonlinear
effective interaction in Ref.~\cite{Niksic02b} where the RHB model
is extended to include density-dependent meson-nucleon couplings.
A density-dependent effective interaction DD-ME1 is proposed in
which the improved isovector properties result in a better
description of charge radii, and especially the calculated values
of neutron thickness are in much better agreement with
experimental data. As what is shown in Fig.~\ref{FigD6}, the
difference between the values calculated with NL3 and DD-ME1
increases with the number of neutrons to about 0.1 fm at $N$ = 82,
but then it remains practically constant for $N >$ 82. While both
interactions reproduce the isotopic trend of the experimental
data, NL3 obviously overestimates the neutron skin. The values
calculated with DD-ME1, on the other hand, are in excellent
agreement with the experimental data. This result presents a
strong indication that the isovector channel of the effective
interaction DD-ME1 is correctly parametrized.

\begin{figure}
\centerline{
\includegraphics[width=8.0cm]{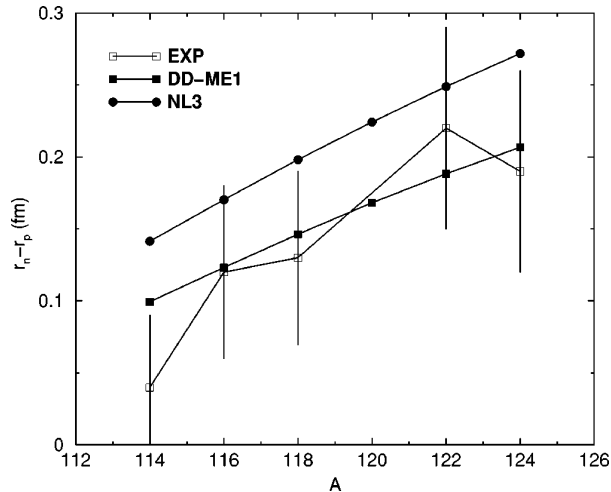}}
\caption{DD-ME1 and NL3 predictions for the neutron skin thickness
of Sn isotopes, compared with experimental data. Taken from
Ref.~\cite{Niksic02b}.} \label{FigD6}
\end{figure}

In Ref.~\cite{Shen05}, the sensitivity of the neutron skin
thickness $S=r_{\mathrm n} - r_{\mathrm p}$ in $^{208}$Pb to the
addition of nucleon-$\sigma$-$\rho$ coupling corrections to a
selection (PK1, NL3, S271, and Z271) of interactions in RMF model
was investigated. With the higher order nucleon-$\sigma$-$\rho$
coupling, the PK1 and NL3 effective interactions lead to a minimum
value of the neutron skin thickness $S$ = 0.16 fm and the S271 and
Z271 effective interactions yield even smaller values of $S$ =
0.11 fm. The addition of these higher order correction terms to
the conventional RMF models can be associated with some kind of
multimeson exchange processes occurring in the inner (higher
density) region of nuclei. In the inner region of a heavy nucleus
such as $^{208}$Pb, one could imagine that multimeson exchange
processes introduced through nucleon-$\sigma$-$\rho$ coupling
corrections could play an important role in modifying the
short-range behavior of the nucleon-nucleon interaction, by
softening the density dependence of the symmetry energy at high
density (the inner part of nuclei) and thus leading to smaller
values of neutron skin thickness $S$ compared to previous
effective interactions.

\subsubsection{Nuclear charge radii: isospin dependence and the $Z^{1/3}$ law}
\label{Z-law}

 Among all the size quantities describing nucleus, nuclear
charge radius has been measured with very high precision by
various techniques and methods experimentally. With its accuracy,
the study of the nuclear charge radii is very important to
understand not only the proton distribution inside the nucleus but
also the exotic phenomena such as halo and skin observed in exotic
nuclei. Particularly if one can get a simple and reliable formula
for nuclear charge radii, it will be very useful to extract the
decoupling of proton and neutron in the exotic nuclei and provide
information for the effective nucleon-nucleon interaction widely
used in all the nuclear models. The available charge radii data
for $A \geq 40$ were examined and its global behavior was studied
in Ref.~\cite{Zhang02b}. Instead of the widely accepted $A^{1/3}$
law, a new $Z^{1/3}$ formula with isospin effect has been
proposed.

Based on the consideration of the nuclear saturation property,
nuclear charge radius $R_c$ is usually described by the $A^{1/3}$
law~\cite{Bohr69,Ring80}
\begin{equation}
 R_c=r_A A^{1/3},
 \label{eqn.A}
\end{equation}
where $A$ is the mass number and $R_c=\sqrt{5/3}~r_c$, with $r_c$
the rms charge radius. For very light nuclei, because of their
small $A$ and large fluctuation in charge distribution due to the
shell effect with short period, it seems that the charge
distribution radius as a bulk property has little meaning. A
detailed analysis of charge radii data for $A\geq 40$ shows that
$r_A$ is by no means a constant, but systematically decreases with
$A$; i.e., $r_A \approx 1.31$ fm for light nuclei $(A\sim 40)$ and
$r_A\approx 1.20$ fm for very heavy nuclei (see upper left panel
in Fig. ~\ref{FigD7}). This fact implies that some physics is
missing in Eq.~(\ref{eqn.A}).

Definite evidences of the violation of $A^{1/3}$ law have been
found in the measurements of isotope shift in rms charge
radii~\cite{Zeng57,Zeng75,Pomorski93}. In particular,
$\delta\langle r^2\rangle_{A+2,A}$ values (associated with an
addition of two neutrons) are often found to be considerably
smaller compared to what is expected from the $A^{1/3}$ law
($\delta \langle r^2\rangle_{A+2,A}=\frac{4}{3A}\langle
r^2\rangle_{A}$). A typical example is that the observed charge
radii of the calcium isotopes $^{40-50}$Ca remain almost the same
(except a very little change induced by deformation or shell
effect), though the mass number $A$ has changed significantly. In
contrast, there is also evidence that the observed $\delta\langle
r^2\rangle_{A+2,A}$ values (associated with the addition of two
protons) are often greater than what is expected from the
$A^{1/3}$ law (e.g., $\delta\langle r^2\rangle$ for
$^{46}$Ti$-^{44}$Ca, $^{50}$Ti$-^{48}$Ca, etc.).

Along the $\beta$-stability line, the ratio $Z/A$ gradually
decreases with $A$, i.e., for light nuclei $Z/A\approx 1/2$, and
for the heaviest $\beta$-stable nucleus $^{238}_{\ 92}$U,
$(Z/A)^{1/3}\approx 0.7285$. It turns out that
$(1/2)^{1/3}/0.7285\approx 1.09$, which is very close to the $r_A$
ratio $1.30/1.20$ shown in the upper left panel of
Fig.~\ref{FigD7}. A naive point of view is that the charge radius
of a nucleus may be more directly related to its charge number
$Z$, rather than its mass number $A$. Therefore, compared to the
$A^{1/3}$ law, a $Z^{1/3}$ dependence for nuclear charge radii may
be more reasonable,
\begin{equation}
R_c = r_Z Z^{1/3}, \label{eqn.Z}
\end{equation}
as noted in Refs.~\cite{Zeng57,Zeng75}. An analysis of the very
limited data of charge radii then available showed that $r_Z$
remains almost a constant, i.e., $r_Z=1.65(2)$ fm for $A\geq 40$.
Similar $Z^{1/3}$ dependent formula for nuclear charge radii was
also proposed in Refs.~\cite{Gambhir86a,Gambhir86b}. With the
physics for such a simple $Z^{1/3}$ law open, one notes here that
the $Z^{1/3}$ law of $R_c$ includes an isospin dependence compared
to the $A^{1/3}$ law as following:
\begin{eqnarray}
 R_c &=& r_Z Z^{1/3}
         = r_Z \left(\frac{A}{2} - \frac{N-Z}{2}\right)^{1/3}
         = \frac{r_Z}{2^{1/3}} A^{1/3} \left(1-\frac{N-Z}{A}\right)^{1/3}
           \nonumber \\
&\approx&  \frac{r_Z}{2^{1/3}} A^{1/3}
           \left(1 - \frac{1}{3}\frac{N-Z}{A}\right).
 \label{eqn.AZ}
\end{eqnarray}
One can get $r_A = r_Z/2^{1/3} \approx 1.31$ fm and 0.33 for the
coefficient of the isospin term. Equation~(\ref{eqn.AZ}) is
approximately equivalent to that obtained in
Ref.~\cite{Pomorski93}, $R_c = 1.25 A^{1/3} \left(1 - 0.2
\frac{N-Z}{A}\right)$, by fitting the data. Obviously
Eq.~(\ref{eqn.Z}) has merits that only one parameter is needed.

The $Z^{1/3}$ dependence of nuclear charge radii was also used to
modify the Coulomb energy term in the semi-empirical nuclear mass
formula~\cite{Zeng80}, and it was found that the agreement between
the calculated and experimental results was improved. Moreover,
the $A^{-1/3}$ law for the nuclear giant (monopole, dipole and
quadrupole) resonance energy ($\propto 1/R$) also could be
improved, if the $A^{-1/3}$ dependence is replaced by a $Z^{-1/3}$
dependence~\cite{Zeng82}.

In the past two decades, a vast amount of new experimental
information on the electromagnetic structure of nuclear ground
states of many nuclei has become available with high accuracy. The
vast amount of improved experimental results have been
investigated by using the $A^{1/3}$ and $Z^{1/3}$
dependence~\cite{Zhang02b}. The measured charge radii for 536
nuclei with $A\geq 40$ are shown in Fig.~\ref{FigD7}. The
dependence of charge radii on the quadrupole deformation $\beta$
has been taken into account for the rare-earth deformed nuclei.

\begin{figure}
\centering
\includegraphics[width=10.0cm]{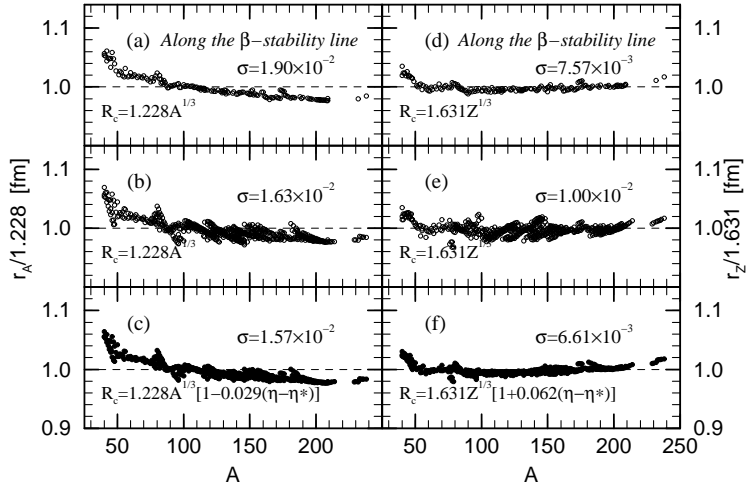}
\caption{Comparison of descriptions of some formulas for nuclear
charge radius $R_c$. The parameters in the first four formulas are
obtained through least square fitting to data on charge radii of
nuclei with $A \geq 40$. Taken from Ref.~\cite{Zhang02b}.}
\label{FigD7}
\end{figure}

In the upper left and right panels of Fig.~\ref{FigD7}, the charge
radii for the most stable 159 nuclei with $A\geq 40$ along the
$\beta$-stability line have been analyzed by using the $A^{1/3}$
and $Z^{1/3}$ dependence. In the middle left and right panels, the
same has been done for the data of 536 nuclei with $A\geq 40$. Two
significant features can be observed: (A) On the one hand, the
agreement between the data and the calculated results using the
$Z^{1/3}$ dependence is much better than that using the $A^{1/3}$
law, i.e., while there exists a global regular decrease of $r_{A}$
with $A$, $r_{Z}$ nearly remains constant ($r_{Z}=1.631(11)$ fm).
The relative rms deviations $\sigma$ for the $Z^{1/3}$ dependence
are much less than those for the $A^{1/3}$ law. (B) On the other
hand, though the rms deviation for the $Z^{1/3}$ dependence is
significantly reduced, an isospin induced scattering of the data
in the middle panels in Fig. \ref{FigD7} can be also observed
compared with that in the top panels.  In
Refs.~\cite{Pomorski93,Pomorski94,Dobaczewski96,Warda98}, the
isospin effects has been considered based on the $A^{1/3}$ law. As
the $Z^{1/3}$ dependence can describe the nuclear charge radii
much better than the $A^{1/3}$ law, the $Z^{1/3}$ dependence is a
more reasonable starting point for describing the isospin
dependence of nuclear charge radii.

To make the isospin dependent $Z^{1/3}$ formula for nuclear charge
radii be based on more strong foundation and be also valid for
exotic nuclei, the charge radii of nuclei far from the
$\beta$-stability line up to drip lines are needed. However, such
data is not well available, an alternative is to require that the
new isospin dependent $Z^{1/3}$ formula should assort with a
reliable and microscopic nuclear model such as the fully
self-consistent and microscopic RCHB theory.

\begin{figure}
\centering
\includegraphics[height=5.0cm]{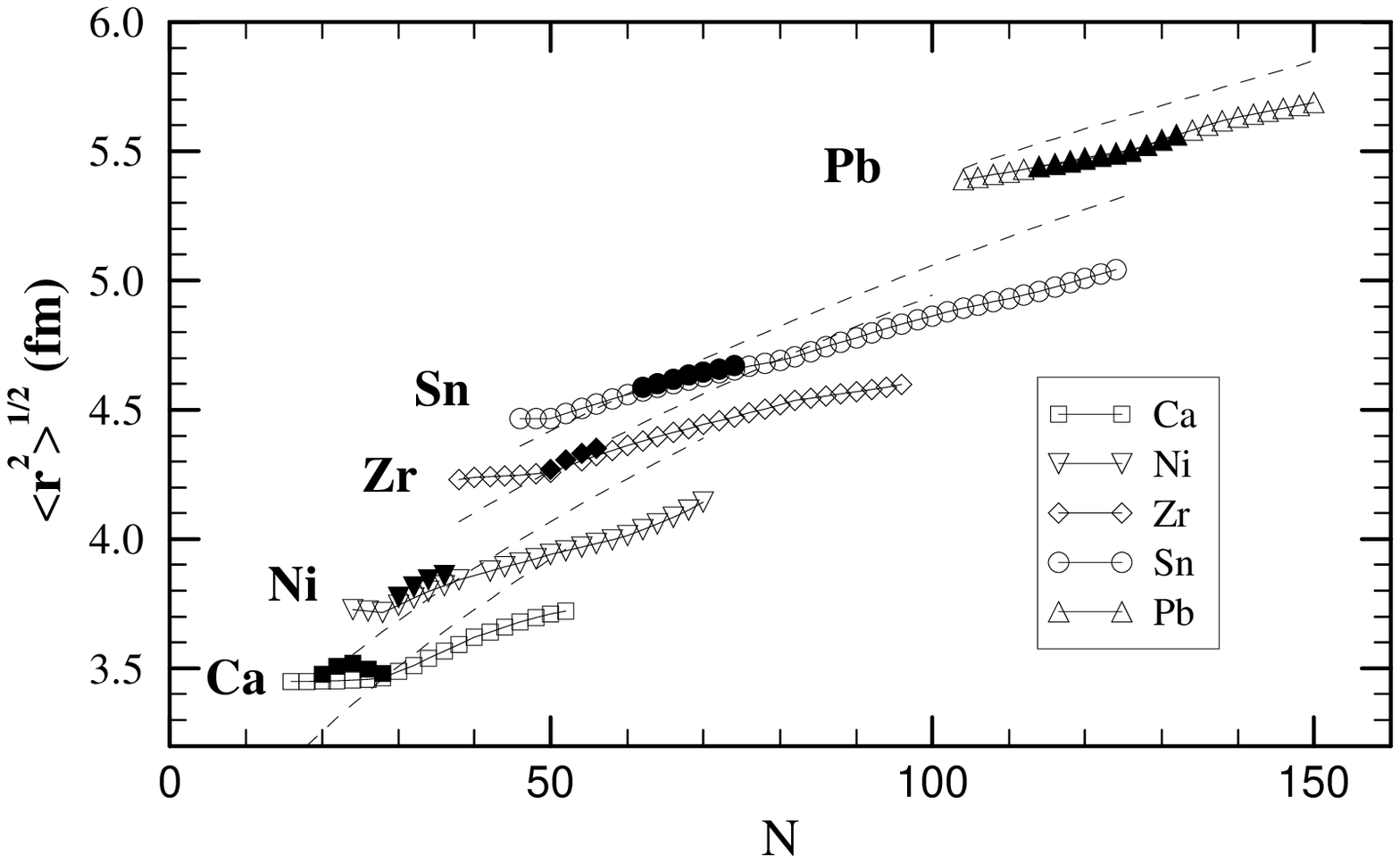}
\includegraphics[height=5.0cm]{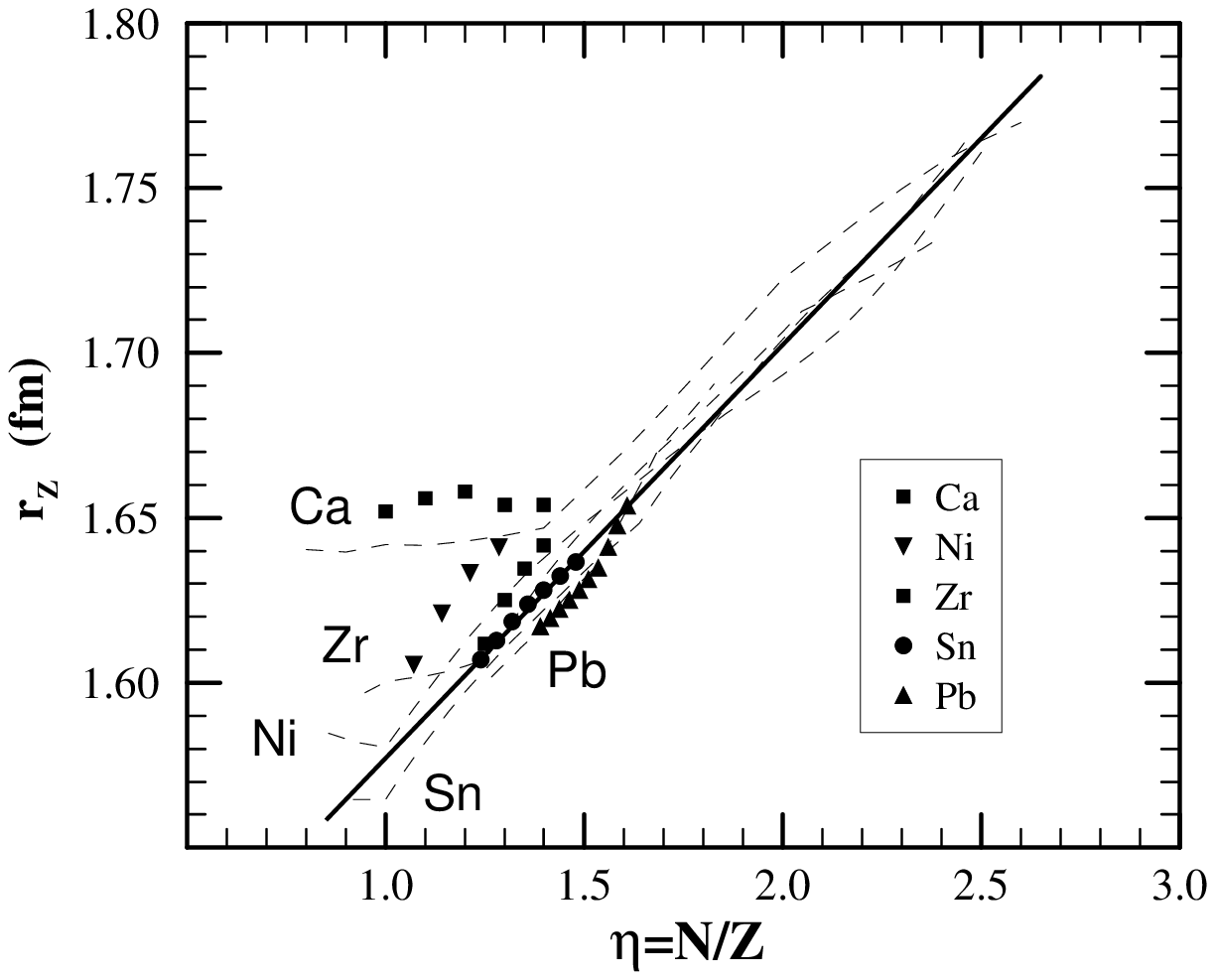}
\caption{Left: The rms charge radii versus the neutron number $N$
for even-even Ca, Ni, Zr, Sn, Pb isotopes. The RCHB calculation
with $\delta$-force is represented by open symbols, while the
corresponding data is denoted by solid symbols. The dashed lines
represent the predictions by the $A^{1/3}$ law with $r_{A}=1.228$
fm. Right: The experimental (solid symbols) and RCHB predicted
(dashed lines) coefficient $r_{Z}=R_c/Z^{1/3}$ for the nuclear
charge radii as a function of isospin quantity $\eta=N/Z$ in
even-even Ca, Ni, Zr, Sn and Pb isotopes. An asymptotic behavior
is drawn as a solid line. Taken from Ref.~\cite{Zhang02b}.}
\label{FigD8}
\end{figure}

For even-even Ca, Ni, Zr, Sn and Pb isotopes, the rms charge radii
$r_c$ obtained from the RCHB theory (open symbols) and the data
available (solid symbols) for the even-even Ca, Ni, Zr, Sn and Pb
isotopes are given in Fig.~\ref{FigD8}. The RCHB calculations
reproduce the data very well. For a given isotopic chain, an
approximate linear $N$ dependence of the calculated rms charge
radii $r_c$ is clearly seen in Fig.~\ref{FigD8}, which shows that
the variation of $r_c$ for a given isotopic chain deviates from
both the simple $Z^{1/3}$ dependence and the simple $A^{1/3}$ law
(denoted by dashed lines in Fig.~\ref{FigD8}). Therefore, a strong
isospin dependence of nuclear charge radii is necessary for nuclei
with extreme $N/Z$ ratio.

In Fig.~\ref{FigD8}, the data and RCHB predicted $r_{Z} = R_c /
Z^{1/3}$ for the proton magic isotopes are also presented as a
function of $\eta=N/Z$. It is clearly seen that the coefficient
$r_{Z}$ increases linearly with $\eta$ (except some deviations due
to deformation or shell effects) and the slopes are nearly the
same for these isotopic chains. The linear $\eta$ (or isospin
$T_Z=(N-Z)/2$) dependence of $r_{Z}$ for an isotopic chain may be
understood as the effect of the first order perturbation
correction of nuclear wave function due to an isospin $T_Z$
dependent interaction~\cite{Wigner57}. Based on the analysis of
data in the middle and upper panels of Fig.~\ref{FigD7} and RCHB
prediction in Fig.~\ref{FigD8}, the following isospin dependent
$Z^{1/3}$ formula for nuclear charge radii has been
proposed~\cite{Zhang02b}:
\begin{equation}
R_c=r_Z\,Z^{1/3}\left[1+b(\eta-\eta^*)\right], \ \ \ \ \eta=N/Z,
\label{eqn.rc}
\end{equation}
where $\eta^*$ is $\eta=N/Z$ for the nuclei along the
$\beta$-stability line which can be directly extracted from the
nuclear mass formula~\cite{Bohr69}, $r_{Z}= 1.631(11)$ fm as
obtained in upper right panel of Fig.~\ref{FigD7}, and
$b=0.062(9)$ obtained from the least square fitting.

The analysis of the available data using Eq.~(\ref{eqn.rc}) with
$r_{Z}$ and $b$ thus obtained is displayed in the lower right
panel of Fig.~\ref{FigD7}. The data are reproduced better by
Eq.~(\ref{eqn.rc}) than by Eq.~(\ref{eqn.Z}). Description of the
data with an isospin dependent $A^{1/3}$ formula,
$R_c=1.228\,A^{1/3}\left[1+0.029(\eta-\eta^*)\right]$ is also
shown in the lower left panel of Fig.~\ref{FigD7} for comparison.
The parameter for the isospin term is also obtained by least
square fitting of the data. It can be seen that this formula
describes the data less successfully than Eq.~(\ref{eqn.rc}).

In Refs.~\cite{Pomorski93,Pomorski94}, a modified $A^{1/3}$
formula with additional $\sim (N-Z)/A$ and $\sim 1/A$ terms made
some successes on describing nuclear charge radius. In addition,
this formua was extended with $\sim A^{-2}$
terms~\cite{Dobaczewski96}, which are considered to account for
the isospin dependence of charge radii in light nuclei. Based on
the fact that $R_p Z^{-1/3}$ is approximately a constant, where
$R_p$ is the rms radius of proton density distribution, a formula
for nuclear charge radius was proposed in
Refs.~\cite{Gambhir86a,Gambhir86b,Dobaczewski96},
$R_c=((r_pZ^{1/3})^2+0.64)^{1/2}$. Detailed analysis of these
expressions show: (i) The $Z^{1/3}$ dependence is a better
approximation to available $R_c$ data than the $A^{1/3}$ one;(ii)
The simple $Z^{1/3}$ dependent formula (\ref{eqn.Z}) gives nearly
the same result as the formula suggested in Refs.
~\cite{Gambhir86a,Gambhir86b}; (iii) The new isospin dependent
$Z^{1/3}$ formula is better than the two-parameter isospin
dependent $A^{1/3}$ formula but less better than the
three-parameter one. A better agreement of the isospin dependence
$Z^{1/3}$ formula of charge radii can be achieved by fitting both
$r_Z$ and $b$ simultaneously. While in Eq.~(\ref{eqn.rc}), the
same $r_Z$ is used as that in Eq.~(\ref{eqn.Z}), which are derived
only from the most stable nuclei. Thus the second term in
Eq.~(\ref{eqn.rc}) mainly corresponds to the isospin dependence of
charge radii for the nuclei away from the $\beta$ stability line.
It is expected that the modified $Z^{1/3}$ formula (\ref{eqn.rc})
will become more useful with more and more data obtained for the
nuclei far from the $\beta$-stability line.

\subsubsection{Density distribution and cross section}
\label{subsec:crosssection}

The measurement of interaction cross sections plays an important
role in exploring the properties of exotic nuclei because it can
provide information of the nuclear density distribution and
nuclear radii. Based on the measurement of interaction cross
section with radioactive beams at relativistic energy, novel and
entirely unexpected features appear: e.g., the neutron halo and
skin as the rapid increase in the measured interaction
cross-sections in the neutron-rich light nuclei. Systematic
investigation of interaction cross sections for an isotope chain
or an isotone chain can provide a good opportunity to study the
density distributions over a wide range of isospin. From the mean
field point of view, the nucleon density distributions in an
atomic nucleus determine the mean potential. Since the single
particle levels and the shell structure are directly determined by
the mean fields, the study of the isospin dependence of shell
structure and the mean potential, which become highly diffuse near
the particle drip line due to the large diffuseness in the density
distribution, is also crucial to understand properties of unstable
nuclei. Particularly the study of the surface diffuseness will
provide another mean to understand exotic nuclei because the
surface diffuseness can be strongly related with the spin-orbital
splitting which is mainly determined by the diffuseness in the
mean field potentials and could be measured experimentally.

\paragraph{The density distribution and shell structure}

Systematic RCHB calculations have been carried out for all the
nuclei in Na isotopes with mass number A ranging from 17 to 45
with parameter set NLSH~\cite{Meng98plb}. The calculated binding
energies $E_B$ and the interaction cross sections with the Glauber
Model are shown in Fig.~\ref{FigD9}. The calculated binding
energies $E_B$ are in good agreement with the empirical
values~\cite{Audi95}. The resonance states of $^{17}$Na and
$^{18}$Na (with a positive Fermi energy) are exactly reproduced.
$^{19}$Na is bound but unstable against the proton emission,
reproducing the experimental observation. The neutron drip line
nucleus has been predicted to be $^{45}$Na. The difference between
the calculations and the empirical values for the stable isotopes
is from the deformation~\cite{Meng98plb}.

\begin{figure}
\centering
\includegraphics[width=6.0cm]{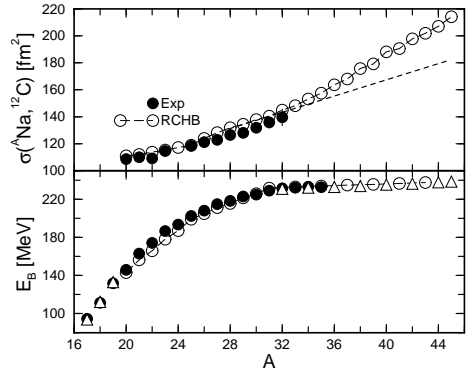}
\caption{Upper: The interaction cross sections $\sigma_I$  of
$^A$Na isotopes on a carbon target at $950 A$ MeV: the open
circles are the result of RCHB and the available experimental data
($A = 20 - 23, 25 - 32$) are given by solid circles with error
bars. The dashed line is a simple extrapolation based on the RCHB
calculation for $^{28-31}$Na. Lower: Binding energies for Na
isotopes, the convention is the same as the upper part, but the
RCHB result for particle unstable isotopes are indicated by
triangle. Taken from Ref.~\cite{Meng98plb}.} \label{FigD9}
\end{figure}

\begin{figure}
\centering
\includegraphics[width=6.0cm]{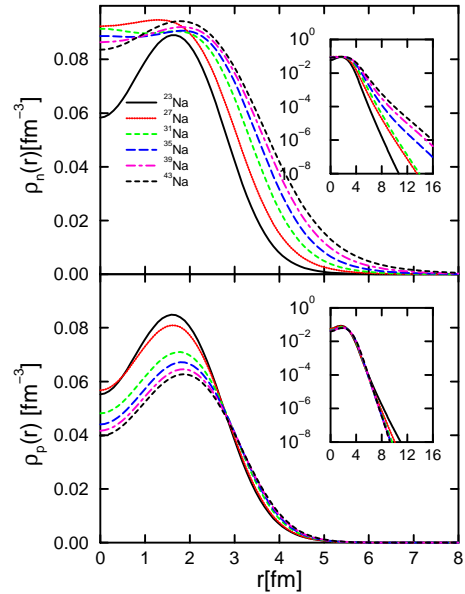}
\caption{The neutron (upper) and proton (lower) density
distributions in Na isotopes. The same figures but in logarithm
scale are given as inserts to show the tail part of the density
distribution. Taken from Ref.~\cite{Meng98plb}.} \label{FigD10}
\end{figure}

The density distributions are examined and the relation between
the development of halo and shell effect within the RCHB theory is
also studied in detail for Na isotopes~\cite{Meng98plb}. For
example, the density distributions for both the proton and the
neutron in Na isotopes are given in Fig.~\ref{FigD10}. As seen in
the upper part of Fig.~\ref{FigD10}, the change of the neutron
density is as follows: for the nucleus with less neutrons (N), the
density at the center is low and it spreads only to some smaller
distance. With the increase of N, the density near the center
increases due to the occupation of the $2s_{1/2}$ level, so does
the development of neutron radius. In the lower part of
Fig.~\ref{FigD10}, the proton density shows different behavior.
The surface is more or less unchanged because of the Coulomb
Barrier, but with the increase of N, the density of the center
decreases due to the slight increasing at the tail. This is due to
the symmetry energy. But as the density must be multiplied by a
factor $4\pi r^2$ before the integration in order to give the
fixed Z, the big change in the center does not influence the outer
part of the proton distribution very much.

\begin{figure}
\centering
\includegraphics[width=6.0cm]{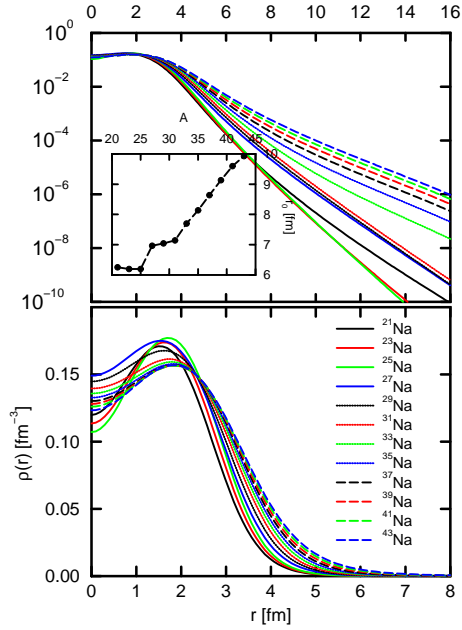}
\caption{The same as Fig.~\ref{FigD10} but for matter density
distribution. The upper part is given in logarithm scale (the
radius is labeled at the top of the figure) and the radius $r_0$
at which $\rho(r_0) = 10^{-4}$ for different isotopes is given as
inserts. Taken from Ref.~\cite{Meng98plb}.} \label{FigD11}
\end{figure}

The matter densities for the even Na isotopes are given in
Fig.~\ref{FigD11} where the shortest tail in the total density
occurs for $^{23-25}$Na, the most stable ones. For either the
proton or the neutron rich sides, the tail density increases
monotonically. The tail ($r > 10$ fm) of the proton rich nuclei is
mainly due to the contribution of the proton and that of the
neutron drip line nuclei is mainly due to the contribution of the
neutron. Compared with the neutron-rich isotopes, the proton
distribution for nucleus with less N has higher density at the
center, lower density in the middle ($2.5 < r < 4.5 $fm), a larger
tail in the outer ($r > 4.5 $ fm) part.

\begin{figure}
\centering
\includegraphics[width=6.0cm,angle=270]{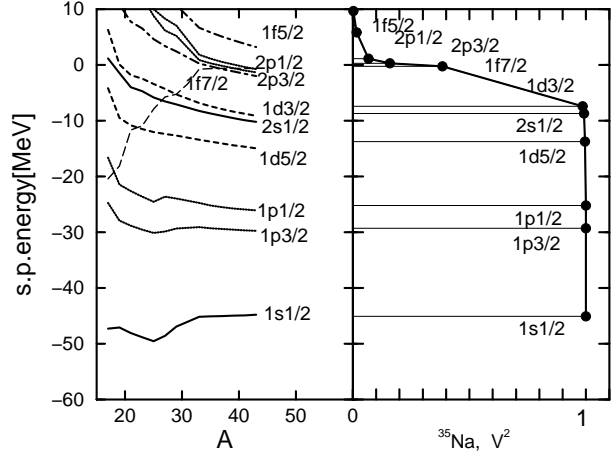}
\caption{Left: Single particle energies for neutrons in the
canonical basis as a function of the mass number. The dashed line
indicates the chemical potential. Right: The occupation
probabilities in the canonical basis for $^{35}$Na. Taken from
Ref.~\cite{Meng98plb}.} \label{FigD12}
\end{figure}

It is found that there is interesting connections between the
matter distribution and the single particle level distribution
(see Fig.~\ref{FigD12}): after $^{25}$Na, as it is a sub-closure
shell for the $1d_{5/2}$, the neutrons are filled in the
$2s_{1/2}$ and $1d_{3/2}$. So the tail of the density for
$^{27}$Na is two order of magnitude larger than $^{25}$Na at $r =
10$ fm, while the tail of the density from $^{27}$Na to $^{31}$Na
is very close to each other. But as more neutrons are filled in,
the added neutrons are filled in the next shell $1f_{7/2}$,
$2p_{3/2}$ and $2p_{1/2}$. So again two order of magnitude's
increase has been seen from $^{31}$Na to $^{33}$Na, and then a
gradual increase after $^{33}$Na. So it becomes clear that the
rapid increase in the cross section is connected with the filling
of neutrons in the next shell or sub shell. In the inserts in
Fig.~\ref{FigD11}: the radius $r_0$ at which $\rho(r_0) = 10^{-4}$
fm$^{-3}$ is given as a function of the mass number to see the
relation between the shell effect and matter distribution more
clearly. It is very interesting to see a slight decrease of $r_0$
from proton drip line to $^{25}$Na. The tail of $^{25}$Na is the
smallest. From $^{26}$Na to $^{31}$Na, one sees a almost constant
$r_0$. After a jump from $^{31}$Na to $^{32}$Na, a rapid
increasing tendency appears again.

In Figs.~\ref{FigD12} , the microscopic structure of the single
particle energies in the canonical basis~\cite{Ring80,Meng98npa}
is given. In the left panel of Fig.~\ref{FigD12}, the single
particle levels in the canonical basis for the isotopes with an
even neutron number are shown. Going from $A=19$ to $A=45$ we
observe a big gap above the $ N=8 $, $ N=20 $ major shell , and $
N=14 $ sub-shell. The $N=28$ shell for stable nuclei fails to
appear, as the $2p_{3/2}$ and $2p_{1/2}$ come so close to
$1f_{7/2}$. When $N \ge 20 $, the neutrons are filled to the
levels in the continuum or weakly bound states in the order of
$1f_{7/2}$, $2p_{3/2}$, $2p_{1/2}$, and $1f_{5/2}$. In the right
part, the occupation probabilities in the canonical basis of all
the neutron levels below  $E = 10$ MeV have been given for
$^{35}$Na to show how the levels are filled in nuclei near the
drip line.

\paragraph{The surface diffuseness and the spin-orbital splitting}

The RCHB theory has also been used to systematically study the tin
isotopes from proton drip line to neutron drip
line~\cite{Meng99npa}. The spin-orbital splitting
\begin{equation}
   E_{ls} = \displaystyle \frac {E_{lj=l-1/2}-E_{lj=l+1/2}} {2l+1}
\label{E-ls}
\end{equation}
versus A for the whole isotope chain is given in Fig.~\ref{FigD13}
for the neutron spin-orbital partners ($1d_{3/2},
1d_{5/2}$),($1g_{7/2}, 1g_{9/2}$), ($1i_{11/2}, 1i_{13/2}$),
($1p_{1/2}, 1p_{3/2}$), ($1f_{5/2}, 1f_{7/2}$), and ($1h_{9/2},
1h_{11/2}$), and the proton spin-orbital partners ($1d_{3/2},
1d_{5/2}$), and ($1f_{5/2}, 1f_{7/2}$). It is very interesting to
see that the spin-orbital splitting for the neutron and proton is
very close to each other, at least for ($1d_{3/2}, 1d_{5/2}$), and
($1f_{5/2}, 1f_{7/2}$) cases. The splitting decreases
monotonically from the proton drip line to the neutron drip line
for all the partners. To see the underlying reason of this
behavior, it is very helpful to examine the origin of the
spin-orbital splitting in the Dirac equation. The equation of
motion for the nucleon moving in scalar and vector potentials
could be de-coupled and reduced for the upper component and the
lower component, respectively. If it is reduced in the lower
component, it will be related with another interesting topic ---
the pseudo-spin symmetry discussed in next subsection. As for the
spin-orbital splitting, the Dirac equation can be reduced for the
upper component as the following:
\begin{eqnarray}
   & &  [ \frac {d^2} {dr^2}  - \frac 1  {E + 2M - V  + S}
        \frac {d(2M-V+S)} {dr} \frac {d} {dr} ] G^{lj}_i(r) \nonumber \\
   & -&  [  \frac { \kappa ( 1 + \kappa ) }  {r^2}
        - \frac 1 {E + 2M - V + S} \frac {\kappa} r
        \frac {d(2M - V + S)} {dr}  ] G^{lj}_i (r) \nonumber \\
  = &-& (E + 2M - V + S ) ( E - V - S ) G^{lj}_i (r),
\label{larspinor4}
\end{eqnarray}
where
\begin{eqnarray}
       \kappa =
        \left\{ \begin{array}{cc}
            -l-1  &  j=l+1/2  ,\\
            l     &  j=l-1/2 .
         \end{array} \right.
\end{eqnarray}
The spin-orbital splitting is due to the corresponding
spin-orbital potential
\begin{equation}
   \displaystyle \frac 1 {E + 2M- V + S} \frac {\kappa} r
        \frac {d(2M - V + S)} {dr}
\label{spp}
\end{equation}
with some proper normalization factor. It is seen that the
spin-orbital splitting in RMF is energy dependent and they depends
on the derivative of the potential $2M- V + S$ as well as the
particle distribution. Therefore one can introduce the so-called
spin-orbital potential: $V_{ls} = \displaystyle \frac {\kappa} r
\frac {d(2M - V + S)} {dr}$. The $V_{ls}$ for $^{110}$Sn,
$^{140}$Sn, and $^{170}$Sn, are given in the left panel of
Fig.~\ref{FigD14}. The $V_{ls}$ for both proton and neutron are
almost the same, as the potential $V-S$ is a big quantity ($\sim
700$ MeV), the isospin dependence in the spin-orbital potential
could be neglected. Therefore the proton and neutron $V_{ls}$ are
the same in the present model. That is the reason why the
spin-orbital splitting for the neutron and proton is very close to
each other in Fig.~\ref{FigD13}. From $^{110}$Sn to $^{170}$Sn,
the amplitude of $V_{ls}$ decreases monotonically due to the
surface diffuseness. Neutron potentials $V - S$ for $^{100}$Sn,
$^{110}$Sn,$^{120}$Sn, $^{130}$Sn, $^{140}$Sn,$^{150}$Sn,
$^{160}$Sn,  and $^{170}$Sn are given in the right panel of
Fig.~\ref{FigD14} and its inserted figure gives the radii $R_0$ at
which $V - S=100$ MeV as a function of the mass number. As seen in
the above equations, the spin-orbital splitting is related with
the derivative of the potential $V - S$ . The surface diffuseness
happens for both the vector and scalar potential, as well as for
$V - S$ and $V + S$.

\begin{figure}
\centering
\includegraphics[width=6.0cm,angle=270]{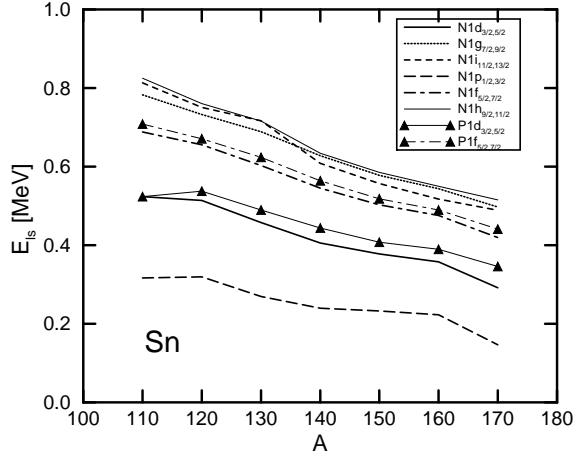}
\caption{The neutron spin-orbit splitting $E_{ls} = \displaystyle
\frac {E_{lj=l-1/2}-E_{lj=l+1/2}} {2l+1}$ versus the mass number
$A$ in tin isotopes for neutron ($1d_{3/2}, 1d_{5/2}$),
($1g_{7/2}, 1g_{9/2}$), ($1i_{11/2}, 1i_{13/2}$), ($1p_{1/2},
1p_{3/2}$), ($1f_{5/2}, 1f_{7/2}$), and ($1h_{9/2}, 1h_{11/2}$)
orbital and proton ($1d_{3/2}, 1d_{5/2}$), and ($1f_{5/2},
1f_{7/2}$) orbital, respectively. Taken from
Ref.~\cite{Meng99npa}.} \label{FigD13}
\end{figure}

For the decline of the spin-orbital splitting, it comes from the
diffuseness of the potential or the outwards tendency of the
potential. The diffuseness of the neutron potentials $V + S $ are
given in the left panel of Fig.~\ref{FigD15} for $^{110}$Sn,
$^{120}$Sn, $^{130}$Sn, $^{140}$Sn, $^{150}$Sn, $^{160}$Sn, and
$^{170}$Sn. It is seen that the depth of the potential decreases
monotonically from the proton drip line to the neutron drip line
and the surface of the potential moves outwards. The inserted
figure gives the radii $R_0$ at which $V + S= -10$ MeV as a
function of the mass number. The proton potentials $V + S$ for
$^{110}$Sn,$^{120}$Sn, $^{130}$Sn,  $^{140}$Sn,$^{150}$Sn,
$^{160}$Sn,  and $^{170}$Sn are given in the right panel of
Fig.~\ref{FigD15}. With the increase of the neutron number, the
proton potentials are pushed towards outside as well due to the
proton-neutron interaction.

\begin{figure}
\centering
\includegraphics[width=6.0cm,angle=270]{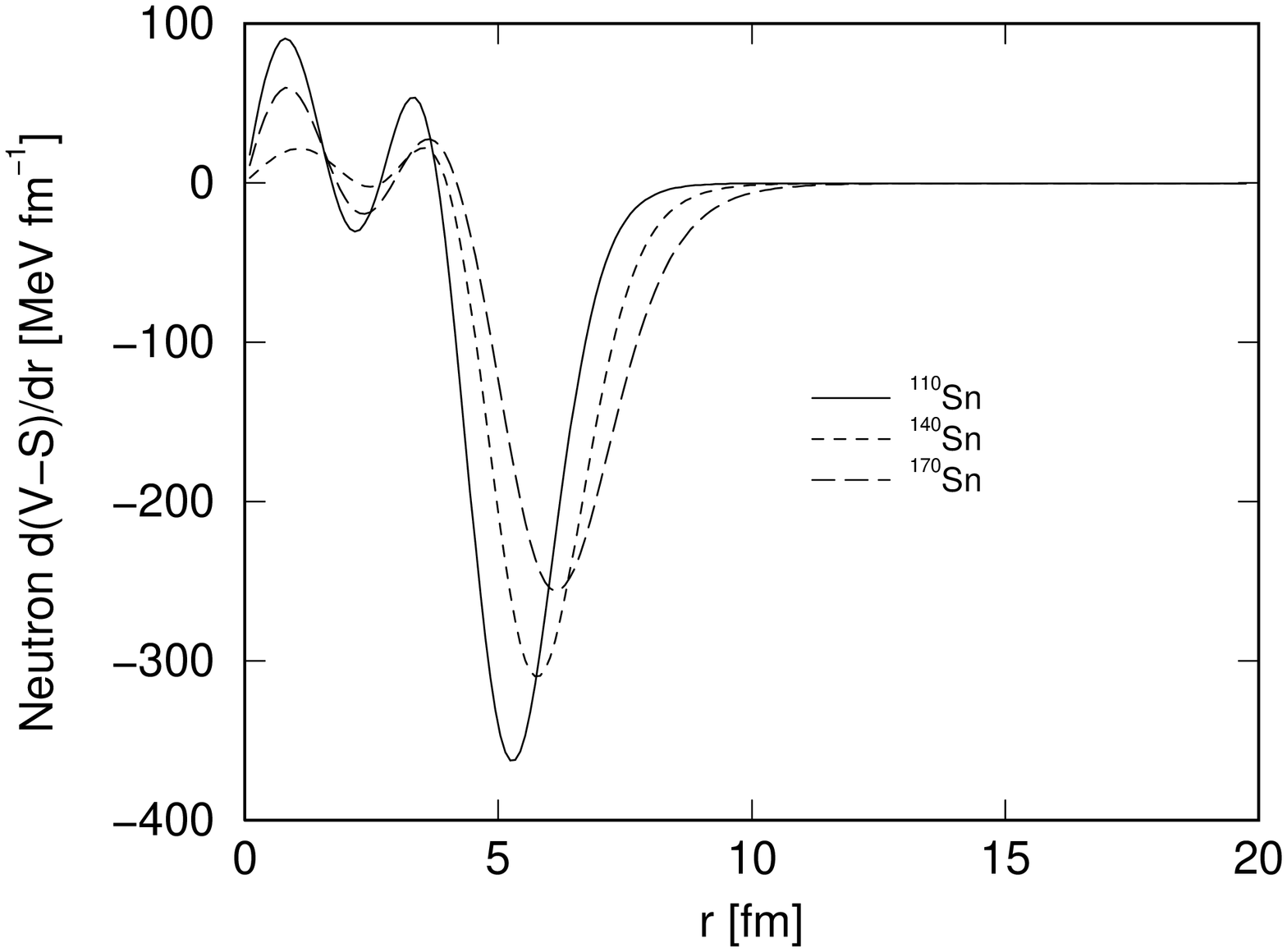}
\includegraphics[width=6.0cm,angle=270]{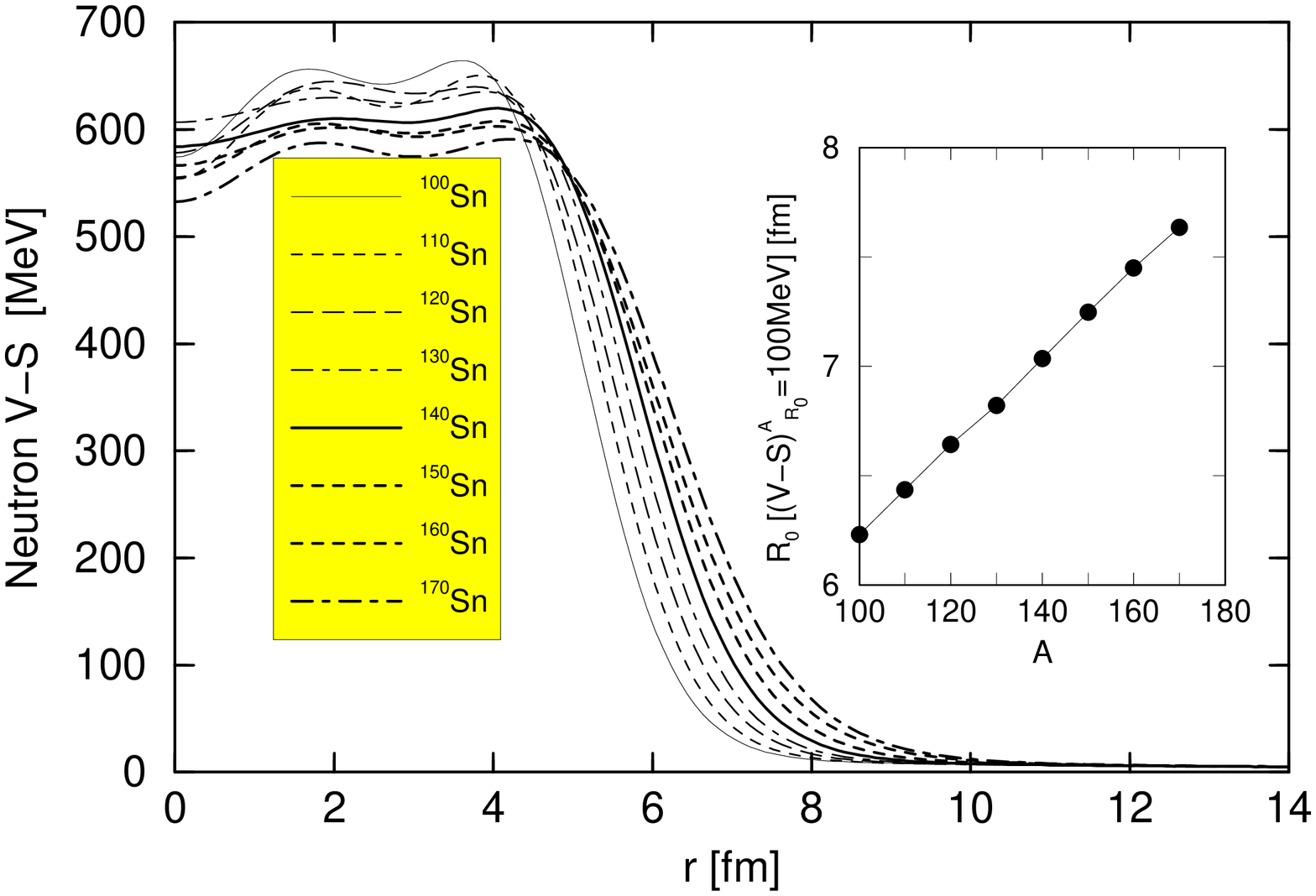}
\caption{Left: The derivative of the neutron potentials $V( r ) -
S( r )$ for $^{110}$Sn, $^{140}$Sn, and $^{170}$Sn; Right: The
neutron potentials $V( r ) - S( r )$ for $^{100}$Sn,
$^{110}$Sn,$^{120}$Sn, $^{130}$Sn,  $^{140}$Sn,$^{150}$Sn,
$^{160}$Sn,  and $^{170}$Sn. In order to examine the surface
diffuseness more clearly, the radii $R_0$ at which $V(R_0) -
S(R_0) = 100$ MeV has been given as an inserted figure. Taken from
Ref.~\cite{Meng99npa}.} \label{FigD14}
\end{figure}

\begin{figure}
\centering
\includegraphics[width=6.0cm,angle=270]{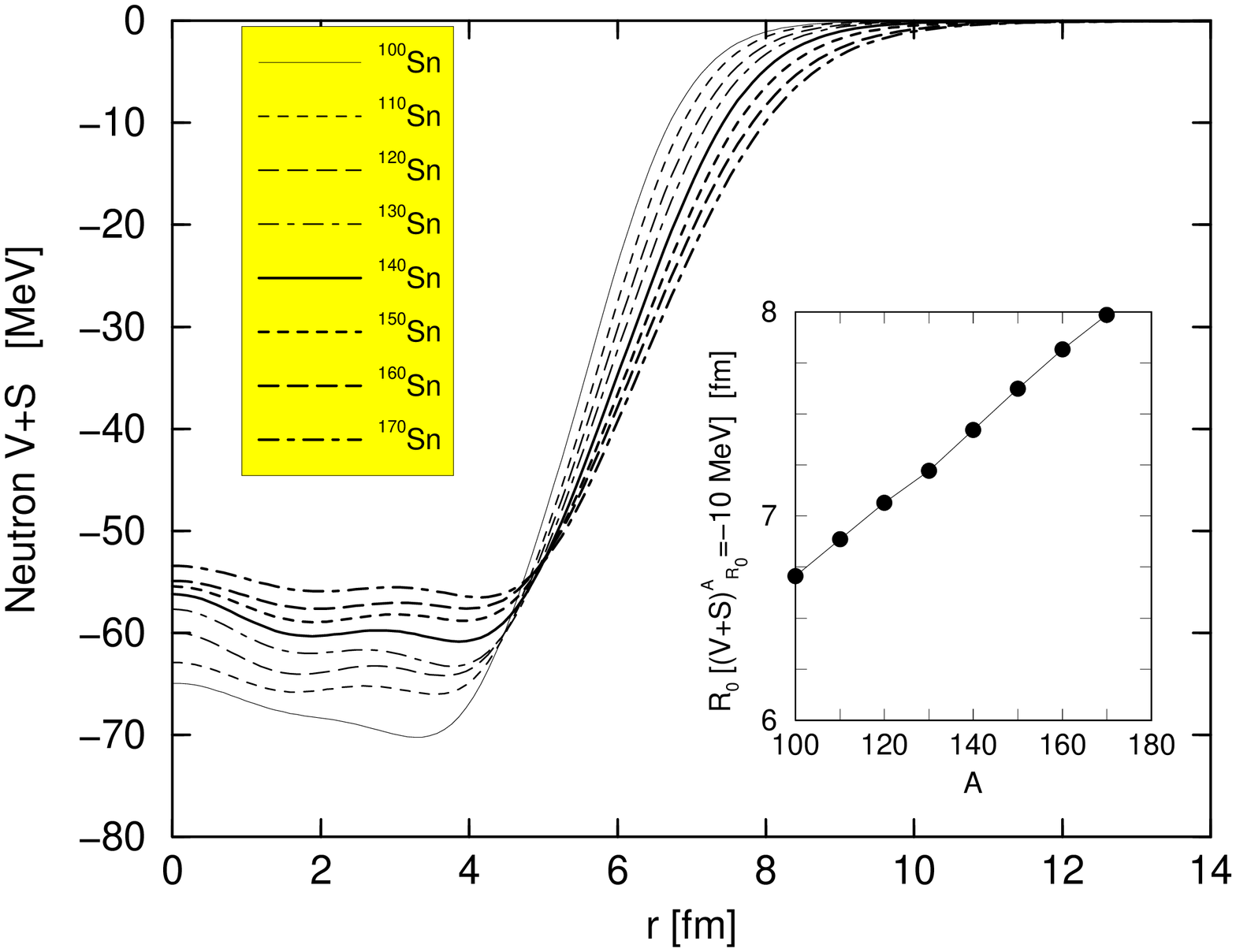}
\includegraphics[width=6.0cm,angle=270]{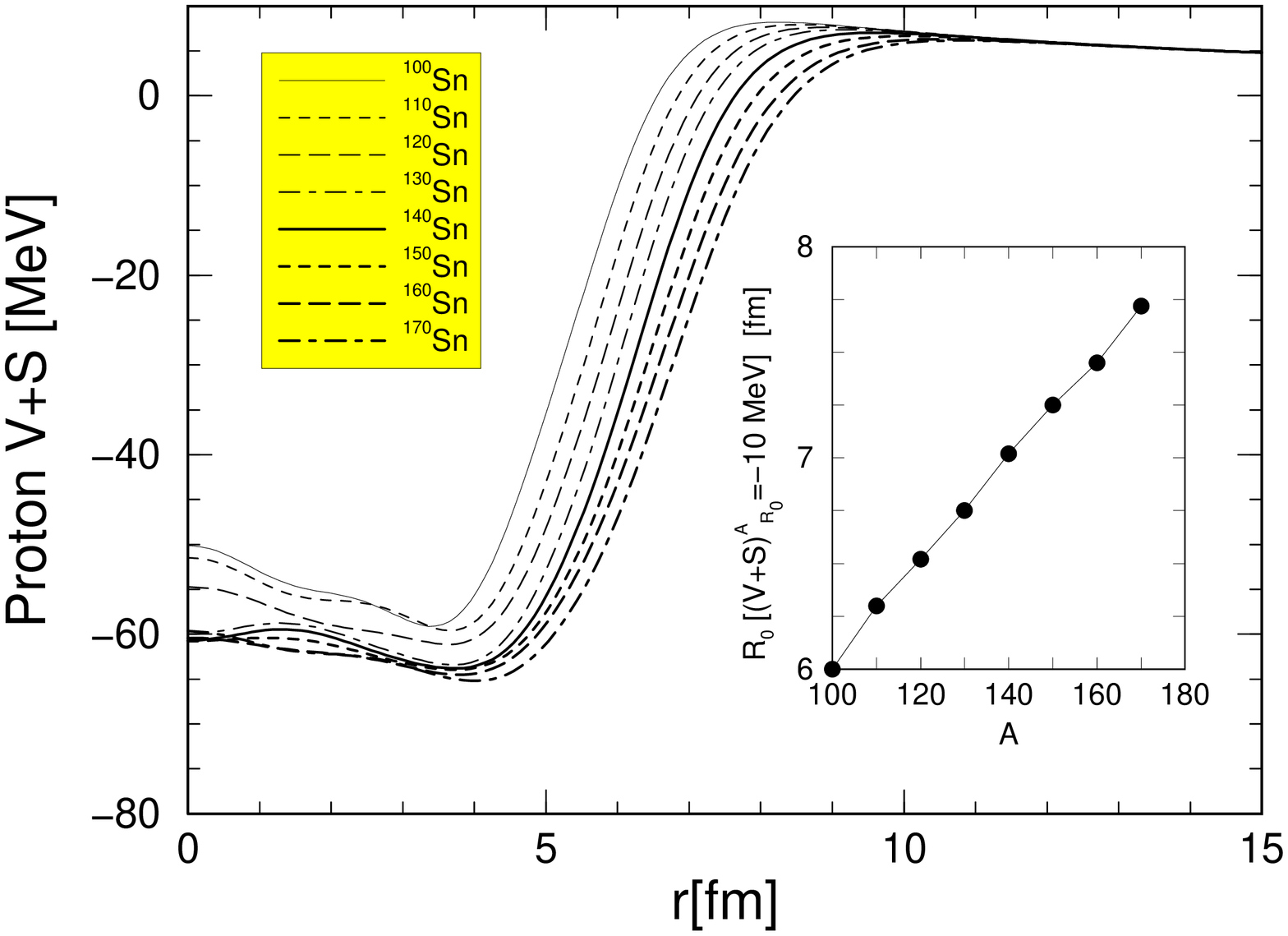}
\caption{Left: The neutron potentials $V( r ) + S( r )$ for
$^{100}$Sn, $^{110}$Sn,$^{120}$Sn, $^{130}$Sn,
$^{140}$Sn,$^{150}$Sn, $^{160}$Sn,  and $^{170}$Sn. In order to
examine the surface diffuseness more clearly, the radii $R_0$ at
which $V (R_0) + S(R_0) = -10$ MeV has been given as an inserted
figure; Right: The proton potentials $V( r ) + S( r )$ for
$^{110}$Sn,$^{120}$Sn, $^{130}$Sn,  $^{140}$Sn,$^{150}$Sn,
$^{160}$Sn,  and $^{170}$Sn.  In order to examine the surface
diffuseness more clearly, the radii $R_0$ at which $V (R_0) +
S(R_0) = -10$ MeV has been given as an inserted figure. Taken from
Ref.~\cite{Meng99npa}.} \label{FigD15}
\end{figure}

\paragraph{Interaction cross sections}

For medium and high energy reactions, the interaction cross
section $\sigma_{\rm int}$ can be calculated with the Glauber
model which is based on the individual nucleon-nucleon collisions
in the overlap volume of the high energy colliding nuclei. In the
framework of this model, the nucleus-nucleus reaction cross
section can be obtained from nucleon-nucleon reaction cross
section. Consider the collision of a projectile nucleus $P$ on a
target nucleus $T$ when the nuclei $P$ and $T$ are situated at an
impact parameter $\bf b$ relative to each other, the total
reaction cross section can be written as
\begin{equation}
\sigma = 2 \pi \int b db \left( 1 - T(b) \right),
\end{equation}
where $T(b)$ is the transparency function describing the
probability that the projectile will pass through the target
without interacting. $T(b)$ is given by
\begin{equation}
T(b) = \exp\left[-\sigma_{NN}\int \rho_P({\bf b}_P, z_P) d{\bf
b}_P dz_P \,\,
 \rho_T({\bf b}_T, z_T) d{\bf b}_T dz_T \,\,
 t({\bf b-b_P+b_T})\right],
 \label{eq:Glauber}
\end{equation}
where $\sigma_{NN}$ is the nucleon-nucleon reaction cross section
suitably averaged over the interacting $n$-$n$, $p$-$p$, and
$n$-$p$ pairs~\cite{Ray79,LiuJY01}. The profile function $t({\bf
b})$ for the NN scattering is taken as a delta function for
zero-range nuclear interaction.

To compare the cross section directly with experimental measured
values, the densities $\rho_{n,p}(r)$ of the target $^{12}$C and
the Na isotopes obtained from RCHB (see Fig.~\ref{FigD10}) were
used in the Glauber model calculation. The cross sections for
reaction of Na isotopes at $950A$ MeV on $^{12}$C have been
compared with the experimental values~\cite{Suzuki95} in the upper
part of Fig.~\ref{FigD9}. The agreement between the calculated
results and measured ones are fine. The cross section below
$^{22}$Na changes only slightly with the neutron number, which
means the proton density has played an important role to remedy
the contribution of less neutron. From $^{25}$Na to the neutron
drip line, a gradual increase of the cross section has been
observed. After $^{32}$Na, although no data exist yet, a
relatively fast increase has been predicted. In short, the
measured cross section shows similar behavior as that of RCHB.

\paragraph{Charge changing cross section}

Systematic investigation of interaction cross sections for an
isotope chain or an isotone chain can provide a good opportunity
to study the density distributions over a wide range of
isospin~\cite{Suzuki95,Meng98plb}. However the contribution from
proton and neutron are coupled in the measurement of interaction
cross section. To investigate possible differences in proton and
neutron density distributions, a combined analysis of the
interaction cross section and other experiment on either proton or
neutron alone are necessary. The charge-changing cross section
which is the cross section for all processes which result in a
change of the atomic number for the projectile can provide good
opportunity for this purpose. In Ref.~\cite{Chulkov00}, the total
charge-changing cross section $\sigma_{\rm cc}$ for the light
stable and neutron-rich nuclei at relativistic energy on a carbon
target were measured. The charge changing cross section
$\sigma_{\rm cc}$ are calculated by using the Glauber Model and
the proton density of projectiles and total density of target
provided by RCHB theory~\cite{Meng02plb}.

\begin{figure}
\centering
\includegraphics[width=8.0cm]{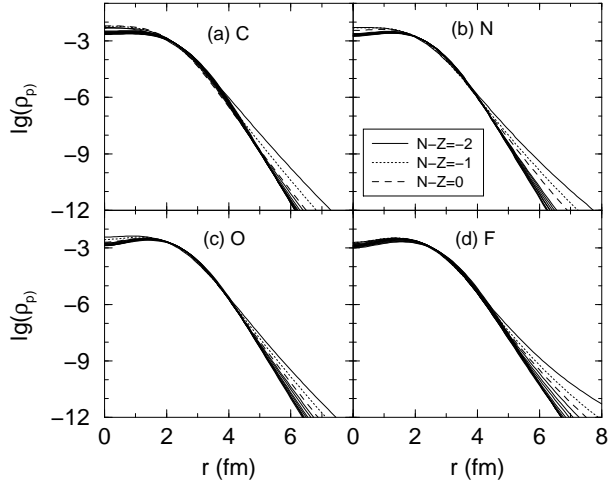}
\caption{The proton density distributions predicted by RCHB for
the nuclei $^{10-22}$C, $^{12-24}$N, $^{14-26}$O and $^{16-25}$F
in logarithm scale. Taken from Ref.~\cite{Meng02plb}.}
\label{FigD16}
\end{figure}

The proton density distributions predicted by RCHB for the nuclei
$^{10-22}$C,$^{12-24}$N,$^{14-26}$O and $^{16-25}$F are given in
Fig.~\ref{FigD16} in logarithm scale. The change in the density
distributions for each isotope chain in Fig.~\ref{FigD16} occurs
only at the tail or in the center part. The density is multiplied
by a factor $4\pi r^2$ before an integration in order to get
proton number or radii, therefore the change of density in the
central region does not matter very much. What is important is the
density distribution in the tail part. Compared with the cases of
neutron-rich isotopes, the proton distribution of nuclei with less
$N$ has higher density in the center, lower density in the middle
part ($2.5 < r < 4.5 $fm), a larger tail in the outer part ($r
> 4.5 $ fm) which gives rise to the increase of $r_{\rm p}$ and
$\sigma_{\rm cc}$ for the proton rich nuclei as will be seen in
following figures.

\begin{figure}
\centering
\includegraphics[width=8.0cm]{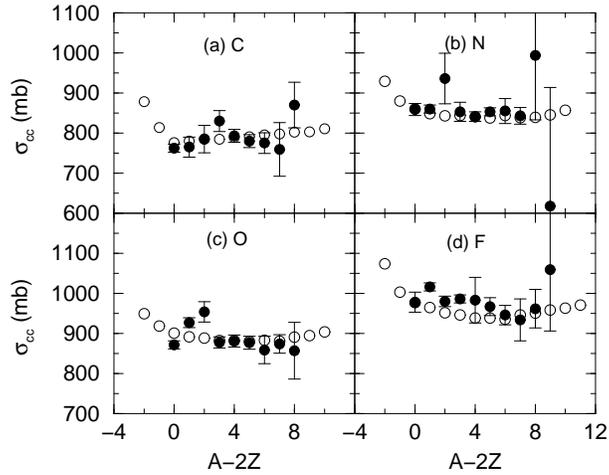}
\caption{The total charge-changing cross sections $\sigma_{\rm
cc}$ of the nuclei $^{10-22}$C,$^{12-24}$N,$^{14-26}$O and
$^{16-25}$F on a carbon target at relativistic energy. The open
circles are the result of RCHB and the available experimental data
are given by solid circles with error bars. Taken from
Ref.~\cite{Meng02plb}.} \label{FigD17}
\end{figure}

With the proton density from RCHB calculations for the projectile
used for $\rho_P({\bf b}_P, z_P)$ in Eq.~(\ref{eq:Glauber}), the
total charge-changing cross sections $\sigma_{\rm cc}$ of the
nuclei $^{10-22}$C, $^{12-24}$N, $^{14-26}$O and $^{16-25}$F on a
carbon target at relativistic energy are calculated and given in
Fig.~\ref{FigD17}. The results of the Glauber model are
represented by open circles and available data~\cite{Chulkov00} by
solid ones with error bars. The agreement between the calculation
and the measurement is very good.

The charge changing cross sections change only slightly with the
neutron number except for proton-rich nuclei. This indicates that
the proton density plays an important role in determining the
charge-changing cross sections $\sigma_{\rm cc}$. A gradual
increase of the cross section can be observed towards the neutron
drip line. However, the big error bars of the data cannot help to
conclude anything here yet. It is shown clearly that the RCHB
theory, when combined with the Glauber model, can provide reliable
description for not only interaction cross section but also charge
changing cross section. From comparison between Fig.~\ref{FigD17}
and Fig.~\ref{FigD4}, one finds similar trends of variations of
proton radii and of charge changing cross sections for each
isotope chain which implies again the important role that proton
plays in determining the charge-changing cross sections.

\subsection{Single particle levels and pseudospin symmetry}

The concept of pseudo-spin is that the single particle orbitals
with $j=l+1/2$ and $j=(l+2)-1/2$ lie very close in energy and can
therefore be labelled as pseudo-spin doublets with quantum number
$\tilde n = n-1$, $\tilde l = l-1$, and $\tilde s = s =1/2$. This
concept is originally found in spherical nuclei~\cite{Arima69,
Hecht69} , but later proved to be a good approximation in deformed
nuclei as well~\cite{Ratna73, Draayer84, Zeng91}. It is shown that
pseudo-spin symmetry remains an important physical concept even in
the case of triaxiality~\cite{Blokhin97}. The origin of
pseudo-spin is proved to be connected with the special ratio in
the strength of the spin-orbit and orbit-orbit
interactions~\cite{Bohr82, Castanos92} and the unitary operator
performing a transformation from normal spin to pseudo-spin space
have been discussed~\cite{Bohr82, Castanos92, Balantekin92}.
However, it is not explained why this special ratio is allowed in
nuclei.

The relation between the pseudo-spin symmetry and the RMF theory
was first noted in Ref.~\cite{Bahri92}, in which Bahri et al found
that the RMF explains approximately the strengthes of spin-orbit
and orbit-orbit interactions found by non-relativistic
calculations. More details have been given in
Refs.~\cite{Blokhin95, Draayer84}, in which it was suggested that
the origin of pseudo-spin is related to the strength of the scalar
and vector potentials. Ginocchio first revealed that
pseudo-orbital angular momentum is nothing but the ``orbital
angular momentum'' of the lower component of the Dirac wave
function~\cite{Ginocchio97}.

In the Dirac equation of the nucleon, when the scalar potential
$S(\mathbf{r})$ and the vector potential $V(\mathbf{r})$ are equal
in amplitudes but opposite in sign, \textit{i.e.}, $S(\mathbf{r})
+ V(\mathbf{r}) = 0$, or more generally, $d[S(\mathbf{r}) +
V(\mathbf{r})]/ dr = 0$, there is an exact pseudospin symmetry in
single particle spectra~\cite{Ginocchio97, Meng98r, Meng99prc}.
Under these conditions, there should be some special relations
between the four components of the Dirac wave functions of the
pseudospin doublets which have been used to test the pseudospin
symmetry in realistic nuclei~\cite{Ginocchio99, Ginocchio01,
Leviatan01, Ginocchio02}.

The pseudo spin symmetry in deformed nuclei has also been
extensively studied in the framework of the RMF model. The
conditions for an exact pseudo spin symmetry and the relationships
for the lower components of the Dirac eigenfunctions were
investigated in Refs.~\cite{Lalazissis98, Tanabe98, Tanabe00,
Tanabe02}. The relationships between the upper and lower
components of the two states in the pseudo spin doublets have been
studied thoroughly for realistic deformed relativistic
eigenfunctions~\cite{Ginocchio02, Ginocchio04}.

There are also many other investigations on the pseudo-spin
symmetry in the framework of RMF theory~\cite{Marcos00, Marcos01,
Lopez03, Marcos03, Marcos04,Alberto01, Chen02, Lisboa04}. A
thorough review is certainly beyond the purpose of the present
paper. In the following, the pseudo-spin symmetry in RCHB theory
will be briefly reviewed.

From the radial Dirac equation for spherical nuclei, similar as
Eq.~(\ref{larspinor4}), one can derive the radial equation for the
lower components:
\begin{eqnarray}
&&
[ \frac {d^2} {dr^2}
     + \frac 1 {E - V - S} \frac {d(V+S)} {dr} \frac {d} {dr}] F^{lj}_i(r)
    +  [ \frac { \kappa ( 1 - \kappa ) }  {r^2}
     - \frac 1 {E - V - S} \frac {\kappa} r \frac {d(V+S)} {dr}] F^{lj}_i(r)
     \nonumber\\
 & = & - ( E + 2M - V + S)  ( E - V - S) F^{lj}_i(r),
\label{smaspinor4}
\end{eqnarray}
where $\kappa ( \kappa - 1 ) = \tilde l (\tilde l +1)$. It is
clear that one can use either Eq.~(\ref{smaspinor4}) or
equivalently Eq.~(\ref{larspinor4}) to get the eigenvalues $E$ and
the corresponding eigenfunctions. In
Section~\ref{subsec:crosssection}, Eq.~(\ref{larspinor4}) has been
used to discuss the spin-orbital splitting in connection with the
corresponding spin-orbital potential. If the
Eq.~(\ref{smaspinor4}) is used instead and the pseudo spin-orbital
potential (PSOP) term, $\displaystyle\frac 1  {E - V - S } \frac
{\kappa} r \frac {d(V+S)} {dr}$, is neglected, then the
eigenvalues $E$ for the same $\tilde l$ will degenerate. This is
the phenomenon of pseudo-spin symmetry observed in
Refs.~\cite{Arima69,Hecht69}. As shown in Ref.~\cite{Meng98prl},
the particle levels for the bound states in canonical basis are
the same as those by solving the Dirac equation with the scalar
and vector potentials from RCHB. Therefore Eqs.~(\ref{larspinor4})
and (\ref{smaspinor4}) remain the same in canonical basis even
after the pairing interaction has been taken into account.

In Eq.~(\ref{smaspinor4}), the term which splits the pseudo-spin
partners is simply the PSOP. The hidden symmetry for the
pseudo-spin approximation is revealed as $d(V+S)/dr = 0$, which is
more general and includes $V + S = 0$ discussed
in~\cite{Ginocchio97} as a special case. For exotic nuclei with
highly diffuse potentials, $d(V+S)/dr \sim 0$ may be a good
approximation and then the pseudo-spin symmetry will be good. But
generally, $d(V+S)/dr = 0$ is not always satisfied in the nuclei
and the pseudo-spin symmetry is an approximation.  However, if $|
\displaystyle\frac 1 {E - V -S} \frac {\kappa} r \frac {d(V+S)}
{dr}| \ll |\displaystyle\frac { \kappa ( 1 - \kappa ) }  {r^2}| $,
the pseudo-spin approximation will be good. Thus, the comparison
of the relative magnitude of the pseudo centrifugal barrier (PCB),
$\displaystyle\frac { \kappa ( 1 - \kappa ) }  {r^2} $, and the
PSOP can provide us with some information on the pseudo-spin
symmetry. In Ref.~\cite{Meng98r}, with the scalar and vector
potentials derived from a self-consistent RCHB calculation, the
pseudo-spin symmetry and its energy dependence have been discussed
in realistic nuclei. Afterwards, the pseudo-spin splitting for
exotic nuclei in Zr and Sn isotopes has been
studied~\cite{Meng99prc}.

\begin{figure}
\centering
\includegraphics[width=6.0cm,angle=270]{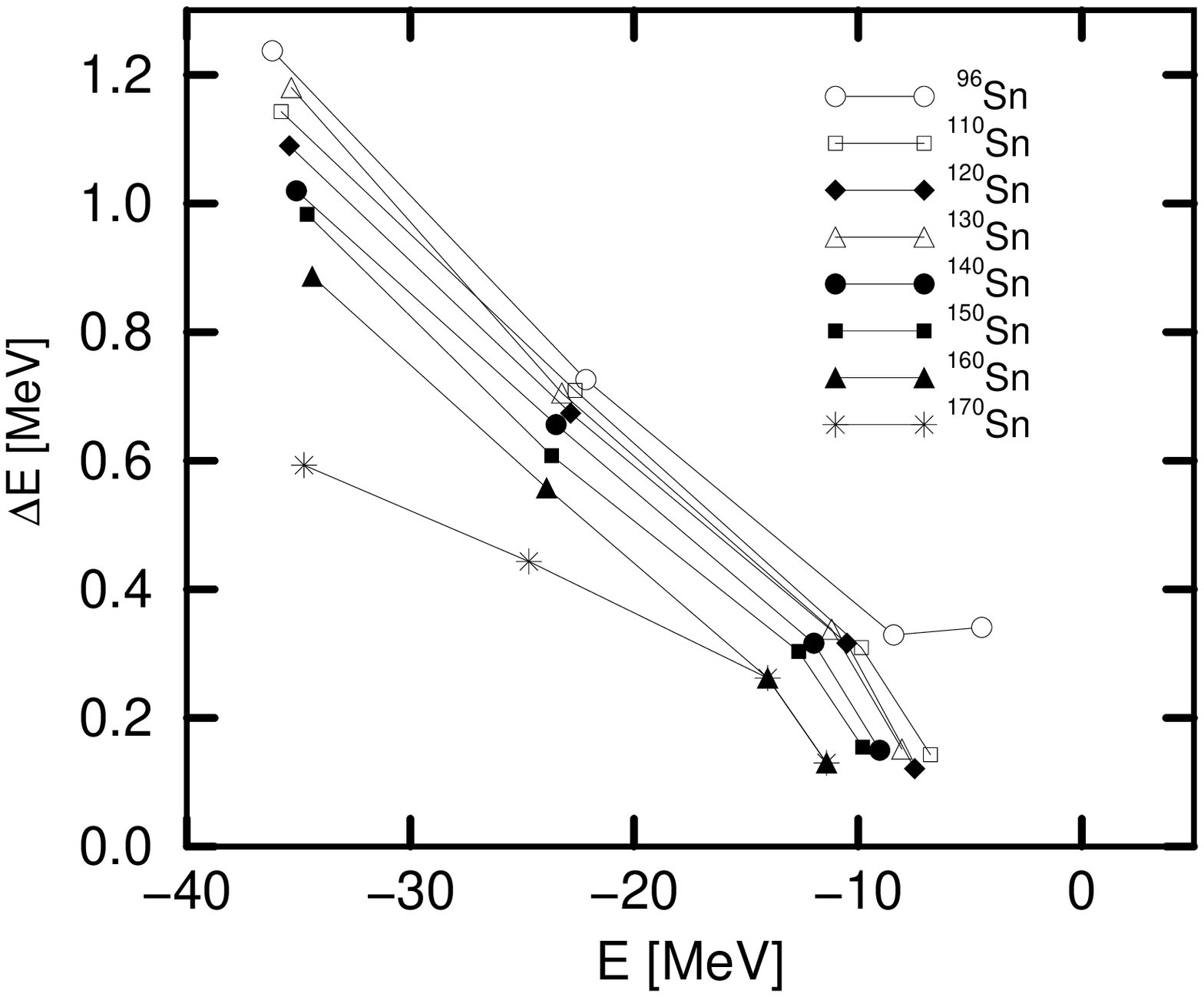}
\includegraphics[width=6.0cm,angle=270]{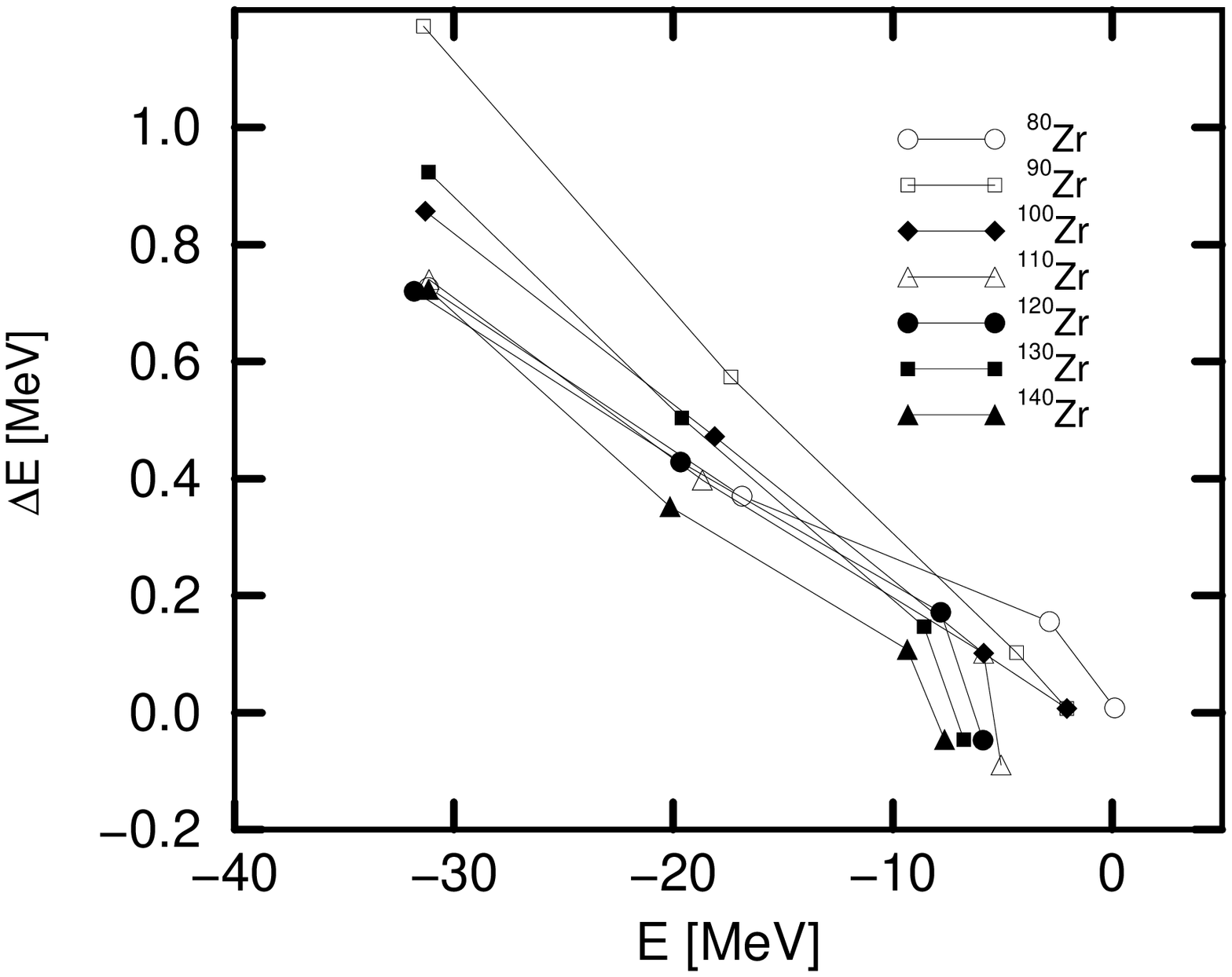}
\caption{The pseudo-spin orbit splitting $\displaystyle\Delta E =
\frac {E_{\tilde lj=\tilde l-1/2}- E_{\tilde lj=\tilde l+1/2}}
{2\tilde l+1}$ versus the binding energy $\displaystyle E = \frac
{ \tilde l E_{\tilde lj=\tilde l + 1/2 } + ( \tilde l + 1 ) E_{
\tilde lj= \tilde l-1/2} } {2 \tilde l+1} $ for Zr and Sn
isotopes. From left to right, the pseudo-spin partners correspond
to ($1d_{3/2}, 2s_{1/2}$), ($1f_{5/2}, 2p_{3/2}$), ($1g_{7/2},
2d_{5/2}$) and ($2d_{3/2}, 3s_{1/2}$), respectively. Taken from
Ref.~\cite{Meng98r}.} \label{FigD18}
\end{figure}

The redefined pseudo-spin orbital splitting $\displaystyle\Delta E
= \frac {E_{\tilde lj=\tilde l-1/2}- E_{\tilde lj=\tilde l+1/2}}
{2\tilde l+1}$ versus the average single particle energy
$\displaystyle E = \frac { \tilde l E_{\tilde lj=\tilde l + 1/2 }
+ ( \tilde l + 1 ) E_{ \tilde lj= \tilde l-1/2} } {2 \tilde l+1}$
for the bound pseudo-spin partners in Sn and Zr isotopes are
plotted in Fig.~\ref{FigD18}. In both isotopes, a monotonous
decreasing behavior with the decreasing binding energy is clearly
seen. The pseudo-spin splitting for $3s_{1/2}$ and $2d_{3/2}$ is
more than $10$ times smaller than that of the $2s_{1/2}$ and
$1d_{3/2}$. As far as the isospin dependence of the pseudo-spin
orbital splitting is concerned, the splitting in Sn isotopes gives
a monotonous decreasing  behavior with the increasing isospin.
Particularly for $2s_{1/2}$ and $1d_{3/2}$ partners, the
pseudo-spin splitting in $^{170}$Sn is only half of that in
$^{96}$Sn. Just as one expects~\cite{Meng98r}, the pseudo-spin
symmetry in neutron-rich nuclei is better. In Zr isotopes,
although the situation is more complicated (e.g., the effect of
the deformation which is neglected here), the pattern is more or
less the same, i.e., a monotonous decreasing  behavior with the
decreasing binding energy and a monotonous decreasing behavior
with the isospin. From these studies, the pseudo-spin symmetry
remains a good approximation for both stable and exotic nuclei. A
better pseudo-spin symmetry  can be expected for the orbital near
the threshold, particularly for nuclei near the particle drip
line.

\begin{figure}
\centering
\includegraphics[width=8.0cm]{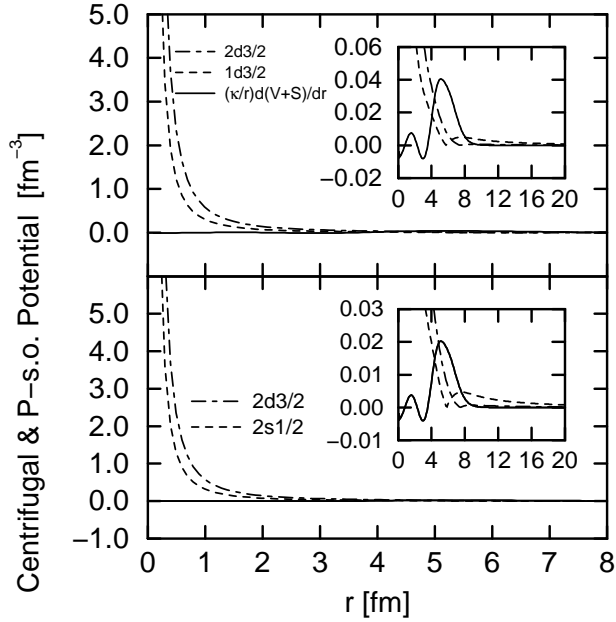}
\caption{The comparison of the effective pseudo centrifugal
barrier (PCB) $\displaystyle (E-V-S) \frac {\kappa(\kappa-1)}
{r^2}$ (dashed lines and dot-dashed lines) and the effective
pseudo-spin orbital potential (PSOP) $\displaystyle \frac {\kappa}
{r} \frac {d(V+S)} {dr}$ (solid line ) in arbitrary scale for
$d_{3/2}$ (upper) and $s_{1/2}$ (lower) in $^{120}$Zr. The dashed
lines are for $1d_{3/2}$ and $2s_{1/2}$, and the dot-dashed lines
are for $2d_{3/2}$ and $3s_{1/2}$. The inserted boxes show the
same quantities, but the ordinate is magnified and the abscissa is
reduced to show the behaviors of the effective PCB and the
effective PSOP near the nuclear surface. Taken from
Ref.~\cite{Meng98r}.} \label{FigD19}
\end{figure}

To understand why the energy splitting of the pseudo-spin partner
changes with different binding energies and why the pseudo-spin
approximation is good in RMF, the PSOP and PCB should be examined
carefully. Unfortunately, it is very hard to compare them clearly,
as the PSOP has a singularity at $E \sim (V+S)$. As one is only
interested in the relative magnitude of the PCB and the PSOP, the
effective PCB, $\displaystyle (E-V-S) \frac {\kappa(\kappa-1)}
{r^2}$, and the effective PSOP, $\displaystyle\frac {\kappa} {r}
\frac {d(V+S)} {dr}$ are introduced for comparison. They
correspond to the PCB and the PSOP multiplied by a common factor
$E-V-S$ respectively.

\begin{figure}
\centering
\includegraphics[width=6.0cm, angle=-90]{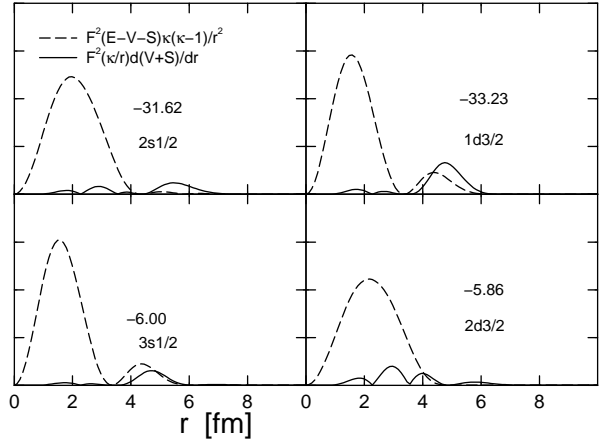}
\caption{The comparison of the effective pseudo centrifugal
barrier (PCB) $\displaystyle (E-V-S) \frac {\kappa(\kappa-1)}
{r^2}$ (dashed lines) and the effective pseudo-spin orbital
potential (PSOP) $\displaystyle\frac {\kappa} {r} \frac {d(V+S)}
{dr}$ (solid line ) multiplied by the square of the wave function
$F$ of the lower components in arbitrary scale for $d_{3/2}$
(upper) and $s_{1/2}$ (lower) in $^{120}$Zr. Taken from
Ref.~\cite{Meng98r}.} \label{FigD20}
\end{figure}

The effective PSOP does not depend on the binding energy of the
single particle level, but depends on the angular momentum and
parity. On the other hand the effective PCB depends on the energy.
Comparing these two effective potentials one could see the energy
dependence of the pseudo-spin symmetry. They are given in
Fig.~\ref{FigD19} for $s_{1/2}$ (lower) and  $d_{3/2}$ (upper) of
$^{120}$Zr in arbitrary scale.

The pseudo-spin approximation is much better for the less bound
pseudo-spin partners, because the effective PCB is smaller for the
more deeply bound states. This is in agreement with the results
shown in Fig.~\ref{FigD18}. The effective PSOP and the effective
PCB are also given as inserts in Fig.~\ref{FigD19} in order to
show their behavior near the nuclear surface.

The effective PCB (dashed lines or dot-dashed lines) and  the
effective PSOP (solid lines) multiplied by the squares of the
lower component wave function $F(r)$ are given in
Fig.~\ref{FigD20}, for $2s_{1/2}$ (upper left), $3s_{1/2}$ (lower
left), $1d_{3/2}$ (upper right), and $2d_{3/2}$ (lower right) of
$^{120}$Zr in arbitrary scale. The pseudo-spin approximation is
much better for the less bound pseudo-spin partners, because the
effective PCB is smaller for the more deeply bound states. This is
in agreement with the results shown above. It is clear that the
contribution of the effective PCB (dashed lines or dot-dashed
lines) is much bigger than that of the effective PSOP (solid
lines). Generally the effective PSOP is two orders of magnitude
smaller than the effective PCB. The energy difference between the
orbital $j = \tilde l + 1/2$ and the orbital $j = \tilde l - 1/2$
is always negative with some exceptions. Further investigations
reveal that the integration of $\dfrac {d(V+S)} {dr} |F|^2$ over
$r$ gives the splitting of the pseudo-spin partners, whose sign
will decide the normal splitting or the reverse.

\subsection{Halos and giant halos}
\label{subsec:halo}

Since the experimental discovery of neutron halo phenomena in
$^{11}$Li \cite{Tanihata85}, the study of exotic nuclei has become
a very challenging topic in nuclear physics. Experimentally more
and more halo nuclei have been observed~\cite{Mueller93,
Tanihata95, Hansen95, Jonson04, Jensen04, Liu01b, Cai02a,
ZhangHY02a, Liu04}. On the theoretical side very different models
have been used to describe these phenomena. Three-body
calculations based on an inert core of $^9$Li and two outside
neutrons treat the translational invariance properly, but they
neglect polarization effects. Full shell model calculations
contain all the configuration mixing, but they are limited to
relatively small configuration spaces in the oscillator model. In
the mean field description, the large radius observed
experimentally in $^{11}$Li has been interpreted by the fact that
filling in more and more neutrons in the nuclear well, the Fermi
surface for the neutrons comes close to the continuum limit. Due
to the small single-neutron separation energy, the tails of the
two wave functions of the last filled orbital (1$p_{1/2}$) reach
very far outside of the nuclear well and a neutron halo is
formed~\cite{Tanihata85}. Although this simple interpretation is
based on the mean field picture, several microscopic theoretical
investigations within self-consistent mean field
models~\cite{Bertsch89, Koepf91b, Sagawa92, Zhu94} have failed. In
fact, it would be a strange accident if the last occupied neutron
level in $^{11}$Li, the 1$p_{1/2}$ orbit, would be so close to the
continuum limit that the tail of the last two uncorrelated neutron
wave functions reaches so far out as would be necessary to
reproduce the large experimental radius observed. Therefore, in
the non-relativistic scheme Bertsch et al \cite{Bertsch89} and
Sagawa \cite{Sagawa92} introduced an artificial modification to
the potential in order to reproduce the small separation energy in
their mean field calculations using Skyrme interactions. In this
way the authors were able to reproduce qualitatively the observed
trends, although some discrepancies still remained.

Koepf et al \cite{Koepf91b} were the first to investigate the
neutron halo in the RMF theory for both spherical and axially
deformed cases. They found large deformations for the lighter Li
isotopes, but a spherical shape for $^{11}$Li and as in the
non-relativistic investigations the binding energy of the
1$p_{1/2}$ was too large so as to reproduce a neutron halo. Zhu et
al \cite{Zhu94} improved the result of the RMF calculations by
applying a similar modification to the potential as in
Ref.~\cite{Sagawa92} in order to adjust the proper size of the
halo.

The first self-consistent microscopic description of the halo in
$^{11}$Li was achieved by using the RCHB theory where the pairing
correlations are taken into account by a density dependent force
of zero range and the halo was reproduced in a self-consistent way
by using the scattering of Cooper pairs to the 2s$_{1/2}$ level in
the continuum~\cite{Meng96}.

In Ref.~\cite{Poschl97}, pairing correlations and the coupling to
particle continuum states are described by finite range two-body
forces and the finite element method are used in the coordinate
space discretization of the coupled system of
Dirac-Hartree-Bogoliubov integro-differential eigenvalue
equations, and Klein-Gordon equations for the meson
fields~\cite{Poschl97b}. The authors of Ref.~\cite{Poschl97}
performed calculations for the isotopic chains of Ne and C nuclei
and found evidence for the occurrence of neutron halo in heavier
Ne isotopes. The RHB theory is also applied to the description of
the properties of light nuclei in isotopic chains of C, N, O, F,
Ne, Na, and Mg with large neutron excess in
Ref.~\cite{Lalazissis98a}. The RHB model predicts the formation of
neutron skin and eventually a neutron halo in Ne and Na due to the
quasi-degeneracy of the triplet of states 1f$_{7/2}$, 2p$_{3/2}$,
and 2p$_{1/2}$ which are close to the Fermi surface in neutron
rich nuclei.

To describe medium-heavy nuclei far from the $\beta$ stability
line, the main problem of nuclear structure models is the
extrapolation of effective forces to nuclei with extreme isospin
values because most of these nuclei are still not accessible
experimentally and therefore one could not compare theoretical
results with empirical data. In order to make predictions for
medium-heavy nuclei at the neutron drip line, one has to test
available effective interactions in detailed calculations of the
properties of neutron-rich nuclei for which a comparison with
experimental data is possible. This has been done in
Ref.~\cite{Lalazissis98e} where the RHB theory is applied in the
description of the ground-state properties of Ni and Sn isotopes.
In a comparison with available experimental data, it is shown that
the NL3 + Gogny D1S effective interaction provides an excellent
description of binding energies, neutron separation energies, and
proton and neutron rms radii. The results indicate that this
choice of model parameters might also be valid for nuclei with
more extreme isospin values, i.e., medium-heavy nuclei at the drip
lines.

We note that the RHB model solved in a discretized coordinate
space is available only for spherical systems up to now. The
deformed RHB theory solved in a harmonic oscillator
basis~\cite{Lalazissis00b} is not applicable for halo nuclei due
to the localization property of the harmonic oscillator potential.
One should work in two or three dimensional spatial lattice or
instead, one can improve the asymptotic behavior of the HO wave
function~\cite{Stoitsov98a, Stoitsov98b} or adopt other basis
which has a correct asymptotic behavior~\cite{Zhou03prc}.

In the following, we will briefly review the RCHB description of
halo phenomena and the prediction of giant halo in light and
medium-heavy nuclei.

\subsubsection{ {Neutron halo in light nuclei}}

The ground state properties of Li isotopes with mass numbers $A=6$
to $A=11$ were investigated by using the RCHB theory with the
effective interaction NL2~\cite{Meng96}. Pairing is neglected for
the three protons and the strength $V_0$ of the pairing force
(\ref{vpp}) for the neutrons is determined by a calculation in the
nucleus $^7$Li adjusting the corresponding pairing energy
-$\displaystyle \frac{1}{2}  \mbox{Tr} \ \Delta\kappa$ to that of
a RHB-calculation in an oscillator basis using the finite range
part of the Gogny force D1S of Ref.~\cite{Berger84} in the pairing
channel. Good agreement with experimental values is found for the
total binding energies and the radii of the isotope chain $^6$Li
to $^{11}$Li. The matter radius shows a considerable increase when
going from the nucleus $^9$Li to $^{11}$Li. In contrast to the
earlier mean field calculations of Refs. \cite{Bertsch89,
Sagawa92, Zhu94} these results are obtained without any artificial
modifications of the potential.

\begin{figure}
\centering
\includegraphics[width=6.0cm,angle=270]{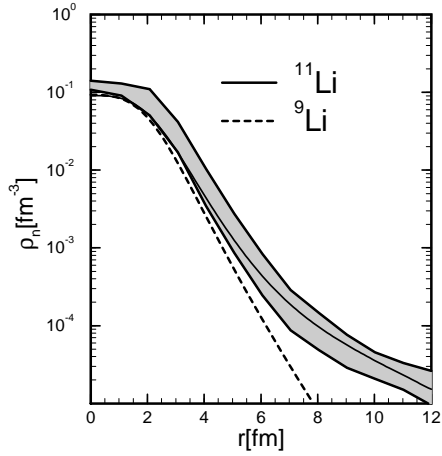}
\caption{Calculated and experimental density distribution in
$^{11}$Li and $^9$Li. The solid line shows the result of $^{11}$Li
while the dashed line corresponds to the calculation of $^9$Li.
The shaded area gives the experimental results with error bars.
Taken from Ref.~\cite{Meng96}.} \label{FigD21}
\end{figure}

In Fig.~\ref{FigD21} the corresponding density distribution for
the neutrons in the isotopes $^9$Li and $^{11}$Li is shown. It is
clearly seen that the increase of the matter radius is caused by a
large neutron halo in the nucleus $^{11}$Li. Its density
distribution is in very good agreement with the experimental
density of this isotope shown with its error bars by the shaded
area.

\begin{figure}
\centering
\includegraphics[width=6.0cm,angle=270]{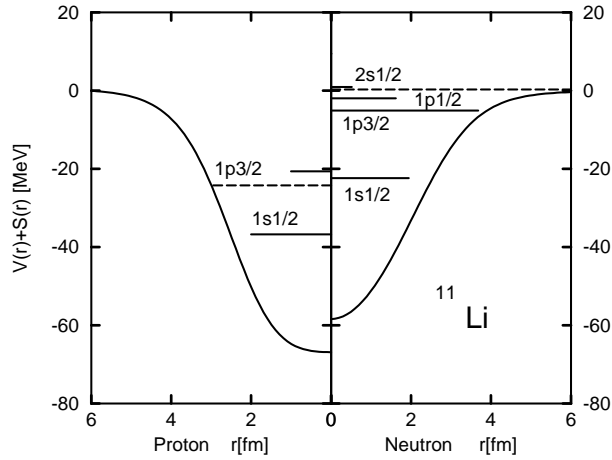}
\caption{The mean field potential $S+V$ for protons (l.h.s.) and
neutrons (r.h.s.). The chemical potential is given by a dashed
line. The energy levels in the canonical basis are indicated by
horizontal lines with various lengths proportional to the
occupation of the corresponding orbit. Taken from
Ref.~\cite{Meng96}.} \label{FigD22}
\end{figure}

In order to understand the microscopic structure of this halo, in
Fig.~\ref{FigD22} the mean field $S(r)+V(r)$ for the protons and
neutrons together with the energy levels in the canonical basis is
shown. The Fermi level for the neutrons is very close to the
continuum limit in close vicinity to the $\nu 1p_{1/2}$ and to the
$\nu 2s_{1/2}$ level. The length of these energy levels in
Fig.~\ref{FigD22} is proportional to the corresponding occupation.
One clearly finds that pairing correlations cause a partial
occupation of both the $\nu 1p_{1/2}$ and the $\nu 2s_{1/2}$
level, i.e. a scattering of Cooper pairs to the continuum.

In contrast to many previous investigations, in the RCHB model,
the halo is not formed by two neutrons occupying the $1p_{1/2}$
level very close to the continuum limit, but is formed by
Cooper-pairs scattered mainly in the two levels $1p_{1/2}$ and
$2s_{1/2}$. This is made possible by the fact that the $2s_{1/2}$
comes down close to the Fermi level in this nucleus and by the
density dependent pairing interaction coupling the levels below
the Fermi surface to the continuum. The RCHB theory provides a
much more general mechanism for halo occurring in nuclei. One
needs only several single particle levels with small orbital
angular momenta and correspondingly small centrifugal barriers
close to, but not necessarily directly at, the continuum limit.

\begin{figure}
\centering
\includegraphics[width=6.0cm,angle=270]{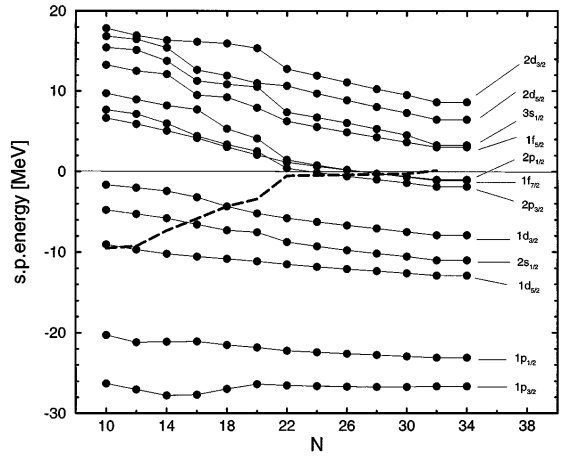}
\caption{ {Canonical basis single-particle neutron levels as
functions of the number of neutrons. The spectrum is calculated
for Ne isotopes. Taken from Ref.~\cite{Poschl97}.}}
 \label{FigD23}
\end{figure}

 In Ref.~\cite{Poschl97}, the isotopic chains of Ne and C
nuclei were studied with the RHB model with finite range pairing
interaction in the coordinate space. The formation of neutron skin
and eventually a neutron halo in Ne was found to be mainly due to
the quasi-degeneracy of the triplet of states 1f$_{7/2}$,
2p$_{3/2}$, and 2p$_{1/2}$. In Fig.~\ref{FigD23}, the neutron
single-particle spectrum in the canonical basis is displayed for
Ne isotopes. The energies of levels in the continuum decrease with
increasing neutron number (cf. Fig.~\ref{FigD28} (a)). The shell
structure dramatically changes at $N \ge$ 22. The triplet of
states 1f$_{7/2}$, 2p$_{3/2}$, and 2p$_{1/2}$ approaches zero
energy, and a gap is formed between these states and all other
states in the continuum. The Fermi level uniformly increases
toward zero for $N$ = 22. Between $N$ = 22 and $N$ = 32, the Fermi
level is practically constant and very close to the continuum. The
addition of neutrons in this region of the drip does not increase
the binding. Only the spatial extension of neutron distribution
displays an increase. At $N$ = 32 the Fermi energy becomes
slightly positive, and heavier isotopes are not bound any more.
The finite range pairing interaction promotes neutrons from the
1f$_{7/2}$ orbital to the 2p levels. Since these levels are so
close in energy, the total binding energy does not change
significantly. Because of their small centrifugal barrier, the
2p$_{3/2}$ and 2p$_{1/2}$ orbitals form the halo.

\subsubsection{Giant halos in Zr isotopes}

In all of the halos observed so far, one has only a very small
number of nucleons, namely one or two outside of the normal core.
In order to study the influence of correlations and many-body
effects it would be very interesting to find also nuclei with a
larger number of neutron distributed in the halo. Following the
successful description of the halo nucleus $^{11}$Li, the RCHB
theory was used to study and predict giant neutron halos for
Zr-isotopes close to the neutron drip line~\cite{Meng98prl}. It is
formed by up to six neutrons outside of the $^{122}$Zr core with
the magic neutron number $N=82$.

\begin{figure}
\centering
\includegraphics[width=6.0cm]{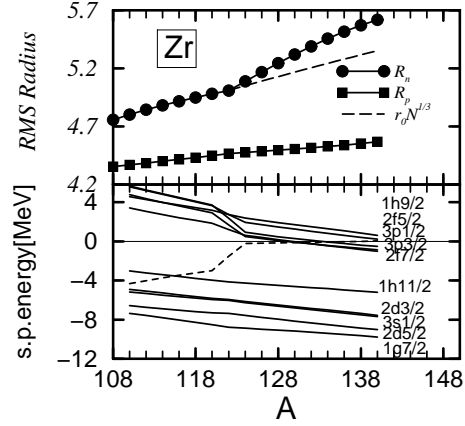}
\caption{Upper: rms radii for neutrons and protons in Zr isotopes
close to the neutron drip line as a function of the mass number
$A$. Lower: single particle energies for neutrons in the canonical
basis. The dashed line indicates the chemical potential. Taken
from Ref.~\cite{Meng98prl}.} \label{FigD24}
\end{figure}

From the upper panel of Fig.~\ref{FigD24} which presents the rms
radii of the protons and neutrons for the Zirconium isotopes, one
finds a kink for the neutron rms radius at the magic neutron
number $N=82$. This kink can be understood more clearly by
considering the microscopic structure of the underlying wave
functions and the single particle energies in the canonical basis
which is shown in the lower panel of Fig.~\ref{FigD24}. Going from
$N=70$ to $N=100$, there is a big gap above the 1$h_{11/2}$ orbit.
For $N>82$, the neutrons are filled to the levels in the continuum
or weakly bound states in the order of $3p_{3/2}$, $2f_{7/2}$,
$3p_{1/2}$, $2f_{5/2}$ and $1h_{9/2}$.

\begin{figure}
\centering
\includegraphics[width=6.0cm]{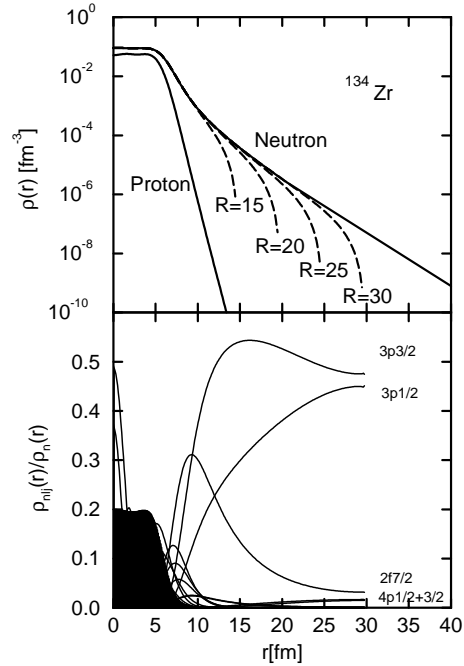}
\caption{Upper: neutron and proton density distribution in
$^{134}$Zr. Dashed lines indicate calculations for different
values of the box size $R$ and the dashed-dotted line give the
neutron distribution for the core $^{122}$Zr. Lower: relative
contributions of the different orbits to the full neutron density
as a function of the radius. The shaded area indicates the total
neutron density in arbitrary units. Taken from
Ref.~\cite{Meng98prl}.} \label{FigD25}
\end{figure}

The neutron chemical potential is given in this Figure by a dashed
line. It approaches rapidly the continuum already shortly after
the magic neutron number $N=82$ and it crosses the continuum at
$N=100$ for the nucleus to $^{140}$Zr. In this region the chemical
potential is very small and almost parallel to the continuum
limit. This means that the additional neutrons are added with a
very small, nearly vanishing binding energy at the edge of the
continuum. The total binding energies $E$ for the isotopes above
$^{122}$Zr are therefore almost identical. This has been
recognized already in Ref.~\cite{Sharma94} in RMF calculations
using the BCS approximation and an expansion in an oscillator
basis, which is definitely not reliable for chemical potentials so
close to the continuum limit.

In the upper panel of Fig.~\ref{FigD25} the density distributions
for neutrons and protons in the nucleus $^{134}$Zr are shown. A
very large box size is needed to describe the halo properly. For
$R=$30 fm the neutron density is reliably reproduced only up to
$r=25$ fm, where it has decreased to 10$^{-6}$ fm$^{-3}$. The full
line in the upper panel of Fig.~\ref{FigD25} is an asymptotic
extension to infinite box size. On the other hand, the density
distribution inside the nucleus is reproduced properly even for
small values of the box size. The relative contributions
$\rho_{nlj}$ of the different orbits characterized by the quantum
numbers $nlj$ with respect to the total neutron density $\rho_n$
are displayed in the lower panel of Fig.~\ref{FigD25}. The shaded
area gives the total neutron density in arbitrary units. The halo
is formed essentially by contributions from three orbits
$3p_{3/2}$, $3p_{1/2}$, and $2f_{7/2}$. The most inner part of the
halo ($7\le r\le 9$ fm) the $2f_{7/2}$ orbit plays the dominant
role. As can be seen in the lower part of Fig.~\ref{FigD25}, this
orbit is slightly below the chemical potential and to the
continuum limit in this nucleus. Further outside ($10\le r\le 15$
fm) its relative contribution is strongly reduced because of the
larger centrifugal barrier felt by the $l=3$ orbit. In this region
the orbit $3p_{3/2}$, which has nearly the same position as the
$2f_{7/2}$ orbit, takes over. Because of the smaller orbital
angular momentum, this orbit feels a reduced centrifugal barrier.
For even larger distances from the center ($r \ge 15$ fm) its
relative contribution is somewhat reduced and the $3p_{1/2}$ orbit
gains importance. The $3p_{3/2}$ and the $3p_{1/2}$ levels feel
the same centrifugal barrier, but the latter is situated directly
at the continuum limit and therefore it is more loosely bound than
the other two orbits.

The occupation probabilities in the canonical basis of all the
neutron levels near the Fermi surface, i.e. in the interval $-20
\le E \le 10 $ MeV are presented in Fig.~\ref{FigD26} for all the
Zr-isotopes between $A=108$ and $A=140$. The chemical potential is
indicated by a vertical line. For the mass numbers $A<122$ the
chemical potential lies several MeV below the continuum limit
($E=0$) and there is only very little occupation in the continuum
($E>0$). The nucleus $^{122}$Zr has a magic neutron number and no
pairing. As the neutron number goes beyond this closed core, the
occupation of the continuum becomes more and more important.
Adding up the occupation probabilities $v^2$ for the levels with
$E>0$ one finds in the continuum 2 particles for $N=84$, 4 for
$N=86$, 6 $N=88$, roughly 3 for $N=90$, roughly 4 for $N=92$,
roughly 3 for $N=94$, roughly 4 for $N=96$, roughly 5 for $N=98$,
and roughly 6 for $N=100$ (where the neutron drip line is
reached).

\begin{figure}
\centering
\includegraphics[width=8.0cm,angle=270]{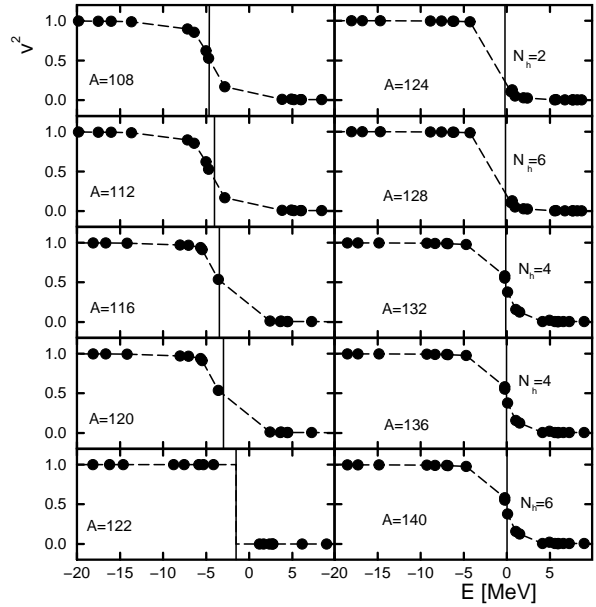}
\caption{The occupation probabilities in the canonical basis for
various Zr isotopes with mass number A as a function of the single
particle energy. The chemical potential is indicated by a vertical
line. For $A>122$ the number $N_h$ of neutrons in the halo is also
shown. Taken from Ref.~\cite{Meng98prl}.} \label{FigD26}
\end{figure}

Considering the very extended neutron rms radii of these systems
and estimating the number of valence neutrons which would fit into
the same volume outside the $^{122}$Zr core (shown in
Fig.~\ref{FigD25} by a dashed dotted line) if they would be packed
with normal neutron density, one finds for $^{134}$Zr 24 neutrons
instead of 12 and for $^{140}$Zr 34 neutrons instead of 18. This
means that the density outside of the $^{122}$Zr core is reduced
by roughly a factor of 2. This phenomenon is therefore clearly a
neutron halo and not a neutron skin. As one sees from Fig. 3 the
tail of the density is proportional to $\exp(-2\sqrt{4mS}r)$ with
$S\approx 0.4\,$MeV. This fact is also emphasized by the extremely
low 2n-separation energies ($<~0.5$ MeV) of these systems. This
phenomenon is called a {\it Giant Halo} because of the large
number of particles in the halo region~\cite{Meng98prl}..

\subsubsection{Giant halos in lighter nuclei}

Although the giant halos were predicted in Zr isotopes, it is very
difficult for experimentalists to confirm them because the exotic
Zr isotopes with $N>82$ are too heavy to be synthesized by the RIB
facilities at present. Thus it is valuable to investigate the
giant halo phenomena in lighter nuclei which are experimentally
accessible today. Ground state properties of all the even-even O,
Ca, Ni, Zr, Sn, Pb isotopes ranging from the proton drip line to
the neutron drip line were investigated~\cite{Meng02r,Zhang03}.
The two neutron sepation energies are shown in Fig.~\ref{FigD1}
where one finds the $S_{2n}$ values for exotic Ca isotopes near
neutron drip line are very close to zero in the large mass region,
i.e., $S_{2n}\approx 2.06,~1.24,~0.76,~0.43,~0.21,~0.04$ MeV for
$A=62,~64,~66,~68,~70,~72$ isotopes, respectively. If one regards
$^{60}$Ca as a core, then the several valence neutrons occupying
levels above the $N=40$ sub-shell for these $A>60$ exotic nuclei
are all weakly bound and can be scattered easily to the continuum
levels due to the pairing interaction, especially for
$^{66-72}$Ca. This case is very similar to $S_{2n}$ in Zr isotopes
with $N>82$ (see Fig.~\ref{FigD1})~\cite{Meng98prl} . Note that
for Sn isotopes in the vicinity of the drip line~\cite{Meng99npa},
however, $S_{2n}$ decrease rather fast with the mass number and
for Ni isotopes \cite{Meng98npa}, $S_{2n}$ are quite large ($\sim
2 $MeV) and undergo a sudden drop to negative value at the drip
line nucleus. Such behavior of $S_{2n}$ in Ca isotopes with $N>40$
indicates that there exist giant halos in these nuclei, just as
what happens in Zr isotopes with $N>82$~\cite{Meng98prl}. From
Fig.~\ref{FigD5}, one also finds similarity between Ca and Zr
isotopes: the neutron radius increases suddenly with neutron
number increasing in the Ca chain above $N=40$ and in Zr chain
above $N=82$. Together with the analysis of nucleon density
distributions, single particle levels and the occupation
probabilities in the canonical basis, it is predicted that giant
halo may exist in Ca isotopes with A$>$60~\cite{Meng02r,Zhang03}.

Compared with giant halos in Zr chain, the halos in Ca isotopes
will be much easier to access with today's radioactive facilities.
This means that the giant halos in the Ca isotopic chain are
challenging but promising for experimentalists. In order to make
the conclusion more solid and exclude the effects from possible
error in $S_{2n}$ ($\sim$ 1 MeV), in Fig.~\ref{FigD27}, the
two-neutron separation energies $S_{2n}$ in drip line region for
even-neutron Ne, Na, Mg, and Al chains (left panel) and for Ar, K,
Ca, Sc and Ti chains (right panel) are  presented. Open symbols
represent the values calculated from the RCHB theory with the NLSH
parameter set while corresponding filled symbols represent the
data available. It should be noted that the calculations are
performed by keeping the shape spherical. Nevertheless it can be
seen from the figure that all these $S_{2n}$ curves are almost
parallel to each other and the zero $S_{2n}$ line. Although the
calculated $S_{2n}$ value may differ from the data by 1 or 2 MeV,
there must be one or more isotope chains in which many nuclei have
the $S_{2n}$ values very close to zero thus are good candidates
for giant halos.

Experimentally much efforts have been made to search the drip line
for heavier elements. New experiments have shown that both
$^{37}$Na and $^{34}$Ne are bound while $^{33}$Ne and $^{36}$Na
are missing \cite{Notani02, Lukyanov02}, as expected from the
predictions in Ref.~\cite{Meng97}. In Fig.~\ref{FigD27}, $^{37}$Na
and $^{34}$Ne lie in the drip line region and approach the
predicted giant halo area. Therefore much more experimental effort
should be devoted to measuring their masses and extending the
isotope chains for the possible giant halos.

\begin{figure}
\centering
\includegraphics[width=6.5cm]{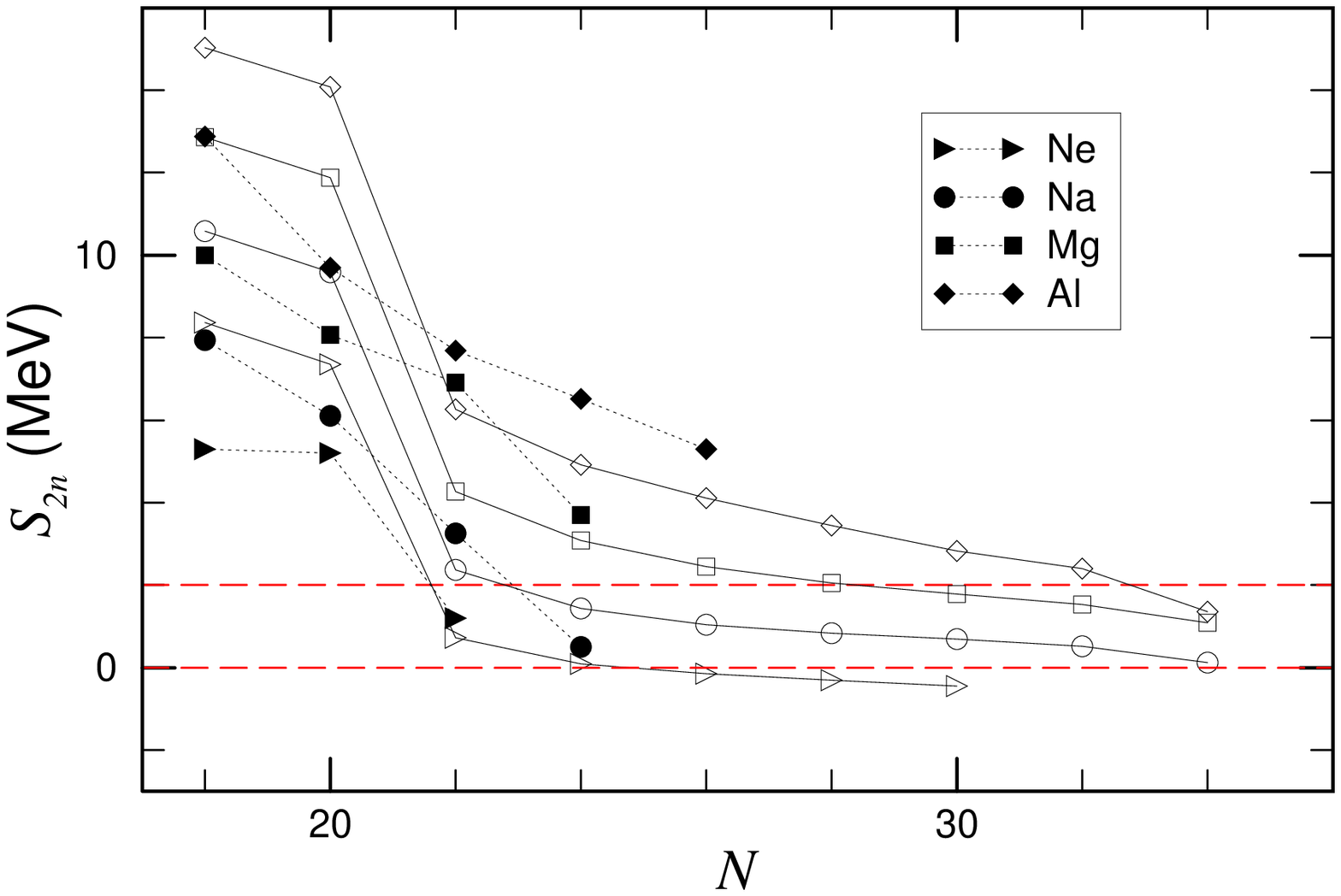}
\includegraphics[width=6.5cm]{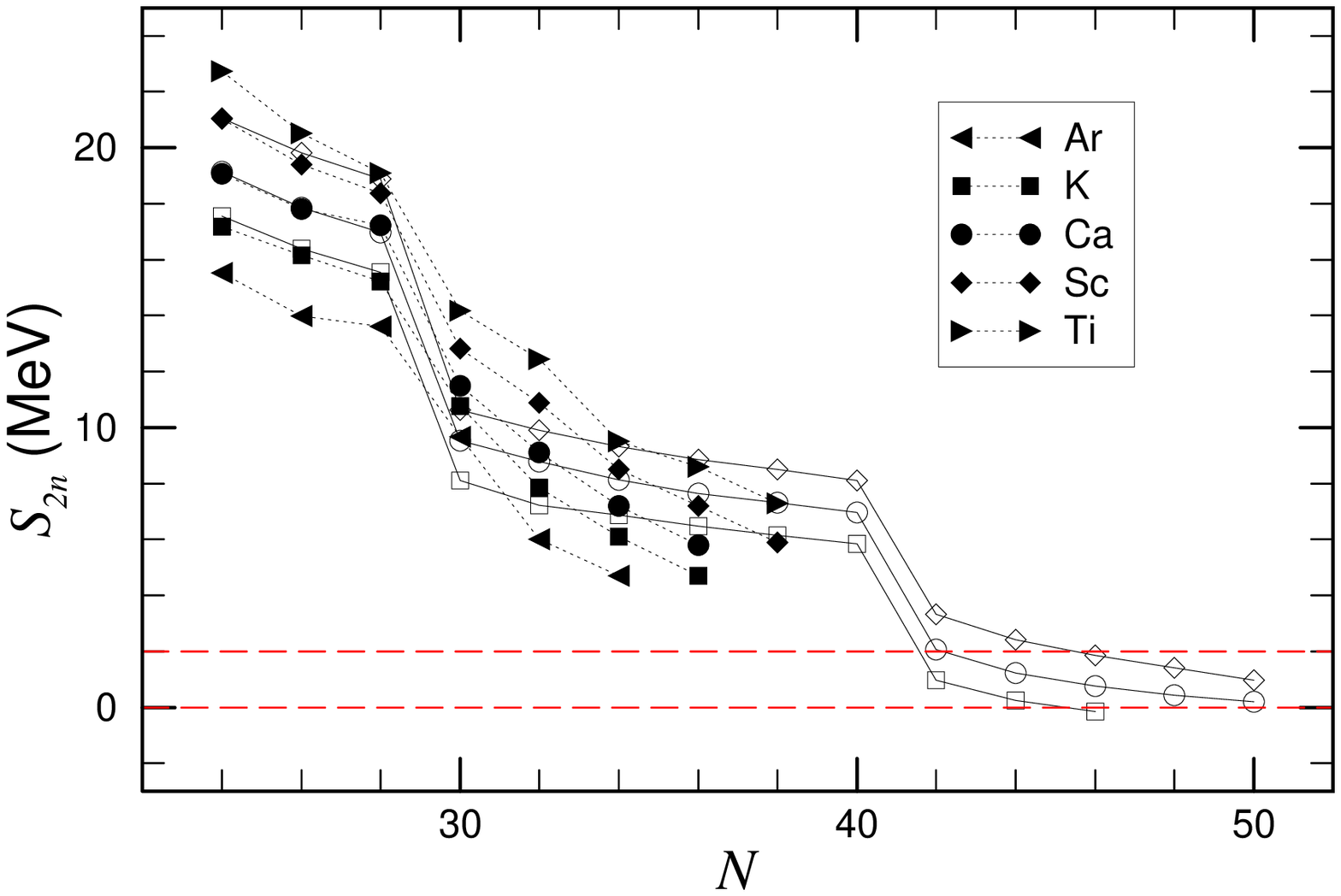}
\caption{Left: The two-neutron separation energies $S_{2n}$ for
even-neutron Ne, Na, Mg, and Al nuclei in the drip line region.
Open symbols represent the values calculated with the RCHB theory
with the NLSH parameter set, while corresponding solid symbols
represent the data available.  The horizontal line at 2MeV denotes
the upper limit for the possible occurrence of halos. Right: for
even Ar, K, Ca, Sc, and Ti nuclei in the drip line region. Taken
from Ref.~\cite{Zhang03}.} \label{FigD27}
\end{figure}

\subsection{Halos in hypernuclei}

\begin{figure}[hbt]
\centering
\includegraphics[width=10.cm]{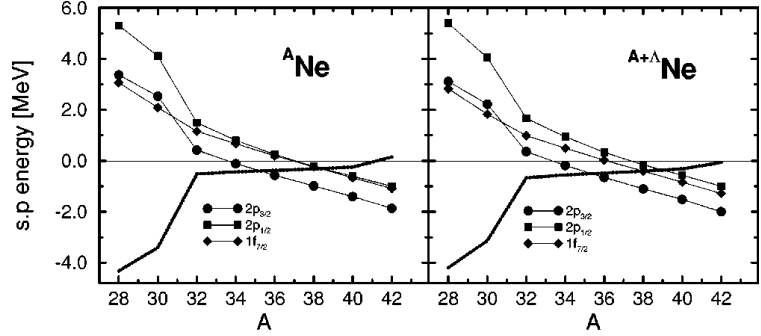}
\caption{ {1f-2p single-particle neutron levels in the canonical
basis for the Ne (a), and Ne+$\Lambda$ (b) isotopes. The dotted
line denotes the Fermi level. Taken from
Ref.~\cite{Vretenar98r}.}} \label{FigD28}
\end{figure}

 The production mechanisms, spectroscopy, and decay modes of
hypernuclear states have been the subject of many theoretical
studies. The relativistic Hartree Bogoliubov model in coordinate
space with finite range pairing interaction~\cite{Poschl97,
Poschl97b} has also been applied to describe $\Lambda$ hypernuclei
with a large neutron excess~\cite{Vretenar98r}. It is found that
the inclusion of the $\Lambda$ hyperon does not produce excessive
changes in bulk properties but shifts the neutron drip by
stabilizing an otherwise unbound core nucleus at drip line. For
example, the nucleus $^{42}$Ne was unbound without the $\Lambda$
(see Fig.~\ref{FigD28} (a)). The presence of the strange baryon
stabilizes the otherwise unbound core and $^{42+\Lambda}$Ne
becomes bound (see Fig.~\ref{FigD28} (b)). The microscopic
mechanism through which additional neutrons are bound to the core
originates from the increase in magnitude of the spin-orbit term
in presence of the $\Lambda$ particle~\cite{Vretenar98r}.

\begin{figure}[hbt]
\centering
\includegraphics[width=8.cm]{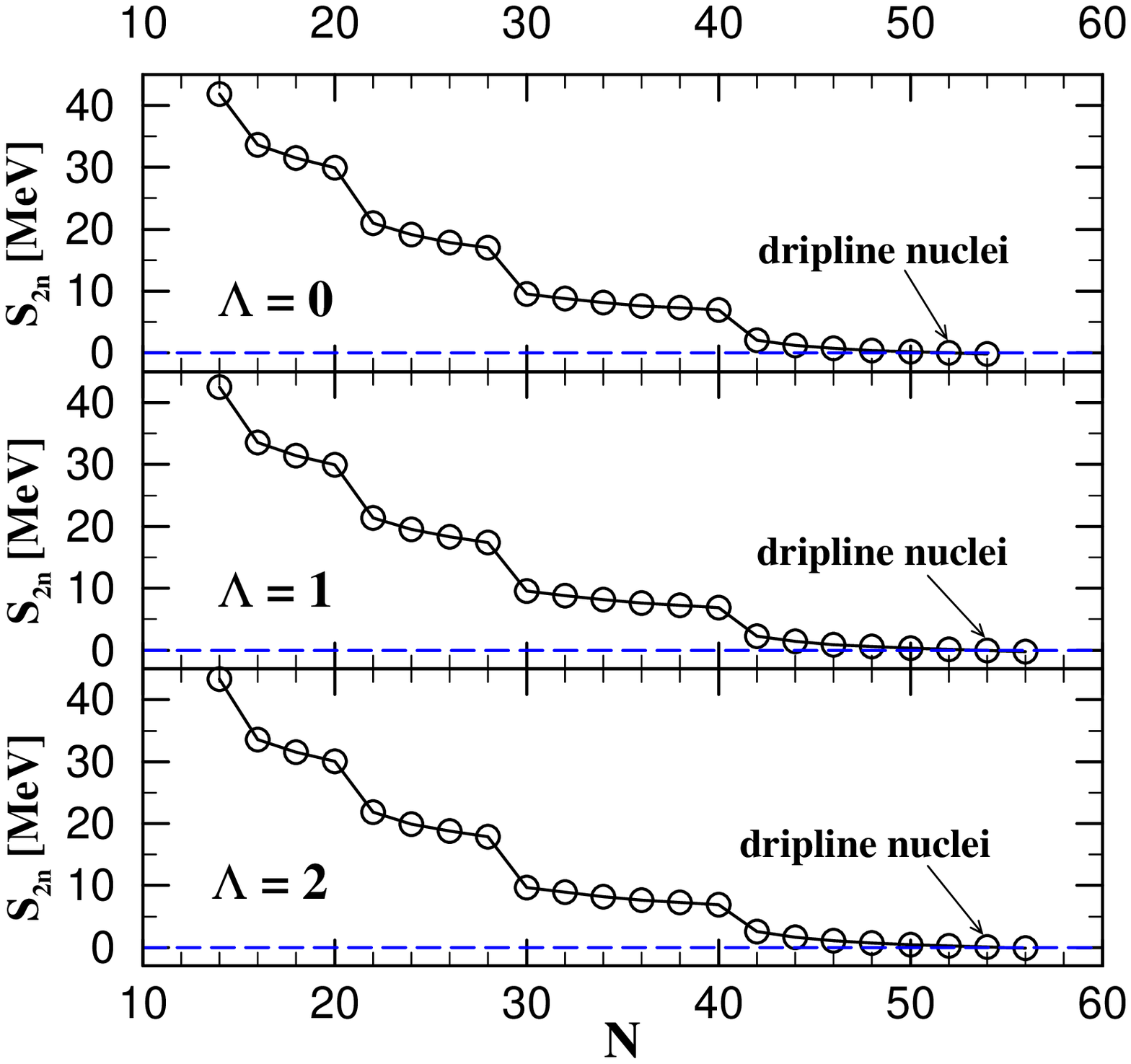}
\caption{The two-neutron separation energies $S_{2n}$ in even-N Ca
isotopes versus the neutron number N. The upper panel is for
ordinary nuclei, the middle one for single-$\Lambda$ hyper nuclei,
and the lower one for double-$\Lambda$ hyper nuclei. Taken from
Ref.~\cite{Lv03}.} \label{FigD29}
\end{figure}

Motivated by experimental knowledge of $\Lambda$-N interaction and
theoretical understanding on giant halo~\cite{Meng98prl, Meng02r,
Zhang03}, it is interesting to observe the possible appearances of
halos in hyper exotic nuclei as what have been done in
Ref.~\cite{Lv03}. In Fig.~\ref{FigD29}, two neutron separation
energies $S_{2n}$ for ordinary nuclei, single-$\Lambda$ hyper
nuclei and double-$\Lambda$ hyper nuclei of Ca isotopes labelled
by $\Lambda = 0$, $\Lambda =1$ and $\Lambda =2$, respectively,
from the proton to neutron drip line are presented. It is clear
that one or two $\Lambda$ hyperons lower the neutron Fermi level a
little bit but keep the neutron shell structure unchanged.
Therefore, the neutron drip line is pushed outside from $N=52$ in
ordinary isotope chain to $N=54$ in hyper isotope chain.
Meanwhile, giant halo due to pairing correlation and the
contribution from the continuum still exist in Ca hyper nuclei
similar as that in ordinary Ca isotopes~\cite{Meng02r, Zhang03}.
This is a slight but rewarding step for exploring the limit of
drip line nuclei on the basis of the giant halo. For more details,
see Ref.~\cite{Lv03}.

\begin{figure}
\centering
\includegraphics[width=10.cm]{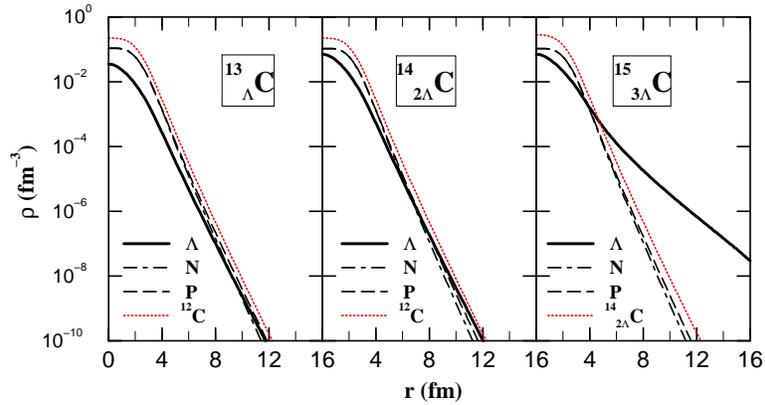}
\caption{Density distributions for $\Lambda$ (solid), neutron
(dot-dashed), proton (long-dashed) and corresponding core (dotted)
in $^{13}_{\Lambda}$C, $^{14}_{2\Lambda}$C and $^{15}_{3\Lambda}$C
in logarithmic scales. Taken from Ref.~\cite{Lv02}.}
\label{FigD30}
\end{figure}

Apart from neutron halos in hyper nuclei, as $\Lambda$ hyperon is
less bound than the corresponding nucleon in nuclei, it is worth
investigating the existence of hyperon halos. In
Ref.~\cite{Rufa90}, a lambda halo was tentatively suggested in a
RMF plus BCS model. However, the normal BCS method will have
unphysical solution by involving baryon gas~\cite{Meng98npa}, such
prediction needs further study. Therefore hyper carbon isotopes
are studied by RCHB theory in Ref.~\cite{Lv02}. In
Fig.~\ref{FigD30}, the baryon density distributions in logarithmic
scales versus the radius $r$ in hyper nuclei $^{13}_{\Lambda}$C,
$^{14}_{2\Lambda}$C and $^{15}_{3\Lambda}$C are presented, where
the solid, dot-dashed, long-dashed and dotted lines represent
density distributions for the lambda, neutron, proton, and
corresponding core, respectively. It can be seen that up to two
$\Lambda$ hyperons added to the core $^{12}$C, the nucleon density
distributions remain the same and hyperon density distributions at
the tail are comparable with those of the nucleons. An intriguing
phenomenon appears in $^{15}_{3\Lambda}$C where the hyperon
density distribution has a long tail extended far outside of its
core $^{14}_{2\Lambda}$C (denoted by dotted line), as shown in the
right panel of Fig.~\ref{FigD30}. This is a signature of hyperon
halo. It is due to the weakly bound state $1p_{3/2}^\Lambda$ in
$^{15}_{3\Lambda}$C which has a small hyperon separation energy
and a density distribution with long tail.


\section{Magic numbers for superheavy nuclei}
\label{sec:she}

One of the fundamental and persistent questions in nuclear science
is the exploration for the limits of charge and mass that a
nucleus can attain and the creation of nuclei with masses and
charges much larger than those we are familiar with, i.e.,
superheavy nuclei. The properties of superheavy nuclei are
important to understand not only the nuclear structure, but also
the structure of the stars and the evolution of the universe.
However, after lots of efforts for both higher luminosities and
better detection efficiencies as well as the impressive progresses
on the synthesis of superheavy nuclei, the borders of the
upper-right corner in the nuclear chart still remain unknown.

Due to fission resulting from the balance of the attractive
nuclear surface tension and the repulsive Coulomb force, the
superheavy nuclei cannot exist in the classical charged
liquid-drop model.  The stability of superheavy nuclei is mainly
determined by shell effects. At the magic proton or neutron
numbers 2, 8, 20, 28, 50, and 82, as well as  N=126 for neutrons,
nuclei have higher stability and abundance compared with their
neighbors. Particularly, the highest stability is observed in the
case of the doubly magic nuclei. Amongst other special properties,
the doubly  magic nuclei are spherical. A semblance of the same
would work also for superheavy nuclei, if there were magic numbers
in this region. Consequently, these nuclei will be guarded against
a faster decay by fission and have more opportunities to be bound.
Therefore, it is important to find out the regions in the ($Z$,
$N$) plane where the shell effects are strong and the long
lifetimes of the superheavy nuclei can be expected.

The pioneering works on the superheavy elements are a series of
macro-microscopic nuclear shell model calculations performed over
more than 30 years to determine nuclear masses and energy
surfaces, which suggested the existence of a group of relatively
stable nuclei separated in neutron and proton numbers from the
known heavy elements by a higher instability when approaching the
closed spherical shells Z=114 and N=184~\cite{Myers66, Meldner67,
Nilsson69, Mosel69}. This group of nuclei become to be known as
the island of superheavy elements and the verification of their
existence is one of the most challenging topics in world-wide
heavy ion research facility.

Experimentally, however, although the superheavy elements up to
$Z$=116 are synthesized or claimed to be
synthesized~\cite{Hofmann95a, Hofmann95b, Hofmann96, Ghiorso95,
Lazarev94, Lazarev95, Lazarev96, Hofmann98, Oganessian99,
Hofmann00, Oganessian00, Oganessian01, Morimoto04, Oganessian04}
by different heavy ion reaction types including the cold, warm and
hot fusion reactions, they are not examples of the originally
thought island of superheavy elements yet.

There is no consensus among different theories with regard to the
center of the island of superheavy nuclei. Based upon
phenomenological models such as finite-range droplet model (FRDM),
the shell closures were predicted at $Z$=114 and
$N$=184~\cite{Moller94}. Additionally, the FRDM also predicts
larger shell gaps at $Z$=104, 106, 108, 110 and at $N$=162,
164~\cite{Moller94}. In Nilsson-Strutinsky scheme, a similar
pattern of deformed nuclei have been predicted about $Z$=108 and
$N$=162 as in FRDM~\cite{Patyk91, Sobiczewski94}. However, the
main obstacle is the question whether the macroscopic approaches
which apply to the region of $\beta$-stability line can be
extrapolated to the superheavy nuclei. Recently, microscopic
calculations~\cite{Wu96, Lalazissis96, Cwiok96, Rutz97, Bender98,
Bender99, Bender00, Meng00, Reinhard02, Long02, Geng03a,
Burvenich04} are also attempted in describing the superheavy
nuclei. In the framework of RHB theory, calculations with a
finite-range pairing force of Gogny interaction D1 and effective
interaction NLSH show that $Z$=114 and $N$=160, $N$=166, $N$=184
exhibit stability compared to their neighbors and indications for
a doubly magic character at $Z$=106 and $N$=160 are also
observed~\cite{Lalazissis96}. The Skyrme Hartree-Fock (SHF) method
with interactions SkP and SLy7 predicts magic numbers at $Z$=126
and $N$=184, and also predicts the increased stability due to the
deformed shell effects at $N$ = 162~\cite{Cwiok96}. Considering
non-relativistic SHF effective interactions SkM*, SkP, SLy6, SkI1,
SkI3, SkI4 and relativistic mean field (RMF) effective
interactions PL-40, NLSH, NL-Z, TM1, the doubly magic spherical
nuclei $_{184}$114, $_{172}$120 and $_{184}$126 based on
two-nucleon gaps $\delta_{2p}$ and $\delta_{2n}$ are
given~\cite{Rutz97}. The uncertainty on the magic numbers lies in
the uncertain strength of the spin-orbit coupling in
non-relativistic approaches or the effective interactions in
relativistic models. The prediction of magic numbers in superheavy
nuclei remains a challenge for the nuclear models, which have to
be rigorously tested by a wide variety of nuclear properties
throughout the periodic table.

As the RCHB formalism allows for the proper description of the
coupling between the bound states and the continuum by the pairing
force and has shown a remarkable success in the description of
nuclei with unusual $N/Z$ ratio ~\cite{Meng96, Meng98prl,
Meng98npa, Meng02plb, Meng02r}, it is suitable to apply the RCHB
theory for superheavy nuclei. In principle, only a calculation in
a large multidimensional deformation space can definitively decide
the appropriate ground-state shape. However, as the traditional
superheavy nuclei are expected to be spherical and are located on
the nuclear chart around a spherical doubly magic nucleus next to
$^{208}_{126}$Pb, the RCHB theory with the assumption of spherical
shape can be applied for the search. Therefore here we will mainly
focus on the magic numbers in spherical superheavy nuclei. Some
crucial observables in searching shell closures and magic numbers
are taken into account, including two-nucleon separation energies,
two-nucleon gaps, the shell correction energies, as well as the
pairing energies and the effective pairing gaps. At the end of
this section, a brief discussion on the stability of the doubly
spherical nuclei against deformation and the main progress on the
investigation of the deformed superheavy nuclei will be given, as
well as the $\alpha-$ decay half-lives of superheavy nuclei.

\subsection{Two nucleon separation energies}

Traditionally, the magic nuclei are characterized by higher
stability than their neighboring ones and the magic numbers can be
revealed by the differences of the binding energy, i.e. the one-
or two-nucleon separation energies~\cite{Bohr69}. Therefore, the
nucleon separation energies are good starting points to
investigate the magic numbers in superheavy region. Due to the
absence of odd-even effects, the two-nucleon separation energy
$S_{2p}$ ($S_{2n}$) is better to quantify shell effects than the
single-nucleon separation energy $S_{p}$ ($S_{n}$). For an
isotonic (isotopic) chain, the $S_{2p}$ ($S_{2n}$) decrease with
the proton (neutron) number and its sudden jump indicates the
occurrence of a proton (neutron) magic number.

\begin{figure}[htbp]
 \centering
 \includegraphics[scale=0.4]{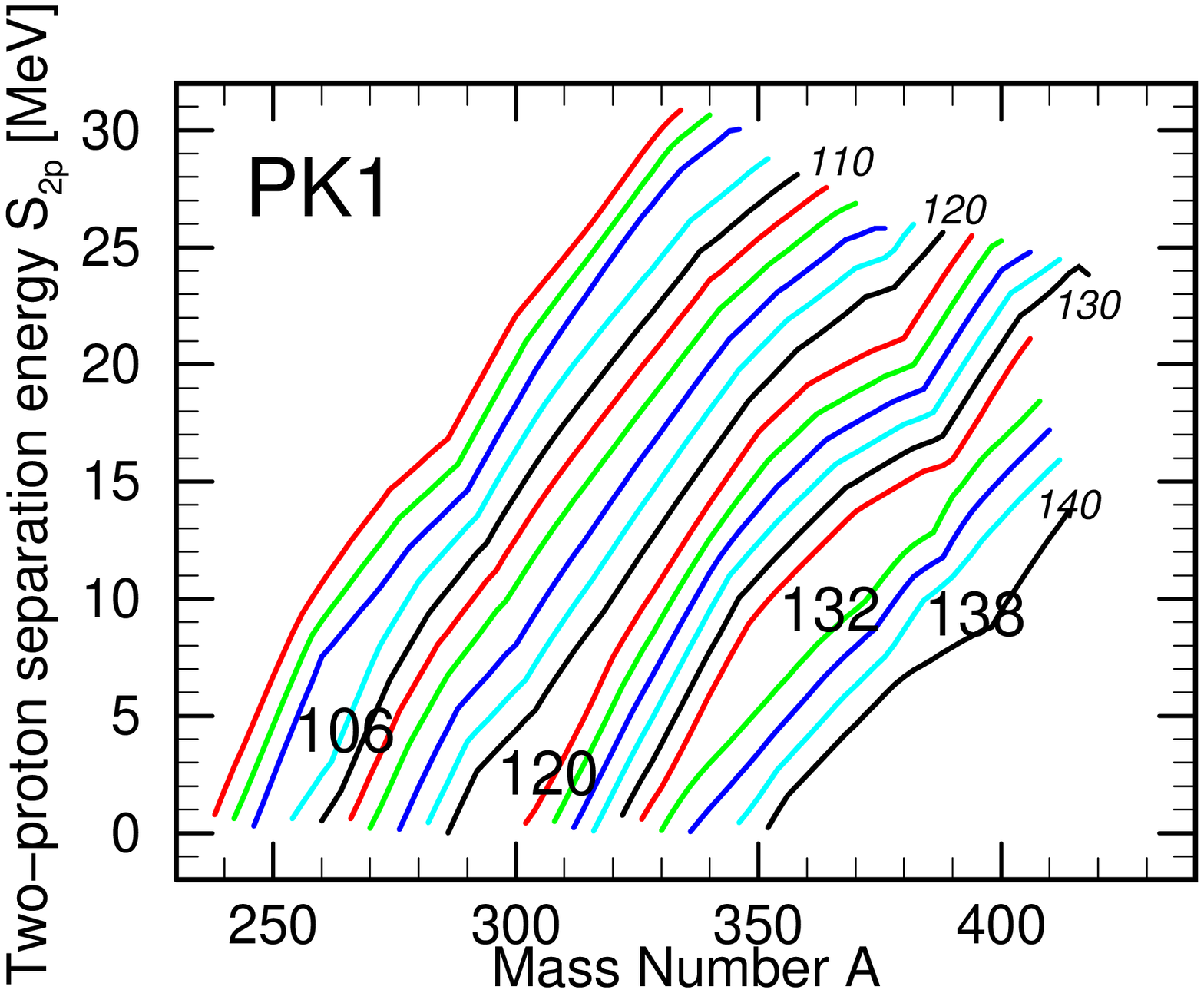}
 \includegraphics[scale=0.35]{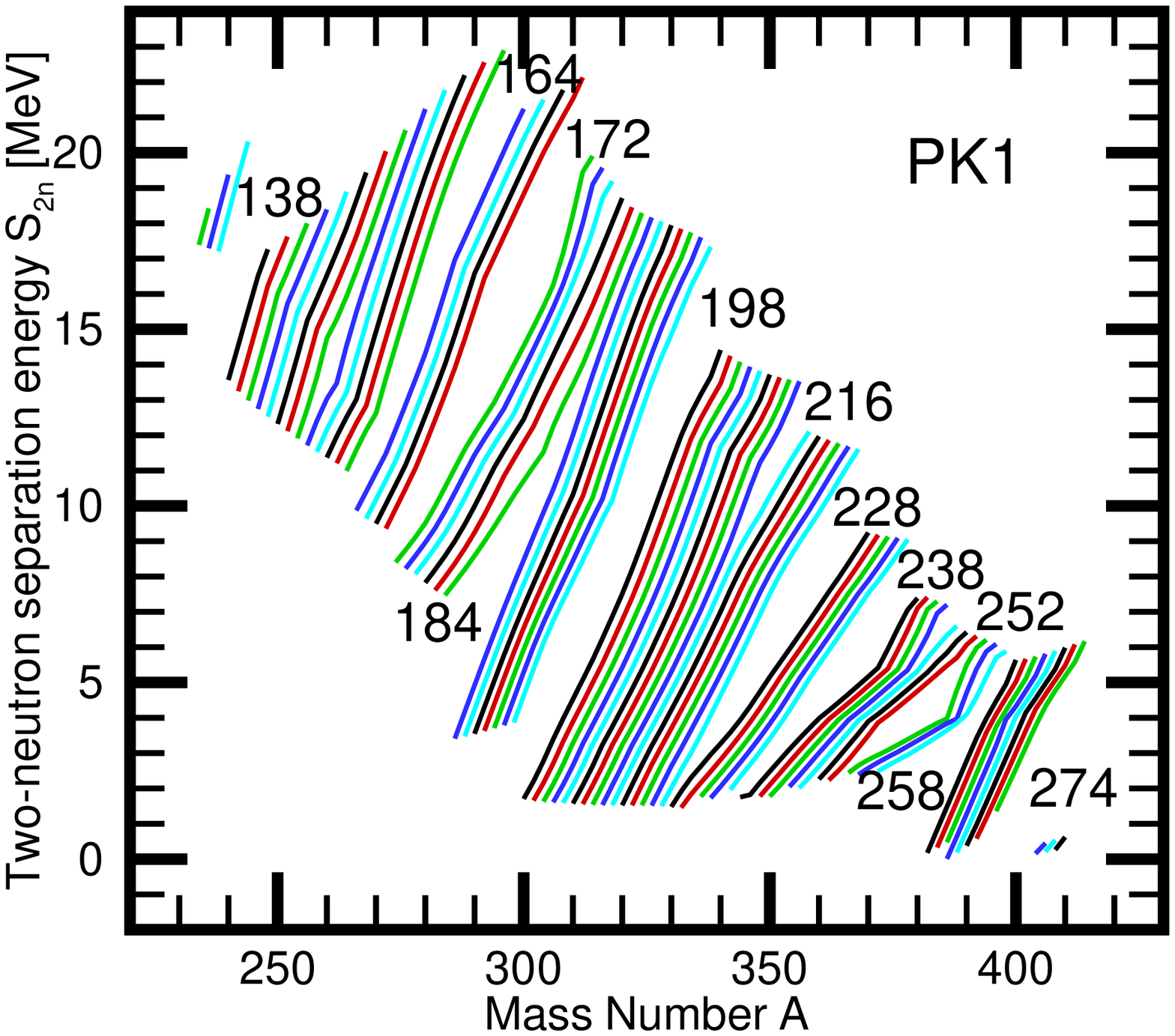}
 \caption{ {The two-proton(left panel) and two-neutron (right panel)
separation energies, $S_{2p}$ and $S_{2n}$, as a function of mass
number $A$ obtained by RCHB calculation with PK1 effective
interaction. Taken from Ref.~\cite{ZhangW05}.}} \label{FigE1}
\end{figure}

 In Ref.~\cite{ZhangW05}, the two-proton and two-neutron
separation energies have been investigated in the superheavy
region with proton number $Z$ ranging from 100 to 140 and neutron
number $N$ from $(Z+30)$ to $(2Z+32)$ by the RCHB theory with NL1,
NL3, NLSH, TM1, TW99, DD-ME1, PK1 and PK1R. As an example, the
results with PK1 are shown as a function of mass number $A$ in
Fig.~\ref{FigE1}. For the proton in left panel, the large gaps
appear at $Z=106, 120, 132, 138$, while the usual $Z$=114 does not
appear.  Furthermore, all the gaps are strongly neutron dependent.
On the neutron side in right panel, the relatively larger gaps can
be seen at $N$=138, 172, 184, 198, 228, 238, 252, 258, and 274,
and also the size and the shape of the gaps differ notably.

 From similar investigations with other effective
interactions~\cite{ZhangW05}, it is found that the magic proton
numbers $Z$=120, 132, and 138 are common while $Z$=106 is observed
only for NL3, NLSH, TW99, PK1, and PK1R. Furthermore the gaps do
not exist for all isotopes. On the neutron side, $S_{2n}$ show
distinguishable gaps at $N$=138, 172, 184, 198, 228, 238, 252,
258, and 274 with all the effective interactions. Apart from
these, there are also other gaps which appear only for some
effective interactions, e.g., $N$ = 164 for NLSH, TW99, DD-ME1,
PK1 and PK1R, and $N$ = 216 for NLSH, TW99, PK1, and PK1R. For the
magic numbers marked either by the common gaps or the
interaction-dependent gaps, the shell quenching phenomena, i.e.,
the gaps at $N$=184, 198 (NL1), 216 (NLSH, TW99, PK1, and PK1R),
228 (NL1, NL3, NLSH, TM1, PK1, and PK1R), 238, 252, and 258 (NLSH,
TM1, PK1, and PK1R) appear only for certain $Z$, are observed.

\subsection{Two-nucleon shell gaps}

The changes of the two-nucleon separation energies can be
quantified by the second difference of the binding energies, i.e.,
the two-nucleon shell gaps:
 \beqn
\delta_{2p}(N,Z) &=& S_{2p}(N,Z)-S_{2p}(N,Z+2)
        =2E_B(N,Z)-E_B(N,Z+2)-E_B(N,Z-2) \\ \nonumber
\delta_{2n}(N,Z) &=& S_{2n}(N,Z)-S_{2n}(N+2,Z)
        =2E_B(N,Z)-E_B(N+2,Z)-E_B(N-2,Z).
 \eeqn
 A pronounced peak of the two-nucleon shell gaps corresponds to a
drastic change of the two-nucleon separation energies and
indicates the shell closure~\cite{Nazarewicz95}. So far, the
two-nucleon shell gaps have been extensively used to be an
indicator for the magic number~\cite{Rutz97, Bender99} and to
analyze the shell quenching phenomenon~\cite{Novikov02}.

\begin{figure}[htbp] \centering
\includegraphics[scale=0.3]{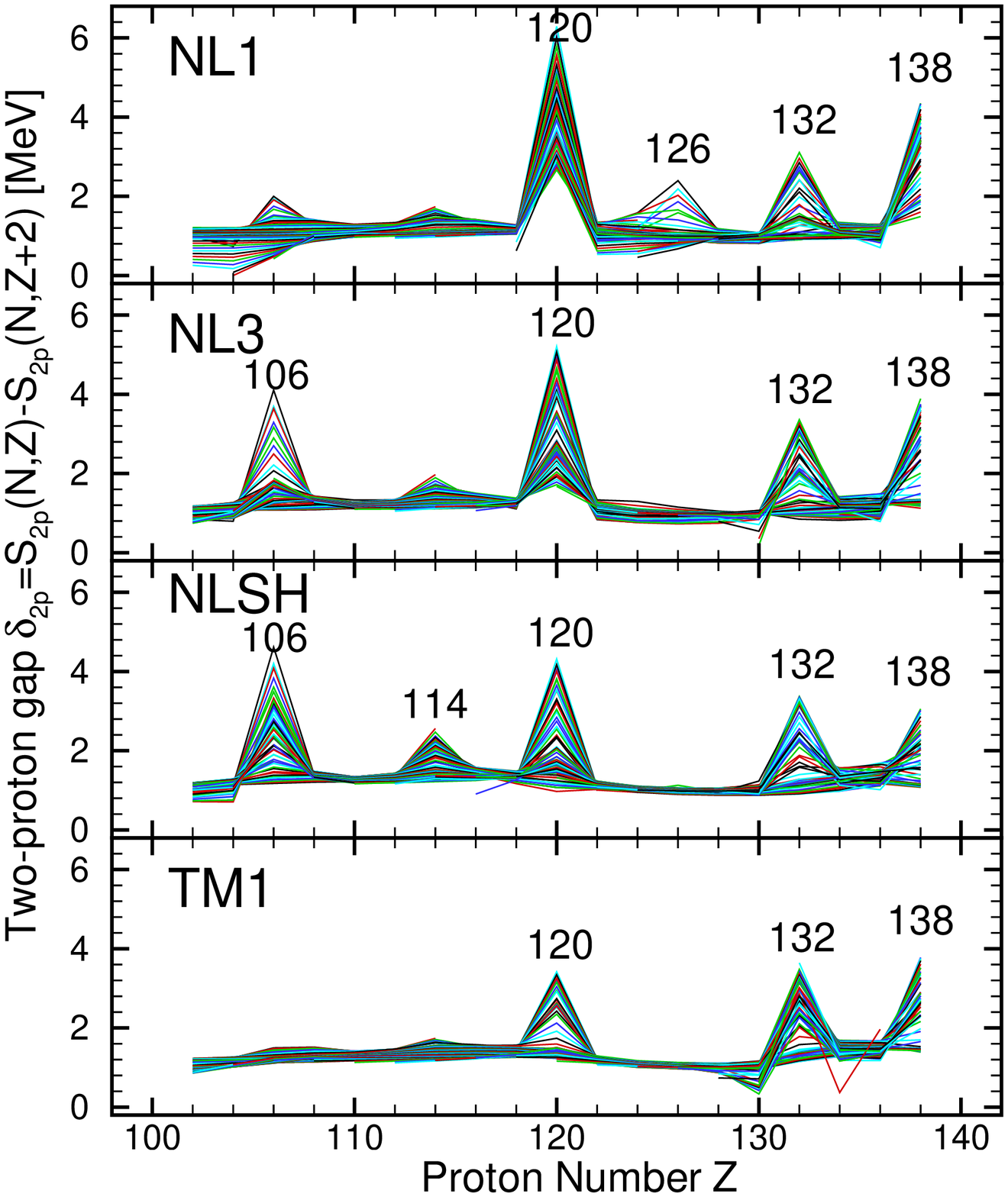}
\includegraphics[scale=0.3]{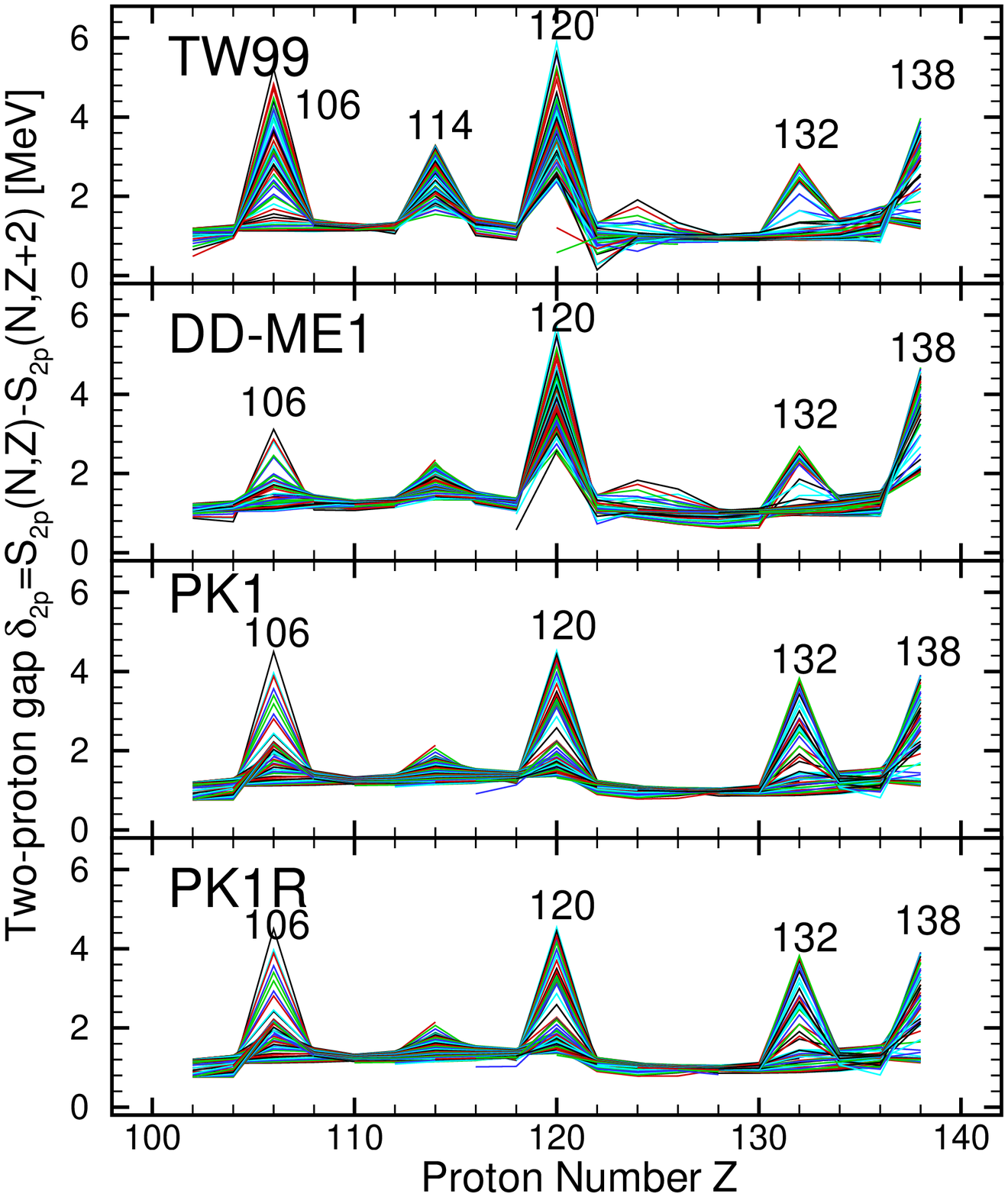}
 \caption{The
two-proton shell gaps, $\delta_{2p}$, as a function of proton
number obtained by RCHB calculation with effective interactions
NL1, NL3, NLSH, TM1, TW-99, DD-ME1, PK1, and PK1R, respectively.
Taken from Ref.~\cite{ZhangW05}.}
 \label{FigE2}
 \end{figure}

\begin{figure}[htbp] \centering
\includegraphics[scale=0.4]{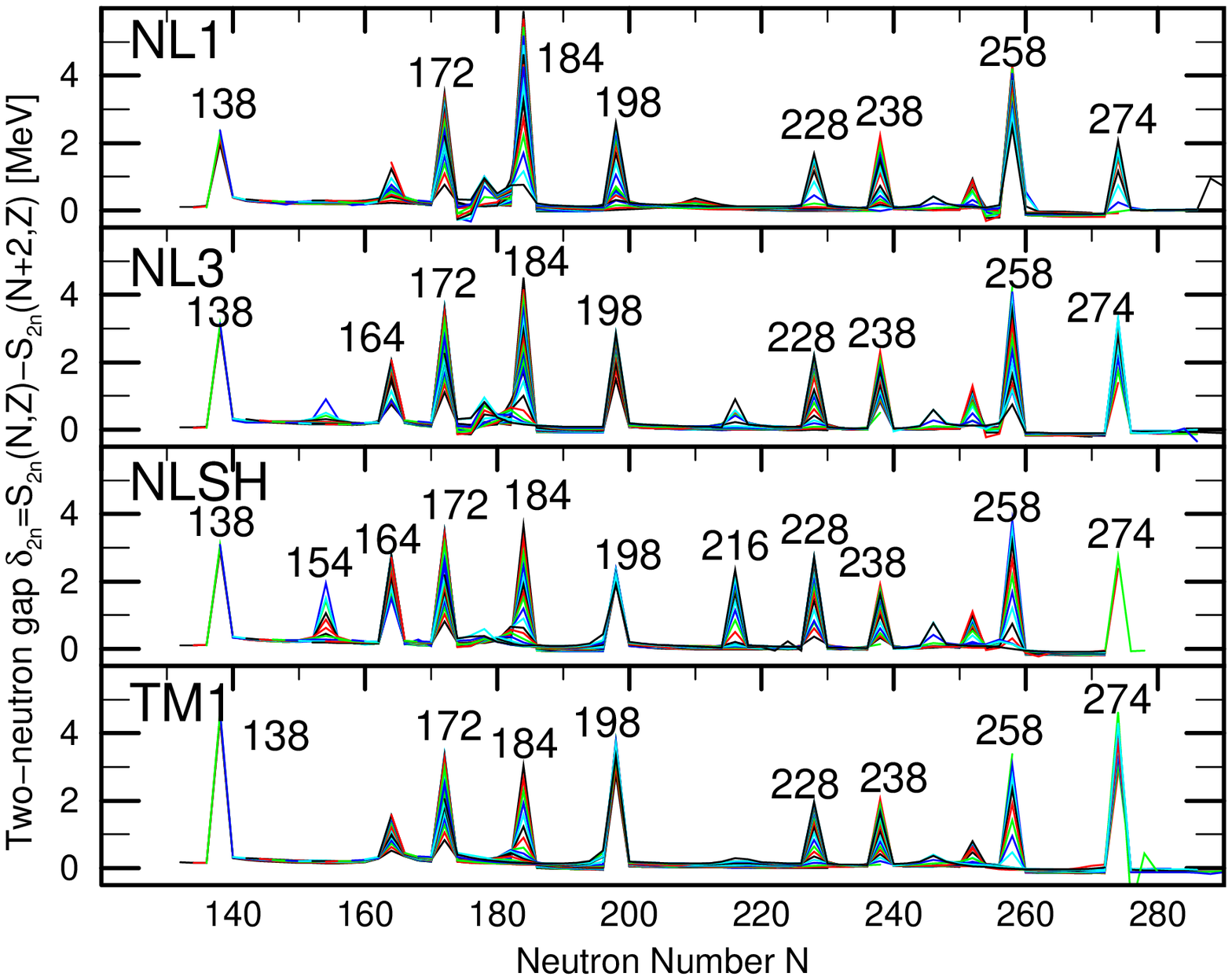}\hspace{0.5cm}
\includegraphics[scale=0.4]{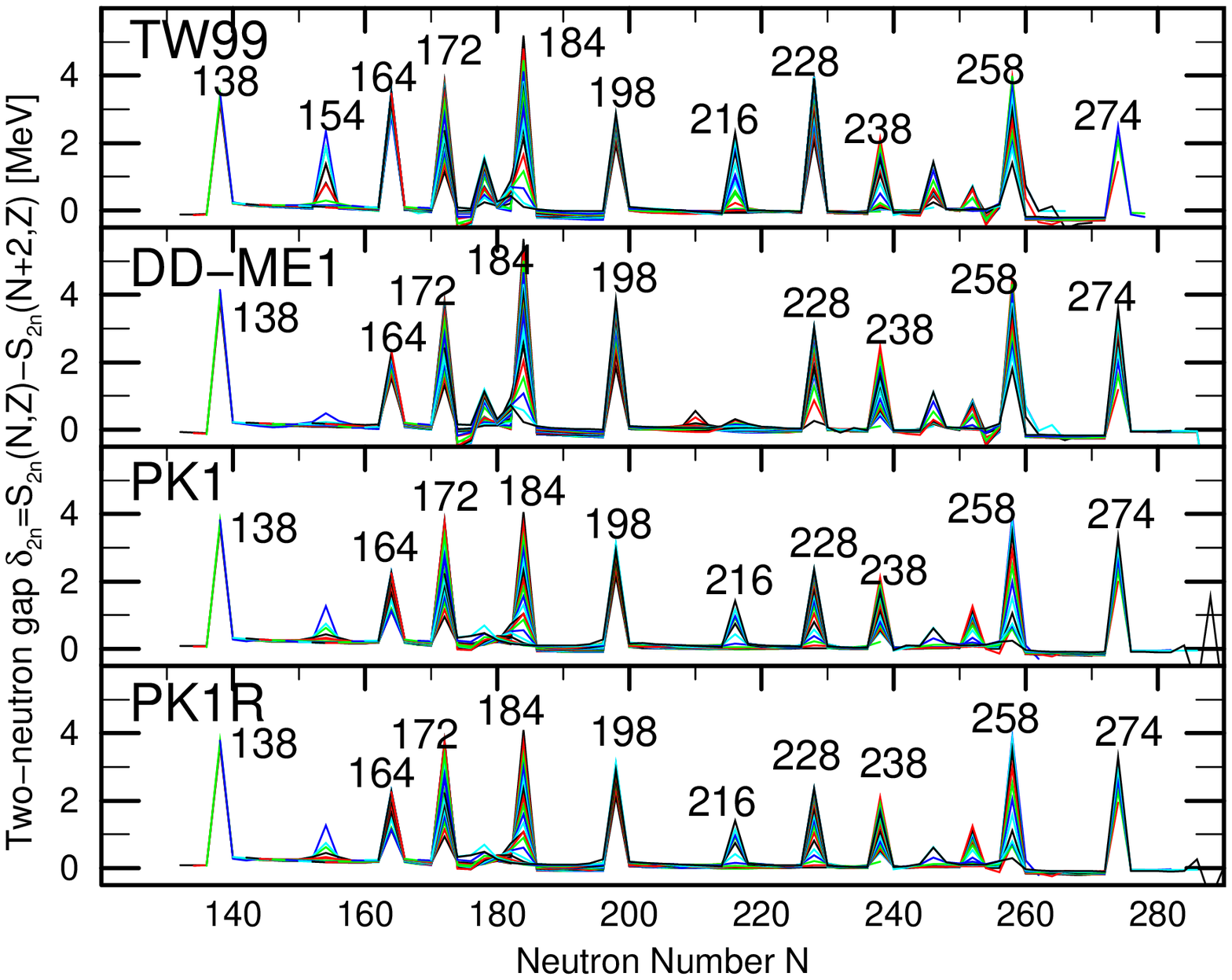}
 \caption{The
two-neutron shell gaps, $\delta_{2n}$, as a function of neutron
number obtained by RCHB calculation with effective interactions
NL1, NL3, NLSH, TM1, TW-99, DD-ME1, PK1, and PK1R, respectively.
Taken from Ref.~\cite{ZhangW05}.}
 \label{FigE3}
 \end{figure}

The two-proton gaps $\delta_{2p}$ from the RCHB calculations for
even-even nuclei with $Z$=102 - 138 as a function of $Z$ are shown
in Fig.~\ref{FigE2}. The two-proton gaps $\delta_{2p}$ with the
same $N$ are connected as a curve. A peak at certain $Z$ in the
curve suggests the existence of magic proton number. The sharpness
of the peaks represent the goodness of the magic numbers while the
quenching effects are associated with the bundle of the curves at
the certain $Z$. The size of the gaps of two-proton separation
energies $S_{2p}$ in Fig. \ref{FigE2} correspond to the magnitude
of the peaks of two-proton gaps $\delta_{2p}$ in Fig. \ref{FigE2}.
It is observed that common magic proton numbers $Z$=120, 132, and
138 exist for all the effective interactions while $Z$=106 is
observed only for NL3, NLSH, TW99, DD-ME1, PK1, and PK1R.
Furthermore, the peak at $Z$=114 for NLSH and TW99 and the peak at
$Z$=126 for NL1 are also observed, though they are not so obvious
as that at $Z$=120.

Similar figure for two-neutron gaps $\delta_{2n}$ are shown in
Fig. \ref{FigE3}. There are interaction-independent peaks of
two-neutron gaps $\delta_{2n}$ at $N$=138, 172, 184, 198, 228,
238, 258, and 274 which are consistent with the conclusion from
the two-neutron separation energies $S_{2n}$ . Moreover, small
peaks are also observed for $N$=154 (NLSH and TW99), $N$=164 (NL3,
NLSH, TW99, DD-ME1, PK1, and PK1R), and $N$=216 (NLSH, TW99, PK1,
and PK1R).

\subsection{Shell correction energies}

The shell correction energy, representing the overall behavior of
the single-particle spectra, is another suitable observable to
identify the shell closure. This quantity, which is derived from
the single-particle spectra by the Strutinsky
procedure~\cite{Strutinsky68}, is defined as the difference
between the total single-particle energy $E$ and the smooth
single-particle energy $\bar {E}$:
 \beq
  E_{shell} = E - \bar {E}
            = \sum\limits_{i = 1}^{N(Z)} {e_i }
            - 2\int\limits_{ -\infty }^{\bar{\lambda}} {e\bar {g}(e)de},
 \eeq
where $N(Z)$ is the particle number, $e_i$ is the single-particle
energy, $\bar{\lambda}$ is the smoothed Fermi level  determined by
the particle number equation
 $N(Z) = 2\int\limits_{ - \infty}^{\bar{\lambda}} {\bar {g}(e)de}$, and
$\bar{g}(e)$ is the smoothed level density \beq
  \bar{g}(e)= \dfrac{1}{\gamma}  \int\limits_{ -\infty }^\infty
             {\left(\sum\limits_{i = 1}^\infty
             {\delta(e'- e_i)}\right) f(\frac{e' - e}{\gamma })de'}
            = \dfrac{1}{\gamma}  \sum\limits_{i = 1}^\infty
             {f(\dfrac{e_i - e}{\gamma }) },
 \eeq
with the smoothing range $\gamma$ , the folding function $f(x) =
\dfrac{1}{\sqrt \pi}e^{ - x^2}P(x)$, and $P(x)$ the associated
Laguerre polynomial $L_s^{1 / 2}(x^2)$. The shell correction
energy provides an indicator about the deviation in the level
structure of nuclei away from uniformly distributed ones. A large
negative shell correction energy corresponds to shell closure at
certain nucleon number.

\begin{figure}[htbp] \centering
\includegraphics[scale=0.3]{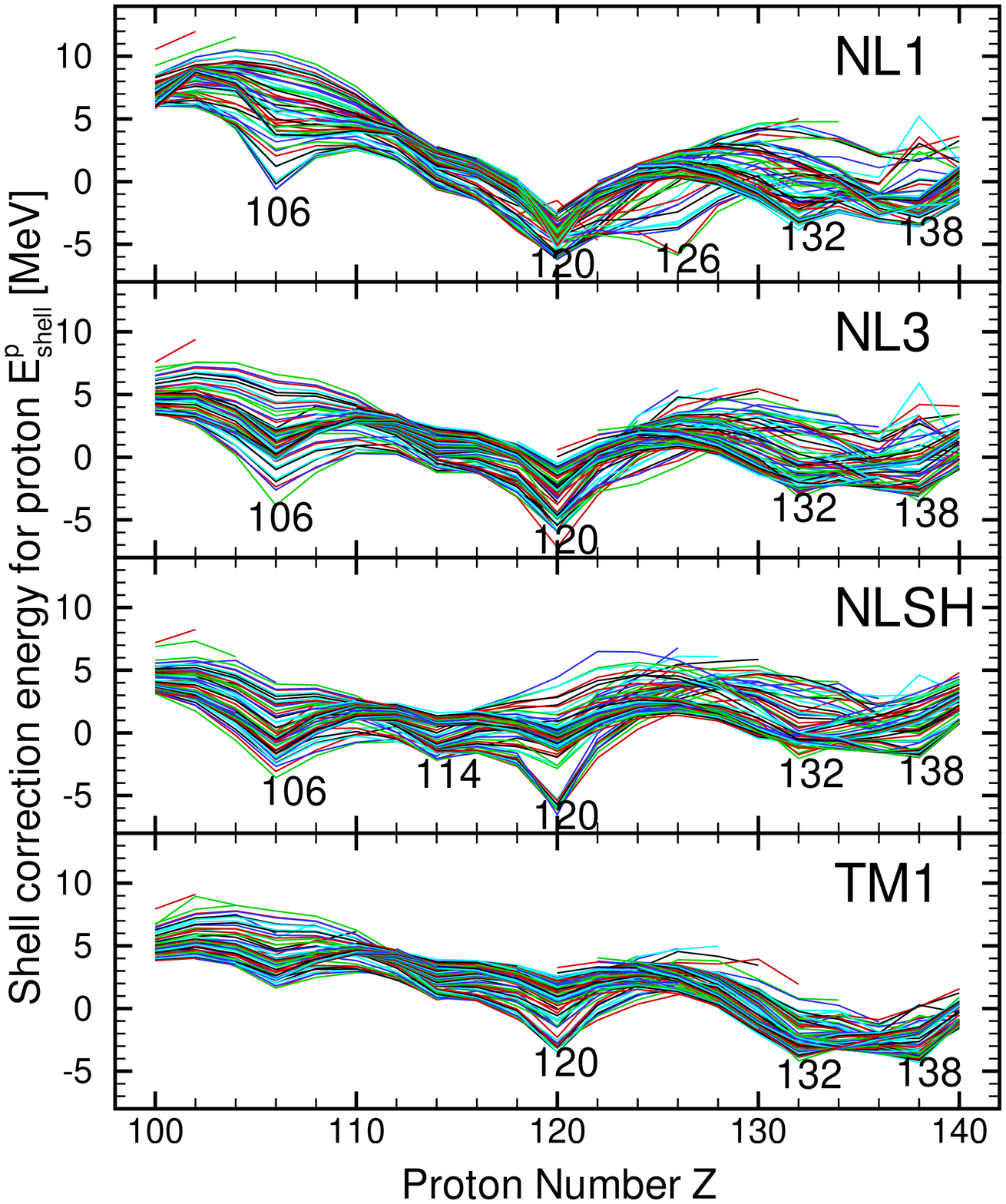}
\includegraphics[scale=0.3]{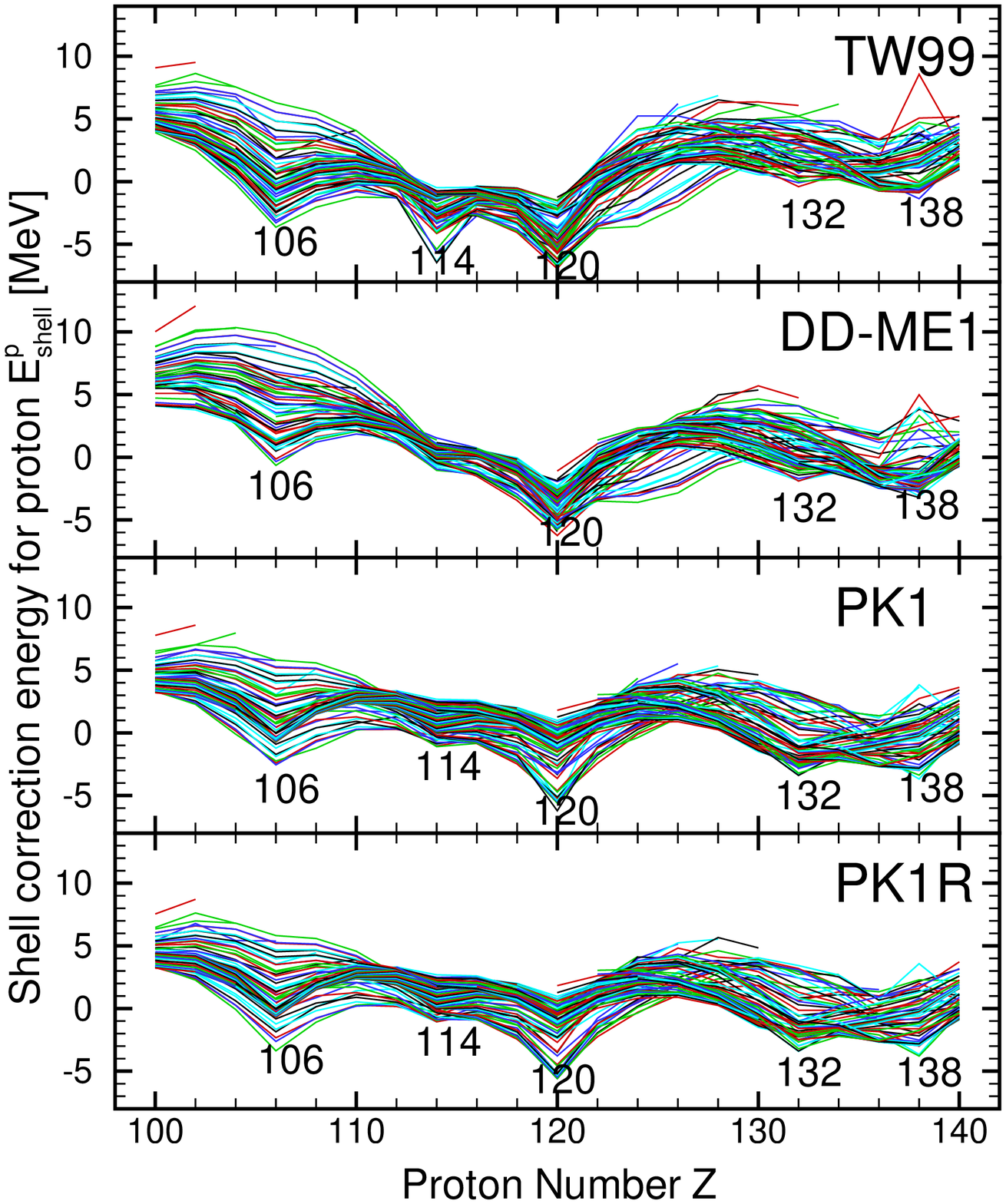}
 \caption{The shell
correction energies for proton, $E_{shell}^{p}$, as a function of
proton number obtained by RCHB calculation with effective
interactions NL1, NL3, NLSH, TM1, TW-99, DD-ME1, PK1 and PK1R,
respectively. Taken from Ref.~\cite{ZhangW05}.} \label{FigE4}
\end{figure}

\begin{figure}[htbp] \centering
\includegraphics[scale=0.3]{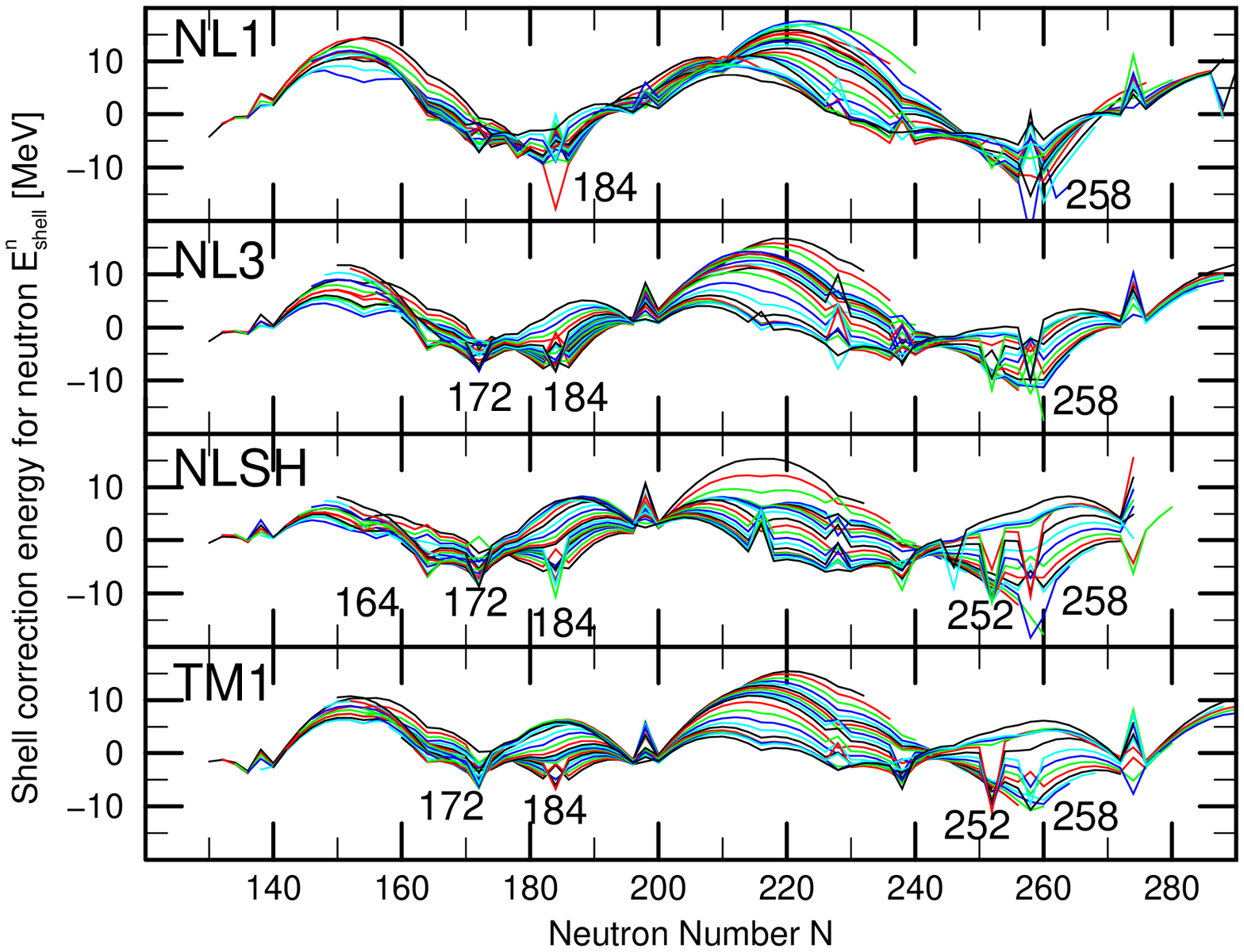}\hspace{0.5cm}
\includegraphics[scale=0.3]{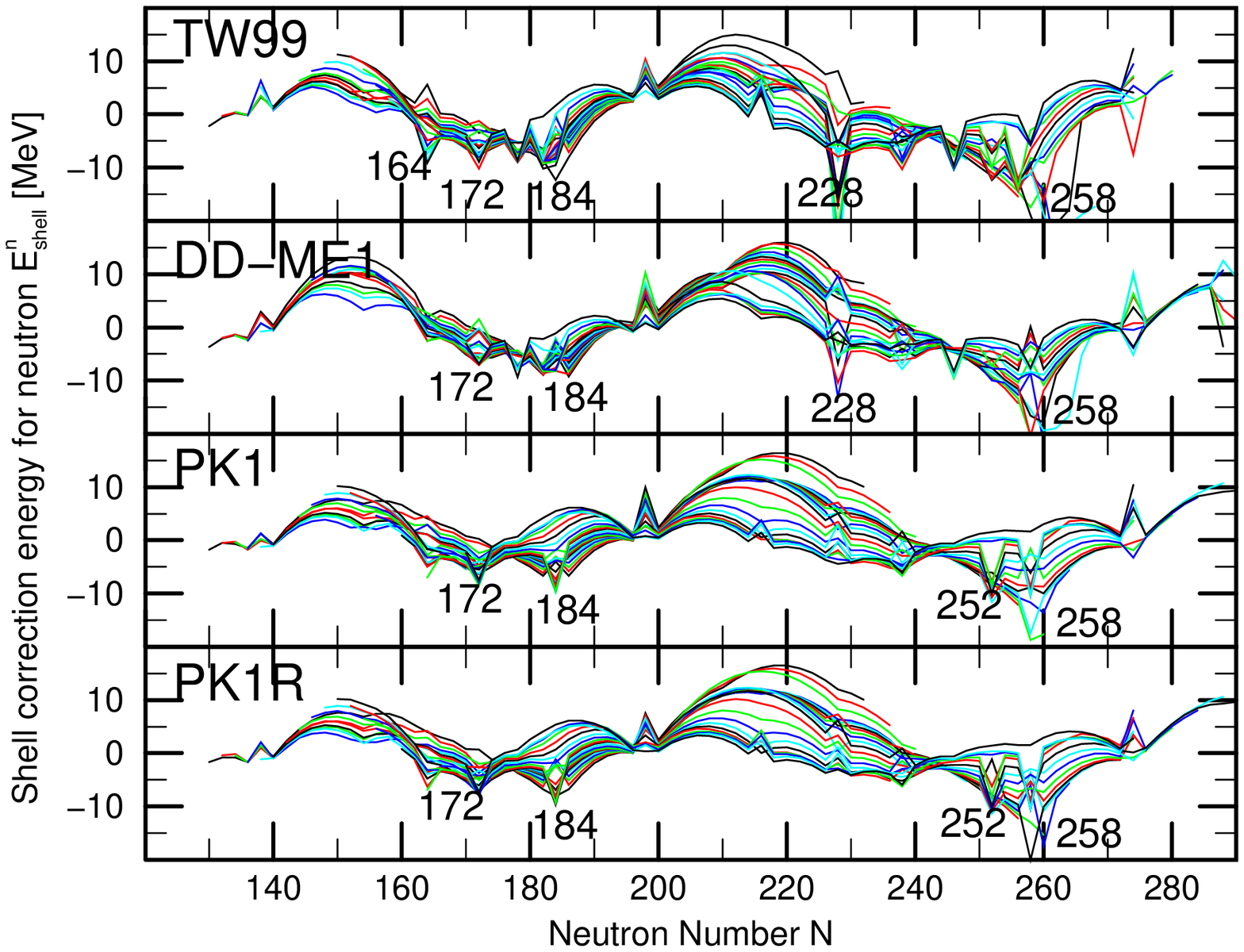}
 \caption{The shell
correction energies for neutron, $E_{shell}^{n}$, as a function of
neutron number obtained by RCHB calculation with effective
interactions NL1, NL3, NLSH, TM1, TW-99, DD-ME1, PK1 and PK1R,
respectively. Taken from Ref.~\cite{ZhangW05}.}  \label{FigE5}
\end{figure}

With the proton single-particle spectra in canonical basis from
the RCHB calculations, the shell correction energies
$E_{shell}^{p}$ for even-even nuclei with $Z$ ranging from 100 to
140 are shown as a function of $Z$ in Fig. \ref{FigE4}. The nuclei
with the same $N$ are connected as a curve. A deep valley at
certain $Z$ on a curve hints the shell closure. The valleys at
$Z$=120, 132, and 138 are conspicuous for all the effective
interactions. Possible shell closure at $Z$=106 for NL1, NL3,
NLSH, TW99, DD-ME1, PK1, and PK1R, at $Z$=114 for NLSH, TW99, PK1,
and PK1R, and at $Z$=126 for NL1 are also observed. The shell
closure from the shell correction energies for proton
$E_{shell}^{p}$  are consistent with the peaks of the two-proton
gaps $\delta_{2p}$, though the magnitude of magicity slightly
differs. Similar to the two-proton gaps $\delta_{2p}$, the spread
of the valleys of the shell correction energies for proton
$E_{shell}^{p}$ indicates the quenching of the magic number. In
view of the shell correction energies for proton $E_{shell}^{p}$,
the quenching phenomena of magic proton numbers in RCHB
calculation are universal. These facts suggest that the magic
proton numbers depend also on neutron number $N$.

The corresponding shell correction energies for neutron
$E_{shell}^{n}$ for even-even nuclei with $N$=130 - 312 are
demonstrated as a function of $N$ in Fig. \ref{FigE5}. As the
neutron numbers extend from $N$=130 to $N$=290, the valleys for
shell correction energies for neutron $E_{shell}^{n}$ in Fig.
\ref{FigE5} are not so obvious as those for shell correction
energies for proton $E_{shell}^{p}$ in Fig. \ref{FigE4}. However,
similar conclusions as the two-neutron separation energies
$S_{2n}$ and the two-neutron gaps $\delta_{2n}$ for magic neutron
numbers can also be drawn. Similar calculation has also been done
in Ref.~\cite{Bender01plb} with SHF and RMF.

The first valley locates in $N$=184 for all the effective
interactions and involves $N$=172 (NL3, NLSH, TM1, TW99, DD-ME1,
PK1, and PK1R), even $N$=164 (NLSH and TW99). And the other valley
is the mixture of the valleys at $N$=228 (TW99 and DD-ME1),
$N$=252 (NLSH, TM1, PK1, and PK1R) and $N$=258. These fine
structures suggest the smearing of magic neutron numbers. Based on
the shell correction energies, the shell closures are smeared
compared with those from the two-nucleon separation energies
$S_{2p}$ and $S_{2n}$ and the two-nucleon gaps $\delta_{2p}$ and
$\delta_{2n}$. However, it is found in Ref.~\cite{ZhangW05} that
the magic numbers at $Z$=120, 132, and 138 are common but strongly
quenched, the shell closures near $N$=172 and $N$=184, and
$N$=228, $N$=252, and $N$=258 are blurred. The other magic numbers
from the shell correction energies, such as $Z$=106, 114, and 126,
appear only for some effective interactions.

In the usual Hartree or HF mean field level, as the nonphysical
particle continuum are involved, much efforts have been made to
avoid the divergence of single particle energy density around the
threshold~\cite{Nazarewicz94, Sandulescu97b, Vertse98, Kruppa00}.
 Based on the Green's-function approach to the level density, a
method of calculating shell corrections has been adopted within
the SHF and RMF theories to a large-scale survey of spherical
shell energies throughout the whole landscape of conceivable
superheavy nuclei~\cite{Bender01plb}. From SHF with SkP, SLy6,
SkI3 and SkI4 and RMF with NL3 and NL-Z2, the total shell
correction energies $E_{shell}$ for spherical superheavy nuclei
around the expected island of stability around $Z$ = 120 and $N$ =
180 are given in Fig.~\ref{FigE6}. Instead of narrow stripes of
large $E_{shell}$ localized around magic numbers for normal
nuclei, a wide area of shell stabilization is found to spreads
over all shell closures predicted. In other words, the predicted
magic shells for superheavy nuclei are much mellowed, which are
consistent with the results from RCHB theory. The disappearance of
a familiar pattern of magic numbers and the appearance of broad
valleys of shell stability are due to the rather large
single-particle level density and the appearance of many low $j$
shells around the Fermi level~\cite{Bender01plb}.

\begin{figure}[htbp]
 \centering
 \includegraphics[scale=1.0,angle=-90]{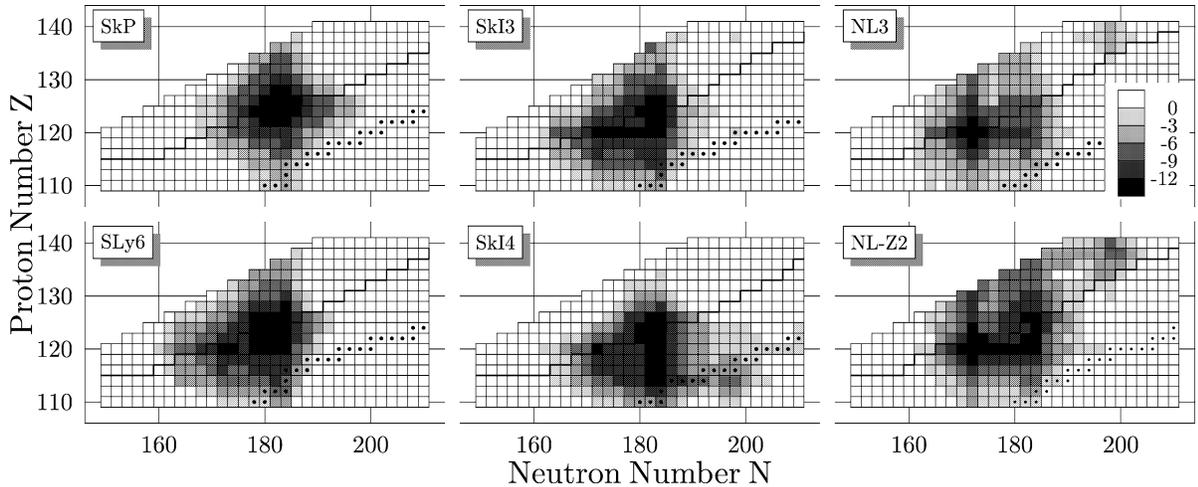}
 \hspace{0.5cm}
\caption{ {Total shell correction energies, calculated from SHF
with SkP, SLy6, SkI3 and SkI4 and RMF with NL3 and NL-Z2, for
spherical even-even superheavy nuclei around the expected island
of stability around $Z$ = 120 and $N$ = 180. The thick solid lines
denote two-particle drip lines. Black squares mark nuclei
calculated to be stable with respect to $\beta$ decay. White color
indicates nuclei with positive shell corrections, black color
denotes nuclei with $E_{shell}$ beyond -12 MeV. Taken from
Ref.~\cite{Bender01}.}}
 \label{FigE6}
\end{figure}

\subsection{ {Magic numbers and deformations}}

\subsubsection{ {Magic numbers for spherical superheavy nuclei}}

 After the detailed analysis of the two-nucleon separation
energies, $S_{2p}$ and $S_{2n}$, the two-nucleon shell gaps,
$\delta_{2p}$ and $\delta_{2n}$, the shell correction energies,
$E_{shell}^{p}$ and $E_{shell}^{n}$, as well as the pairing
effects, the magic numbers for superheavy nuclei in the RCHB
theory are found to be~\cite{ZhangW05}: $Z=120,132,138$ for proton
and $N=172,184,198,228,238,258$ for neutron. As many theoretical
methods have been adopted to probe the superheavy magic numbers
and there is no consensus among them with regard to the center of
the island of superheavy nuclei. It is interesting to briefly
review the up-to-date studies of the superheavy magic numbers
including the present one:

\begin{itemize}
  \item {\bf $Z$=106:} As we have seen, this shell closure is strongly
supported by the effective forces NL3, NLSH, TW-99, PK1, and PK1R,
partially by NL1 and DD-ME1, but not at all by TM1. We also noted
that the shell closure at $Z$=106 depends sensitively on the
neutron number and is visible only around $N$ $\sim$ 154 and
$\sim$ 218, while the later can only be clearly seen in the
results calculated with TW-99 and NLSH. The RHB calculations with
the effective force NLSH and the Gogny interaction D1 for the
pairing correlation predicts a doubly magic character at $Z$=106
and $N$=160~\cite{Lalazissis96}. In addition, the shell closures
at $Z$=104, 106, 108, and 110 and $N$=162, and 164 which
correspond to prolate deformations are predicted by the
FRDM~\cite{Moller94} and the deformed shell closure occurs at
$Z$=108 in Nilsson-Strutinsky scheme~\cite{Patyk91,
Sobiczewski94}.

  \item {\bf $Z$=114:} The shell closures are predicted at $Z$=114 and
$N$=184 in the phenomenological shell
models~\cite{Nilsson69,Mosel69, Patyk91, Moller94}. However, as we
have noted, in the RCHB theory the magicity of $Z$=114 is not at
all obvious, in agreement with other RMF
investigations~\cite{Rutz97,Lalazissis96,Furnstahl96,Serot97,Sil04}.
By adopting a new effective interaction NL-RA1, the nucleus
$^{298}_{184}114$ is predicted to be the next spherical doubly
magic nucleus in a RMF plus BCS calculation~\cite{Rashdan01}.
Within the SHF approach, only SkI4 prefers the doubly magic
nucleus at $Z$=114 and $N$=184, which is not supported by other
Skryme forces such as SkM*, SkP, SkI1, SkI3, SLy6, and
SLy7~\cite{Rutz97,Bender99}.

  \item {\bf $Z$=120:} The shell closure at $Z$=120 is strongly
supported in the RCHB calculations. The doubly magic
characteristic of $^{292}_{172}120$ has been addressed within the
RMF theory~\cite{Rutz97, Bender99, Rashdan01, Kruppa00} as well as
the effective field theory~\cite{Sil04}. Most Skyrme HF
investigations also suggest that this nucleus is doubly magic
except for SkP~\cite{Rutz97, Bender99}. Due to the strong shell
quenching effect in the $Z$=120 isotopic chain, other doubly magic
candidates with $N$=184, 198, 228, 238, 252, 258, and 274 are
interaction-dependent and need to be investigated in more detail.

   \item {\bf $Z$=126:} Several Skyrme HF investigations with the effective
interactions such as SkP, SkM*, SLy6 and SLy7 predict doubly magic
system at $Z$=126 and $N$=184~\cite{Cwiok96, Rutz97, Bender99}.
However, this conclusion is not supported by the RMF calculations
although $N$=184 might be a neutron magic number.

  \item {\bf $Z$=132 and 138:} Little attention has been paid to
the possible proton shell closures at $Z>130$ due to their large
Coulomb repulsive potentials. However, the shell closures at
$Z$=132 with $N \sim$ 210-240, and at $Z$=138 with $N\sim$ 180-200
and $N \sim$ 258 appear in most relativistic and non relativistic
mean field models~\cite{Rutz97, ZhangW05}, .
\end{itemize}


Normally the magic numbers correspond to the large energy gaps
between the single particle levels and the doubly magic nuclei are
those with protons and neutrons filled up to the gap. It is
therefore interesting to study the single-particle level structure
in these doubly magic nuclei, or vice versa.

\begin{figure}[htbp]
 \centering
 \includegraphics[scale=1.0]{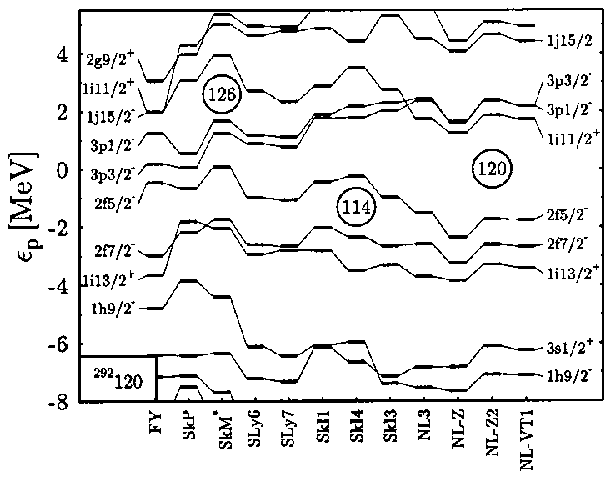}
 \includegraphics[scale=1.0]{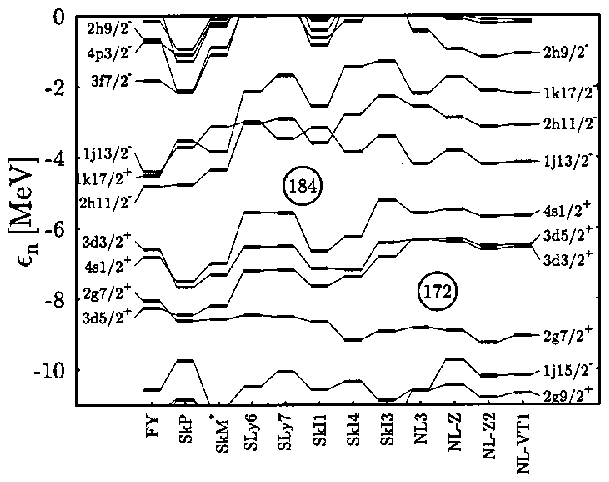}
 \hspace{0.5cm}
\caption{ {Spherical single particle spectra in $^{292}_{172}$120
for protons (left) and neutrons in various mean-field calculations
as indicated. Taken from Ref.~\cite{Bender99}.}}
 \label{FigE7}
 \end{figure}

 As examples, the spherical single particle spectra in
$^{292}$120 for protons (left) and neutrons in various mean-field
calculations are given in Fig.~\ref{FigE7}~\cite{Bender99}. The
occurrence of the shell closure at $Z=120$ depends on the
amplitude of the spin-orbit splitting of the 3p states above the
Fermi level and the 2f levels below the Fermi energy. It has also
been pointed out that the shell closures may be coupled to the
density profiles (and vice versa) based on the systematic analysis
of density distribution of superheavy nuclei. For $^{292}$120, a
significant central depression or central semibubble is obtained
in the spherical case~\cite{Bender99, Decharge99}. Recently, the
influence of the central depression in the density distribution of
spherical superheavy nuclei on the shell structure is studied in
some detail within the RMF theory~\cite{Afanasjev05}. It is found
that a large depression leads to the shell gaps at the proton $Z$
= 120 and neutron $N$ = 172 numbers, whereas a flatter density
distribution favors $N$ = 184 and leads to the appearance of a $Z$
= 126 shell gap and to the decrease of the size of the $Z$ = 120
shell gap. On the other side, the deformation effect has been
found to be significant in the calculation of the density
distribution in this nucleus, namely only a weak central
depression in the deformed case for $^{292}$120 compared to the
predicted semibubble at the spherical
shape~\cite{Pei05,Afanasjev05}.

\subsubsection{ {Stability of the doubly magic superheavy nuclei
against deformation}}

The magic numbers discussed above are mainly based on the
assumption of spherical geometrical configuration. However, such
assumption is not always true for superheavy
nuclei~\cite{Moller94,Patyk91, Sobiczewski94,Ren02c}. It is
worthwhile to investigate the potential energy surfaces to see the
validity of spherical configuration.

In Fig. \ref{FigE8}, the potential energy surfaces for
$^{292}$120, $^{304}$120, $^{318}$120, $^{348}$120, $^{358}$120,
and $^{378}$120 (solid lines)  {obtained from constrained RMF
calculations with the pairing correlations treated by BCS
approximations} are displayed. Although these nuclei have the same
proton number, their potential energy surfaces are quite different
from each other as shown in Fig. \ref{FigE8}. For nuclei
$^{292}$120, $^{304}$120 and $^{378}$120, there is an obvious
local minimum with the spherical configuration $\beta_2 \sim 0$,
while another local minimum with large deformation $\beta_2 \sim
0.6$ can be also clearly seen. For $^{318}$120, the spherical
local minimum is very shallow, and for $^{348}$120 and
$^{358}$120, the spherical minimum is hardly seen and a local
minimum with $\beta_2 \sim 0.25$ appears instead. There are also
local minimum around $\beta_2 \sim -0.4$ with oblate deformation.
In addition to the local minima discussed above, we would like to
make a few remarks on the absolute minima for these nuclei. For
$^{292}$120 and $^{378}$120, the two minima with different
deformations have almost the same energies, i.e. the so-called
``shape coexistence" may exist. In particular, the spherical
minimum is indeed the absolute minimum for $^{292}$120 ($E_B=$
-2064.3 MeV with $\beta_2 \sim 0$ vs. $E_B=$ -2063.6 MeV with
$\beta_2 \sim 0.6$), while the ground state of $^{378}$120 is
quite delicate with $E_B=$ -2398.7 MeV for $\beta_2 \sim 0$ and
$E_B=$ -2398.7 MeV for $\beta_2 \sim 0.6$. For $^{304}$120, the
absolute minimum at $\beta_2\sim 0.6$ is much deeper ($\sim$ 5.5
MeV) than the spherical configuration. In addition for
$^{348}$120, the absolute minimum lies at $\beta_2 \sim 0.2$, and
it is a well-deformed nucleus according to the present
calculation. From the RCHB and the constrained RMF+BCS
calculations, it is shown that spherical doubly magic nuclei may
exist, e.g., $^{292}$120 and $^{378}$120~\cite{ZhangW05}.

 The role of shell effects in superheavy nuclei can be also
seen in Fig. \ref{FigE8}, in which the corresponding macroscopic
energy, defined as the difference between the binding energy and
the total shell correction energy, $E_B-E_\mathrm{shell}$, is
plotted with dashed lines. It can be seen that without the shell
effects, the superheavy nuclei hardly exist. As noted long time
ago, the shell effects play an essential role to stabilize the
superheavy nuclei against the fission.

\begin{figure}[htbp] \centering
\includegraphics[scale=0.25]{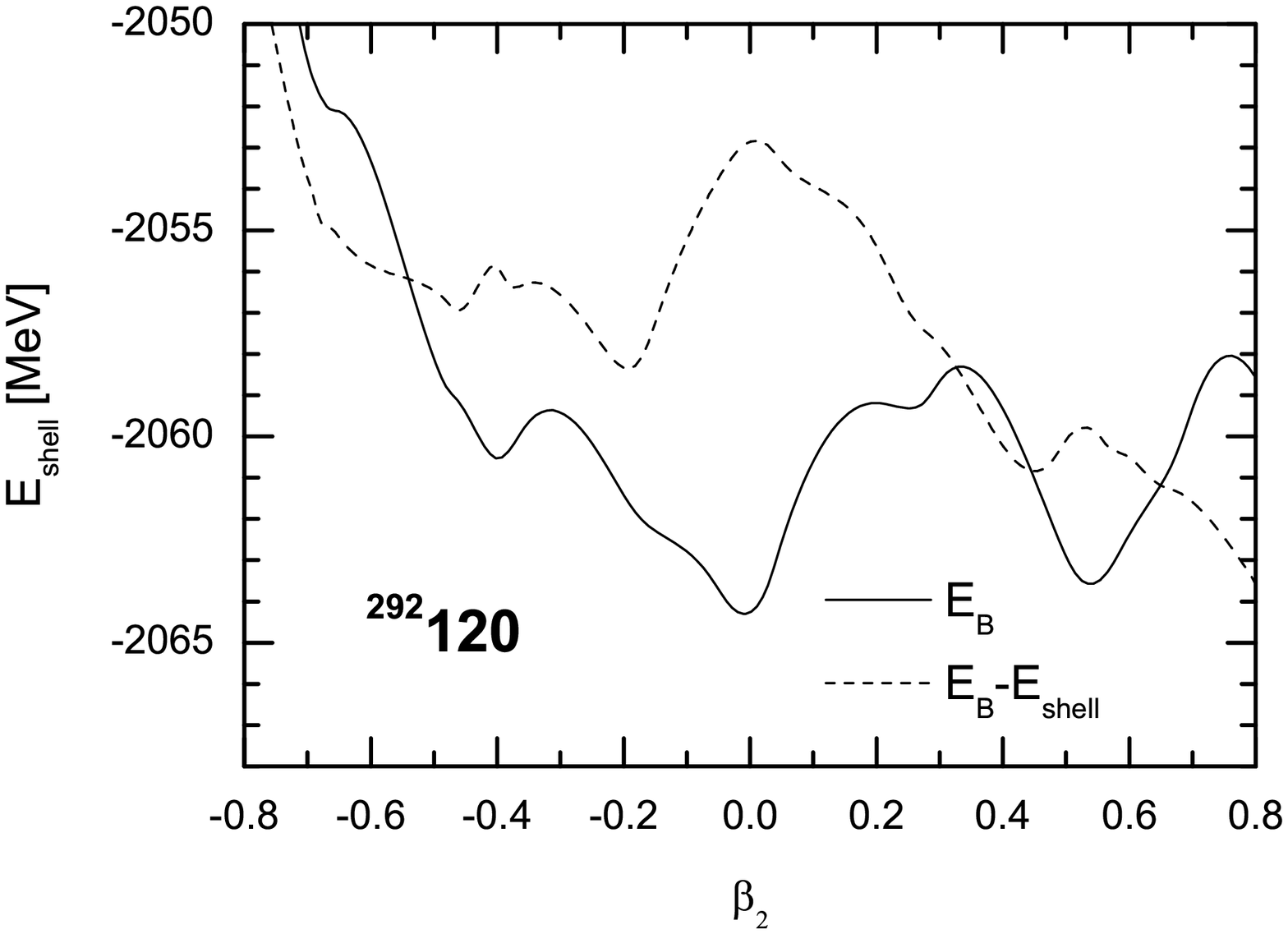}
\includegraphics[scale=0.25]{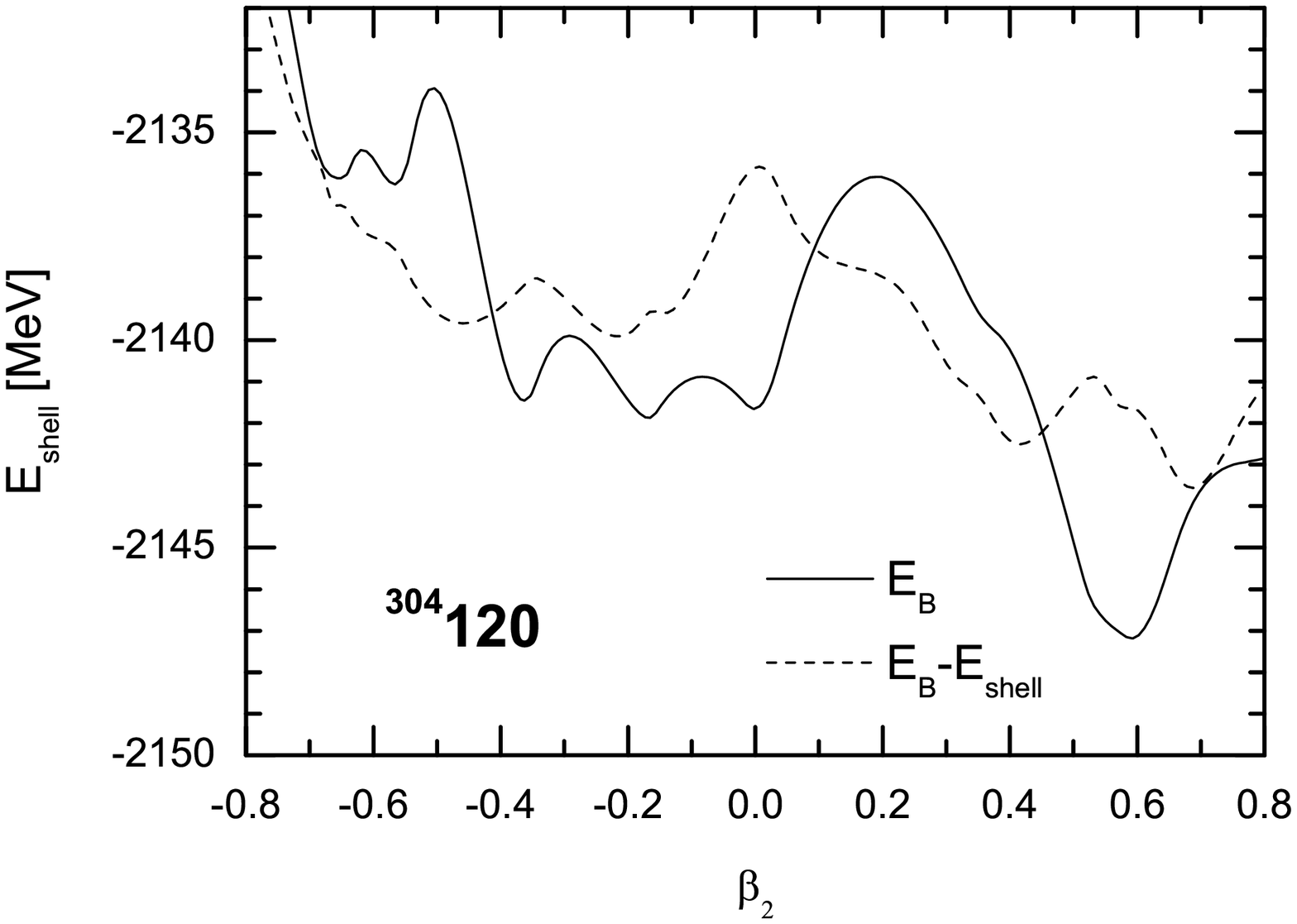}
\includegraphics[scale=0.25]{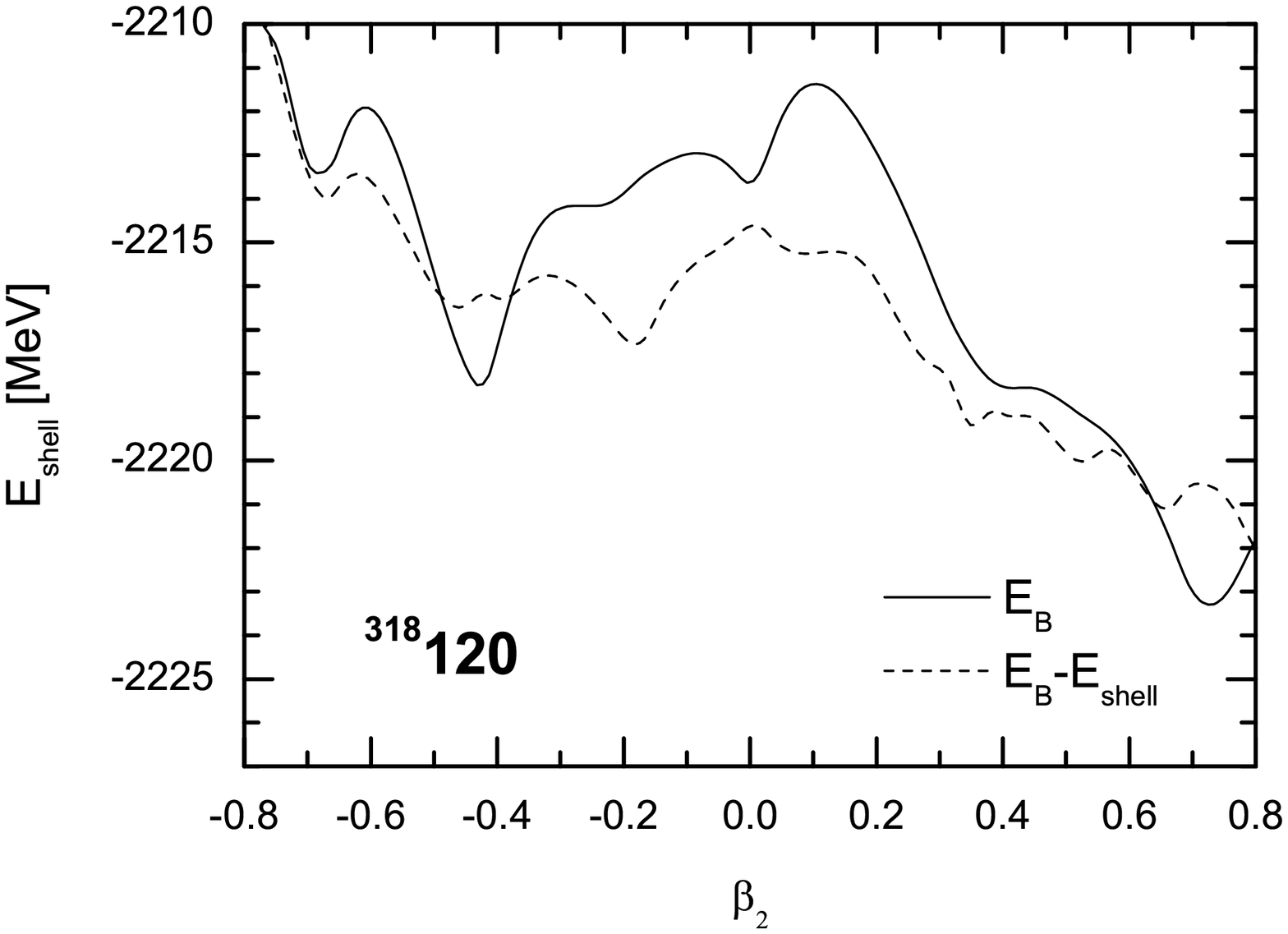}
\includegraphics[scale=0.25]{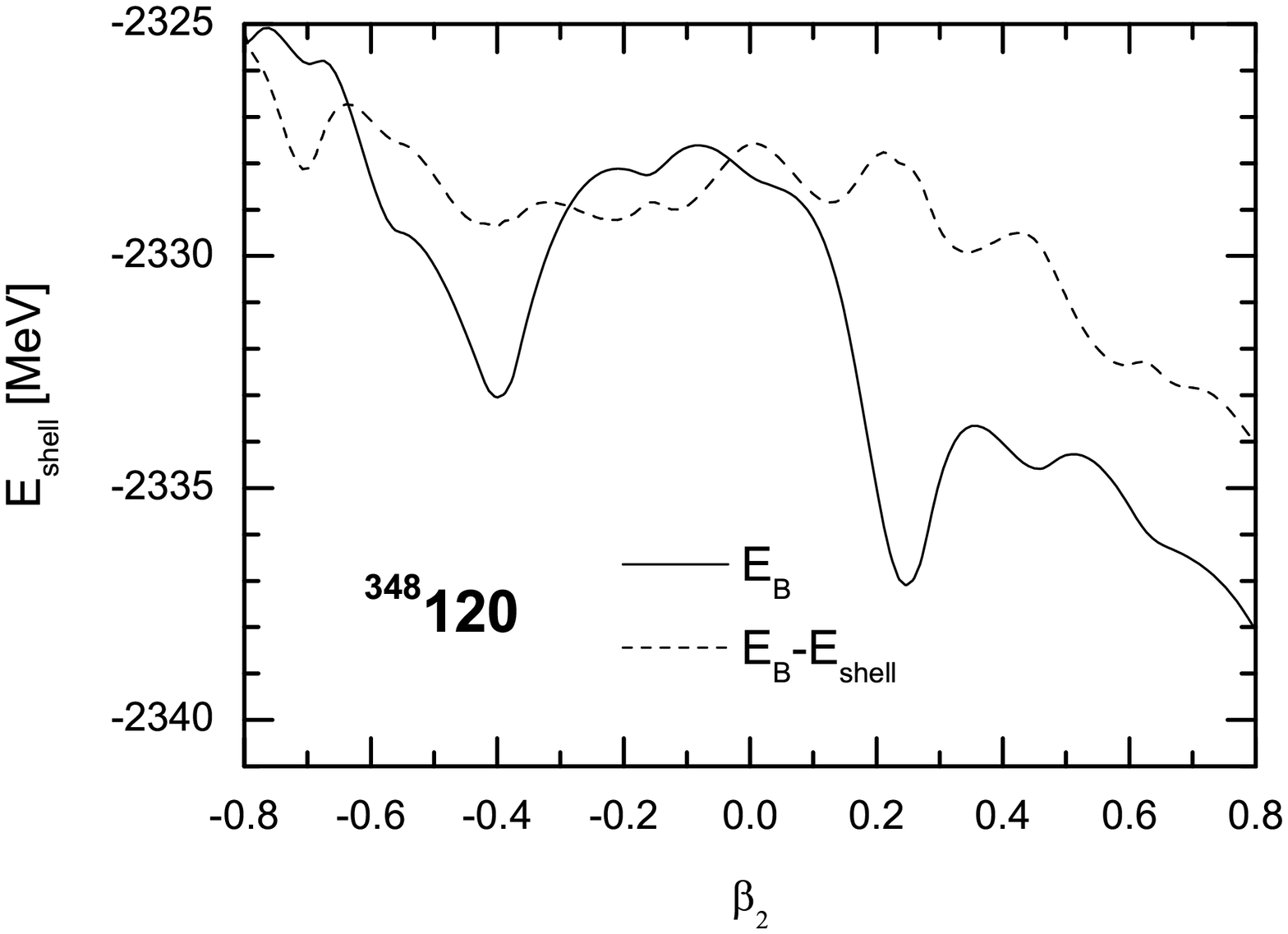}
\includegraphics[scale=0.25]{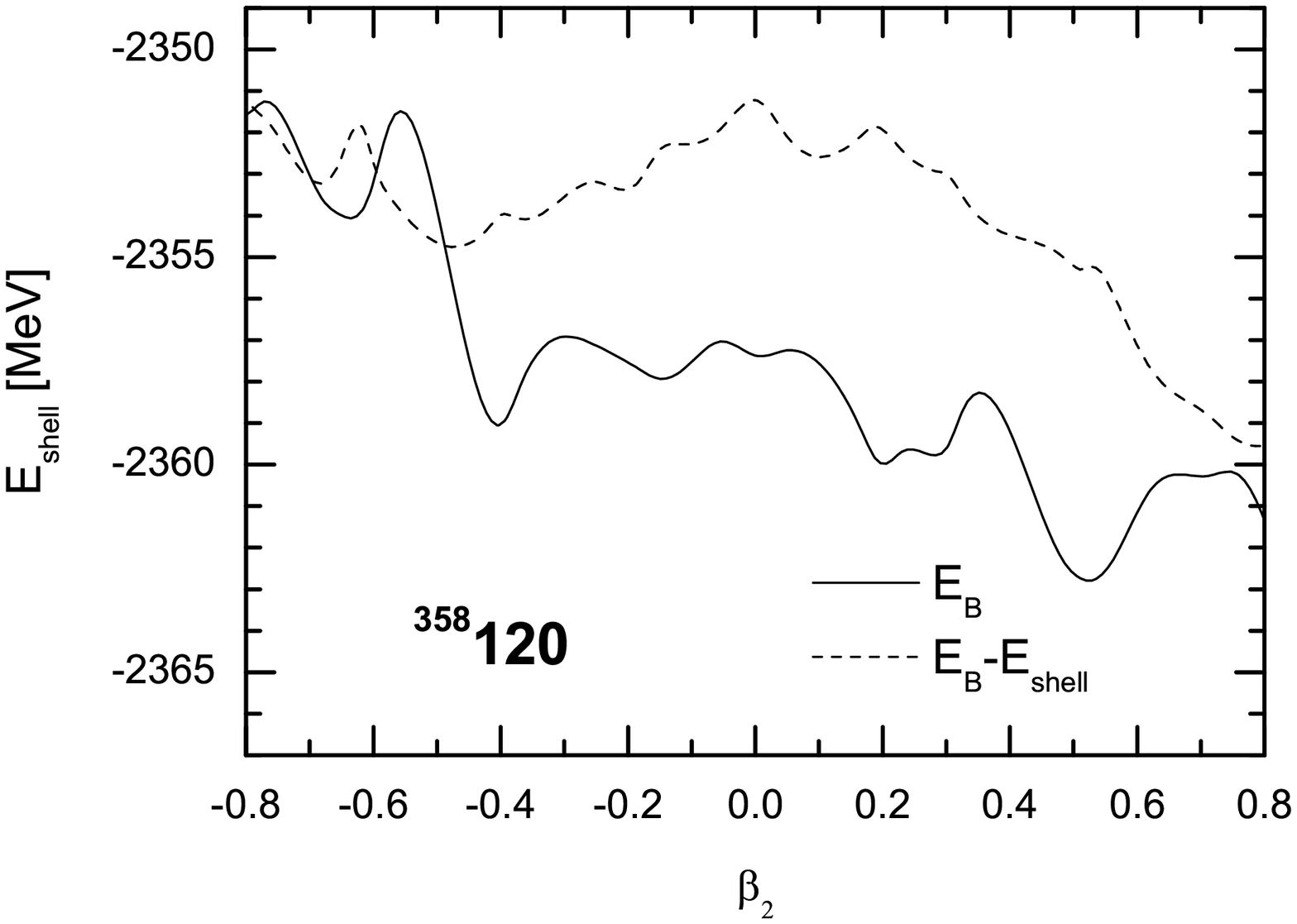}
\includegraphics[scale=0.25]{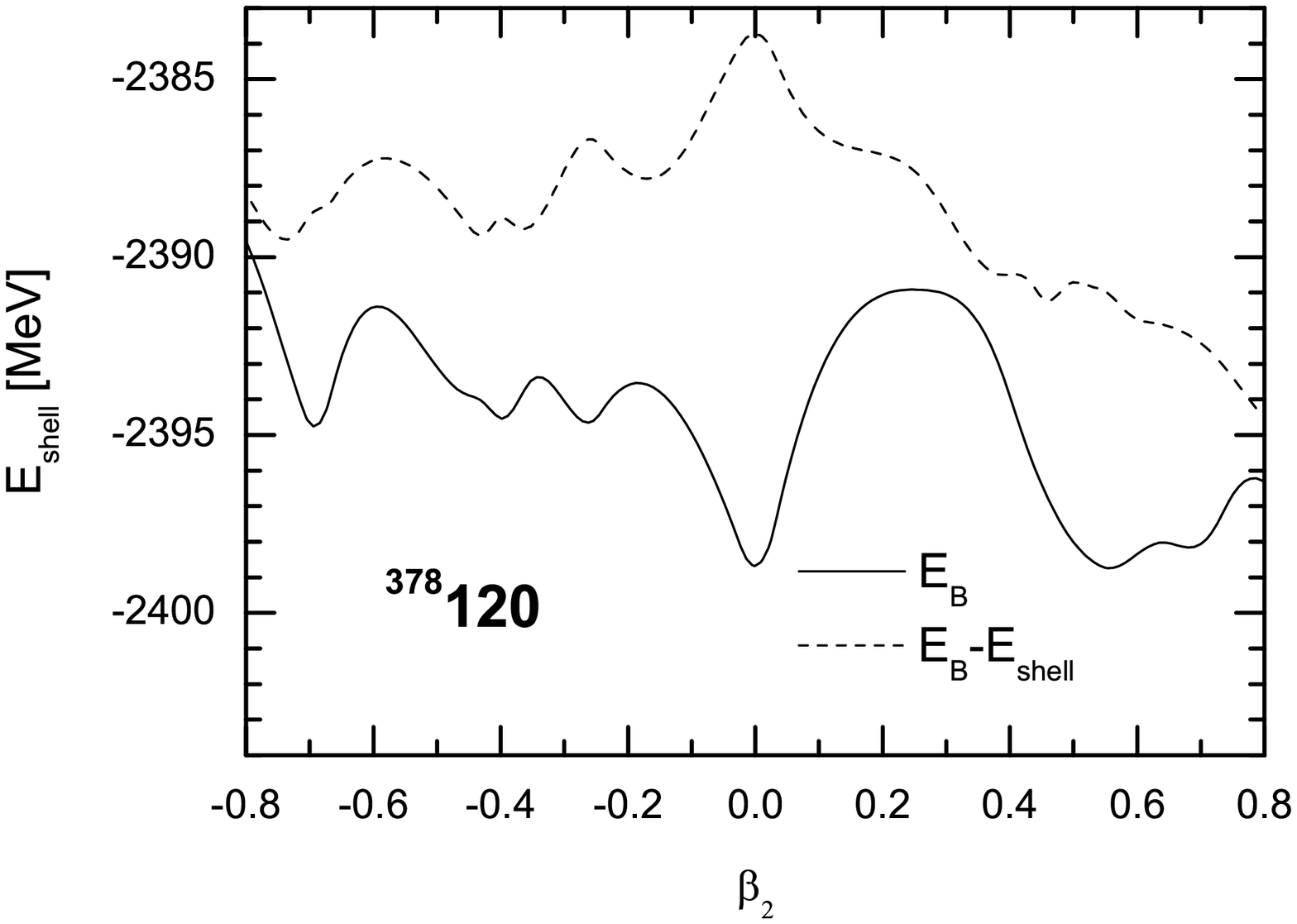}
 \caption{The total energy, $E_B$, and the equivalent macroscopic energy,
$E_B-E_\mathrm{shell}$, of $^{292,304,318,348,358,378}$120
calculated in the constrained RMF theory with effective
interaction NL3. Taken from Ref.~\cite{ZhangW05}. } \label{FigE8}
\end{figure}


Apart from axial symmetric case, the influence of other
deformation such as triaxiality and reflection asymmetry
deformations has been investigated as well, e.g., see
Ref.\cite{Bender03, Burvenich04}. Figure~\ref{FigE9} presents an
example of potential energy surface of the typical actinide
nucleus $^{240}$Pu, which has often served as a benchmark for
mean-field models, obtained with the Skyrme force SLy6. A typical
doubled-humped fission barrier is well shown. The inner barrier
explores triaxial degrees of freedom, which reduce the axial
barrier by about 3 MeV, while the outer barrier explores
reflection asymmetric shape~\cite{Burvenich04}. The similar
behavior is held for heavier nuclear system~\cite{Bender98}. It is
also noted that though the further deformation degrees of freedom
reduce the fission barriers obviously, they hardly change the
energies of mimina. Therefore, the constraint RMF model with axial
symmetry remains valid to confirm the spherical configuration of a
superheavy nucleus.

\begin{figure}[htbp]
\centering
\includegraphics[scale=1.0, angle=-90]{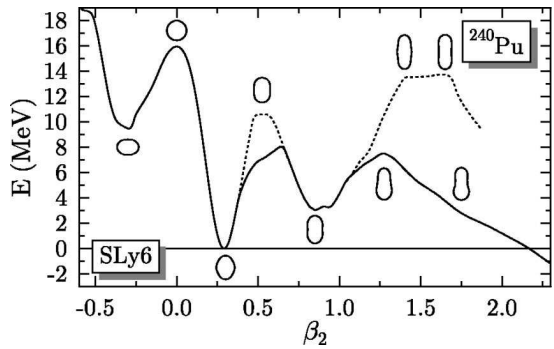}
\caption{ {Double-humped fission barrier of the typical actinide
nucleus $^{240}$Pu. The dotted line denotes an axial and
reflection-symmetric calculation, the full line denotes a triaxial
(inner barrier) and axial and reflection-asymmetric calculation
(outer barrier). The various shapes along the axial paths are
indicated by the contours of the total density at $r_0=0.07$
fm$^{-3}$. Taken from Ref.\cite{Burvenich04}}}
 \label{FigE9}
\end{figure}

\subsection{Alpha decay and the half-lives}
\label{subsec:shealpha}

The preferred decay mode of shell stabilized superheavy nuclei is
$\alpha$ decay. A key quantity for $\alpha$ decay, the $Q_\alpha$
value is defined as
 \begin{equation}
 Q_\alpha(N, Z) = E(N, Z) -  E(N - 2, Z - 2) - E(2, 2),
 \end{equation}
 which is the energy of the $\alpha$-particles emitted by
radioactive heavy and superheavy nuclei. It is interesting to
compare the $Q_\alpha$ data of the discovered superheavy nuclei
with predictions from mean-field models. Such works have been done
in various of mean field framework, e.g. Refs.~\cite{Meng00,
Reinhard02,Long02, Geng03a,Bender03,Cwiok05}. In the recent
experiments designed to synthesize the element 115 in the
$^{243}$Am+$^{48}$Ca reaction at Dubna in Russia, three similar
decay chains consisting of five consecutive $\alpha$-decays, and
another different decay chain of four consecutive $\alpha$-decays
are observed, and the decay properties of these synthesized nuclei
are claimed to be consistent with consecutive $\alpha$-decays
originating from the parent isotopes of the new element 115,
$^{288}115$ and $^{287}115$, respectively \cite{Oganessian04}. The
deformed RMF+BCS calculation with a density-independent
delta-function interaction in the pairing channel and the
effective interaction TMA is made to study these newly synthesized
superheavy nuclei $^{288}115$, $^{287}115$, and their
$\alpha$-decay daughter nuclei~\cite{Geng03a}. The calculated
$\alpha$-decay energies and half-lives agree well with the
experimental values and with those of the macroscopic-microscopic
FRDM+FY and YPE+WS models.

For an area of enhanced stability, the $\alpha$-decay half-lives
are expected to be longer than its neighbors. The half-lives of
$\alpha$-decay can be obtained with $Q_\alpha$ value by
phenomenological Viola and Seaborg systematics \cite{Viola66}:
 \beq
    log_{10}T_{\alpha}
        =(1.66175 Z - 8.5166) Q_{\alpha}^{-1/2}-(0.20228 Z + 33.9069)
 \label{eq:vs}
 \eeq
where $Z$ is the proton number of the  parent nucleus,
$T_{\alpha}$ in second and the parameters taken from
Ref.~\cite{Sobiczewski89}. It should be noted that
Eq.~(\ref{eq:vs}) is based on the WKB approximation, and provides
only a rather crude estimation of $T_{\alpha}$ since it disregards
many structure effects such as deformation, and configuration
changes, etc.

\begin{figure}[htbp]
 \centering
 \includegraphics[scale=0.4]{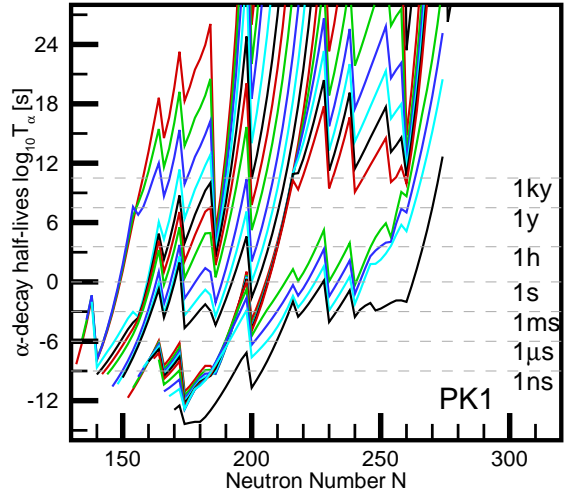}
 \caption{The $\alpha$-decay half-lives $T_{\alpha}$
as a function of neutron number obtained by RCHB calculation with
PK1 effective interaction. Taken from Ref.~\cite{ZhangW05}.}
 \label{FigE10}
\end{figure}

In Fig. \ref{FigE10}, the half-lives $T_{\alpha}$ from the RCHB
calculation in logarithm scale are plotted as a function of
neutron number $N$ with the effective interaction PK1. The
half-lives corresponding to 1 ns, 1 $\mu$s, 1 ms, 1 s, 1 h, 1 y,
and 1 ky are marked by the dashed lines. Each curve in this figure
corresponds to an isotopic chain in the region with $Z$ ranging
from 102 to 140. In Fig. \ref{FigE10}, the half-lives
$log_{10}T_{\alpha}$ increase with $N$. The jumps of the curves
correspond to magic proton numbers $Z$=106, 120, 132, and 138,
which depend on the effective interactions. The peaks of each
curve correspond to magic neutron numbers, i.e., $N$=172, 184,
228, 238 and 258. It can be seen that the magic numbers suggested
by two-neutron separation energies $S_{2n}$(two-neutron gaps
$\delta_{2n}$) or two-proton separation energies $S_{2p}$
(two-proton gaps $\delta_{2p}$) can also be found here. It should
be noted that the half-lives $T_\alpha$ shown here are obtained
from the spherical RCHB calcualtions. In fact, both the pairing
correlation and deformation are essential to reproduce the
experiment $Q_\alpha$ values~\cite{Meng00, Long02, Xu04}.
Calculations along this line are necessary.


\section{Summary and perspectives}
\label{sec:summary}

This article has been concerned with the relativistic description
of the exotic nuclei particularly with the Relativistic Continuum
Hartree-Bogoliubov (RCHB) theory and its applications. The last
two decades have witnessed the development of the relativistic
mean field (RMF) theory and its successful application in nuclear
physics. At the very beginning, the question on the clear signal
of relativistic effects in atomic nuclei has puzzled us for quite
a long time. Now after lots of efforts, it become evident that
there are at least the following points which do favor the
relativistic description of nuclei:
\begin{itemize}
  \item One may say that the kinetic energy of a nucleon moving in
        an nucleus is much
        smaller than its mass. However, as far as the
        mass is concerned, we have to remember that the mass
        of a nucleus is smaller than the sum of the mass of its consistent
        nucleons, which is naturally obtained in a relativistic
        model. Furthermore there are always a large scalar and a
        large
        vector potentials which are both comparable to the mass of the
        nucleons in nucleus.
  \item Relativistically the spin-orbit potential comes out
        naturally. Furthermore the long existing problems on
        the origin of pseudo-spin symmetry were solved.
        A very well developed spin symmetry in single
        anti-neutron and single anti-proton spectra is found.
        This spin symmetry in anti-particle spectra and the
        pseudo-spin symmetry in
        particle spectra have the same origin.
  \item In the framework of the relativistic approach,
        the nucleons interact via the exchanges of mesons and
        photons. From this point of view, the relativistic approach
        is more microscopic and fundamental in the sense of
        describing the nuclear system at the meson level.
  \item In practice, it turns out that the RMF theory can
        reproduce better the nuclear saturation properties
        (the Coester line) in nuclear matter
        and the isotopic shifts in the
        Pb-region, and is more reliable for nuclei far away from
        the line of $\beta$-stability.
\end{itemize}
Altogether the relativistic mean field theory is a reliable and
effective model for nuclear structure. Therefore, with the fast
progress in producing the exotic nuclei, it is natural to describe
the exotic nuclei based on the RMF theory.

The formalism, the numerical solutions and the effective
interactions for the RMF theory as well as its application for
nuclear matter and neutron stars are presented in Section 2. The
numerical solutions of the Dirac equation for spherical nuclei,
particularly on the Woods-Saxon basis, are discussed in detail.
The new effective interactions with nonlinear self-coupling meson
fields and density dependent meson-nucleon couplings, and their
influences on properties of nuclear matter and neutron stars are
also discussed.

Pairing correlation plays important roles in open shell nuclei.
The conventional BCS method can be combined easily with the RMF
theory, but it is not justified for exotic nuclei because it could
not include properly the contribution of continuum states. The
formalism and related discussions of the RCHB theory, in which the
coupling between the bound states and continuum could be included
self-consistently, are given in Section 3. Discussion on the
continuum states, the resonant BCS method and several methods for
obtaining single particle resonant states are briefly reviewed.

The study of exotic nuclei has attracted world wide attention due
to their large $N/Z$ ratios (isospin) and interesting properties
such as halo and skin. In Section 4 we reviewed the application of
the RCHB theory to exotic nuclei, the binding energies, particle
separation energies, the radii and cross sections, the single
particle levels, shell structure, the restoration of the
pseudo-spin symmetry, the halo and giant halo, and halos in hyper
nuclei, etc. For examples, (1) the Li isotopes were investigated
and the halo phenomenon in $^{11}$Li was found to be the result of
the scattering of particle pairs into continua; (2) the giant
halos were predicted in Zr, Ca and even lighter nuclei when
neutron drip line is approached based on the analysis of two
neutron separation energies, single particle energy levels, the
orbital occupation, the contribution of continuum and nucleon
density distribution. These predictions are supported to some
extent by the recent experimental information; (3) exotic
phenomena in hyper-nuclei were also studied; etc.

The RCHB theory has been applied to study superheavy nuclei which
is reviewed in Section 5. The possible doubly magic superheavy
nuclei have been searched within the RCHB theory using a variant
of the effective interactions. Base on the detailed analysis on
the two-nucleon separation energies $S_{2p}$ and $S_{2n}$,
two-nucleon shell gaps $\delta_{2p}$ and $\delta_{2n}$, the shell
correction energies $E_{shell}^{p}$ and $E_{shell}^{n}$, as well
as the pairing energies $E_{pair}^{p}$ and $E_{pair}^{n}$, the
effective pairing gaps $\Delta_{p}$ and $\Delta_{n}$, the proton
and neutron shell closures have been predicted by using the RCHB
theory with the effective interactions NL1, NL3, NLSH, TM1, TW-99,
DD-ME1, PK1 and PK1R. Proton numbers $Z$=120, 132, and 138 and
neutron numbers $N$=172, 184, 198, 228, 238, and 258 are supported
by all effective interactions to be magic. The
deformation-constrained RMF calculations have been done for these
doubly magic nuclei candidates and discussions on the deformation
of these nuclei are also presented.

Among many perspectives of the application of the RMF and the RCHB
theories, we mention briefly the following:

\begin{enumerate}

\item Attempts have been made by extending the present RCHB theory
for the axially deformed nuclei. Although the coupled Dirac
equations and Hartree-Bogoliubov equations with fixed pairing
potentials have been solved in coordinate space, the simultaneous
self-consistent solution of axially deformed RCHB equations is
proved to be very difficult~\cite{Meng03a}. Therefore the
Woods-Saxon basis has been suggested to solving the RMF theory in
order to generalize it to study exotic nuclei by solving either
the Schr\"odinger equation or the Dirac equation~\cite{Zhou03prc}.
It combines the advantages of the efficiencies in harmonic
oscillator basis and the appropriateness for exotic nuclei in
coordinate space. For spherical case, the Woods-Saxon basis has
been proved to be very successful~\cite{Zhou03prc}. The
development of deformed RCHB in the Woods-Saxon basis is in
progress.

\item The Shell-model-Like
APproach (SLAP) method has been suggested to treat pairing
correlation in RMF theory~\cite{Guo04}. With the single particle
bound states from the RMF calculation and resonant states obtained
from RMF theory combined with the Analytic Continuation of
Coupling Constant (ACCC) method~\cite{Yang01,ZhangSS04b} or the
phase shift scattering method~\cite{Sandulescu03,Geng03b}, the
SLAP for the pairing can also be used to describe exotic nuclei as
well.

\item Since the pion does not contribute in the Hartree level, we
did not mention its contribution throughout the paper. However, if
the Fock term is considered, one must include pion in the
Lagrangian. This would bring out some complication in solving the
equations as well as new and interesting physics.

\item Many symmetries have been broken in the mean field
theory. Restoration of some symmetries could be achieved by making
projections or corrections. The microscopic center of mass
correction for the broken translation invariance in the RMF model
has been discussed~\cite{Long04}. The single particle states of a
deformed mean field do not have good total angular momentum, so is
the total intrinsic wave function of the deformed nucleus. Thus
one must project out a state with good total angular momentum from
the intrinsic wave function. This can be done by the full angular
momentum projection method~\cite{Ring80} or approximately in a
simple way~\cite{Price87}.  Recent works on the restoration of
rotation symmetry~\cite{Li04a,Li04b} are very interesting.
Possible extensions and applications for chiral doublet bands and
nuclear isomeric states are
expected~\cite{Madokoro00,Frauendorf97,Peng03a,Peng03b,Sun04,Frauendorf01,Ward01}
.

\item In most of the RMF calculations, the no-sea approximation is
applied. Therefore the anti-nucleon degrees of freedom are not
considered. It would be quite interesting to investigate these
degrees of freedom and the polarization of the vacuum as well as
abnormal nuclei containing
anti-baryons~\cite{Mao99,Lv02,Zhou03prl,Lv03,Mishustin04}.

\end{enumerate}

{\bf Acknowledgments}\\

The authors would like to thank their colleagues, collaborators
and students, in particular, A. Arima, S. F. Ban, T. S. Chen, J.
Dobaczewski, H. Geissel, J. Ginocchio, J. Y. Guo, G. C. Hillhouse,
Soojae Im, H. Y. Jia, T. T. S. Kuo, A. Leviatan, J. Li, G. L.
Long, H. F. L\"u, H. Madokoro, M. Matsuzaki, G. M\"unzenberg, A.
Ozawa, J. Peng, P. Ring, N. Sandulescu, J. P. Sang, G. Shen, W. Q.
Shen, K. Sugawara-Tanabe, S. Sugimoto, B. H. Sun, B. X. Sun, Y.
Sun, N. Takigawa, I. Tanihata, N. Van Giai, S. J. Wang, R. Wyss,
S. Yamaji, S. C. Yang, J. Y. Zeng, H. Q. Zhang, J. Y. Zhang, W.
Zhang, S. S. Zhang, E. G. Zhao, who contribute directly or
indirectly to the investigations presented here. This work was
partly supported by the Major State Basic Research Development
Program Under Contract Number G2000077407 and the National Natural
Science Foundation of China under Grant No. 10025522, 10435010,
10475003 and 10221003, the Doctoral Program Foundation from the
Ministry of Education in China and the Knowledge Innovation
Project of Chinese Academy of Sciences under contract No.
KJCX2-SW-N02.

\end{document}